# Synthesis of Multifunctional Charge Transfer Agents: Towards Single Walled Carbon Nanotubes with Defined Covalent Functionality and Preserved π System

Alphonse Fiebor[1], Antonio Setaro[1,4,*], Andreas J. Achazi[2], Georgy Gordeev[1], Manuela Weber[2], Daniel Franz[2], Beate Paulus[2], Mohsen Adeli[1,3,*], and Stephanie Reich[1]

*[1] Department of Physics, Free University Berlin, Berlin, 14195, Germany*
*[2] Department of Chemistry, Biochemistry & Pharmacy, Free University Berlin, Berlin, 14195, Germany*
*[3] Department of Chemistry, Faculty of Science, Lorestan University, Khorramabad, 68151-44316, Iran*
*[4] Pegaso University, Naples, 80132, Italy*

[*] *Corresponding authors: AS: setaro@physik.fu-berlin.de, antonio.setaro@unipegaso.it - MA: adeli.m@lu.ac.ir*

## Abstract

The attachment of well-defined charge transfer agents to the surface of nanomaterials is an efficient strategy to control their charge density and also to tune their optical, electrical, and physicochemical properties. Particularly interesting are charge transfer agents that either donate or withdraw electrons depending on the arrangements of their building units and that promise a non-destructive attachment to delicate nanomaterials like $sp^2$ compounds. In this work, we rationally synthesize molecular moieties with versatile functionalities. A reactive anchor group allows to attach them to carbon nanotubes as defined charge transfer agents while preserving the tube π-conjugation. The charge transfer agents were synthesized through the stepwise nucleophilic substitution of either one (monosubstituted series) or two chlorine (disubstituted series) atoms of cyanuric chloride by aniline derivatives containing one, two or three methoxy groups in the para and meta positions. Variation in the number and position of methoxy as an electron transferring group help us to manipulate the electronic and optical properties of the molecular probes and their charge transfer to the SWNTs systematically. The correlation between the optical properties of these molecular probes and their functionality was investigated by experiments and quantum chemical calculations. While the optoelectronic properties of the conjugated charge transfer agents were dominated by the aniline segments, the triazine warrants the ability to nondestructively attach to the surface of SWNTs. This study is one step ahead towards the production of SWNTs with desired optical and electrical properties by covalent π-preserving functionalization.

## Introduction

Controlling and altering the nature and concentration of charge carriers has shaped the development of modern solid-state electronics. The demand for faster and energy efficient systems has led to the miniaturization of devices, reaching single computational units of a few nanometers. At this scale, the injection of charge carriers into nanocrystals affects their thermoelectric efficiency[1-4], photocatalytic activity[5,6], and plasmonic response[7,8] in addition to their transport properties. A powerful control over the charge density is achieved by functionalizing materials with charge transfer agents (CTAs)[9]. CTAs are simple molecular compounds that donate or extract electrons thereby altering the number of charges in a material in a controlled way. Example of extensively investigated CTAs are the electron accepting tetracyanoquinodimethane (TCNQ) and the electron donating tetrathiafulvene (TTF)[10].

The critical parameters to control the electron transfer process between a CTA and a nanostructure are the structure of CTA and the way in which it was interfaced with the nanosystems. To understand the delicate interplay between the two parameters it is extremely interesting to systematically vary the electron donating or withdrawing efficiency of a CTA while keeping its overall backbone and functionalization identical. An interesting system to achieve this are molecules with methoxy groups attached in different positions. A methoxy substituent attached to a benzene ring in the para position donates electrons, whereas it withdraws them at the meta position[11]. This offers us the advantage of implementing CTA compounds with opposing transferring characters just by starting with the same building units but differentiating the way they are brought together. As a model nanosystem on which to apply our CTA compounds, we chose to work with single walled carbon nanotubes (SWNTs). Among the many novel nanosystems of technological interest, carbon nanotubes stand out for their ability to bridge the quantum and macroscopic world. They are quantum ballistic channels few nanometers thin that can get long up to the millimeters scale[12]. Their room-temperature stable emission is also a direct signature of their quantum nature[12-15]. At the same time, SWNTs are very sensitive to molecular attachment that normally breaks the conjugation of its $\pi$ electrons[16,19]. This is one reason why controlling their charge carrier concentration has remained challenging so far. Over the years, our groups have gained experience on the covalent and non-covalent functionalization of SWNTs[20-30]. In our set of CTA probes, we chose to attach the methoxy groups at different positions of an aniline species instead of a pure benzene ring. The last building block of our CTA probes is the bridge between the charge transfer moieties (the methoxy-aniline derivatives) and the SWNTs. For this role we picked up cyanuric chloride, whose three chlorine atoms can be replaced one by one in a controlled way by fine-tuning the reaction conditions[31-33]. We substitute one or two of the chlorine atoms with the methoxy-aniline groups. The third chlorine provides the anchor the CTA onto the SWNTs: We convert it into a reactive nitrene derivative, which we have shown to efficiently attach to the sidewall of SWNTs[20,21]. We chose a covalent attachment protocol to ensure long-term stability of the attachment of the CTA onto the SWNTs.

In this work we produce triazine based CTAs and conjugate them to SWNTs. Charge transfer agents with different functionalities in terms of their numbers and positions but the same backbone are synthesized through stepwise nucleophilic substitution of the chlorine atoms of cyanuric chloride by aniline derivatives. We show in this work how the energy of the HOMO-LUMO band gap and the optoelectronic properties of our charge transfer agents can be tuned by proper positioning of the methoxy groups in our functional aniline segments. The ability of our molecular probes to create electron deficient nitrenes qualifies them for the nondestructive covalent attachment to the surface of SWNTS through [2+1] cycloaddition reaction followed by $sp^3$-$sp^2$ rehybridization[20]. We attached those charge transfer agents to SWNTs and demonstrated that they do not disintegrate their π system. The nondestructive feature of this covalent functionalization together with defined charge transfer agents is a rational architecture that opens new ways to construct nanodevices with controlled optoelectronic properties.

## Experimental

*Materials*

Cyanuric chloride, *p*-Anisidine, 3,5-dimethoxyaniline, 3,4,5-trimethoxyaniline, sodium hydroxide, sodium azide, single walled carbon nanotube (produced by the CoMoCAT® method, batch SG76, diameters ranging between 0.7 nm and 1.2 nm and length up to 1 μm), acetone, ethanol, hexane, tetrahydrofuran, dichloromethane, trichloromethane, N-Methyl-2-pyrrolidone, toluene and methanol were all purchased from Sigma-Aldrich and Carl Roth (Germany). All materials, except the SWNTs, were used as supplied. The purchased single walled carbon nanotubes were washed with $HNO_3$ and annealed at 800 °C before usage to remove residual metallic impurities and amorphous carbon.

*Methods*

$^{1}H$ NMR and $^{13}C$ NMR spectra were acquired at room temperature by a JEOL ECP 500 and Bruker AVANCE 500. Tetramethyl silane (TMS) was used to calibrate the chemical shift δ, which is reported in parts per million (ppm). Elemental analysis (CHN) was performed using ELEMENTAR apparatus with three columns and detectors for carbon, nitrogen, and hydrogen. UHPLC-ESI-IMS-MS/MS was used to measure the mass of our molecular agents equipped with a DAD detector and a standard separation column (RP, C18). We acquired the melting points of charge transfer agents at a ramp rate of 5 °C/min using Stuart Melting Point SMP30. Ultraviolet–visible–infrared absorption spectra were recorded using a Perkin-Elmer Lambda 950 spectrophotometer at room temperature. Fluorescence maps of molecular probes were obtained by a Nanolog spectrofluorometer from Horiba, equipped with a Xenon lamp. Single excitation lines from HgXe short arc lamp were selected using a monochromator. A thermoelectrically cooled CCD detector was used to record the molecular spectra, a nitrogen-cooled InGaAs detector for the SWNTs. The IR spectra were recorded using an Alpha II FT-IR spectrometer (111717, ATR platinum Diamond 1). The Raman spectra of the synthesized molecular probes and functionalized SWNTs were acquired with a Horiba XploRA spectrometer. The signals were recorded in the backscattering configuration, acquired with a 10x objective (N.A.=0.33) and a thermoelectrically cooled charge coupled devices (CCD) detector. The molecular spectra were acquired with an excitation line at 785 nm and a 600 lines/mm grating, the functionalized nanotubes with a 638 nm pump line and a 1200 lines/mm grating. We calibrated our setup for the molecular probes with 4-aminobenzonitrile and that of the functionalized SWNTs with a neon lamp. The UV/Vis/IR absorption spectra as well as the emission spectra were acquired from a 10 µM concentrated THF solution of the probes, the Raman and IR spectra were acquired in powder form. The emission spectra of the pristine and functionalized nanotubes were collected from a suspension of the tubes in water (density 0.1 g/l) and sodium cholate (1 wt %). The solution was sonicated with a tip-sonicator (Bandelin Sonopuls HD 2070) for 1 h at 60 W and centrifuged (Hettich Mikro 220 R centrifuge) at 30000 g for 1 h, following the standard routine described in our past works[20,22]. The collected supernatant was used for optical measurements.

All geometry optimizations were performed with the density functional PBE0-D3(BJ)[34-40], the Resolution-of-Identity (RI) approximation[41,42], the "multiple grid" m4 for the quadrature of exchange-correlation terms[43], and with and without the dielectric continuum solvent model COSMO[44,45] (simulating THF, relative permittivity $\varepsilon_r$=7.4, refractive index $n_D$=1.4040) in the Turbomole 7.3 software package[46,47]. The HOMO-LUMO gaps are reported for the gas phase.

To get further insights the electronic transitions were calculated with time-dependent density-functional theory (TDDFT)[48] and PBE0-D3(BJ)/def2-TZVP and the COSMO model (simulating THF).

*Synthesis of the molecular probes*

The molecular probes were synthesized according to the reported procedures for the reaction between aniline derivatives and cyanuric chloride[49].

*Synthesis of 4,6-dichloro-N-(4-methoxyphenyl)-1,3,5-triazin-2-amine*

*p*-Anisidine (1.002 g, 8.13 mmol) was added gradually at 0 °C to a cloudy solution of cyanuric chloride (1.50 g, 8.13 mmol) in acetone (25 mL). The mixture was stirred for 30 minutes then it was basified (pH ≥9) by 1N NaOH (6 mL). The mixture was then stirred at 0 °C for additional 2 h. The reaction mixture was filtered, and the product was washed with deionized water (100 mL). The solid off-white product was dried overnight by a lyophilizer to obtain the mono-substituted probe which we abbreviated A1M in 82% yield. Similarly, the probes A2M (yield 87%) and A3M (yield 79%) were synthesized using the same route with 3,5-dimethoxyaniline and 3,4,5-trimethoxyaniline precursors, respectively. The chemical structures of A1M, A2M and A3M probes can be found in Figure 1.

*Synthesis of 6-chloro-N2, N4-bis(4-methoxyphenyl)-1,3,5-triazine-2,4-diamine*

Cyanuric chloride (1.0 g, 5.42 mmol) in acetone (50 mL) was heated at 50 °C and stirred for 30 minutes. The solution was then cooled down to 0-2 °C and *p*-Anisidine (1.34 g, 10.85 mmol) was gradually added to it. The resulted mixture was stirred for 1 h at 0-2 °C and then it was basified by 1N NaOH (10 mL, pH≥9). This mixture was then further stirred for 2 h at 50 °C. The reaction mixture was filtered, and the product was washed with acetone (10 mL) and deionized water (150 mL). The white solid product was dried by lyophilization overnight to obtain the di-substituted probes abbreviated A1D in 74% yield. The probes A2D (yield 72%) and A3D (yield 76%) were similarly synthesized using this protocol but 3,5-dimethoxyaniline, 3,4,5-trimethoxyaniline were used as precursors, respectively. The chemical structures of A1D, A2D, and A3D can be found in figure 1.

*Functionalization of SWNTS*

We followed the same routine to functionalize SWNTs with each of our six CTA probes[20]. Cleaned pristine SWNTs (30 mg, refer to the ESI for details of the cleaning procedure) were added to *N*-methyl-2-pyrrolidone (50 ml), sonicated with a tip sonicator at 60 W with sonic pulses of 0.5 s for 1 h, and then stirred at room temperature for an additional 1.5 h. The molecular probe 6-azido-N2,N4-bis(4-methoxyphenyl)-1,3,5-triazine-2,4-diamine (A1D, see ESI for reaction details and for mass spectrometry characterization) in *N*-methyl-2-pyrrolidone (30 ml) was added to the mixture and allowed to stir at 70 °C for 72 h. The reaction mixture was centrifuged (3700 g. for 10 min) and the crude products were dispersed in NMP and centrifuged again under the same condition. Dispersion and centrifugation of the product was repeated in toluene, methanol, acetone, ethanol, and water to obtain the purified hybrid compounds. The products were lyophilized to obtain 40 mg black solid compound of functionalized SWNT. The other molecular probes were conjugated to the surface of SWNTs using the same protocol.

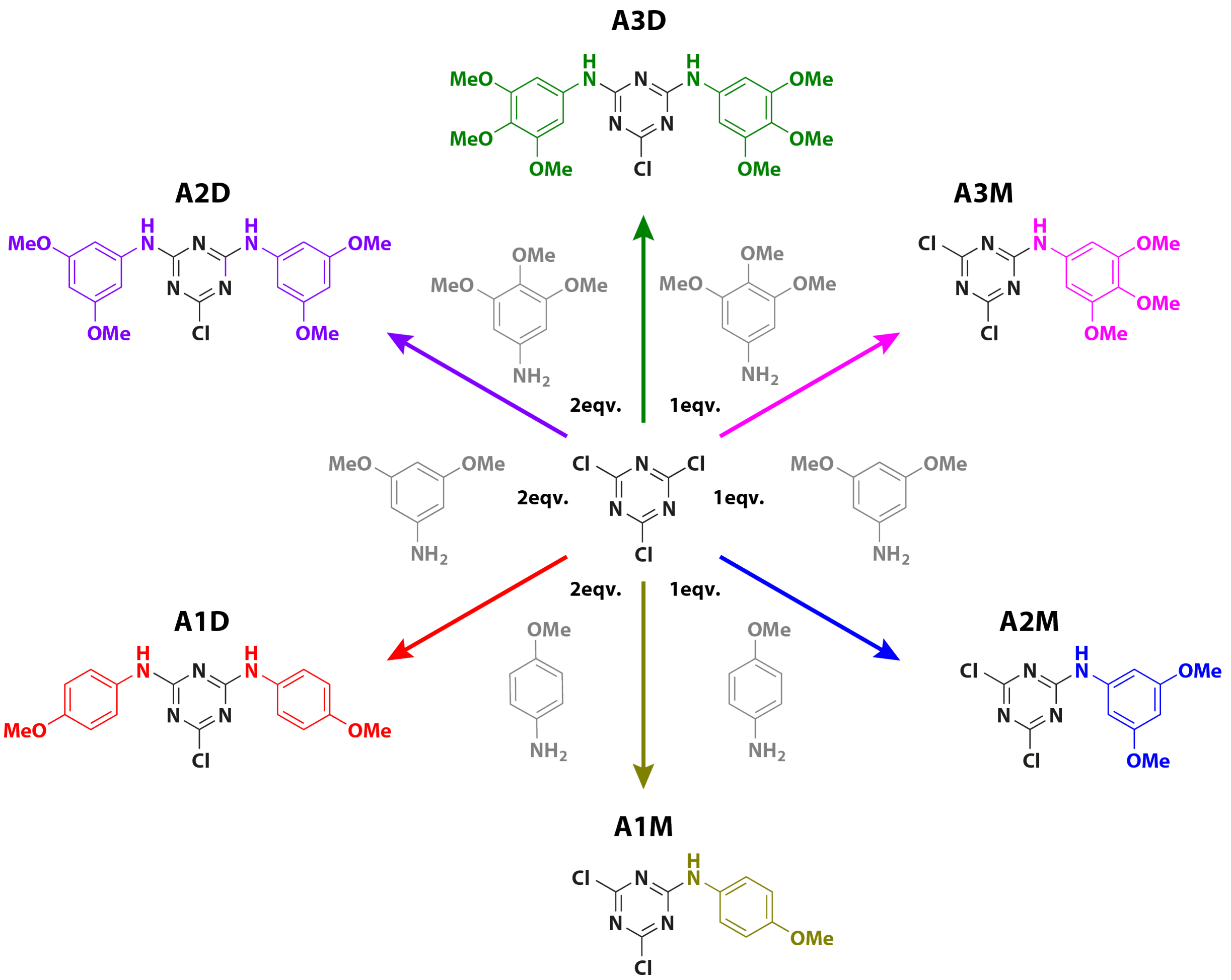

**Figure 1.** Schematic representation of the synthesis of the mono- and di-substituted molecular probes. A1M, A2M, A3M, A1D, A2D and A3D represent mono- and di-substituted probes with 1, 2, and 3 methoxy groups at different positions, respectively.

## Results and discussion

To investigate the key factors affecting the covalent bond-mediated electron transfer between SWNTs and CTAs, triazine-pivot molecular probes with different and defined functionalities were synthesized and characterized by different spectroscopic methods and elemental analysis. The stepwise nucleophilic substitution of the chlorine atoms of the cyanuric chloride by the aniline derivatives resulted in molecular probes with different functionalities in terms of number and structure of their aniline segments (Figure 1). Two series of molecular probes, the mono- and di-substituted triazine structures series, were synthesized by varying the number and position of the methoxy substituents. The mono- and di-substituted series were abbreviated as AxM and AxD, respectively, where x stands for the number of methoxy functional groups.

The crystal structures of A1M and A3M revealed the attachment of the N atom from the aniline derivatives to the C atom from cyanuric chloride, proving the successful nucleophilic substitution of the chlorine atom and the synthesis of our molecular probes (Figures 2a.). The bond length and angles in the crystal structures were matched with the expected chemical formulas. In the crystal structure of A3M, the distance between the hydrogen H5 and the nitrogen N4 was 2.298 Å and the C5-H5-N4 bond angle was 120°, in agreement with the expected chemical structure (Figure 2a). These selected bond lengths and angles were consistent with the reported values in literature[50-53]. The crystal structure refinement parameters are summarized in the ESI Table S1-S14.

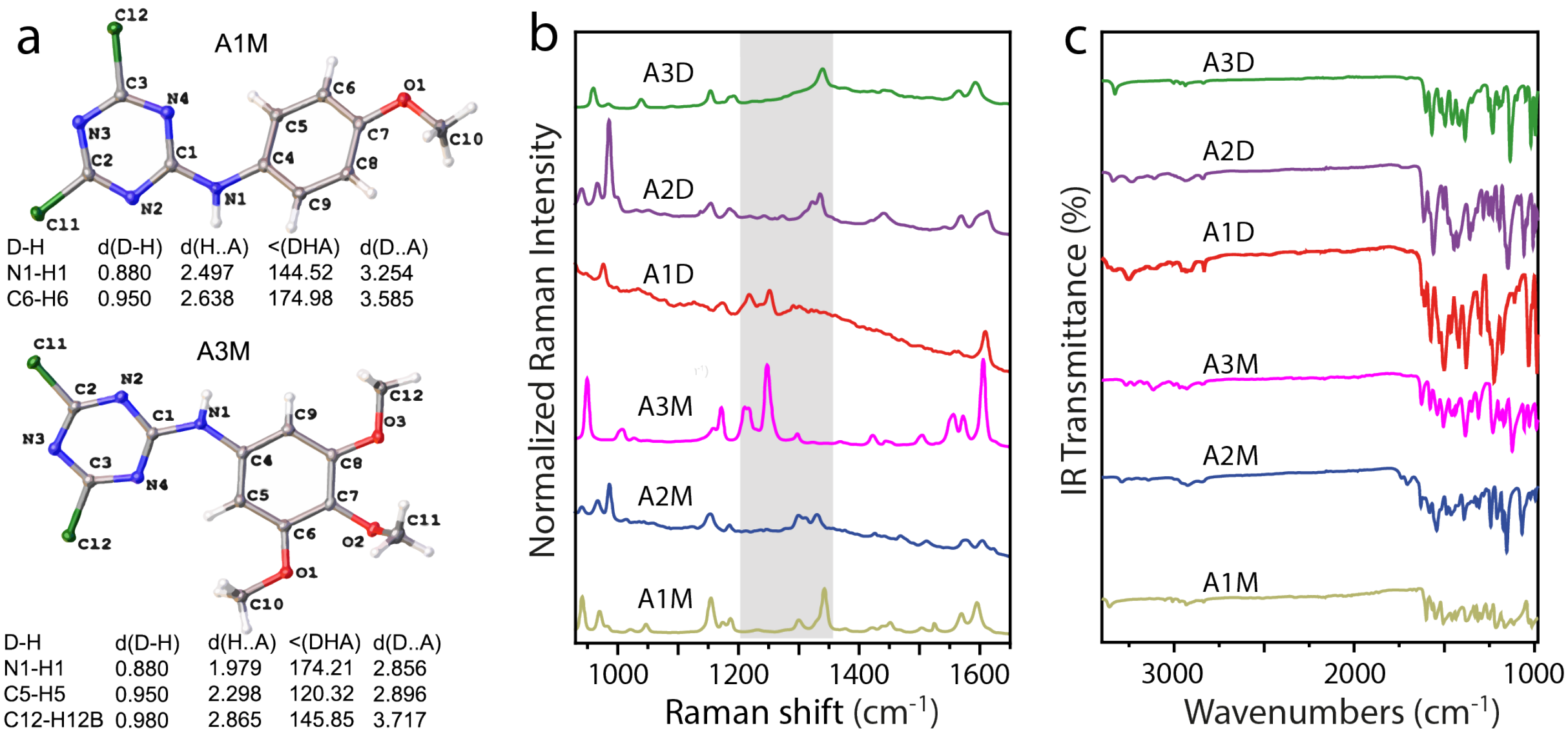

**Figure 2.** (a) Single crystal XRD depicting the chemical structure of the A1M and A3M compounds with selected bond lengths and angles. (b) Raman and (c) IR spectra of the synthesized mono- and di-substituted molecular probes.

The structure of the synthesized materials was also investigated by different spectroscopy methods. IR and Raman spectra of the molecular probes showed the vibration features of the aniline and triazine segments (Figure 2b and 2c). The Raman spectra of the molecules (Figure 2b and ESI Figure S34) displayed different signatures. For example, the distinctive peaks between 1609 $cm^{-1}$ and 1593 $cm^{-1}$ for all probes indicated the ν(CC) and ν(CN) of the phenyl and triazine rings. The β(NH) plus ν(CN) vibration modes also appeared at 1569 $cm^{-1}$ and 1566 $cm^{-1}$ in the Raman spectra of A3M and A3D, at 1575 $cm^{-1}$ and 1568 $cm^{-1}$ for A2M and A2D, and at 1570 $cm^{-1}$ and 1562 $cm^{-1}$ for A1M and A1D, respectively. We also noticed the β(CH) band of the bending vibration of the phenyl ring at 1172 $cm^{-1}$ and at 1153 $cm^{-1}$ in A3M and A3D spectra, respectively. For A1M, A1D, A2M and A2D, the β(CH) bending vibration of the phenyl ring appeared at 1154 $cm^{-1}$, 1173 $cm^{-1}$, 1152 $cm^{-1}$ and 1153 $cm^{-1}$, respectively. These vibrational frequencies agree with reported data in the literature[54,55].

Generally, the stretching frequency of the aryl C-O with electron-donating groups at the meta positions is higher due to the increased double bond feature of the C-O bond than their counterparts at the ortho and para positions[55,56]. Conversely, the stretching vibration is higher at the ortho and para positions than at the meta positions for electron-withdrawing groups[55,56]. We observed that the aryl C-O stretching vibrations (highlighted in gray, Figure 2b) of the synthesized molecular probes are affected by the number and position of the methoxy groups as well as the triazine ring. The trend of the vibration energy of the C-O bonds in the di-substituted probes, A1D<A2D<A3D, showed that they are dominated by the number and position of the methoxy groups. Accordingly, the bands of the C-O bonds in A3D (with three methoxy groups at para and meta positions), A2D (with two meta methoxy groups), and A1D (with one para methoxy group) appeared at 1343 $cm^{-1}$, 1336 $cm^{-1}$ and 1253 $cm^{-1}$, respectively. In the mono-substituted probes, however, the triazine played a key role and the observed trend was the opposite of the one of the di-substituted series. A3M (with three methoxy groups at the meta and para position) showed the lowest C-O vibration, appearing at 1248 $cm^{-1}$. This vibration was shifted to 1329 $cm^{-1}$ and 1344 $cm^{-1}$ for A2M and A1M (with two meta methoxy and one para methoxy groups, respectively). This trend could be ascribed to the transfer of electrons from the methoxy groups to the triazine ring mediated by the aniline segment. The electron-poor triazine ring in the AxM series facilitates electron transfer from the para methoxy groups by resonance effect. Electron transfer results in the higher double bond character for the C-O bonds in A1M and causes a shift of C-O vibration to a higher energy compared to A2M and A3M. Electron transfer from methoxy to phenyl and then to the triazine rings in the AxM series elongates their π-conjugated systems and decreases their band gap energies, as it will be discussed in the next sections of this manuscript.

The IR spectra of the molecular probes showed weak bands for the secondary amine (N-H) at 3367 $cm^{-1}$ (Figure 2c and ESI Figure S33). This band indicates attachment of the aniline derivatives to the triazine rings trough the amine groups. Multiple C-H stretching bands from 2850 to 3100 $cm^{-1}$ were assigned to the C-H bonds of the methoxy groups and the phenyl ring[54-56]. Also, the C=C and C=N stretching bands of the phenyl and triazine rings appeared at 1610 $cm^{-1}$ and 1570 $cm^{-1}$, respectively[54]. The C-O stretching vibrations were also observed at 1486 $cm^{-1}$. These results showed that the triazine and phenyl moieties are integrated and form a new conjugated structure. Moreover, the ring breathing mode of the triazine-pivot appeared between 985 $cm^{-1}$ and 1000 $cm^{-1}$ in both the IR and Raman spectra of the synthesized probes. This is consistent with the reported ring breathing mode of the triazine in literature[54-59].

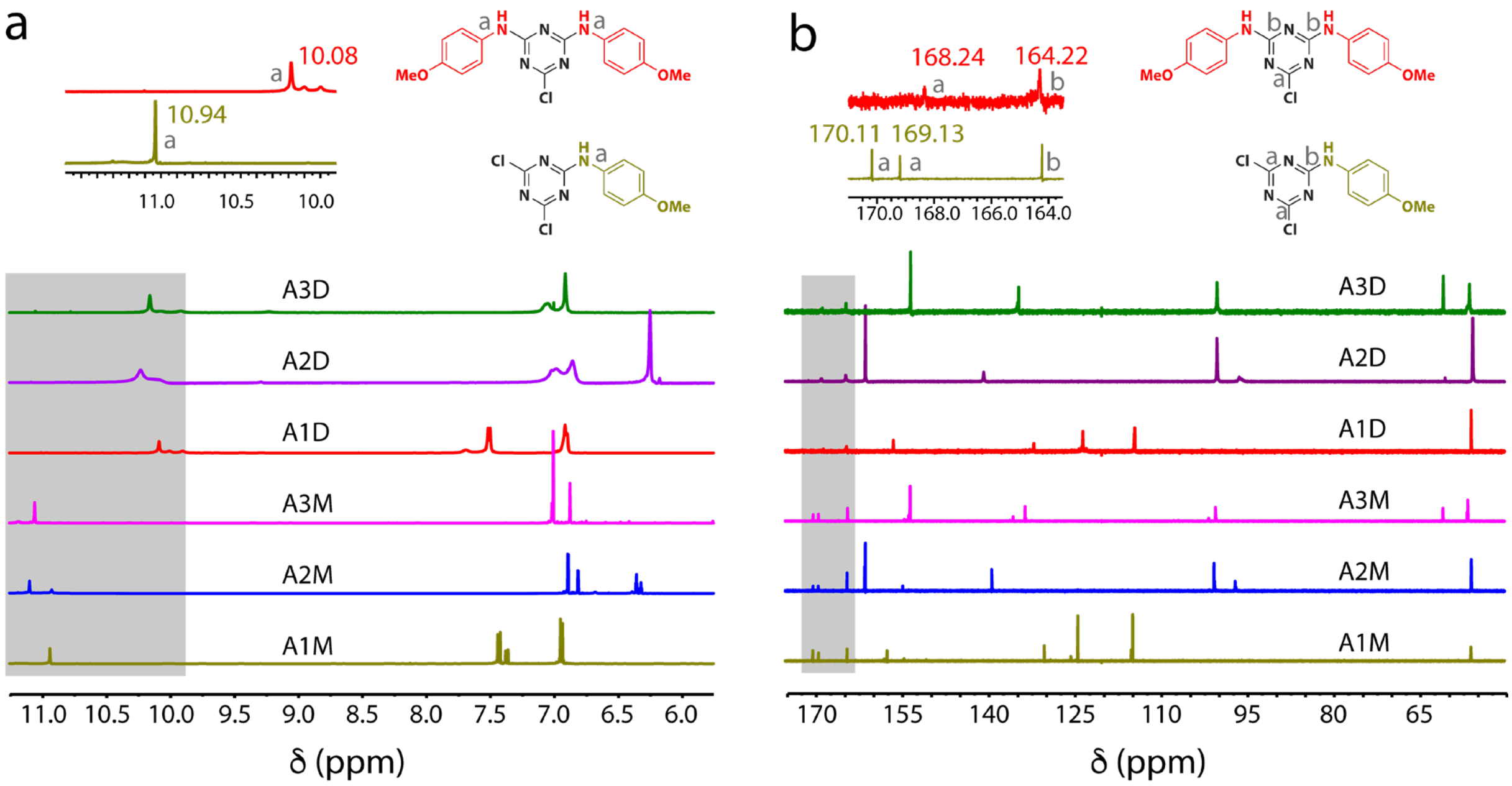

**Figure 3.** (a) $^{1}$H and (b) $^{13}$C NMR spectra of the mono- and di-substituted molecular probes in DMSO-d6 at 25°C. Signals of different protons and carbons of each molecular probe are shown on the corresponded spectra.

The structure of the synthesized molecular probes was also investigated by $^{1}$H and $^{13}$C NMR spectroscopies (Figure 3a and 3b.). Different signals are assigned to the corresponding protons and carbons of these compounds on their NMR spectra (see ESI fig. S1 – S12, pages S12 – S18 for magnified and detailed spectra). In general, the signal of the N-H protons of the AxM probes (highlighted in gray) showed a larger downfield shift in comparison to their di-substituted counterparts, exemplified by A1M and A1D at 10.94 ppm and 10.08 ppm, respectively (Figure 3a). This is assigned to the more electron poor nature of the dichlorotriazine derivatives in comparison to their monochlorotriazine analogs. In the $^{13}C$ NMR spectra (Figure 3b), the carbon atoms of the triazine ring shifted upfield upon attachment to the aniline derivatives, indicating successful substitution of chlorine atoms of the cyanuric chloride and synthesis of the molecular probes. This trend of chemical shift can also be explained by the first term of the Ramsay-Karplus-Pople equation: $\sigma_\pi \simeq \frac{1}{\Delta E}\langle r^{-3}\rangle \sum Q_{ij}$[60,61], where the paramagnetic term $\sigma_\pi$, which is responsible for the chemical shift, inversely depends on the HOMO-LUMO energy gap $\Delta E$. To illustrate this point, we compared the calculated DFT HOMO-LUMO gaps of the molecular probes to the chemical shifts of the coordinating atoms (N-H protons in $^{1}H$ NMR). Here, a smaller energy gap corresponds to a bigger chemical shift. Accordingly, the AxM probes series with larger chemical shifts showed a relatively smaller calculated DFT HOMO-LUMO gaps whereas the AxD series with lower chemical shifts displayed bigger band gaps as shown in Figure 4a. Along the same line, the composition of the molecular probes was investigated by elemental analysis. Figure 4b illustrates the agreement between the experimental and the calculated elemental contents of the molecular probes and confirms their assigned chemical formulas. Further physicochemical properties of the synthesized materials including melting points, solubility in different solvents and mass were also investigated, and the results are shown in ESI. Temperature played a significant role for the solubility of A2M, A2D, and A3D in the non-polar and polar aprotic solvents.

The effect of the functionality of the molecular probes on their physicochemical properties was further investigated by evaluating their optical properties. UV spectra of the molecular probes showed an absorption band at 4.96 eV (250 nm) – 3.76 eV (330 nm) as depicted in Figure 4c. The maximum absorption of AxD molecular probes showed a blue shift in comparison to their AxM counterparts. The transformation of the smaller molecular structure (AxM) to the bigger structure (AxD) means higher number of electrons, whose repulsion leads to an increase of their energies. Thus, the blue shift indicates an excellent conjugation/delocalization of the electrons over the molecules that pushes the unoccupied orbitals even higher in energy. DFT calculations agree with the experimental data, showing a similar trend between the AxM and AxD series with higher energy gaps for AxD series (Figure 4d).

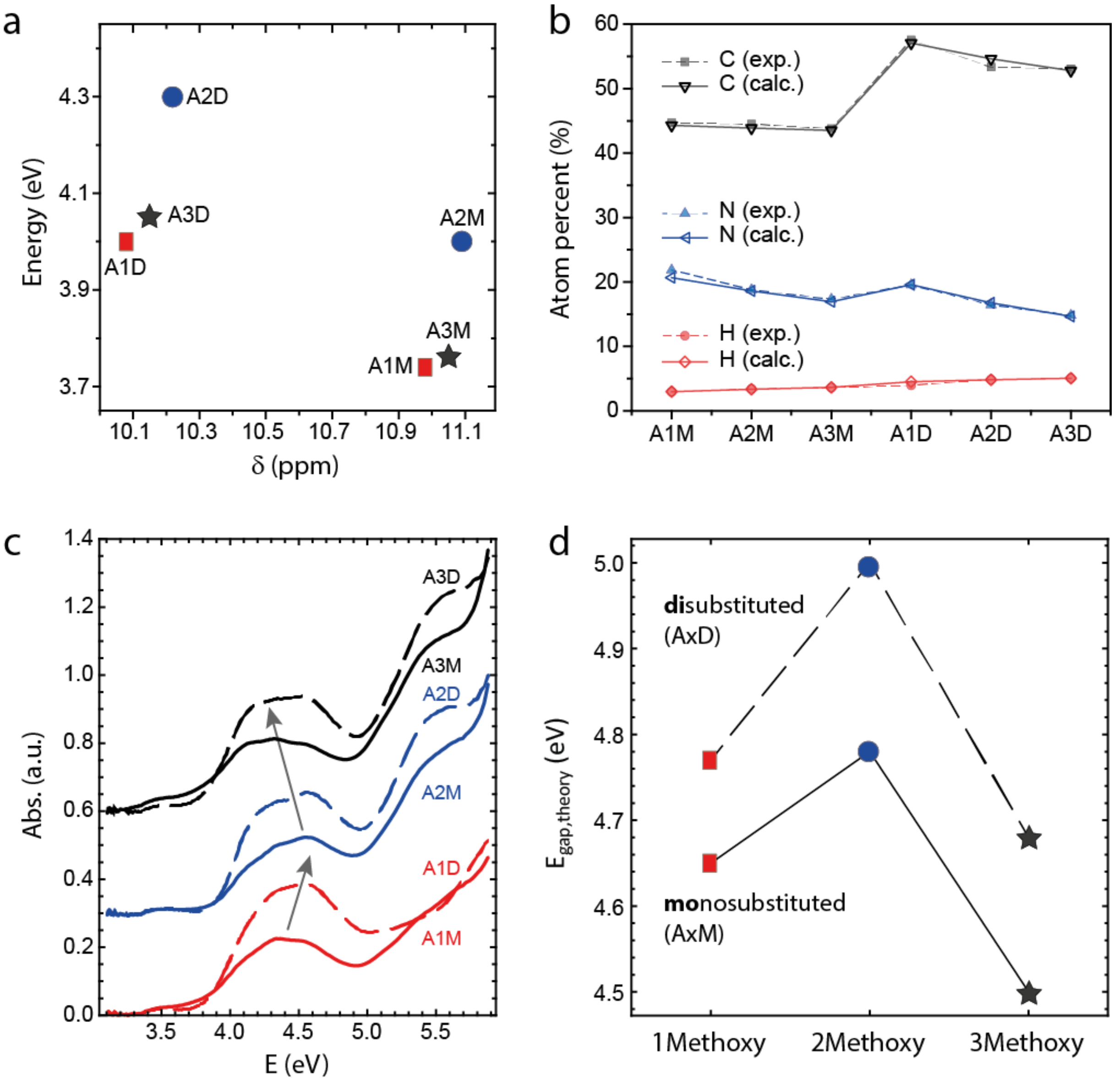

**Figure 4.** (a) Relation between the calculated DFT HOMO-LUMO gaps and the $H^1$ NMR chemical shifts for the mono- and di-substituted molecular probes of the coordinating NH bond. (b) Experimental and calculated elemental contents of the synthesized molecular probes. (c) Absorption spectra of the mono- and disubstituted molecular probes. (d) Comparison between the calculated DFT HOMO-LUMO gap values of the mono- and the disubstituted molecular probes series.

Also, the position and number of the methoxy functional groups affected the maximum absorption and HOMO-LUMO gap of the molecular probes (see arrows in Figure 4c). The change in the electronic and optical properties of the molecular probes is well correlated with their Hammett values. The Hammett value for the methoxy group at meta ($\sigma_m$= 0.12) [11,62] is bigger than the one at para ($\sigma_p$= -0.27) [11,62] position, indicating more efficient electron transfer from this group to the phenyl ring at the para position (Figure 5a).

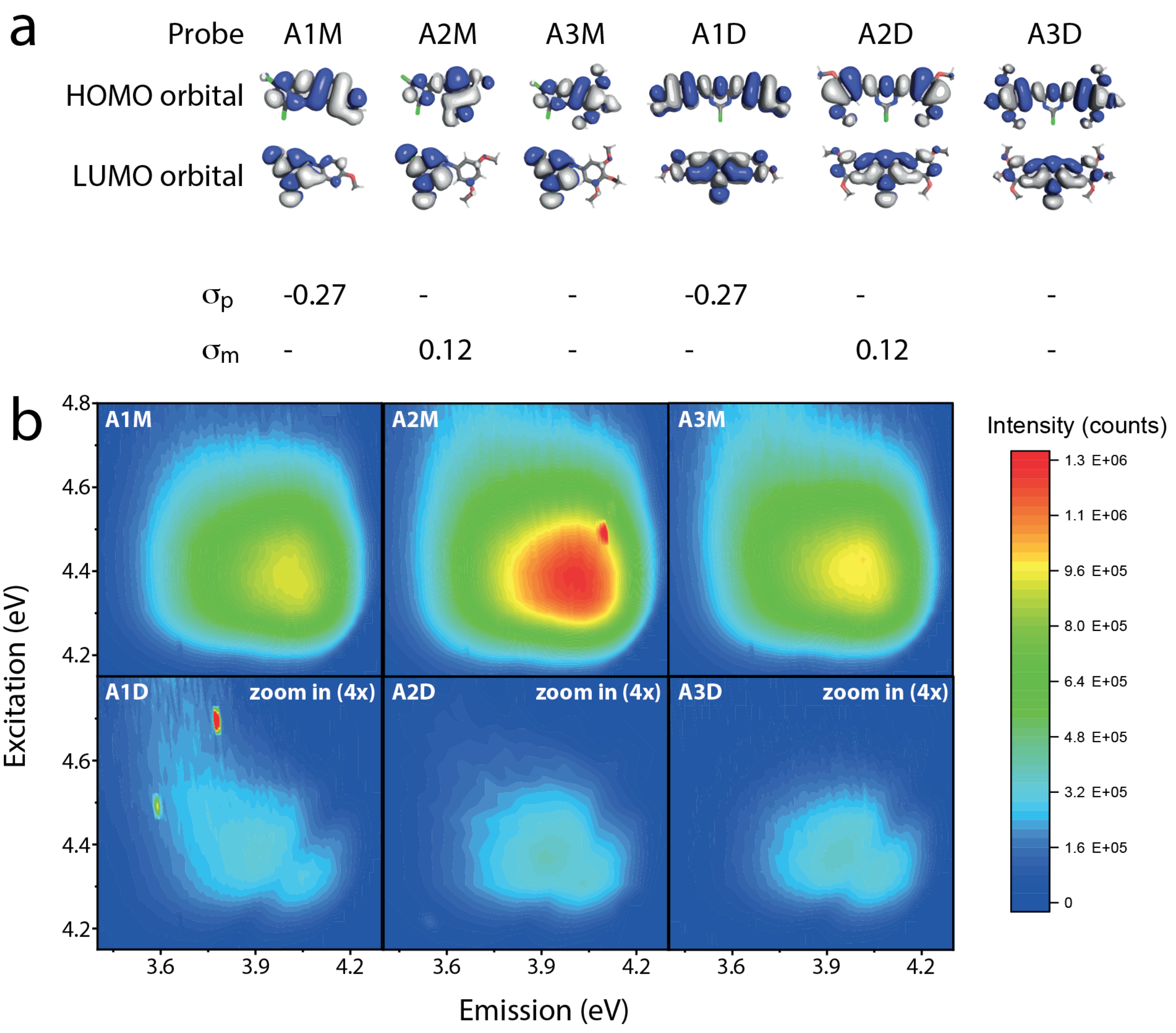

**Figure 5.** (a) HOMO-LUMO orbitals of the mono- (AxM) and di-substituted (AxD) molecular probes, together with their Hammett constants (taken from Ref. 11,62) (b) Excitation-emission maps of the monosubstituted (AxM) and disubstituted (AxD) probes series. The intensity of AxD series is amplified four times.

The methoxy group in A1M and A1D is in the para position and well-integrated with the π-conjugated system of probe and result in a red shift of the absorption bands. When the methoxy groups are in the meta position, they contribute less efficiently to electron transfer and a blue shift of the bands with larger energy band gap is observed. This is in accordance with the literature, where a linear correlation between HOMO-LUMO gap and Hammett constant of substituents is reported[62].

To complement the UV absorption experiments and gain more intuitive understanding of the optoelectronic properties of our probes, we optically excited them in 4.2 eV – 5 eV range and recorded their emission spectra. The optical excitation of the molecular probes resulted in emission of photons in the 3.5 eV – 4.3 eV photon energy range, which indicated an integrated π system in their backbone. The 2D excitation-emission photoluminescence maps (PLE) showed a higher intensity of emission for the mono-substituted probes in comparison with their di-substituted counterparts (Figure 5b). Among the monosubstituted AxM series, A2M showed the highest fluorescent intensity which was assigned to the less electron donating effect of methoxy groups in the meta position and its effect on the energy gap of this molecule through Hammett constant.

To better understand how the fluorescence intensity of the probes relates to the methoxy groups, the electronic transitions were investigated by TD-DFT calculations. The connection between geometry and different parameters including orbitals, energy gaps, oscillator strength and excitation wavelength on the fluorescence emission of the probes was investigated (see ESI, tables S16-S24 for more details). The calculations show that all molecules (AxM and AxD), upon excitation, will be promoted into higher excited states ($S_2$, $S_3$ or $S_4$) but not into the first excited state $S_1$. Furthermore, the $S_1$ state is nonfluorescent (de-excites without emission of photons) for all molecules (AxM and AxD). In case of AxD, the higher excited states are very close in energy to the first excited state $S_1$. Due to this factor and the larger number of vibrational modes in AxD (compared to AxM), the higher excited states ($S_2$, $S_3$ or $S_4$) in the AxD series transform very fast to the nonfluorescent state $S_1$. A possible, fast transition pathway between the excited states in A1D is shown and discussed in the ESI (Figure S23-S25). In contrast, the transition from higher excited states to the nonfluorescent state $S_1$ for AxM needs a large change in the molecular geometry (ESI Figure S26-S32), which takes time. Therefore, in a rare exception from Kasha's rule[63], AxM de-excite with the emission of photons from the higher excited states ($S_2$, $S_3$ or $S_4$) to the ground state $S_0$ as reported in literature for some π-conjugated compounds[64,65]. The calculations also showed (ESI Figure S30) that in A2M an additional barrier has to be overcome for the higher states to go to the nonfluorescent $S_1$ state which explains why A2M has the strongest emission intensity.

**Table 1.** Elemental analysis of pristine and functionalized SWNTs.

| Sample | Carbon (%) | Hydrogen (%) | Nitrogen (%) |
|---|---|---|---|
| Pristine Tube | 93.8 | 0.12 | 0.03 |
| A1M-CNT | 86.57 | 2.61 | 5.45 |
| A2M-CNT | 87.88 | 2.36 | 5.16 |
| A3M-CNT | 86.65 | 2.26 | 2.91 |
| A1D-CNT | 91.69 | 2.36 | 3.01 |
| A2D-CNT | 89.94 | 2.08 | 2.85 |
| A3D-CNT | 89.90 | 2.42 | 2.24 |

After characterization of our molecular probes and investigation of their optical and physicochemical properties, we conjugated them to SWNTs by nondestructive covalent functionalization. The presence of the chlorine atoms in the structure of AxM and AxD series provided more possibilities to modify these molecular probes into reactive precursors. The substitution of one of their chlorine atoms by azide resulted in new probes with the ability to produce nitrenes upon elevating the temperature. [2+1] cycloaddition reaction between the azidated probes and SWNTs followed by $sp^3$-$sp^2$ rehybridization resulted in the covalent nondestructive functionalization of SWNTs (Figure 6a). Elemental analysis of the functionalized tubes confirms the successful functionalization of SWNTs by the molecular probes (Table 1). The CHN data showed increasing content of nitrogen and hydrogen with respect to the pristine SWNTs, indicating the attachment of the molecular probes to the surface of the nanotubes.

To prove the nondestructive character of the covalent functionalization of SWNTs by molecular probes, the Raman spectra of nanotubes were analyzed after functionalization. We did this by exciting our functionalized tubes (AxM-CNT and AxD-CNT) with a 638 nm laser pump to estimate and quantify the amount of disorder induced in the crystalline structure of the SWNTs. The Raman spectra in a relevant region for the assessment of the defectivity and crystalline integrity of the SWNTs are shown in Figure 6b, where we find a weak D mode at 1320 $cm^{-1}$ and two G modes, $G^+$ at 1600 $cm^{-1}$ and $G^-$ at 1560 $cm^{-1}$respectively. ESI Figure S35 zooms in the spectral region of the D band for better visualization while ESI Figure S36 provides the overview Raman spectra of our functionalized SWNTs. The ratio between integrated areas of the D and $G^+$ mode is used to estimate the defect concentration is SWNTs; we refer to it briefly as the D/G ratio. The negligible change in the D/G ratio confirmed the ability of these molecular probes to functionalize SWNTs nondestructively and covalently by reintegrating their π conjugated system upon opening of the aziridine ring and rehybridization of $sp^3$ to $sp^2$ carbon atoms[21,20]. Table 2 is a summary of the D/G ratio of our SWNTs samples. The data confirm that the number of defects even after covalent attachment of our probes is almost negligible, remaining within the limit of defectivity in the SWNTs from their growth process.

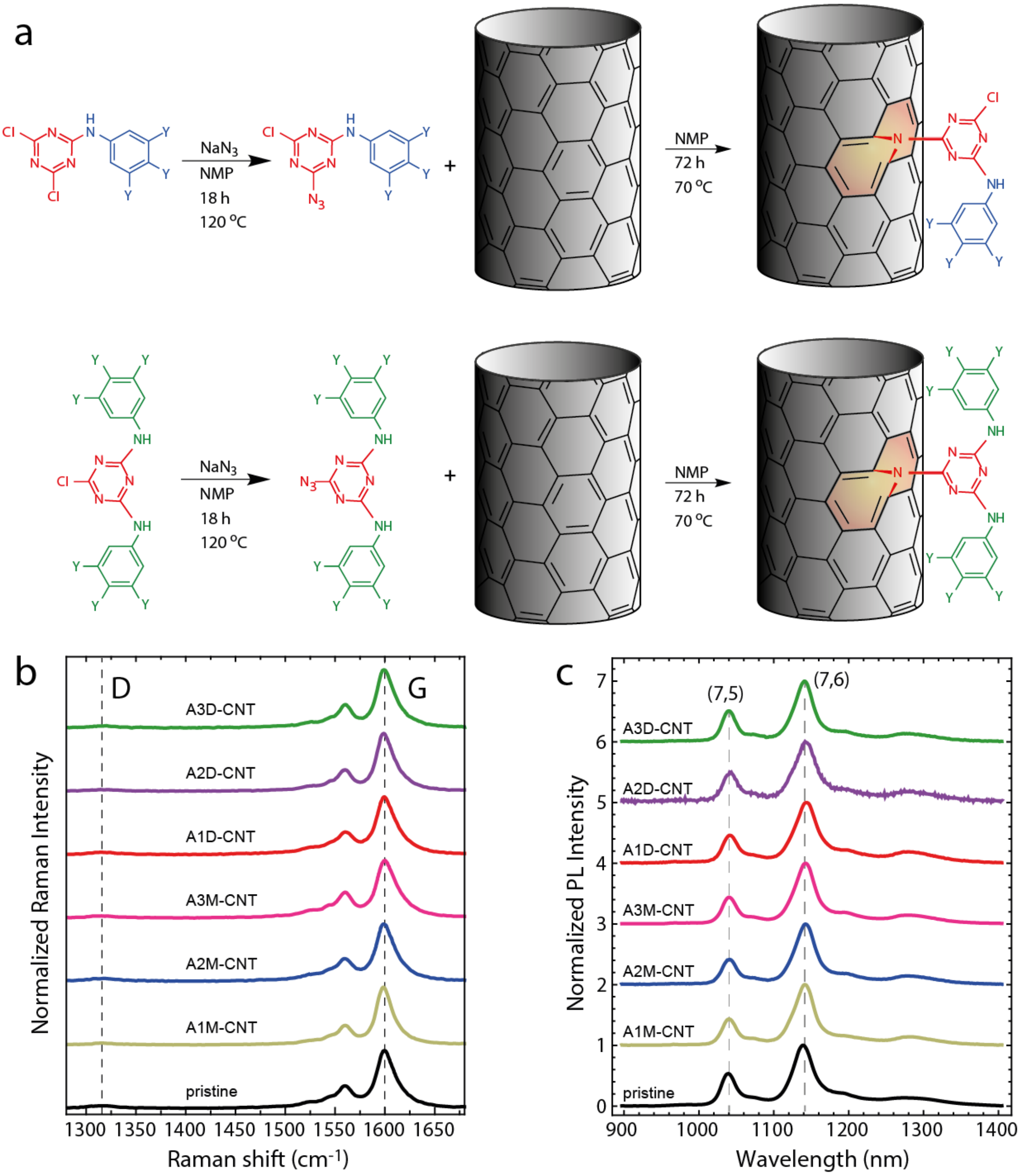

**Figure 6.** (a) Schematic representation of the covalent attachment of molecular probes to the surface of SWNTs via nitrene [2+1] cycloaddition reaction. (b) Representative Raman spectra of the pristine and functionalized SWNTs at 638 nm laser excitation showing the D and G mode bands. (c) Emission spectra of the pristine and functionalized SWNTs excited at 650 nm, showing the characteristic emission lines from the (7,5) and (7,6) tubes species.

To ensure that the π-conjugation of the tubes is not compromised by the addition of our functional probes, we then addressed their emission, which is an extremely sensitive indicator of their unperturbed conjugation. The emission from the (7,5) and (7,6) SWNTs species can be clearly observed in both the pristine and functionalized SWNTs, as shown in Figure 6c. The conjugated $\pi$-network of our functionalized tubes was kept intact as it is in the pristine tubes[21,20]. Those results confirm the ability of our molecular probes to covalently attach onto the tubes without destroying their conjugation. The detailed effect of the functionalization on the physical properties of the nanotubes goes beyond the scope of the present manuscript and will be the subject of future publications.

**Table 2.** Amplitude of the Raman D and G bands, together with their ratio.

| Sample | $I_D$ (counts/sec/mM) | $I_G$ (counts/sec/mM) | Area D/Area G |
|---|---|---|---|
| Pristine | 815 | 23856 | 0.03 |
| A1M-CNT | 1417 | 31867 | 0.04 |
| A2M-CNT | 1887 | 22555 | 0.08 |
| A3M-CNT | 1893 | 36337 | 0.05 |
| A1D-CNT | 2243 | 33839 | 0.07 |
| A2D-CNT | 1413 | 29016 | 0.05 |
| A3D-CNT | 944 | 19365 | 0.05 |

## Conclusion

In summary, a set of defined molecular probes were synthesized and characterized using different spectroscopic tools, and covalently attached onto single walled carbon nanotubes to produce new systems with the preserved and integrated π conjugated systems. While the optoelectronic properties of the charge transfer agents strongly depended on their aniline segments and the number and position of their methoxy groups, their ability to integrate with the π system of SWNTs was based on the triazine segments. Depending upon the number of methoxy groups attached to the aniline derivatives our custom moieties absorbed light between 4.5 eV and 5.0 eV and emitted between 3.5 eV and 4.3 eV. Regardless of the number of aniline segments, our molecular probes showed a systematical blue shift between the di- and the monosubstituted series. In both series, the moieties bearing methoxy groups in the meta positions showed the highest values of the HOMO-LUMO band gaps. Quantum chemical calculations and experiments showed the highest emission yield for the species bearing two methoxy groups in the meta positions of the mono-substituted series and an overall lower emission yield of the di-substituted series than the mono-substituted one. Integration of the π-conjugated systems of defined charge transfer agents and SWNTs is a step ahead for the deep understanding of the mechanism of covalent bond mediated electron transfer in nanomaterials with a huge aromatic backbone. We showed detailed and excellent correlation between experimental and calculated data based on the structural and optical properties of our probes. We also illustrated that the number and position of methoxy substituents and triazine play a significant role in their electrical and optical features. The low fluorescent intensity in our AxD probes is related to the low oscillator strength for the transition between the ground and the first excited state, whereas the high fluorescent intensity in the AxM is a special case considering Kasha's rule. We also observe no increase of the SWNTs defectivity after attaching our molecular probes onto them based on their Raman and emission spectra. Future studies will focus on the injection of charges and their effects on the properties of the tubes.

**Supporting Information**

The supporting information includes: Experimental details on the reaction conditions and product purification procedures. Spectroscopy by single crystal XRD, NMR, and mass spectrometry. Data obtained from the DFT calculations, including cartesian coordinates, oscillator strengths, ground and excited states energies and transitions, potential energy surface of the molecules. IR and Raman spectra.

**Ackowledgements**

This work was supported by the Focus Area NanoScale at Freie Universität Berlin. M.A. and S.R. acknowledge funding by the European Research Council (ERC) under grant 772108. The authors would like to thank the HPC Service of ZEDAT, Freie Universität Berlin, for computing time.

**Electronic Supplementary Information**

# Synthesis of Multifunctional Charge Transfer Agents: Towards Single Walled Carbon Nanotubes with Defined Covalent Functionality and Preserved π System

Alphonse Fiebor[1], Antonio Setaro[1,4], Andreas J. Achazi[2], Georgy Gordeev[1], Manuela Weber[2], Daniel Franz[2], Beate Paulus[2], Mohsen Adeli[1,3], and Stephanie Reich[1]

*[1]Department of Physics, Free University Berlin, Arnimallee 14, 14195 Berlin, Germany*
*[2]Department of Chemistry, Biochemistry & Pharmacy, Free University Berlin, Arnimallee 22, 14195 Berlin, Germany*
*[3]Department of Chemistry, Faculty of Science, Lorestan University, Khorramabad, Iran*
*[4]Department of Engineering, Pegaso University, Naples, Italy*

## Table of contents

## Experimental

Triazine (cyanuric chloride) is an electron poor moiety that has been extensively used as a precursor in various synthetic protocols. The three different yet identical carbons bearing chlorine atoms make it easy to be tuned and modified[1-4]. This property enables it to undergo successive nucleophilic substitutions although its reactivity is decreased after the first two substitutions. This is due to the acquisition of pi electrons from the incoming nucleophiles, hence the need for elevated temperatures (above 75 °C) for usually the third substitution[5-8]. We exploited this characteristic of cyanuric chloride and the concept of mole equivalence to synthesize the mono- and di-substituted molecular probes.

## Functionalization of SWNT with probes

*Cleaning of pristine SWNTs*

Pristine SWNTs in 37% HCl was sonicated for 1 h in a bath sonicator at a temperature of 60 °C. This is followed by extensive rinsing with distilled water. This step is to neutralize the sample and can be done using vacuum filtration setup and polycarbonate membranes with pore size of 0.5 µm. The obtained SWNT sample was dried by lyophilizer for 2 days. The dried SWNTs was then annealed in high vacuum at 800 to 1000 °C. We anneal to remove defects and non-nanotube carbon materials[9].

*Azidation of molecular probes*

We converted molecular probes (both mono and di-substituted) to an azide derivative to be able to attach on the surface of SWNTs by nitrene [2+1] cycloaddition reaction. This was done by reacting molecular probes with sodium azide in a 1:1 ratio at 120 °C overnight (18 h) in *N*-methyl-2-pyrrolidone (50 ml). The reaction was quenched with DI water and extracted with dichloromethane (3 × 30 ml). Organic phase was dried over $MgSO_4$, concentrated *in vacuo* and the crude azido product was used without further purification for functionalization of SWNTs. Mass spectrometry (fig. S19-S22, page *S22-S23*) shows successful azidation of the molecular probes.

All reactions were carried out under inert conditions unless otherwise stated. All reactions leading to molecular probes were monitored by thin-layer chromatography (TLC) using silica gel coated aluminium plates (0.2 mm). Plates were visualized with UV light or under iodine staining.

## Christallographic characterization

**Table S1.** Crystal data and structure refinement for A1M.

| | |
|---|---|
| Identification code | A1M |
| Empirical formula | $C_{10}H_8Cl_2N_4O$ |
| Formula weight | 271.10 |
| Temperature/K | 100(2) |
| Crystal system | monoclinic |
| Space group | $P2_1/c$ |
| a/Å | 7.37860(5) |
| b/Å | 20.15672(12) |
| c/Å | 8.09134(5) |
| α/° | 90 |
| β/° | 114.5191(2) |
| γ/° | 90 |
| Volume/Å$^3$ | 1094.891(12) |
| Z | 4 |
| $\rho_{calc}$g/cm$^3$ | 1.645 |
| μ/mm$^{-1}$ | 5.254 |
| F(000) | 552.0 |
| Crystal size/mm$^3$ | 0.27 × 0.11 × 0.05 |
| Radiation | CuKα (λ = 1.54178) |
| 2Θ range for data collection/° | 8.774 to 140.064 |
| Index ranges | $-8 \leq h \leq 8$, $-24 \leq k \leq 24$, $-9 \leq l \leq 8$ |
| Reflections collected | 35714 |
| Independent reflections | 2076 [$R_{int}$ = 0.0320, $R_{sigma}$ = 0.0114] |
| Data/restraints/parameters | 2076/0/155 |
| Goodness-of-fit on $F^2$ | 1.062 |
| Final R indexes [I>=2σ (I)] | $R_1$ = 0.0227, $wR_2$ = 0.0619 |
| Final R indexes [all data] | $R_1$ = 0.0228, $wR_2$ = 0.0619 |
| Largest diff. peak/hole / e Å$^{-3}$ | 0.22/-0.21 |

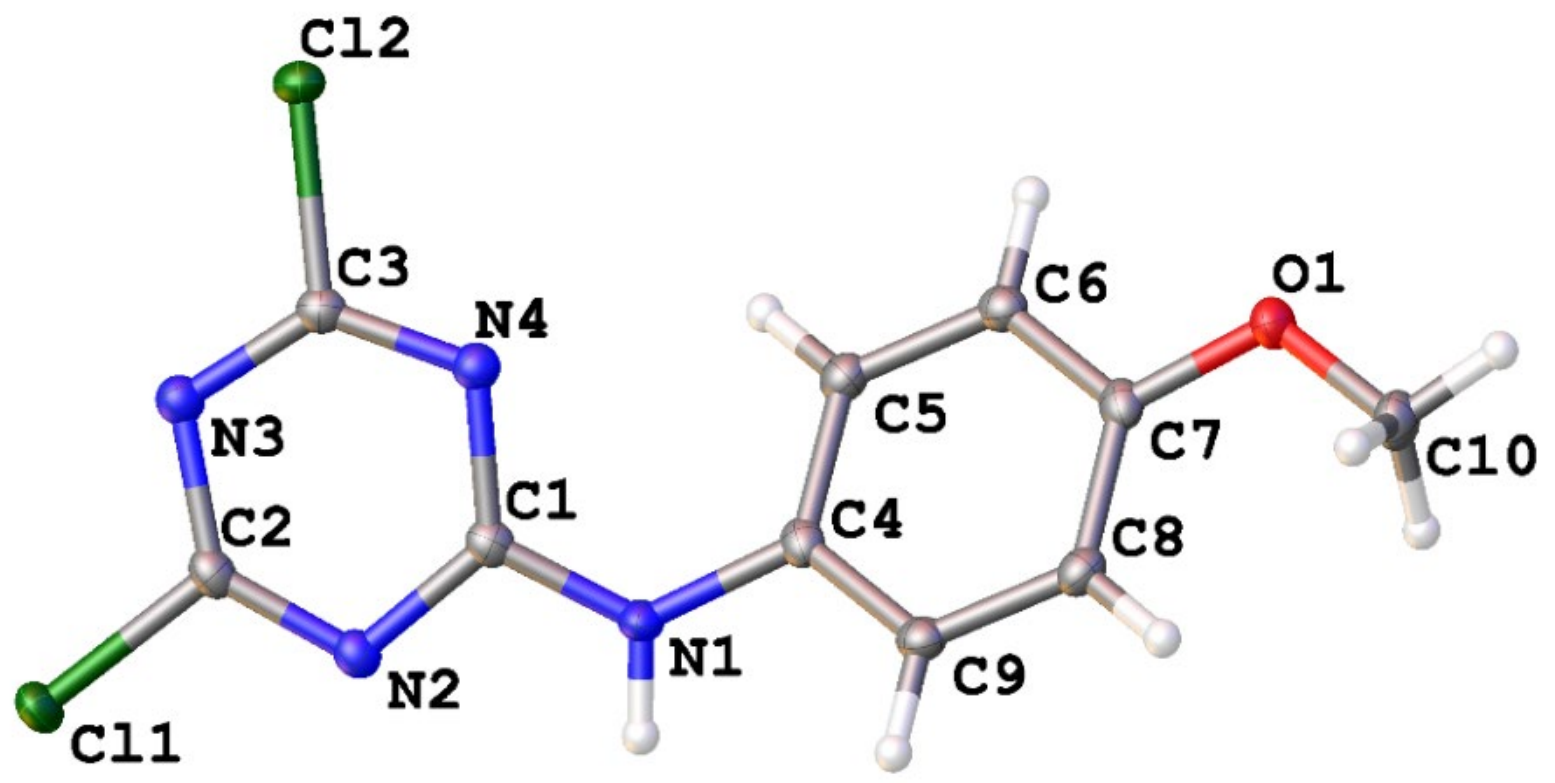

**Table S2.** Fractional Atomic Coordinates (×10$^4$) and Equivalent Isotropic Displacement Parameters (Å$^2$×10$^3$) for A1M-final. $U_{eq}$ is defined as 1/3 of the trace of the orthogonalized $U_{IJ}$ tensor.

| **Atom** | ***x*** | ***y*** | ***z*** | **U(eq)** |
|---|---|---|---|---|
| Cl1 | 9234.6(5) | 8751.4(2) | 8354.2(4) | 21.62(10) |
| Cl2 | 4898.9(4) | 9470.6(2) | 1697.2(4) | 19.02(10) |
| O1 | -471.4(14) | 6037.2(5) | -2526.3(12) | 19.2(2) |
| N1 | 4799.8(16) | 7208.2(5) | 3918.3(14) | 15.8(2) |
| N2 | 6895.6(15) | 7949.6(5) | 5930.6(14) | 15.2(2) |
| N3 | 7007.6(15) | 9068.2(5) | 4999.2(14) | 16.3(2) |
| N4 | 4991.5(15) | 8268.6(5) | 2829.7(14) | 15.2(2) |
| C1 | 5557.6(18) | 7820.6(6) | 4200.1(16) | 14.1(2) |
| C2 | 7528.6(18) | 8561.5(6) | 6180.8(17) | 15.4(3) |
| C3 | 5739.7(18) | 8862.4(6) | 3368.9(17) | 14.7(2) |
| C4 | 3417.4(18) | 6919.4(6) | 2268.3(16) | 15.0(3) |
| C5 | 1866.6(18) | 7275.6(6) | 959.3(17) | 16.0(3) |
| C6 | 599.4(18) | 6964.6(6) | -626.1(17) | 16.0(3) |
| C7 | 848.3(19) | 6293.3(6) | -909.2(17) | 15.6(3) |
| C8 | 2364.9(19) | 5930.4(6) | 419.0(17) | 16.4(3) |
| C9 | 3639.4(19) | 6249.1(6) | 2000.5(17) | 15.8(3) |
| C10 | -245(2) | 5348.9(6) | -2830.4(18) | 19.9(3) |

**Table S3.** Anisotropic Displacement Parameters ($Å^2 \times 10^3$) for A1M-final. The Anisotropic displacement factor exponent takes the form: $-2\pi^2[h^2a^{*2}U_{11}+2hka^*b^*U_{12}+\ldots]$.

| **Atom** | **$U_{11}$** | **$U_{22}$** | **$U_{33}$** | **$U_{23}$** | **$U_{13}$** | **$U_{12}$** |
|---|---|---|---|---|---|---|
| C11 | 25.71(18) | 18.60(17) | 13.09(17) | -1.59(11) | 0.61(13) | -1.19(11) |
| Cl2 | 21.25(17) | 16.13(16) | 16.23(16) | 4.19(10) | 4.35(12) | 0.30(11) |
| O1 | 22.7(5) | 16.5(5) | 14.7(4) | -2.1(3) | 4.1(4) | -2.3(4) |
| N1 | 19.2(5) | 14.7(5) | 11.5(5) | 1.8(4) | 4.2(4) | -1.2(4) |
| N2 | 15.7(5) | 16.5(5) | 13.0(5) | -0.7(4) | 5.5(4) | 0.7(4) |
| N3 | 17.2(5) | 15.8(5) | 15.2(5) | -0.2(4) | 6.2(4) | 0.6(4) |
| N4 | 16.6(5) | 15.3(5) | 13.8(5) | 0.3(4) | 6.6(4) | 0.3(4) |
| C1 | 14.1(6) | 16.1(6) | 13.4(6) | -0.3(5) | 7.0(5) | 1.8(5) |
| C2 | 14.5(6) | 18.4(6) | 12.8(6) | -1.7(5) | 5.3(5) | 0.8(5) |
| C3 | 14.7(6) | 16.0(6) | 14.5(6) | 1.5(5) | 7.2(5) | 1.8(5) |
| C4 | 15.5(6) | 17.1(6) | 13.5(6) | -0.5(5) | 7.2(5) | -3.4(5) |
| C5 | 17.3(6) | 14.4(6) | 17.6(6) | 0.7(5) | 8.6(5) | -0.7(5) |
| C6 | 15.5(6) | 16.7(6) | 15.9(6) | 2.5(5) | 6.5(5) | 0.1(5) |
| C7 | 16.3(6) | 18.2(6) | 13.9(6) | -1.1(5) | 7.8(5) | -4.3(5) |
| C8 | 20.0(6) | 13.6(6) | 17.7(6) | 0.6(5) | 9.9(5) | -0.9(5) |
| C9 | 16.8(6) | 16.0(6) | 15.0(6) | 2.8(5) | 6.9(5) | 0.1(5) |
| C10 | 23.7(7) | 17.6(6) | 18.4(6) | -4.6(5) | 8.7(5) | -3.8(5) |

**Table S4.** Bond Lengths for A1M.

| Atom | Atom | Length/Å | Atom | Atom | Length/Å |
|---|---|---|---|---|---|
| Cl1 | C2 | 1.7280(13) | N3 | C2 | 1.3417(17) |
| Cl2 | C3 | 1.7380(12) | N4 | C3 | 1.3146(17) |
| O1 | C7 | 1.3672(15) | N4 | C1 | 1.3546(16) |
| O1 | C10 | 1.4308(15) | C4 | C9 | 1.3887(18) |
| N1 | C1 | 1.3350(16) | C4 | C5 | 1.3931(18) |
| N1 | C4 | 1.4253(16) | C5 | C6 | 1.3847(18) |
| N2 | C2 | 1.3046(17) | C6 | C7 | 1.3972(18) |
| N2 | C1 | 1.3622(16) | C7 | C8 | 1.3943(18) |
| N3 | C3 | 1.3278(16) | C8 | C9 | 1.3912(18) |

**Table S5.** Bond Angles for A1M.

| Atom | Atom | Atom | Angle/° | Atom | Atom | Atom | Angle/° |
|---|---|---|---|---|---|---|---|
| C7 | O1 | C10 | 116.30(10) | N4 | C3 | Cl2 | 114.76(9) |
| C1 | N1 | C4 | 128.60(11) | N3 | C3 | Cl2 | 115.44(9) |
| C2 | N2 | C1 | 113.58(11) | C9 | C4 | C5 | 119.71(11) |
| C3 | N3 | C2 | 110.21(11) | C9 | C4 | N1 | 117.30(11) |
| C3 | N4 | C1 | 113.01(11) | C5 | C4 | N1 | 122.97(11) |
| N1 | C1 | N4 | 120.76(11) | C6 | C5 | C4 | 119.79(12) |
| N1 | C1 | N2 | 115.18(11) | C5 | C6 | C7 | 120.41(12) |
| N4 | C1 | N2 | 124.05(11) | O1 | C7 | C8 | 124.38(11) |
| N2 | C2 | N3 | 129.12(11) | O1 | C7 | C6 | 115.64(11) |
| N2 | C2 | Cl1 | 115.47(9) | C8 | C7 | C6 | 119.98(12) |
| N3 | C2 | Cl1 | 115.40(9) | C9 | C8 | C7 | 119.12(11) |
| N4 | C3 | N3 | 129.79(12) | C4 | C9 | C8 | 120.95(12) |

**Table S6.** Torsion Angles for A1M.

| A | B | C | D | Angle/° | A | B | C | D | Angle/° |
|---|---|---|---|---|---|---|---|---|---|
| C4 | N1 | C1 | N4 | 1.86(19) | C1 | N1 | C4 | C9 | 145.43(12) |
| C4 | N1 | C1 | N2 | -178.43(11) | C1 | N1 | C4 | C5 | -35.67(19) |
| C3 | N4 | C1 | N1 | 174.34(11) | C9 | C4 | C5 | C6 | -2.19(18) |
| C3 | N4 | C1 | N2 | -5.34(17) | N1 | C4 | C5 | C6 | 178.93(11) |
| C2 | N2 | C1 | N1 | -176.67(10) | C4 | C5 | C6 | C7 | 0.96(18) |
| C2 | N2 | C1 | N4 | 3.03(17) | C10 | O1 | C7 | C8 | 0.25(17) |
| C1 | N2 | C2 | N3 | 1.83(19) | C10 | O1 | C7 | C6 | -179.55(11) |
| C1 | N2 | C2 | Cl1 | -179.88(8) | C5 | C6 | C7 | O1 | -179.41(11) |
| C3 | N3 | C2 | N2 | -3.42(18) | C5 | C6 | C7 | C8 | 0.79(18) |
| C3 | N3 | C2 | Cl1 | 178.28(9) | O1 | C7 | C8 | C9 | 178.92(11) |
| C1 | N4 | C3 | N3 | 3.58(19) | C6 | C7 | C8 | C9 | -1.29(18) |
| C1 | N4 | C3 | Cl2 | -175.07(8) | C5 | C4 | C9 | C8 | 1.69(18) |
| C2 | N3 | C3 | N4 | 0.39(18) | N1 | C4 | C9 | C8 | -179.37(11) |
| C2 | N3 | C3 | Cl2 | 179.03(8) | C7 | C8 | C9 | C4 | 0.05(18) |

**Table S7.** Hydrogen Atom Coordinates (Å×$10^4$) and Isotropic Displacement Parameters (Å$^2$×$10^3$) for A1M.

| Atom | *x* | *y* | *z* | U(eq) |
|---|---|---|---|---|
| H1 | 5208.08 | 6948.47 | 4878.59 | 19 |
| H5 | 1679.14 | 7730.44 | 1153.85 | 19 |
| H6 | -445.61 | 7209.22 | -1526.98 | 19 |
| H8 | 2525.96 | 5471.67 | 246.43 | 20 |
| H9 | 4676.35 | 6004.7 | 2910.15 | 19 |
| H10A | -1258.46 | 5218.59 | -4022.98 | 30 |
| H10B | -404.99 | 5084.03 | -1883.94 | 30 |
| H10C | 1083.39 | 5272.63 | -2792.47 | 30 |

**Table S8.** Crystal data and structure refinement for A3M.

| | |
|---|---|
| Identification code | A3M |
| Empirical formula | $C_{12}H_{12}Cl_2N_4O_3$ |
| Formula weight | 331.16 |
| Temperature/K | 100(2) |
| Crystal system | monoclinic |
| Space group | $P2_1/n$ |
| a/Å | 10.39984(6) |
| b/Å | 10.68405(7) |
| c/Å | 12.63319(8) |
| α/° | 90 |
| β/° | 90.8071(2) |
| γ/° | 90 |
| Volume/Å$^3$ | 1403.565(15) |
| Z | 4 |
| $\rho_{calc}$g/cm$^3$ | 1.567 |
| μ/mm$^{-1}$ | 4.323 |
| F(000) | 680.0 |
| Crystal size/mm$^3$ | 0.37 × 0.28 × 0.15 |
| Radiation | CuKα (λ = 1.54178) |
| 2Θ range for data collection/° | 10.846 to 140.23 |
| Index ranges | $-11 \leq h \leq 12$, $-13 \leq k \leq 13$, $-15 \leq l \leq 15$ |
| Reflections collected | 37176 |
| Independent reflections | 2658 [$R_{int}$ = 0.0289, $R_{sigma}$ = 0.0118] |
| Data/restraints/parameters | 2658/0/193 |
| Goodness-of-fit on $F^2$ | 1.063 |
| Final R indexes [I>=2σ (I)] | $R_1$ = 0.0225, $wR_2$ = 0.0590 |
| Final R indexes [all data] | $R_1$ = 0.0226, $wR_2$ = 0.0591 |
| Largest diff. peak/hole / e Å$^{-3}$ | 0.24/-0.22 |

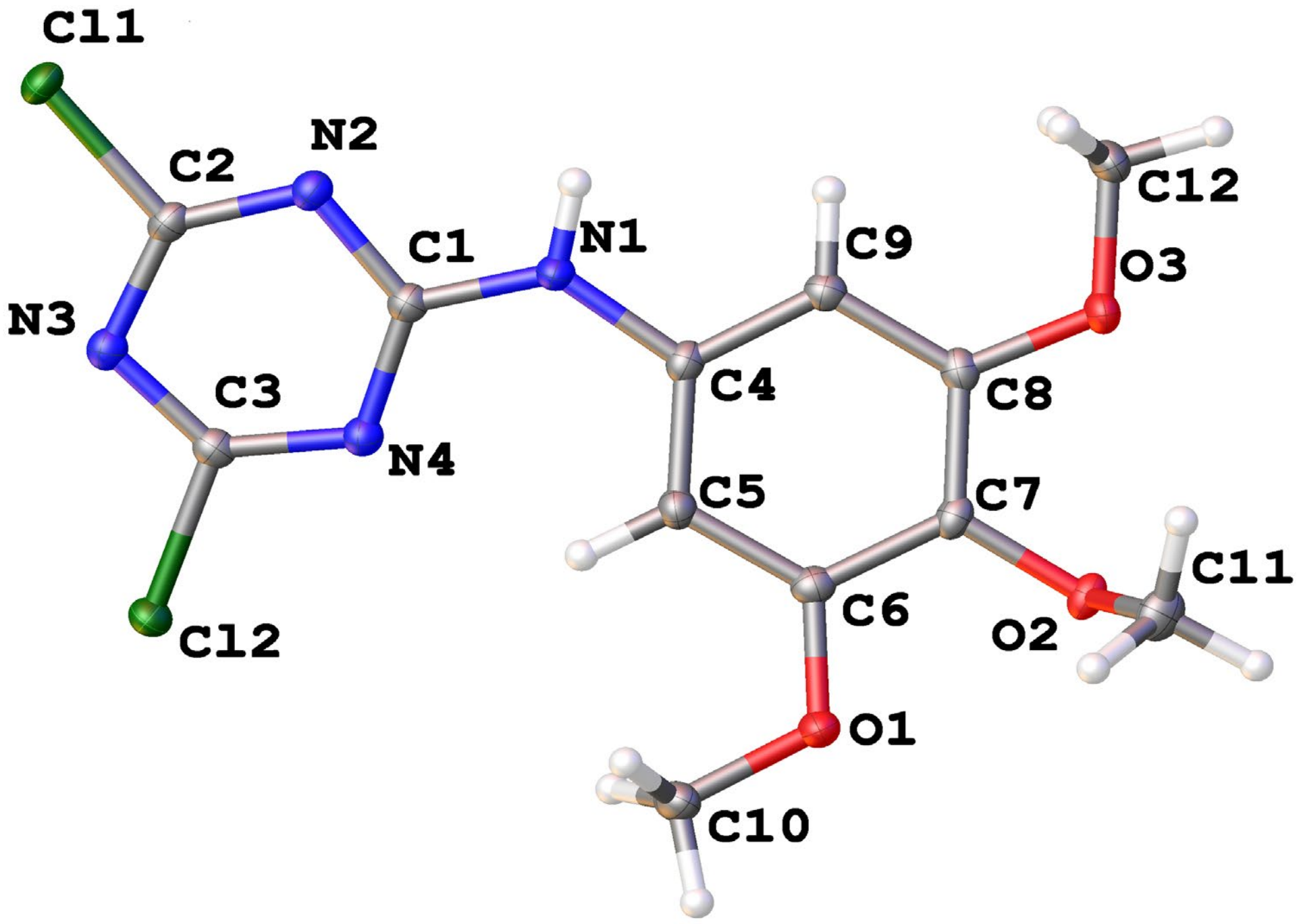

**Table S9.** Fractional Atomic Coordinates (×10[4]) and Equivalent Isotropic Displacement Parameters (Å$^2$×10$^3$) for A3M-final. $U_{eq}$ is defined as 1/3 ofthe trace of the orthogonalised $U_{IJ}$ tensor.

| **Atom** | ***x*** | ***y*** | ***z*** | **U(eq)** |
|---|---|---|---|---|
| Cl1 | 10582.5(3) | 7755.3(3) | 2629.3(2) | 17.98(9) |
| Cl2 | 7647.1(3) | 3867.2(3) | 2077.6(2) | 17.96(9) |
| O1 | 5003.7(9) | 2722.4(8) | 6223.9(7) | 18.6(2) |
| O2 | 5501.2(8) | 2867.4(8) | 8303.5(7) | 16.04(19) |
| O3 | 7250.3(8) | 4496.8(8) | 9072.6(7) | 18.07(19) |
| N1 | 8376.9(10) | 5816.8(10) | 5565.7(8) | 15.4(2) |
| N2 | 9381.3(10) | 6685.3(10) | 4154.2(8) | 15.8(2) |
| N3 | 9092.3(10) | 5806.2(10) | 2428.9(8) | 15.9(2) |
| N4 | 8048.5(10) | 4875(1) | 3908.2(8) | 15.5(2) |
| C1 | 8590.1(11) | 5772.7(11) | 4522.5(9) | 14.2(2) |
| C2 | 9576.2(11) | 6624.0(11) | 3134.3(10) | 15.0(2) |
| C3 | 8345.9(11) | 4970.7(11) | 2901.1(9) | 14.6(2) |
| C4 | 7611.6(11) | 5037.7(11) | 6212.6(9) | 14.2(2) |
| C5 | 6683.7(12) | 4218.8(11) | 5811.9(9) | 15.6(2) |
| C6 | 5955.6(11) | 3527.0(11) | 6524.8(10) | 15.2(2) |
| C7 | 6176.5(11) | 3624.0(11) | 7615.3(9) | 14.5(2) |
| C8 | 7106.3(11) | 4454.9(11) | 7999.6(9) | 14.6(2) |
| C9 | 7821.2(11) | 5173.9(11) | 7300.0(9) | 14.5(2) |
| C10 | 4668.9(13) | 2684.8(12) | 5119.5(10) | 20.0(3) |
| C11 | 4389.7(12) | 3465.6(13) | 8749.5(10) | 21.1(3) |
| C12 | 8124.9(13) | 5417.9(13) | 9478.8(10) | 20.6(3) |

**Table S10.** Anisotropic Displacement Parameters ($Å^2 \times 10^3$) for A3M-final. The Anisotropic displacement factor exponent takes the form: $-2\pi^2[h^2a^{*2}U_{11}+2hka^*b^*U_{12}+\ldots]$.

| **Atom** | **$U_{11}$** | **$U_{22}$** | **$U_{33}$** | **$U_{23}$** | **$U_{13}$** | **$U_{12}$** |
|---|---|---|---|---|---|---|
| Cl1 | 18.45(15) | 17.55(16) | 18.05(15) | 2.98(10) | 3.69(11) | -3.8(1) |
| Cl2 | 21.46(16) | 17.58(16) | 14.85(15) | -3.16(10) | 1.00(11) | -3.11(11) |
| O1 | 21.4(4) | 19.0(5) | 15.5(4) | 0.3(3) | 0.2(3) | -6.7(3) |
| O2 | 16.7(4) | 14.8(4) | 16.8(4) | 3.9(3) | 4.9(3) | 1.8(3) |
| O3 | 22.6(4) | 20.1(5) | 11.5(4) | 1.9(3) | -0.3(3) | -3.4(4) |
| N1 | 18.4(5) | 15.0(5) | 13.0(5) | -1.3(4) | 1.0(4) | -4.1(4) |
| N2 | 16.1(5) | 15.6(5) | 15.7(5) | 1.0(4) | 1.2(4) | -1.0(4) |
| N3 | 15.8(5) | 17.0(5) | 15.1(5) | 0.8(4) | 1.6(4) | 0.1(4) |
| N4 | 16.8(5) | 15.5(5) | 14.4(5) | -0.3(4) | 1.4(4) | -1.0(4) |
| C1 | 13.9(5) | 13.6(6) | 15.1(6) | 0.8(5) | 0.8(4) | 1.2(4) |
| C2 | 12.9(5) | 14.9(6) | 17.3(6) | 3.2(5) | 1.5(4) | 0.6(4) |
| C3 | 13.8(5) | 14.4(6) | 15.6(6) | -1.1(5) | -0.3(4) | 2.2(4) |
| C4 | 15.4(6) | 12.6(6) | 14.6(6) | 1.3(4) | 3.3(4) | 2.0(4) |
| C5 | 18.1(6) | 15.8(6) | 13.0(5) | 0.1(5) | 1.1(5) | -0.5(5) |
| C6 | 15.3(6) | 12.8(6) | 17.7(6) | -0.5(5) | 0.6(5) | 0.4(5) |
| C7 | 15.7(6) | 12.6(6) | 15.5(6) | 2.8(5) | 3.6(4) | 2.4(5) |
| C8 | 16.7(6) | 14.1(6) | 12.9(6) | 0.7(4) | 0.6(4) | 4.6(5) |
| C9 | 14.7(6) | 13.2(6) | 15.5(6) | -0.3(5) | 0.0(4) | 1.0(4) |
| C10 | 22.1(6) | 20.0(7) | 17.8(6) | 1.1(5) | -3.6(5) | -4.3(5) |
| C11 | 18.4(6) | 21.3(7) | 23.7(7) | 2.3(5) | 6.9(5) | 4.1(5) |
| C12 | 25.9(7) | 22.3(7) | 13.6(6) | 0.2(5) | -2.5(5) | -4.3(5) |

**Table S11**. Bond Lengths for A3M.

| Atom | Atom | Length/Å | Atom | Atom | Length/Å |
|---|---|---|---|---|---|
| Cl1 | C2 | 1.7268(12) | N2 | C1 | 1.3620(16) |
| Cl2 | C3 | 1.7261(12) | N3 | C3 | 1.3298(16) |
| O1 | C6 | 1.3612(15) | N3 | C2 | 1.3409(16) |
| O1 | C10 | 1.4337(15) | N4 | C3 | 1.3176(16) |
| O2 | C7 | 1.3855(14) | N4 | C1 | 1.3516(16) |
| O2 | C11 | 1.4424(15) | C4 | C5 | 1.3925(17) |
| O3 | C8 | 1.3625(14) | C4 | C9 | 1.3954(17) |
| O3 | C12 | 1.4304(15) | C5 | C6 | 1.3964(17) |
| N1 | C1 | 1.3404(16) | C6 | C7 | 1.3974(17) |
| N1 | C4 | 1.4186(15) | C7 | C8 | 1.3948(18) |
| N2 | C2 | 1.3089(16) | C8 | C9 | 1.3937(17) |

**Table S12.** Bond Angles for A3M.

| Atom | Atom | Atom | Angle/° | Atom | Atom | Atom | Angle/° |
|---|---|---|---|---|---|---|---|
| C6 | O1 | C10 | 117.07(9) | N3 | C3 | Cl2 | 115.57(9) |
| C7 | O2 | C11 | 113.75(9) | C5 | C4 | C9 | 121.41(11) |
| C8 | O3 | C12 | 116.10(9) | C5 | C4 | N1 | 123.42(11) |
| C1 | N1 | C4 | 130.37(10) | C9 | C4 | N1 | 115.14(11) |
| C2 | N2 | C1 | 113.77(11) | C4 | C5 | C6 | 118.52(11) |
| C3 | N3 | C2 | 110.81(10) | O1 | C6 | C5 | 123.58(11) |
| C3 | N4 | C1 | 113.43(10) | O1 | C6 | C7 | 115.53(11) |
| N1 | C1 | N4 | 121.07(11) | C5 | C6 | C7 | 120.89(11) |
| N1 | C1 | N2 | 114.82(11) | O2 | C7 | C8 | 120.53(11) |
| N4 | C1 | N2 | 124.11(11) | O2 | C7 | C6 | 119.84(11) |
| N2 | C2 | N3 | 128.68(11) | C8 | C7 | C6 | 119.60(11) |
| N2 | C2 | Cl1 | 115.58(9) | O3 | C8 | C9 | 124.11(11) |
| N3 | C2 | Cl1 | 115.74(9) | O3 | C8 | C7 | 115.67(11) |
| N4 | C3 | N3 | 129.18(11) | C9 | C8 | C7 | 120.23(11) |
| N4 | C3 | Cl2 | 115.24(9) | C8 | C9 | C4 | 119.30(11) |

**Table S13.** Torsion Angles for A3M.

| A | B | C | D | Angle/° | A | B | C | D | Angle/° |
|---|---|---|---|---|---|---|---|---|---|
| C4 | N1 | C1 | N4 | -0.7(2) | C10 | O1 | C6 | C7 | 174.37(11) |
| C4 | N1 | C1 | N2 | 179.52(11) | C4 | C5 | C6 | O1 | 178.39(11) |
| C3 | N4 | C1 | N1 | 179.45(11) | C4 | C5 | C6 | C7 | -1.88(18) |
| C3 | N4 | C1 | N2 | -0.79(17) | C11 | O2 | C7 | C8 | 85.12(13) |
| C2 | N2 | C1 | N1 | 179.98(10) | C11 | O2 | C7 | C6 | -96.95(13) |
| C2 | N2 | C1 | N4 | 0.21(17) | O1 | C6 | C7 | O2 | 4.04(16) |
| C1 | N2 | C2 | N3 | 0.93(18) | C5 | C6 | C7 | O2 | -175.71(11) |
| C1 | N2 | C2 | Cl1 | -179.58(8) | O1 | C6 | C7 | C8 | -178.01(10) |
| C3 | N3 | C2 | N2 | -1.24(18) | C5 | C6 | C7 | C8 | 2.24(18) |
| C3 | N3 | C2 | Cl1 | 179.27(8) | C12 | O3 | C8 | C9 | 5.14(17) |
| C1 | N4 | C3 | N3 | 0.41(18) | C12 | O3 | C8 | C7 | -175.06(11) |
| C1 | N4 | C3 | Cl2 | -178.53(8) | O2 | C7 | C8 | O3 | -2.61(16) |
| C2 | N3 | C3 | N4 | 0.49(18) | C6 | C7 | C8 | O3 | 179.46(10) |
| C2 | N3 | C3 | Cl2 | 179.43(8) | O2 | C7 | C8 | C9 | 177.20(10) |
| C1 | N1 | C4 | C5 | -15.5(2) | C6 | C7 | C8 | C9 | -0.73(18) |
| C1 | N1 | C4 | C9 | 166.54(12) | O3 | C8 | C9 | C4 | 178.72(11) |
| C9 | C4 | C5 | C6 | 0.03(18) | C7 | C8 | C9 | C4 | -1.08(17) |
| N1 | C4 | C5 | C6 | -177.84(11) | C5 | C4 | C9 | C8 | 1.43(18) |
| C10 | O1 | C6 | C5 | -5.88(17) | N1 | C4 | C9 | C8 | 179.48(10) |

**Table S14.** Hydrogen Atom Coordinates (Å×$10^4$) and Isotropic Displacement Parameters (Å$^2$×$10^3$) for A3M-final.

| Atom | *x* | *y* | *z* | U(eq) |
|---|---|---|---|---|
| H1 | 8775.71 | 6427.04 | 5903.62 | 19 |
| H5 | 6549.07 | 4132.72 | 5070.17 | 19 |
| H9 | 8444.63 | 5750.68 | 7560.63 | 17 |
| H10A | 4457.55 | 3530.92 | 4873.29 | 30 |
| H10B | 3922.92 | 2136.74 | 5011.4 | 30 |
| H10C | 5397.37 | 2360.97 | 4719.09 | 30 |
| H11A | 3768.49 | 3661.95 | 8183.1 | 32 |
| H11B | 4653.67 | 4239.96 | 9107.07 | 32 |
| H11C | 3991.86 | 2901.3 | 9261.37 | 32 |
| H12A | 8986.26 | 5258.54 | 9204.26 | 31 |
| H12B | 8150.26 | 5373.97 | 10253.64 | 31 |
| H12C | 7837.56 | 6252.31 | 9256.7 | 31 |

Ref. 10-12

## $^1$H and $^{13}$C NMR spectra

In $^1$H NMR spectra of molecular probes, protons of aniline segment are not magnetically equivalent. Therefore, their signals are splitted and they are not singlet. Also, in the $^{13}$C NMR spectra of AxM series, two carbons of triazine bearing chlorine atoms are not magnetically equivalent and show two signals.

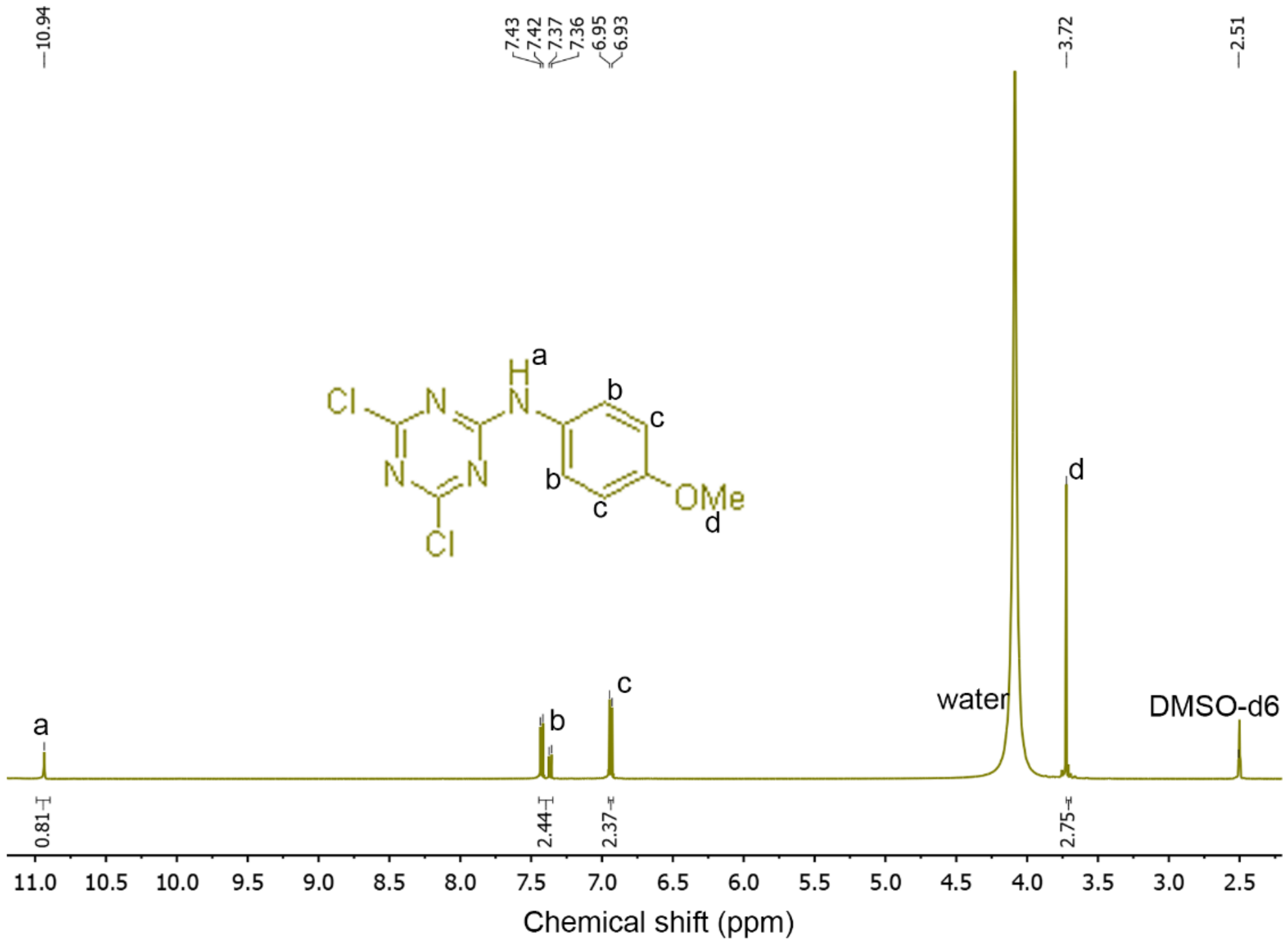

**Figure S1.** $^1$H NMR spectrum of A1M.

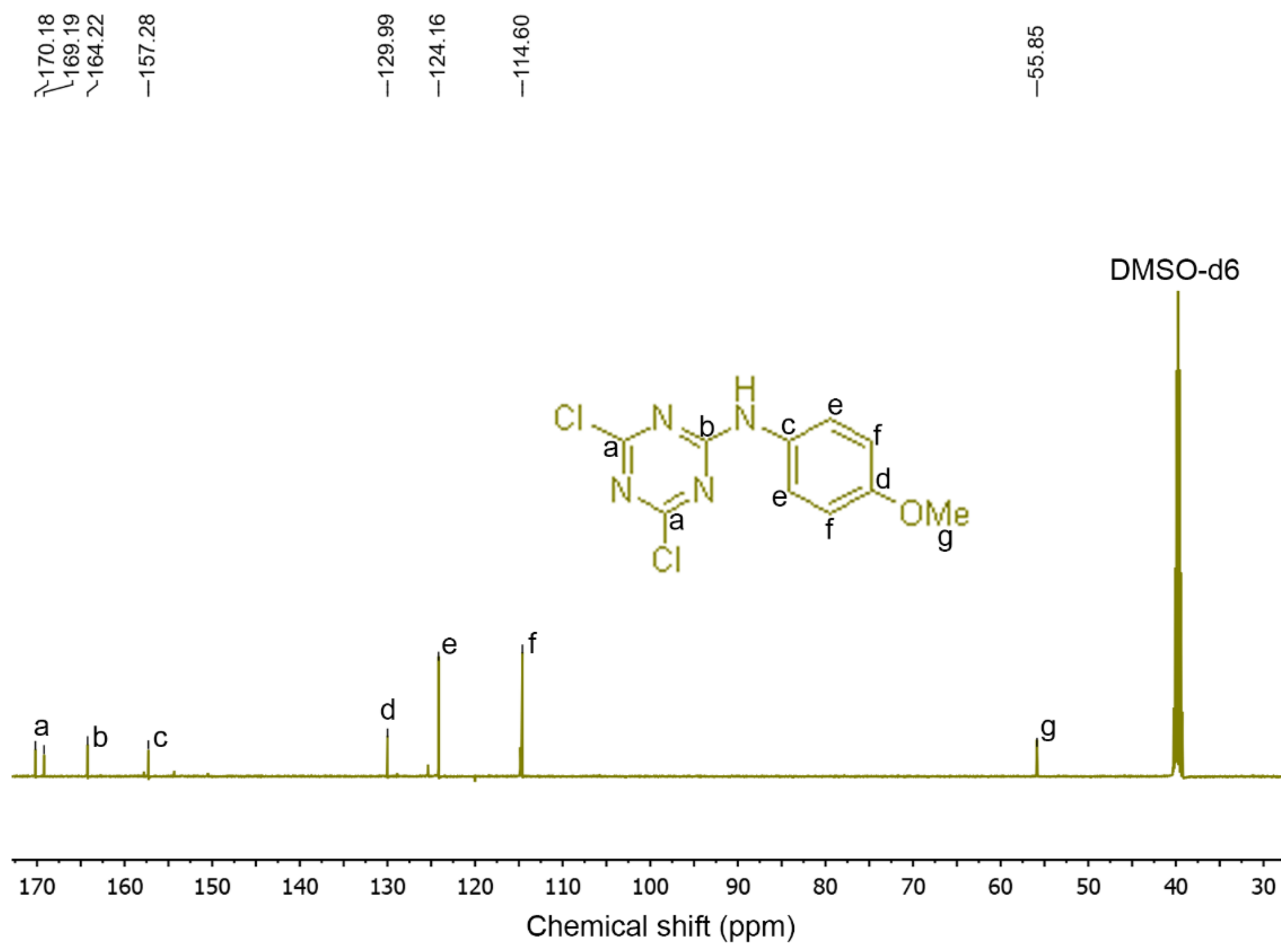

**Figure S2.** $^{13}C$ NMR spectrum of A1M.

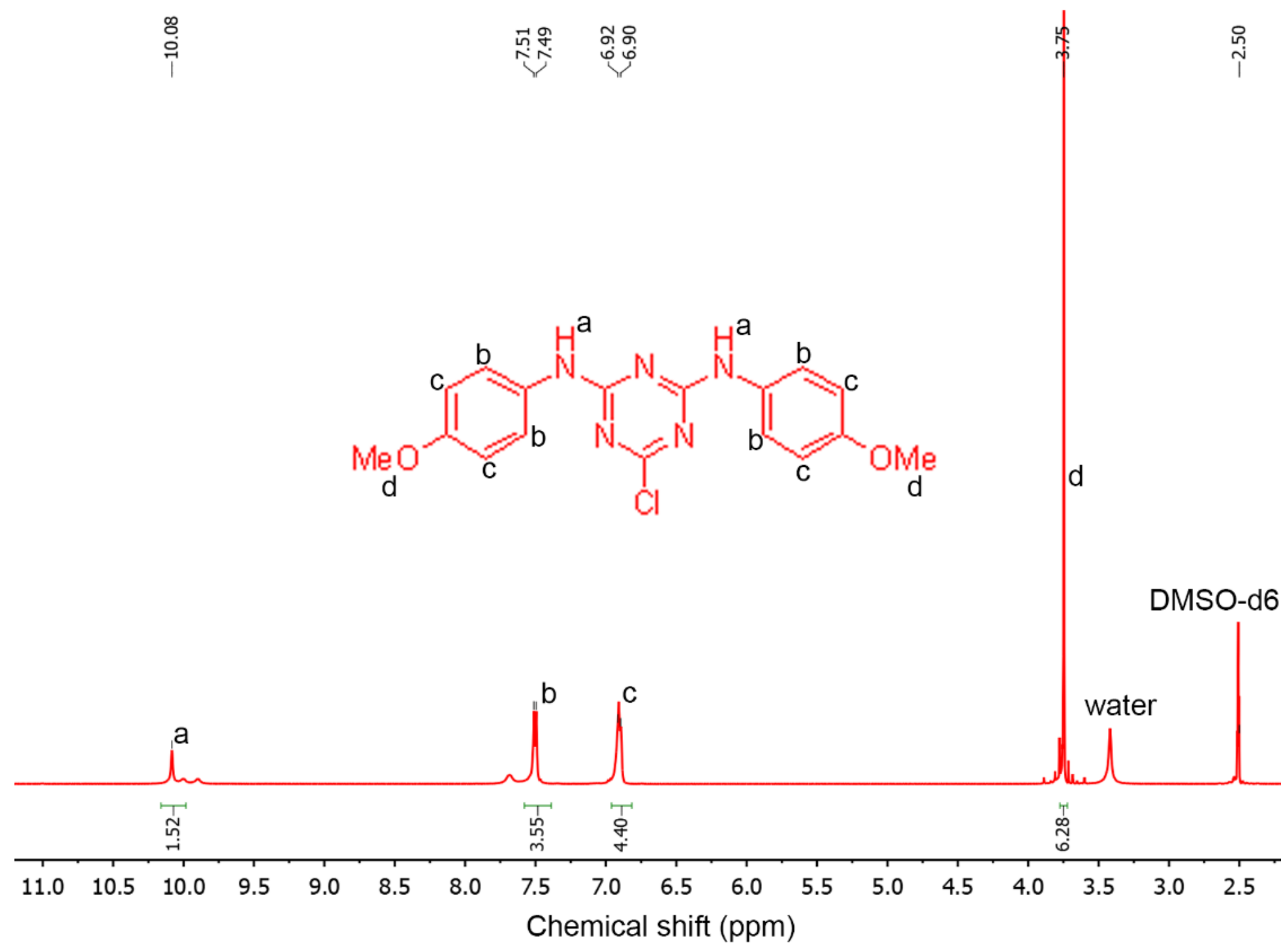

**Figure S3.** $^{1}H$ NMR spectrum of A1D.

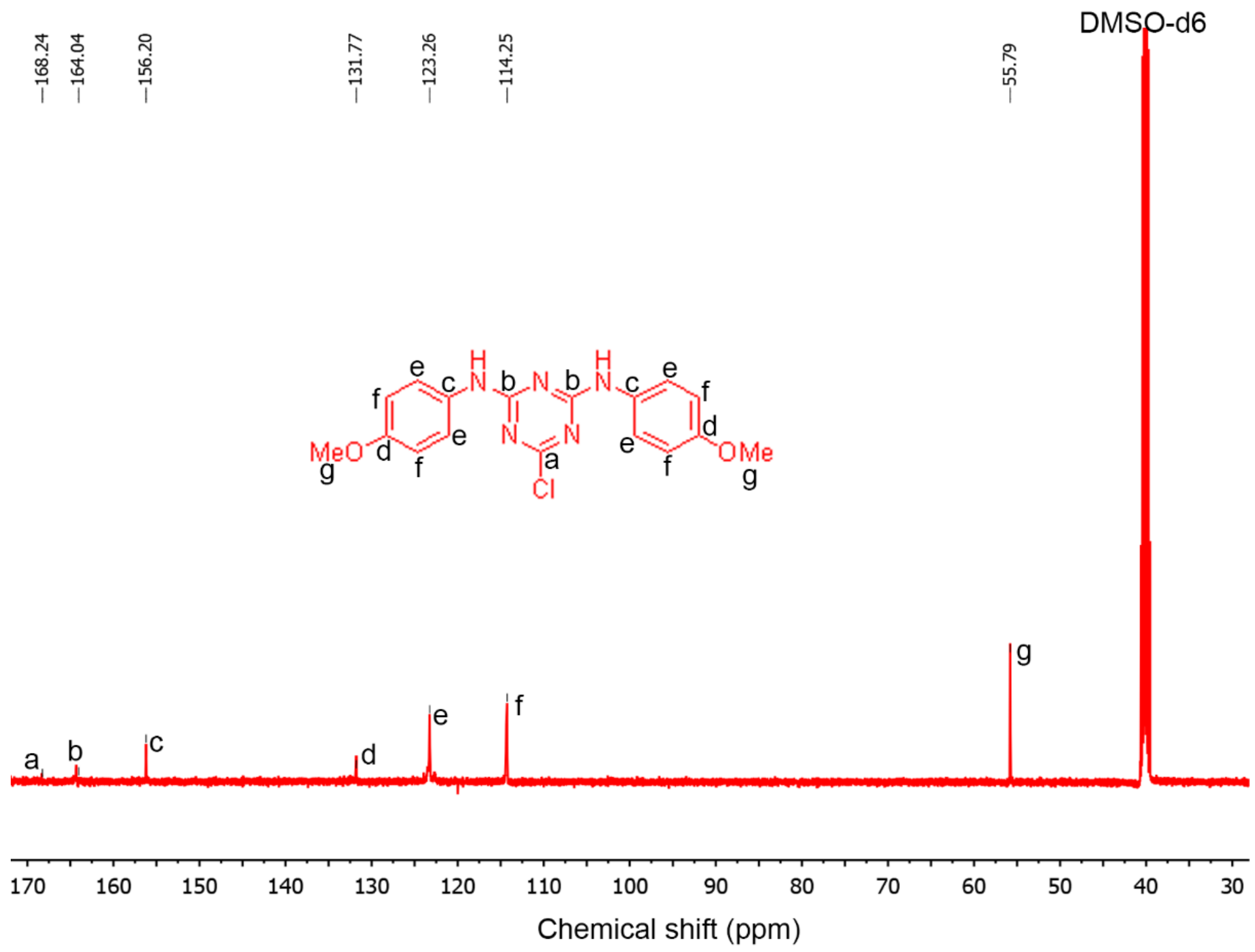

**Figure S4.** $^{13}$C NMR spectrum of A1D.

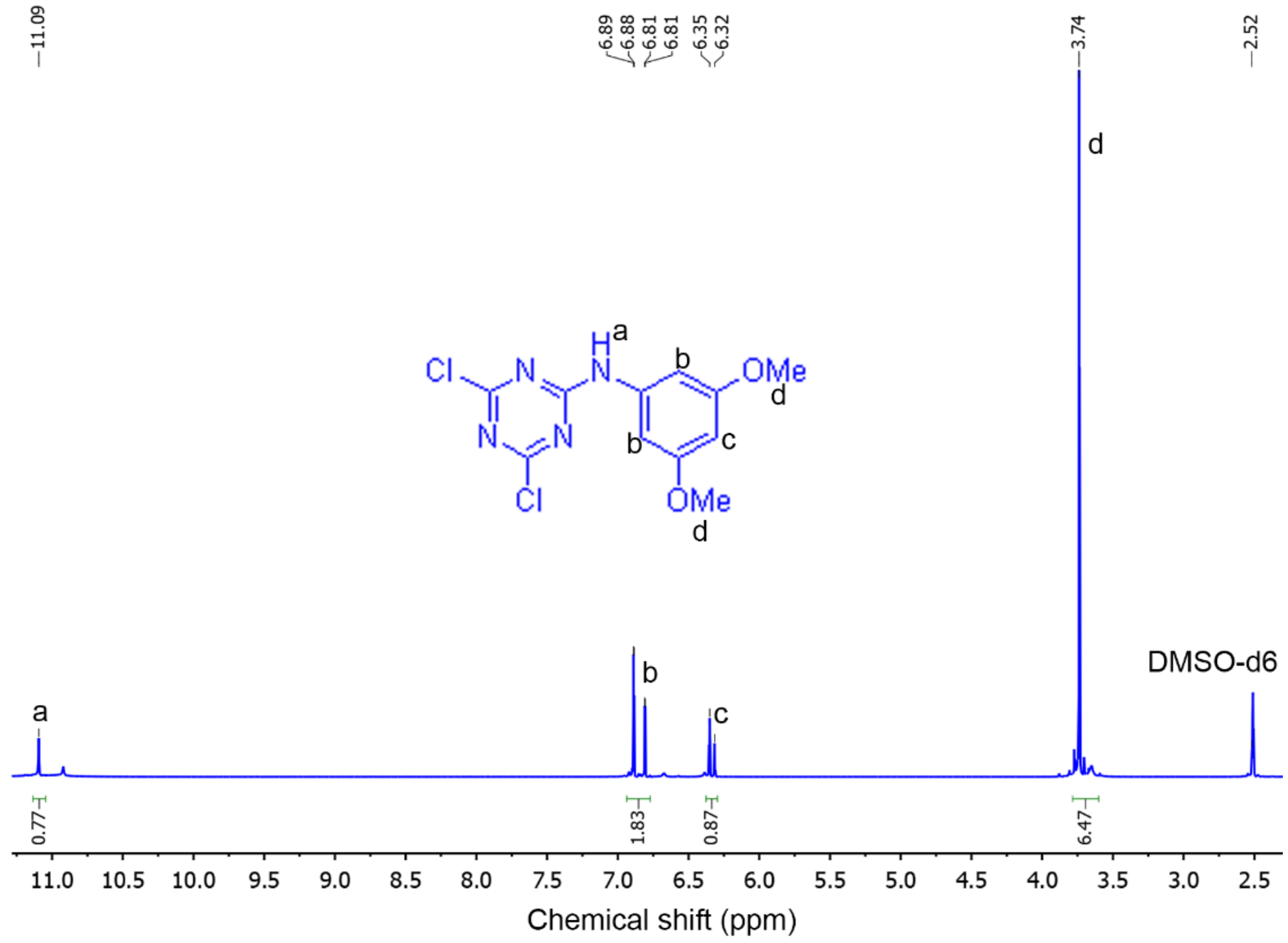

**Figure S5.** $^{1}$H NMR spectrum of A2M.

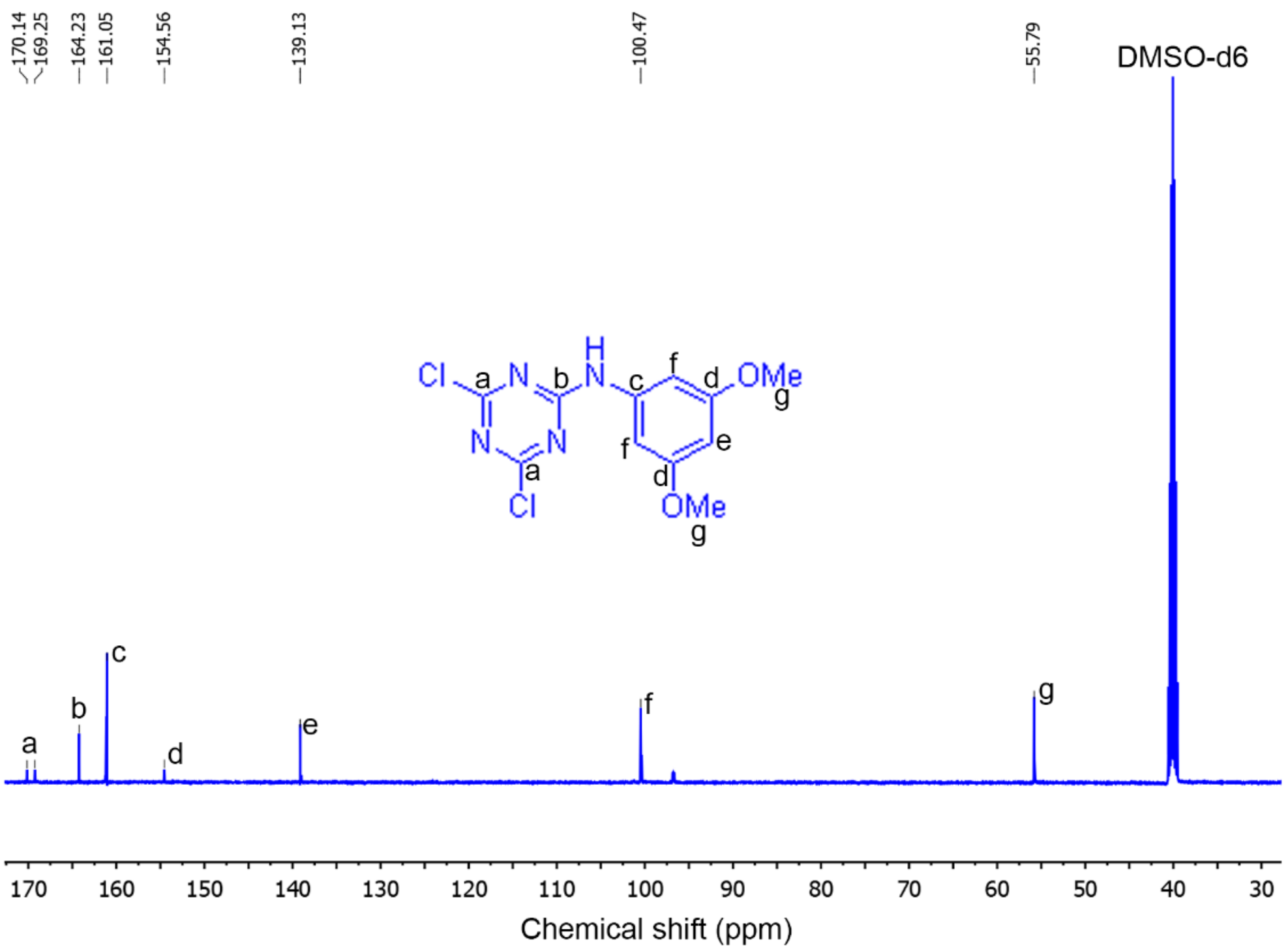

**Figure S6.** $^{13}C$ NMR spectrum of A2M.

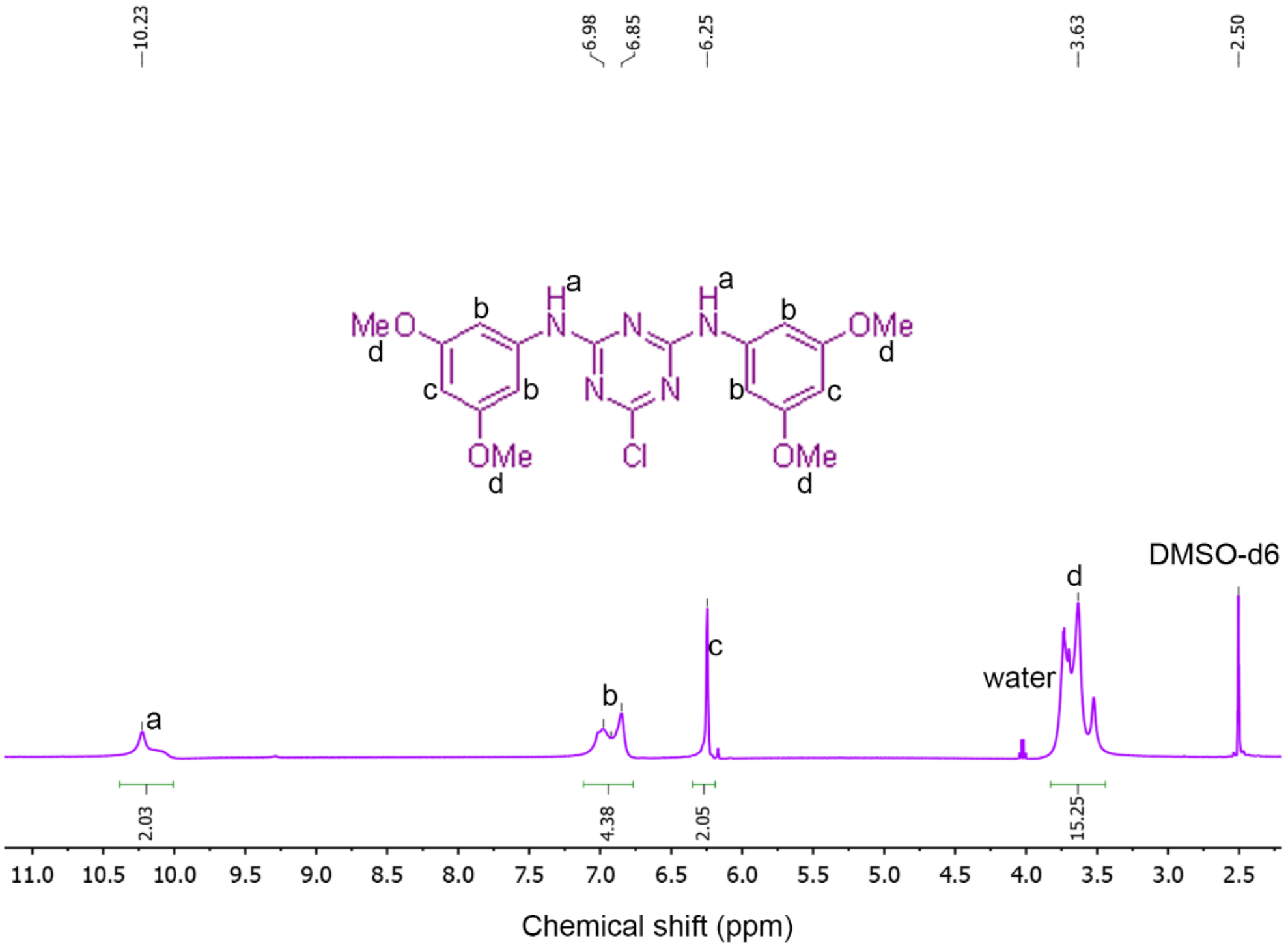

**Figure S7.** $^{1}H$ NMR spectrum of A2D.

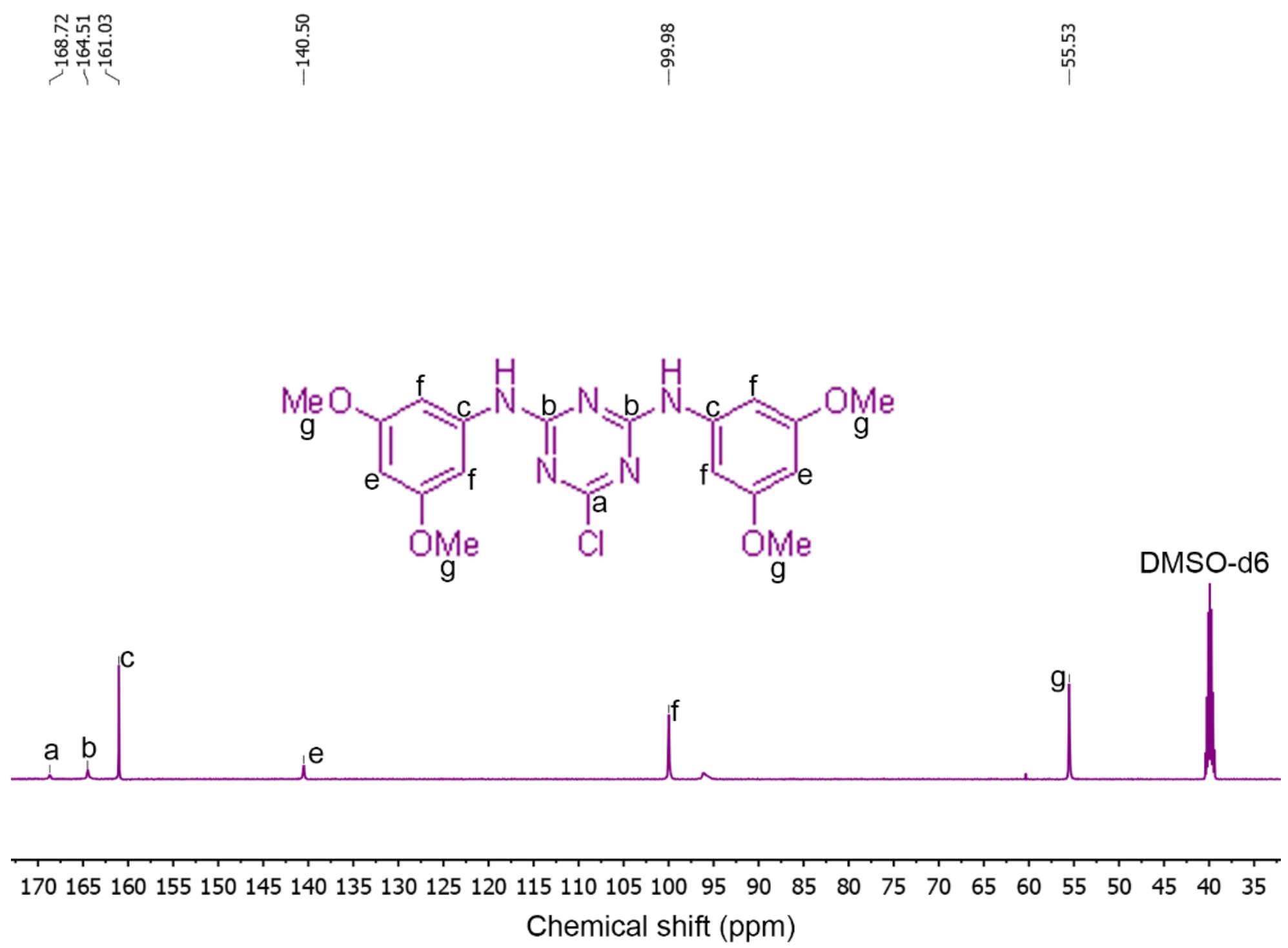

**Figure S8.** $^{13}$C NMR spectrum of A2D.

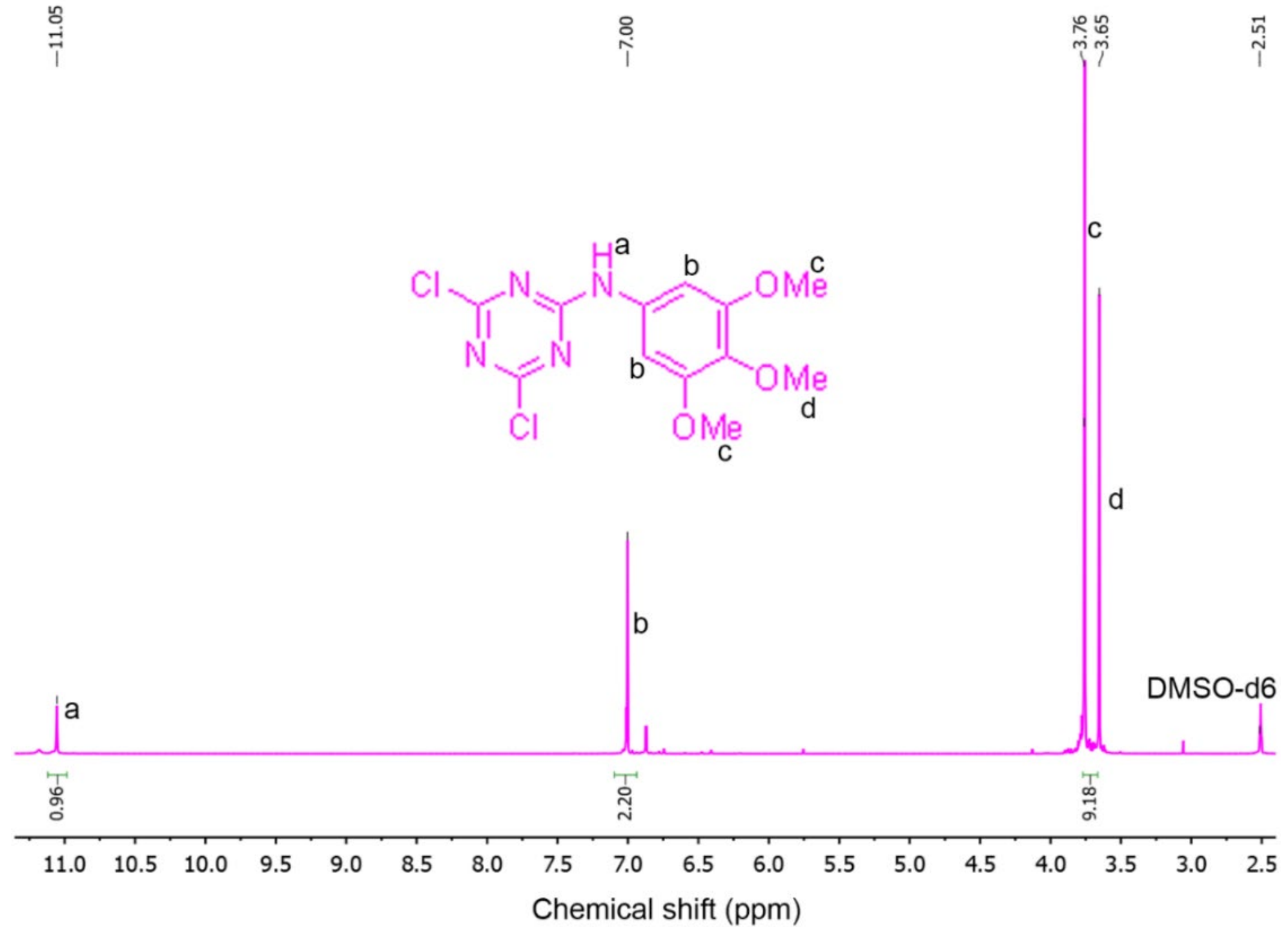

**Figure S9.** $^{1}$H NMR spectrum of A3M.

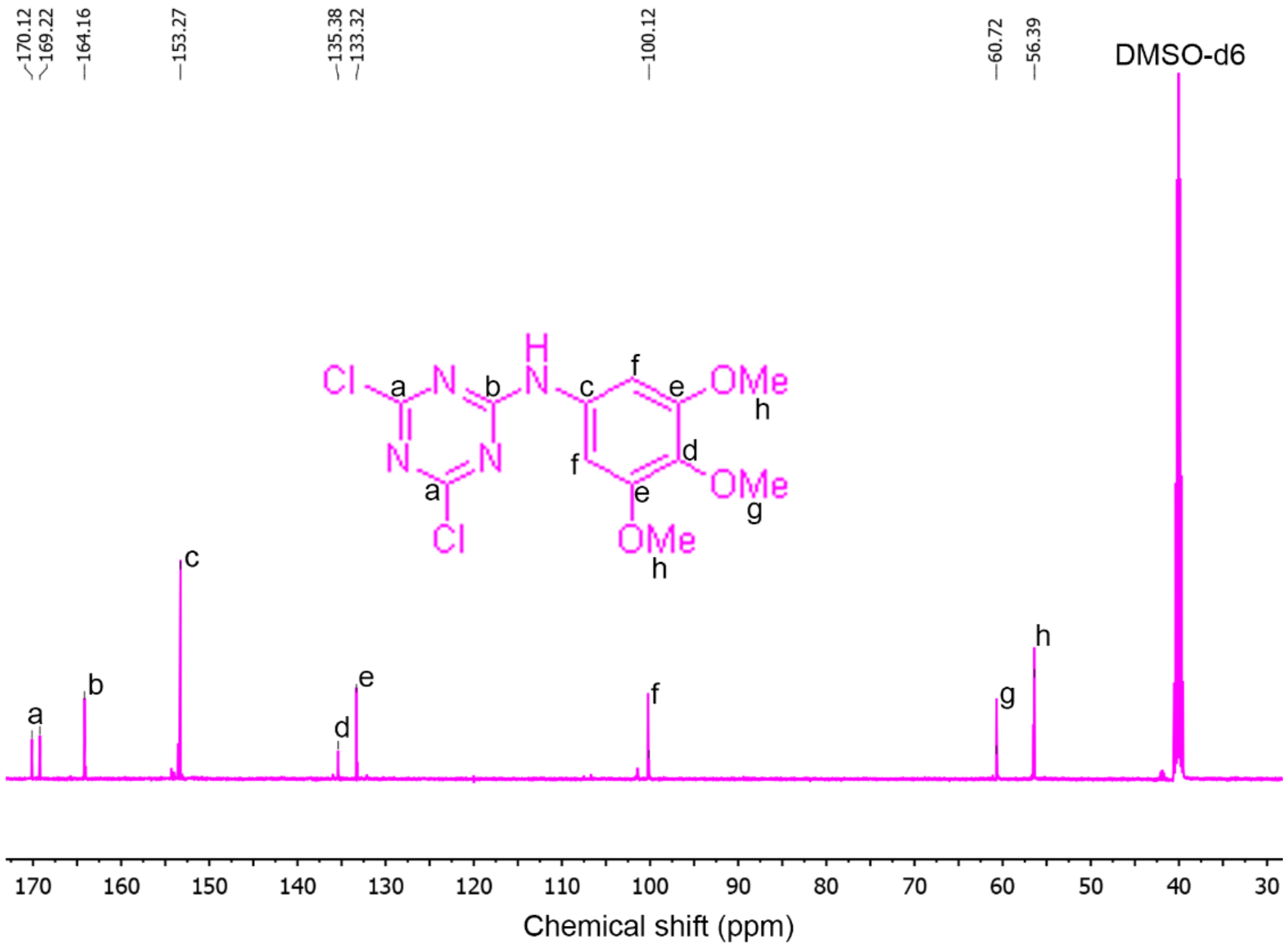

**Figure S10.** $^{13}$C NMR spectrum of A3M.

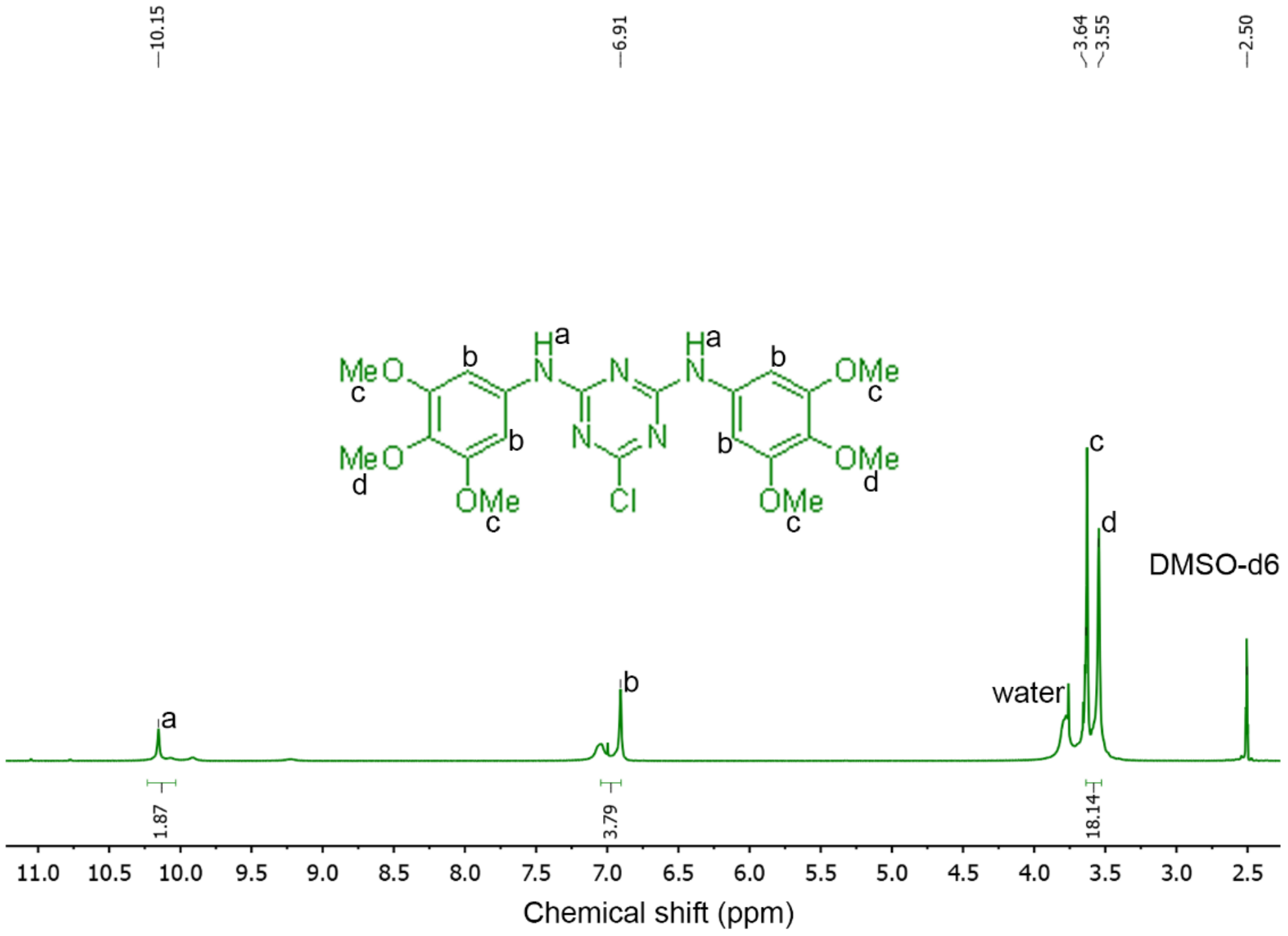

**Figure S11.** $^{1}$H NMR spectrum of A3D.

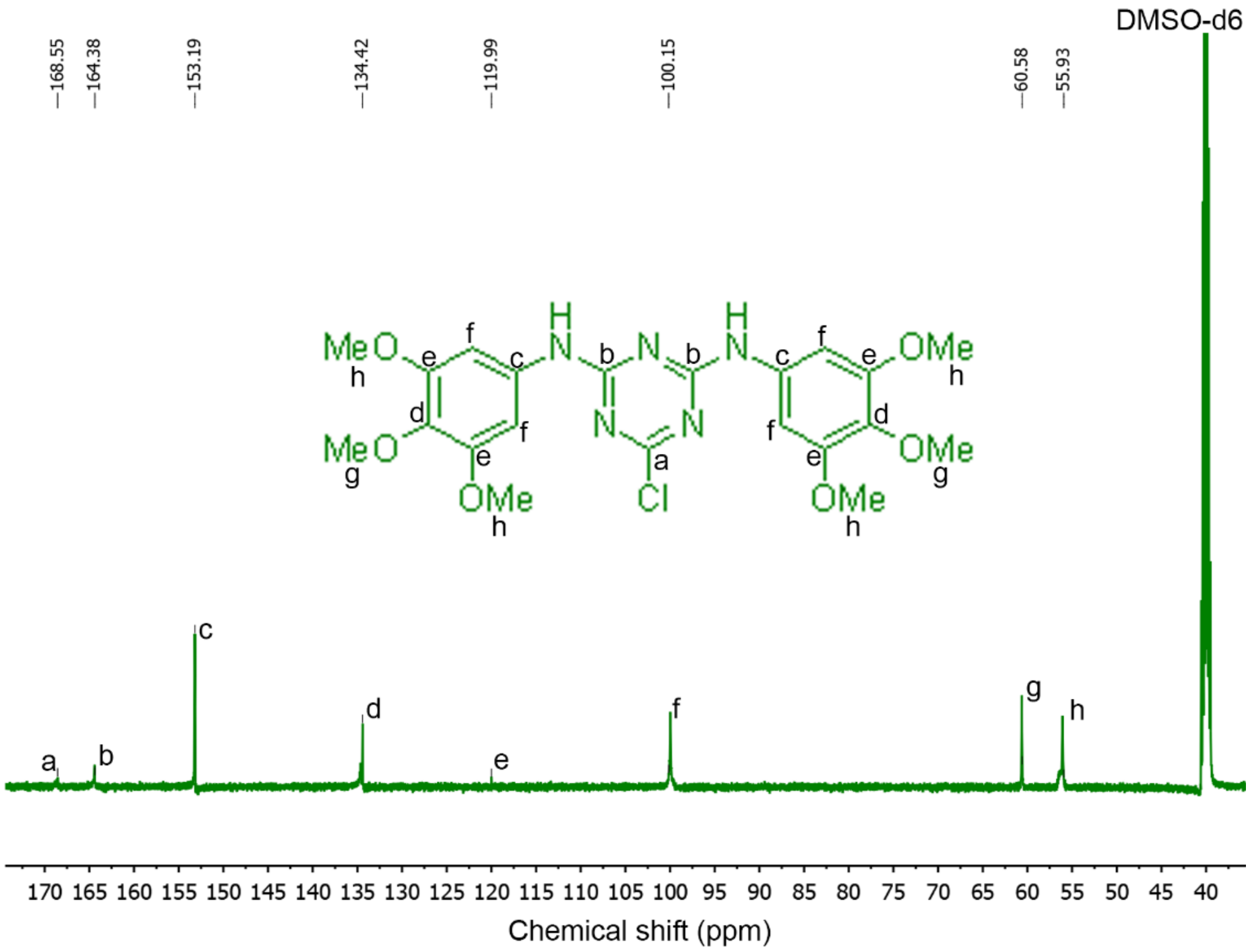

**Figure S12.** $^{13}$C NMR spectrum of A3D.

## Solubility Test

**Table 15.** Solubility test and melting point for the synthesized molecular probes.

| Probes | $CHCl_3$ | Acetone | DMF | $H_2O$ | NMP | DMSO | THF | m.p. (°C) |
|---|---|---|---|---|---|---|---|---|
| A1M | * | * | * | - | * | * | * | 165-166 |
| A2M | ** | ** | * | - | * | * | ** | |
| | *** | *** | | | | | *** | 191.7-192.2 |
| A3M | * | * | * | - | * | * | * | 184.2-185.7 |
| A1D | * | * | * | - | * | * | * | 202.7-203.4 |
| A2D | *** | ** | * | - | * | * | ** | |
| | | *** | | | | | *** | 171.3-172 |
| A3D | ** | ** | * | - | * | * | ** | |
| | *** | *** | | | | | *** | 242.2-243 |

*soluble at room temperature; **sparingly/partially soluble at room temperature; ***soluble after 60 °C heating; -insoluble. Solubility was based on a 0.1 w/v % concentration. Stirring time = 10 – 30 minutes. m.p = melting point.

## Mass Spectra

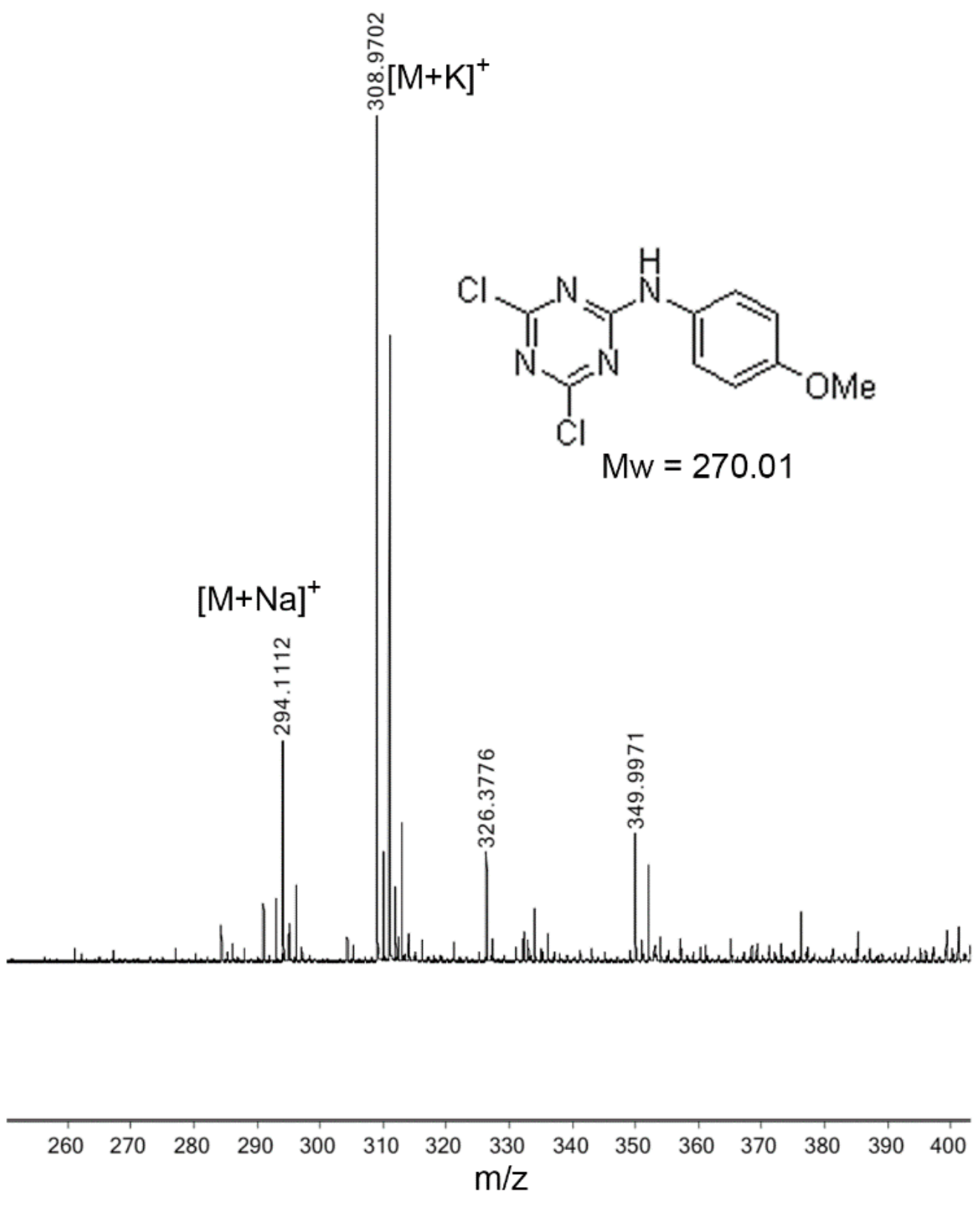

**Figure S13.** Mass spectra of A1M.

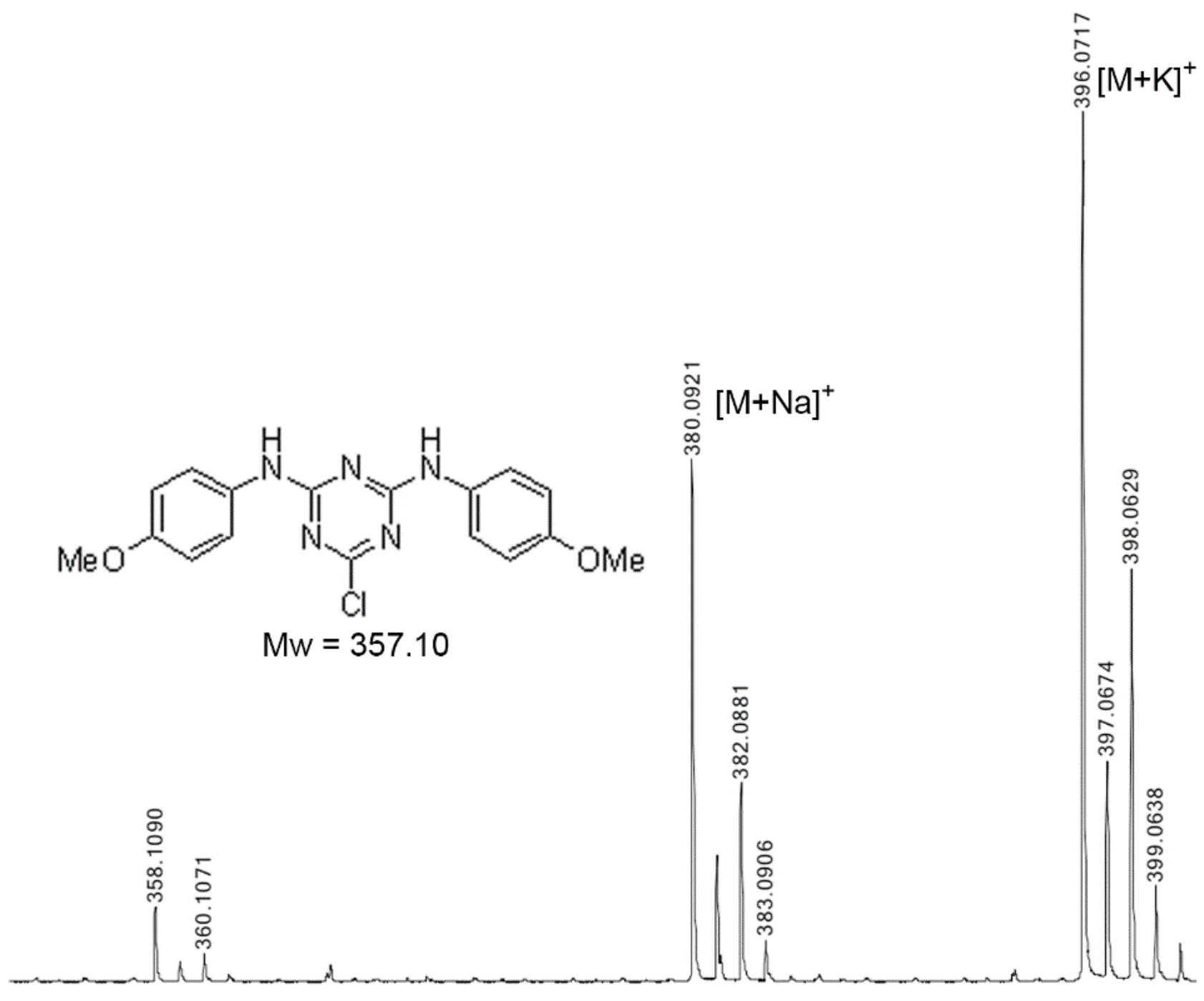

**Figure S14.** Mass spectra of A1D.

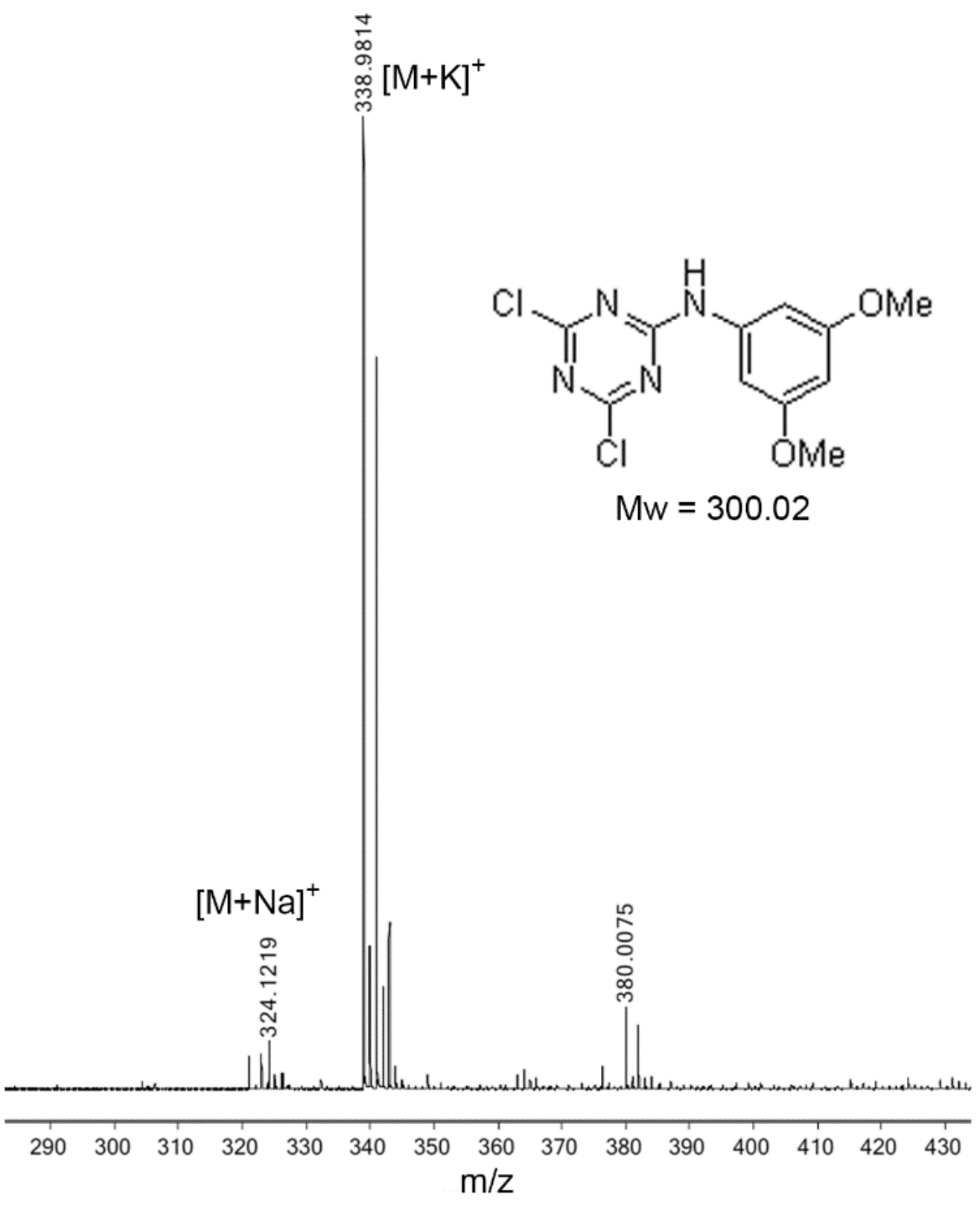

**Figure S15.** Mass spectra of A2M.

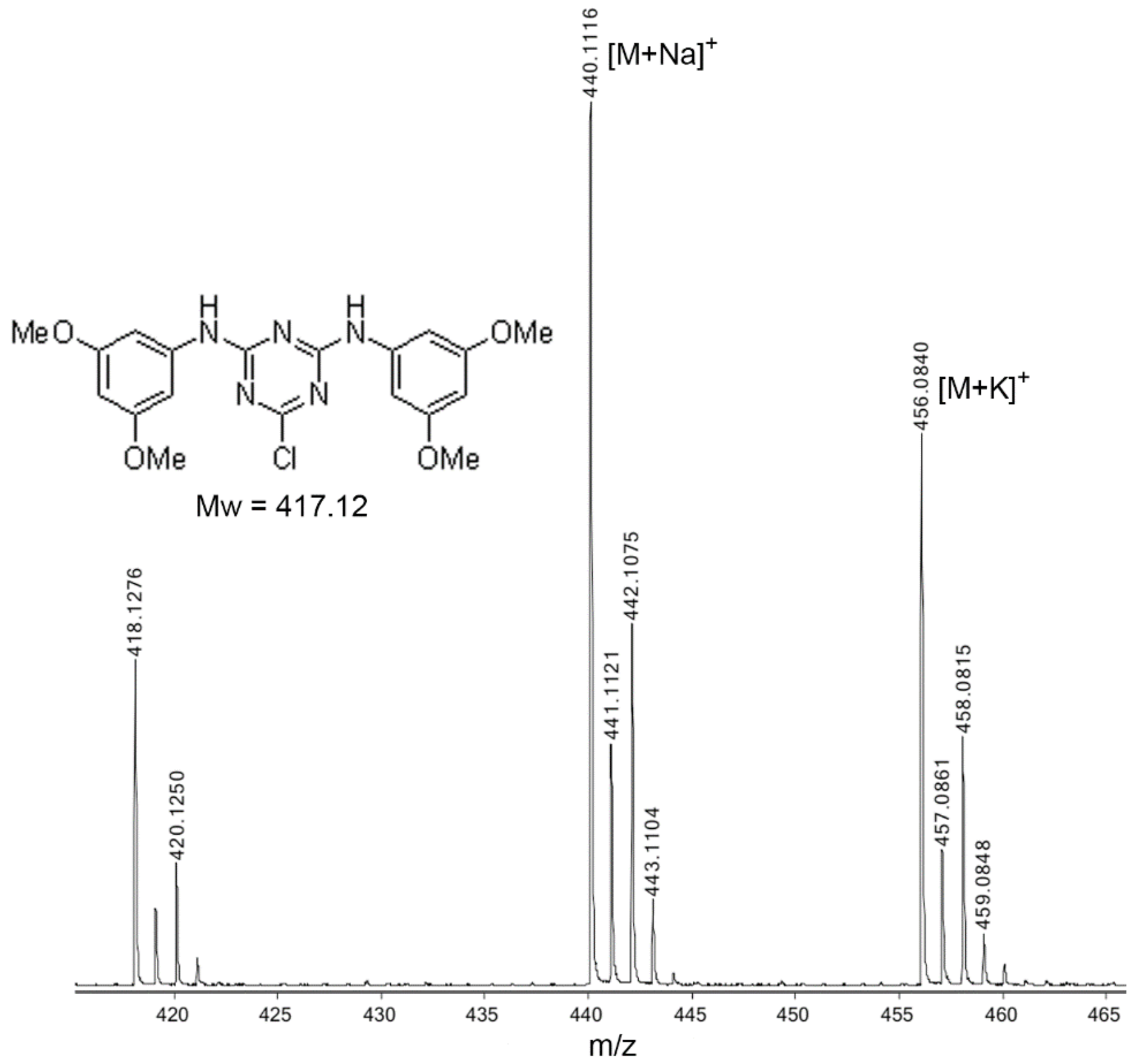

**Figure S16.** Mass spectra of A2D.

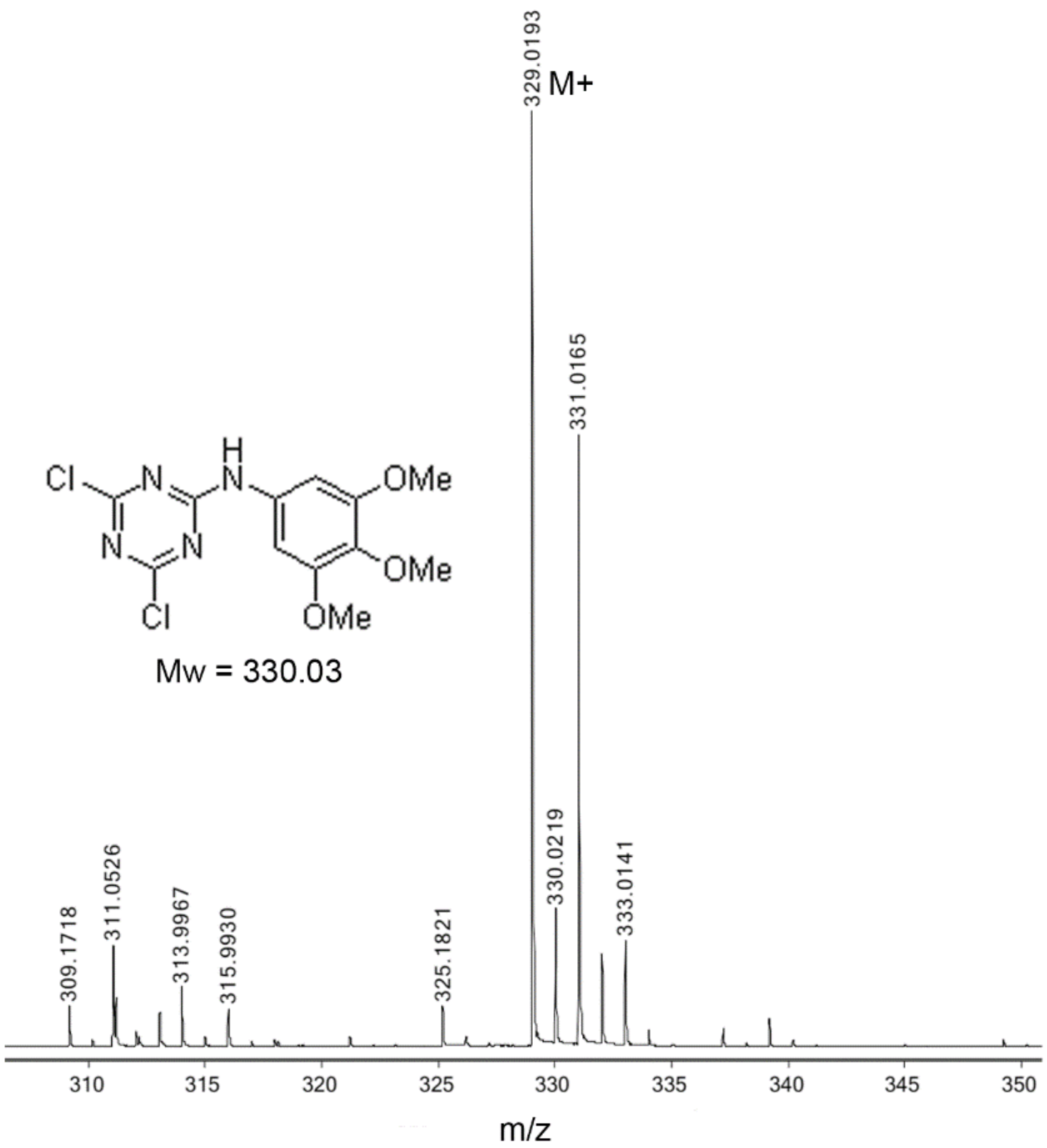

**Figure S17.** Mass spectra of A3M.

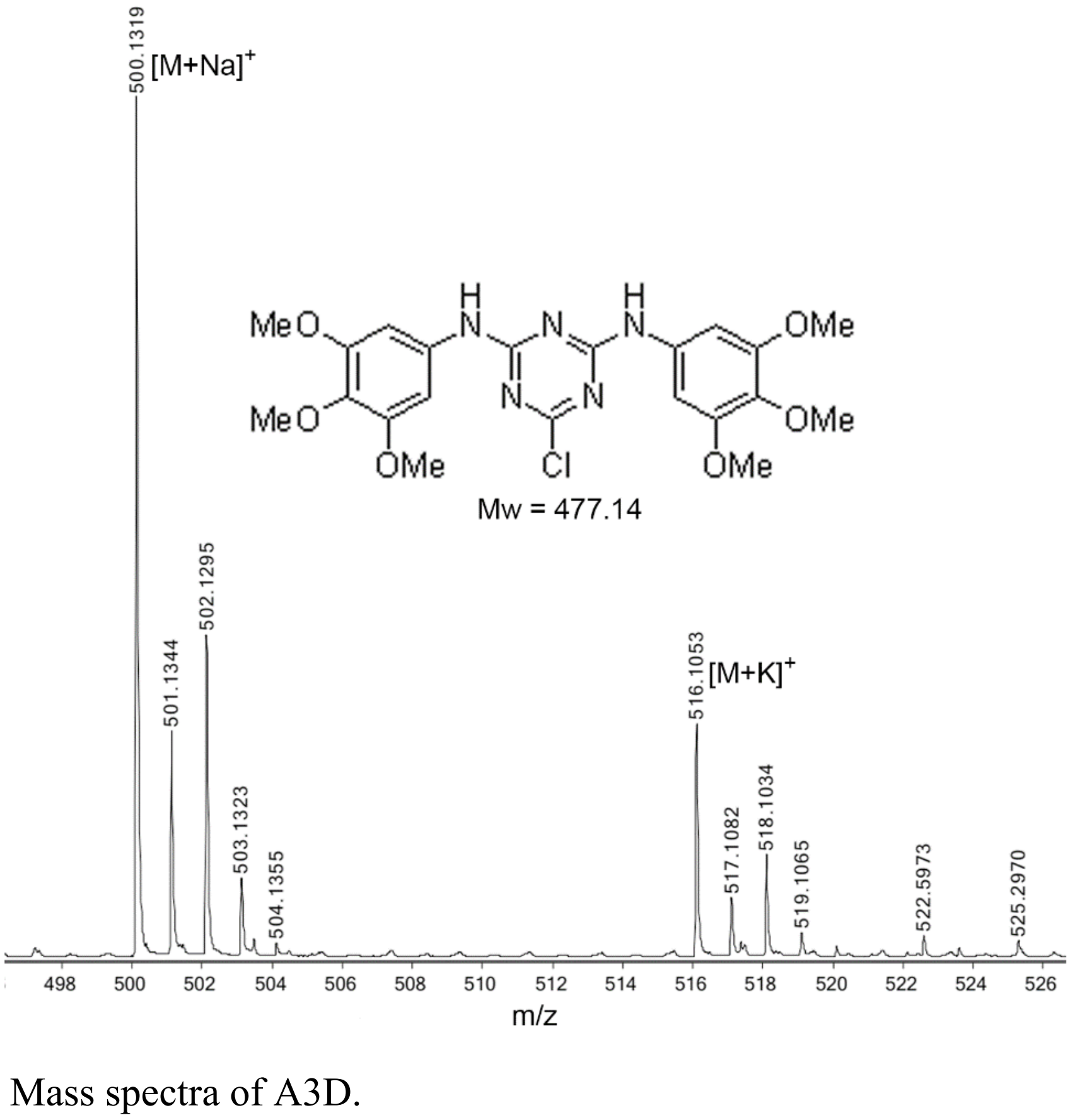

**Figure S18.** Mass spectra of A3D.

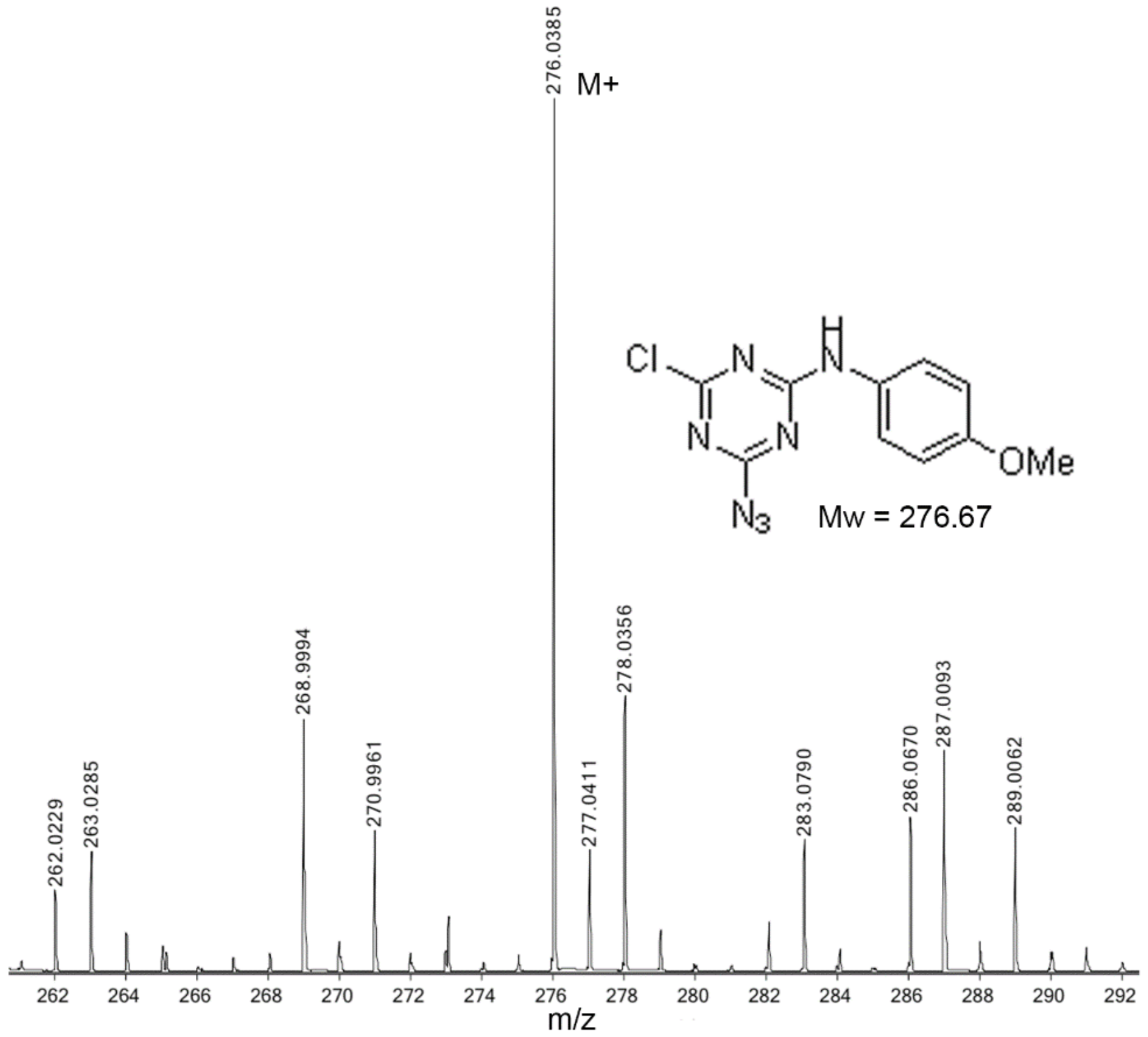

**Figure S19.** Mass spectra of A1M bearing azide group.

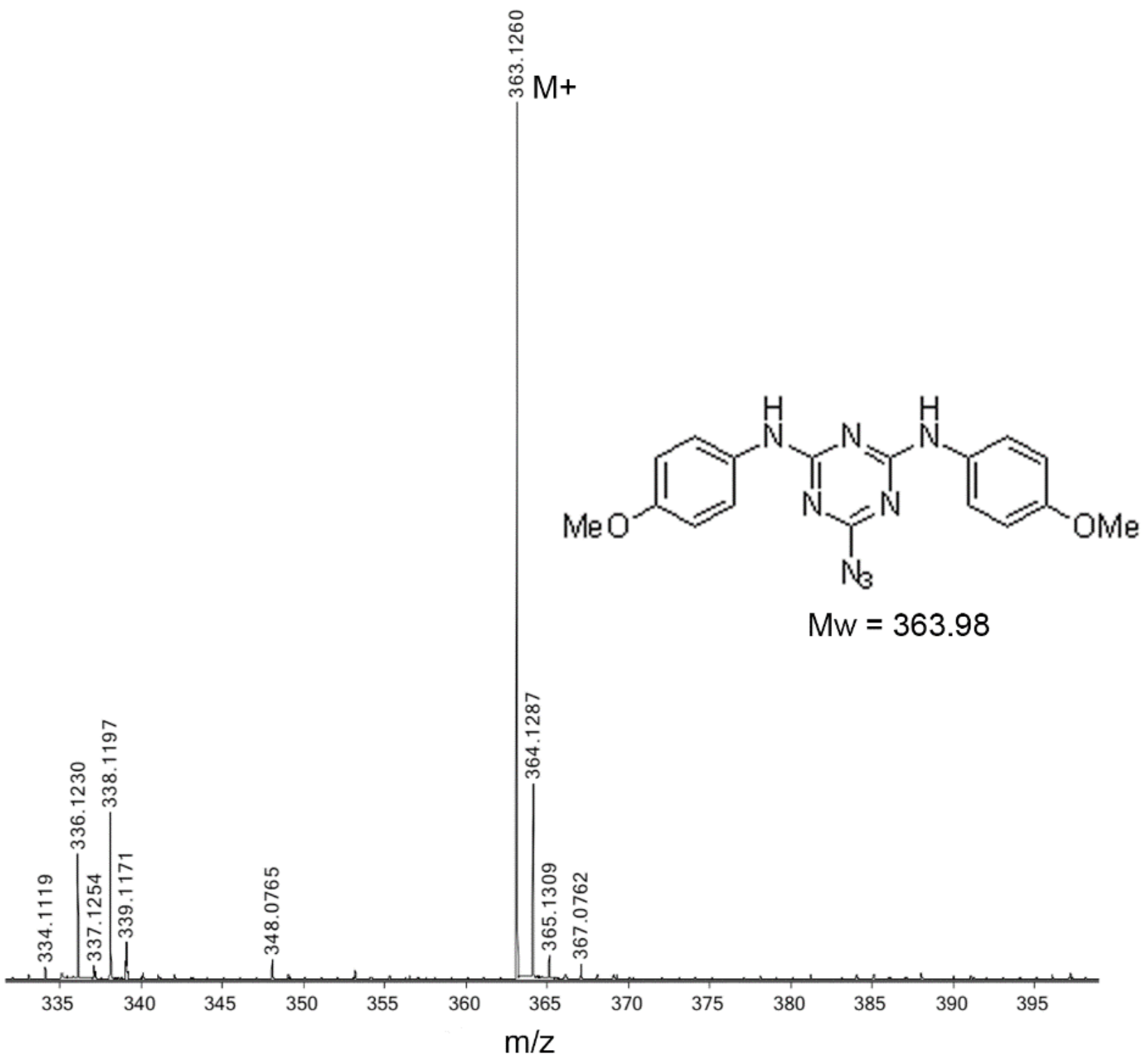

**Figure S20.** Mass spectra of A1D bearing azide group.

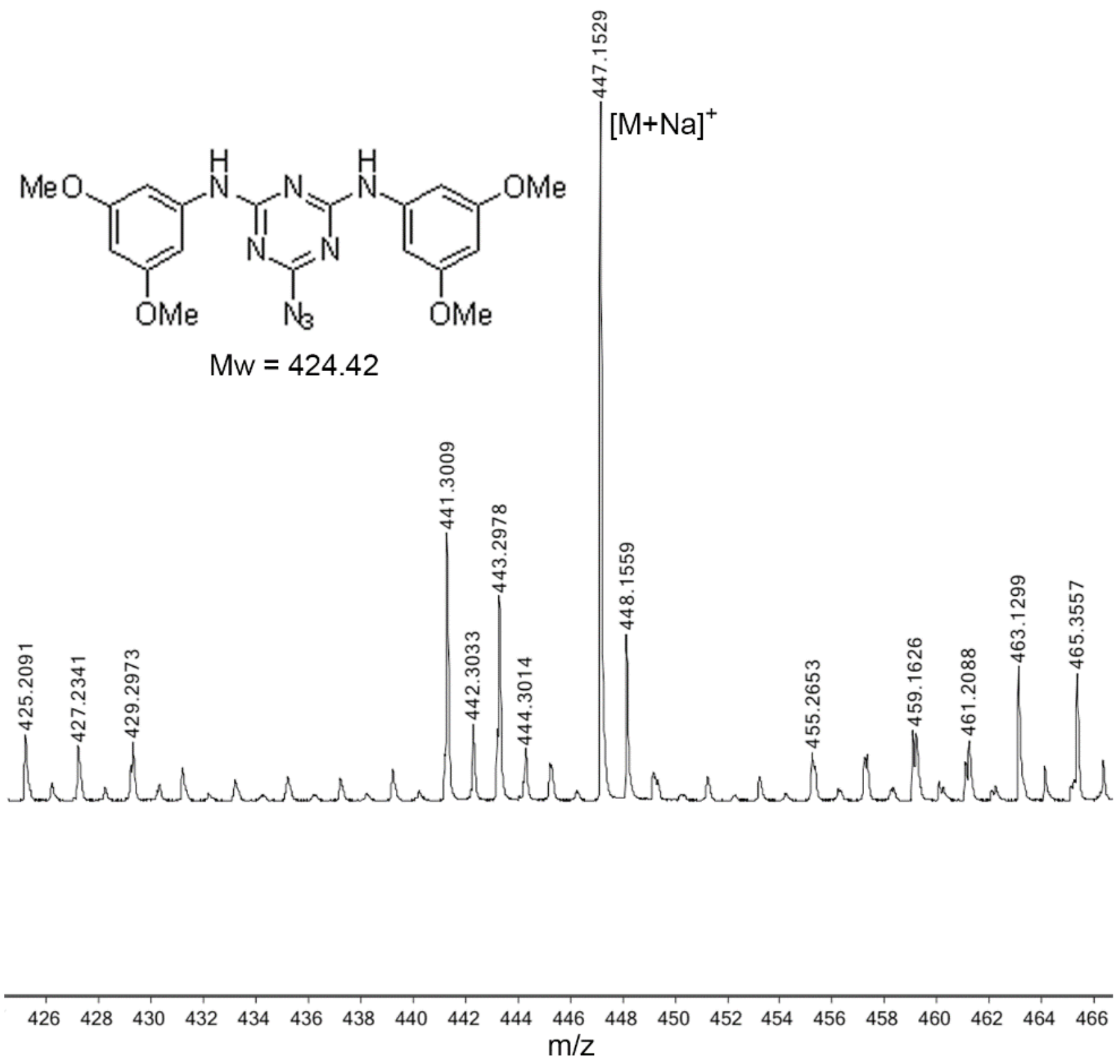

**Figure S21.** Mass spectra of A2D bearing azide group.

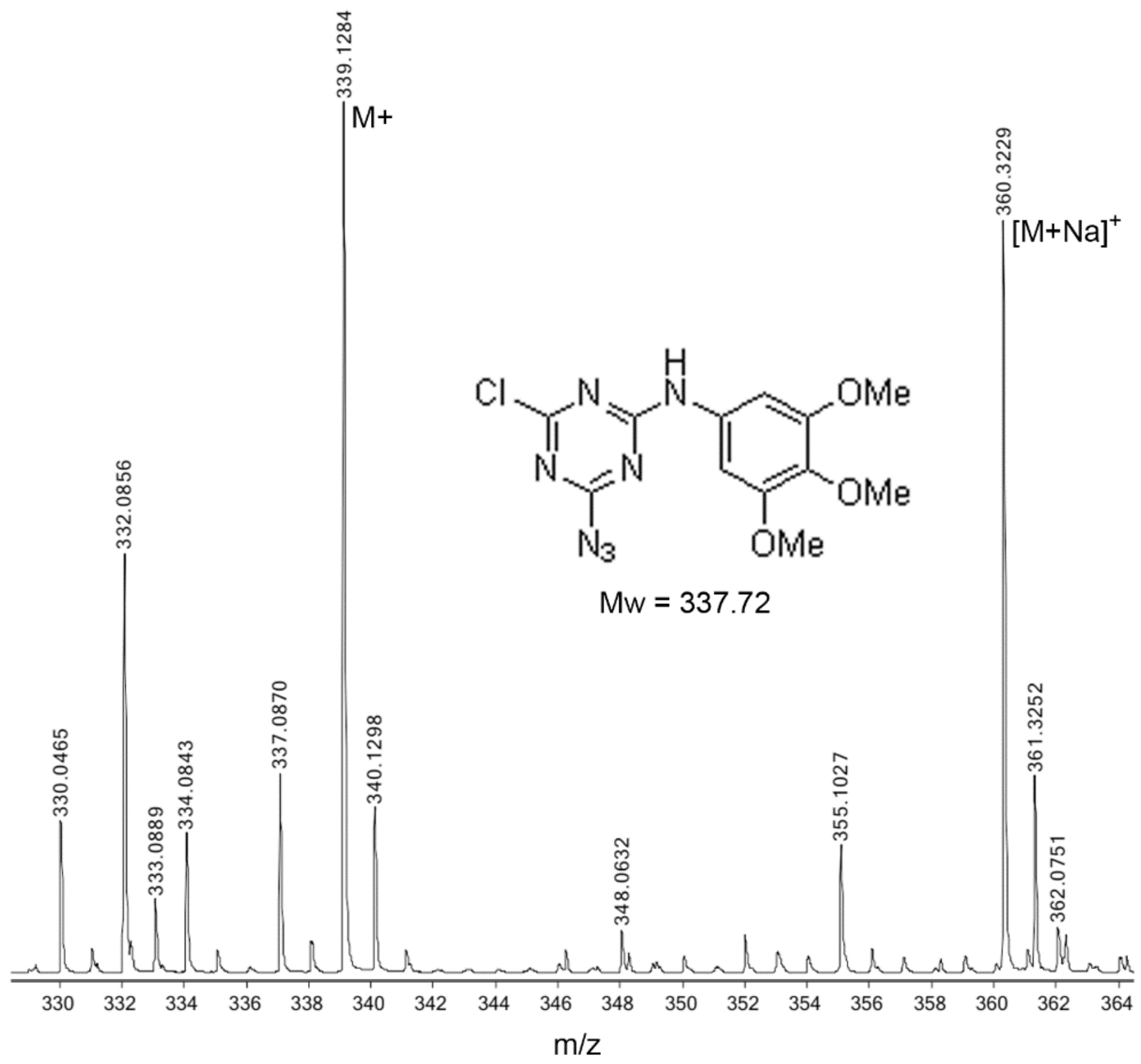

**Figure S22.** Mass spectra of A3M bearing azide group.

## Simulation studies based on TD-DFT PBE0-D3(BJ)/def2-TZVP COSMO(THF) [13-27]

**Table S16**. Electronic energy and oscillator strength of different excited states of A1M in the equilibrium geometry of $S_0$ (of A1M).

| Excited State | Orbital | Energy in eV | Wavelength in nm | Oscillator strengths (length) |
|---|---|---|---|---|
| $S_1$ | HOMO → LUMO | 3.74 | 331.7 | 0.0066 |
| $S_2$ | HOMO → LUMO+1 | 4.20 | 294.9 | 0.7221 |
| $S_3$ | HOMO → LUMO+2 85%<br>HOMO-1 → LUMO+1 10.3% | 4.27 | 290.3 | 0.0830 |
| $S_4$ | HOMO-3 → LUMO+1 84%<br>HOMO → LUMO+1 12% | 4.46 | 277.8 | 0.0026 |

**Table S17.** Electronic energy and oscillator strength of different excited states of A2M in the equilibrium geometry of $S_0$ (of A2M).

| Excited State | Orbital | Energy in eV | Wavelength in nm | Oscillator strengths (length) |
|---|---|---|---|---|
| $S_1$ | HOMO-1 → LUMO 13%<br>HOMO → LUMO 84% | 4.00 | 310.3 | 0.0081 |
| $S_2$ | HOMO-1 → LUMO 78%<br>HOMO → LUMO 14% | 4.21 | 294.3 | 0.0060 |
| $S_3$ | HOMO-1 → LUMO 77%<br>HOMO-1 → LUMO+1 12%<br>HOMO → LUMO+1 6% | 4.27 | 290.3 | 0.1678 |
| $S_4$ | HOMO-1 → LUMO+1 84%<br>HOMO → LUMO+1 12% | 4.46 | 277.8 | 0.5481 |

**Table S18.** Electronic energy and oscillator strength of different excited states of A3M in the equilibrium geometry of $S_0$ (of A3M).

| Excited State | Orbital | Energy in eV | Wavelength in nm | Oscillator strengths (length) |
|---|---|---|---|---|
| $S_1$ | HOMO → LUMO | 3.76 | 329.5 | 0.0053 |
| $S_2$ | HOMO → LUMO+1 | 4.15 | 298.7 | 0.5764 |
| $S_3$ | HOMO-1 → LUMO | 4.32 | 286.81 | 0.0781 |
| $S_4$ | HOMO-1 → LUMO+1 88.2%<br>HOMO → LUMO+2 6.3% | 4.43 | 280.00 | 0.0165 |

**Table S19.** Electronic energy and oscillator strength of different excited states of A1D in the equilibrium geometry of $S_0$ (of A1D).

| Excited State | Orbital | Energy in eV | Wavelength in nm | Oscillator strengths (length) |
|---|---|---|---|---|
| $S_1$ | HOMO → LUMO | 4.00 | 310.0 | 0.0087 |
| $S_2$ | HOMO-1 → LUMO | 4.13 | 300.0 | 0.3042 |
| $S_3$ | HOMO → LUMO+1 | 4.28 | 290.5 | 1.2866 |

**Table S20.** Electronic energy and oscillator strength of different excited states of A2D in the equilibrium geometry of $S_0$ (of A2D).

| Excited State | Orbital | Energy in eV | Wavelength in nm | Oscillator strengths (length) |
|---|---|---|---|---|
| $S_1$ | HOMO → LUMO | 4.30 | 288.5 | 0.0004 |
| $S_2$ | HOMO-1 → LUMO 85%<br>HOMO-3 → LUMO 7% | 4.31 | 287.6 | 0.2484 |
| $S_3$ | HOMO → LUMO+1 | 4.43 | 279.9 | 1.2520 |

**Table S21.** Electronic energy and oscillator strength of different excited states of A3D in the equilibrium geometry of $S_0$ (of A3D).

| Excited State | Orbital | Energy in eV | Wavelength in nm | Oscillator strengths (length) |
|---|---|---|---|---|
| $S_1$ | HOMO → LUMO | 4.05 | 306.3 | 0.0019 |
| $S_2$ | HOMO-1 → LUMO | 4.14 | 299.6 | 0.3673 |
| $S_3$ | HOMO → LUMO+1 | 4.23 | 293.1 | 1.1402 |

## *Transition in A1D*

Figure S23 shows the potential energy surface of the excited states $S_2$ and $S_3$ in A1D along a small geometric distortion. In this geometric distortion atoms shift on average from position 0 to position 5 (see Figure S24 and S25) by only 0.0188 angstrom (this is also called root-mean-square deviation (RMSD) between the coordinates). In position 3 (Figure S23) is a conical intersection. In the ground state $S_0$ it takes only 7 kJ/mol to shift A1D from position 0 to the point of the conical intersection (position 3 in Figure S23) and atoms shift on average by only 0.0113 angstrom (RMSD). In the excited state $S_3$ the shift from position 0 to position 3 is favorable. Thus, the transition from $S_3$ to $S_2$ should be very fast and the energy difference between $S_2$ and $S_1$ is also small. Hence, it can be assumed that the subsequent transition from $S_2$ and $S_1$ is fast as well. This indicates that A1D is following Kasha's rule[28]. A2D and A3D will probably behave similar.

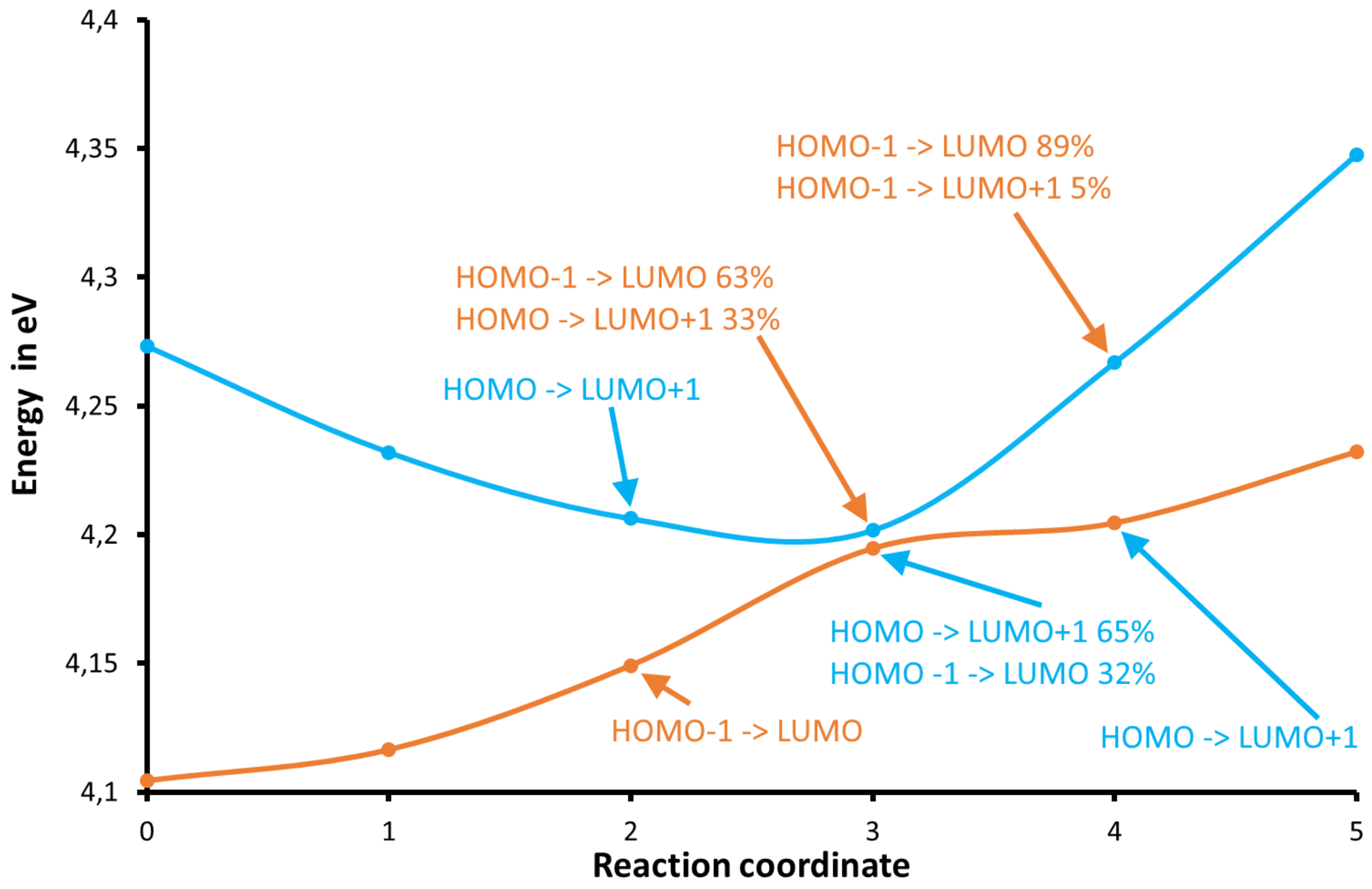

**Figure S23**. PES for the excited states $S_2$ (orange) and $S_3$ (blue) of A1D. Conical intersection at position 3.

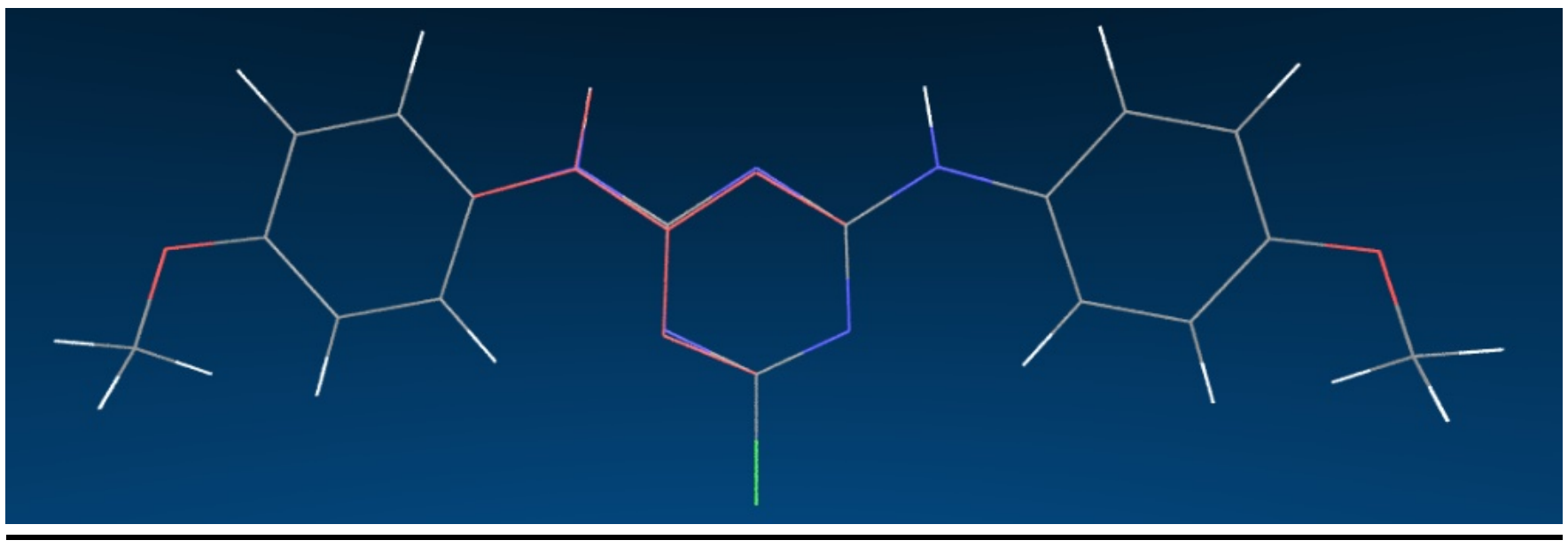

**Figure S24.** Structure of A1D at position 0 in fig. S23. The red lines show the structure of A1D at position 5. The RMSD between the coordinates of the atoms in position 0 and position 5 is 0.0188 Angstrom.

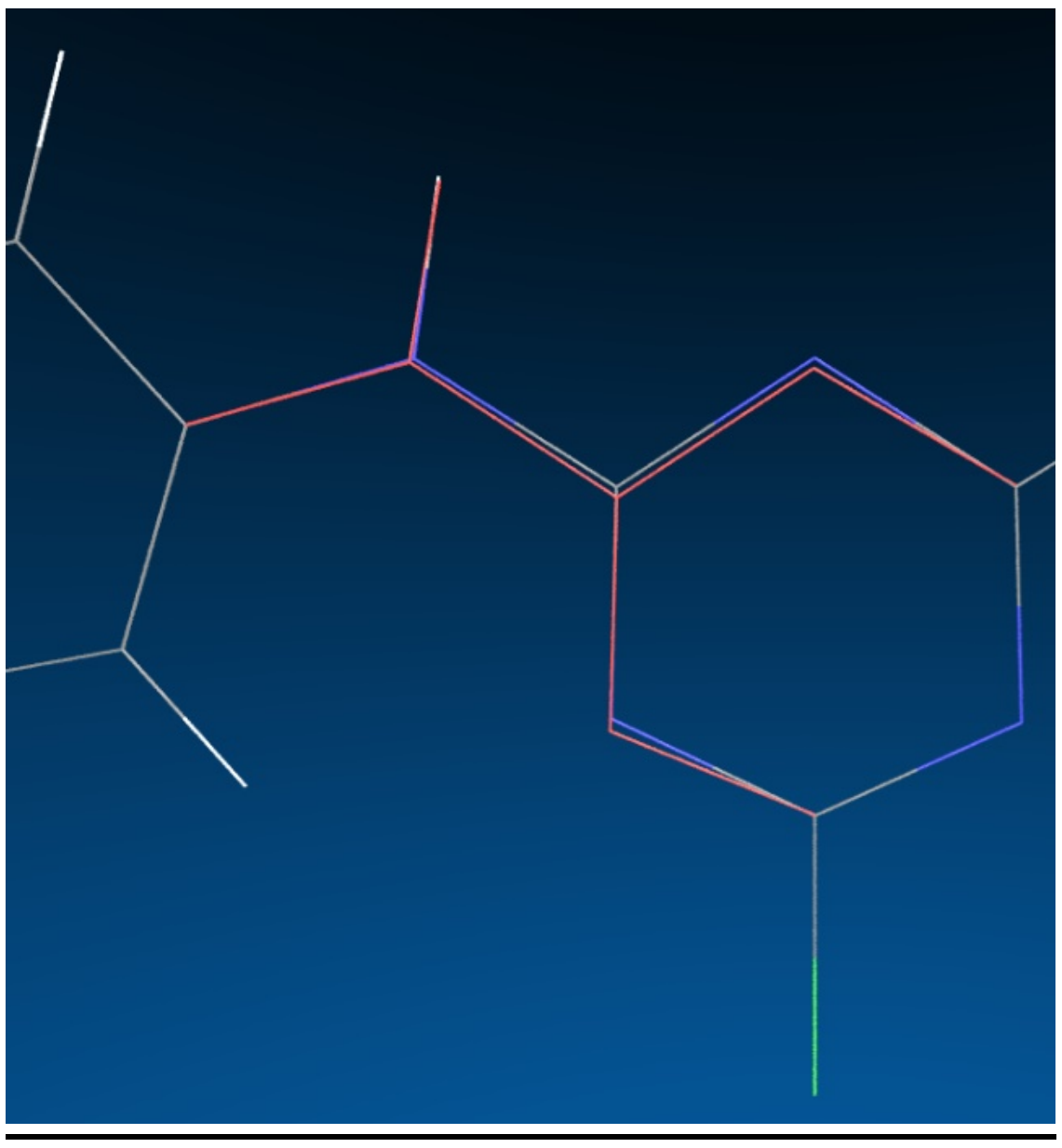

**Figure S25.** Enhanced figure S24. Structure of A1D at position 0 in figure S23. The red lines show the structure of A1D at position 5.

### *Transition in A1M, A2M and A3M*

The geometry optimization of A1M in its second excited state $S_2$ showed that the dichloro-triazine ring is rotating out-of-plane. This rotation is expressed in the internal coordinates of A1M by changing of the dihedral angle shown in figures S26 and S27. In order to investigate the correlation between the geometry and energy of the excited states, the equilibrium structure (in the $S_0$ state) of A1M, A2M and A3M was used and only the dihedral angle shown in figure S26 was changed. Figure S28 and table S22 show the potential energy surface (PES) of A1M at $S_1$ and $S_2$ and different angles. The excitation goes from $S_0$ to $S_2$. Therefore, A1M in the $S_2$ state is initiallly in the equilibrium geometry of the ground state $S_0$. In this equilibrium geometry the dichloro-triazine ring is in-plane and the dihedral angle is 179.94°. The equilibrium geometry of the $S_0$ is a maximum for the $S_2$ state. The minimum energies are observed at 132.5° and 227.5°. Also from 132.5° over 179.94° to 227.5°, a large oscillator strength can be seen (Table S22) indicating that the $S_2$ state can be fluorescent in this region. However, conical intersections take place at 132.5° and 227.5° and based on figure S28 and table S22, the actual minimum will be 90° and 270° for the actual $S_2$ state. At these points the oscillator strength is very low. Alternatively, at 132.5° and 227.5° the $S_2$ state could transition into the $S_1$ state. The $S_1$ state has the minimum at 179.94° like $S_0$. So, for A1M in the $S_2$ state the dichloro-triazin ring is rotating out-of-plane until 132.5° (or 227.5°). There $S_2$ is transition into $S_1$ and then the dichloro-triazin ring will rotate back to be in-plane again. The $S_1$ state has a very low oscillator strength around 179.94° (=in-plane). Thus, the A1M should not be fluorescent. It is shown that the geometry in A1D has to change only by 0.0113 Angstrom per atom to allow a transition between the excited states. The chlorine atoms in the dichloro-triazine ring of A1M on the other hand have to move 2.07 angstrom for A1M in state $S_2$ in the initial position (where the dihedral angle shown in figures S26 is 179.94°) to reach the conical intersection. This change in structure is favorable and without barrier as can be seen in figure S28. Therefore, the structural change should be very fast. However, for the calculation of the PES it is assumed that at every step the solvent molecules (THF) instantly adjust to the solute (A1M). In reality it would take time to move the solvent molecules and it would probably be necessary to overcome an activation barrier. Overall, a huge change in the geometry of A1M is needed for the $S_2$ state to be able to either transition into $S_1$ or to change to geometry where $S_2$ has a low oscillator strength. This huge change in the geometry takes probably so much time that the deexcitation of the $S_2$ state can happen by emitting a photon (=fluorescence) in exception from Kasha's rule. Figure S29 shows not only the PES of $S_1$ and $S_2$ of A1M but also the PES ground state $S_0$. While in $S_2$ it is favorable for the dichloro-triazin ring to rotate out-of-plane, it is unfavorable for $S_0$ making the energy gap between $S_0$ and $S_2$ smaller. In the experimental spectrum this is called Strokes shift and the reduction of the energy gap was around 0.4 eV in the experiments. A reduction of 0.4 eV required dihedral angles 145° and 215° and is well before the conical intersection is reached (figure S28).

The results for A3M are shown in figure S31 and table S24 and are basically identical to the results for A1M.

A2M shows in the experiments a much stronger fluorescence than A1M and A3M. In A1M and A3M the main excitation (in an absorption spectrum) is $S_0$ to $S_2$. In A2M, however, the main excitation is $S_0$ to $S_4$. Similar to A1M and A3M, the dichloro-triazin ring is rotating out-of-plane in the $S_4$ state of A2M. The PES for the rotation of the dichloro-triazin ring for the excited states of A2M are shown in figure S30 and table S23. In this case conical intersections are at 150° and 210°. From 90° to 150° and 210° to 270°, the $S_4$ state is following what is marked as $S_3$ state in figure S30 and table S23. The energy stays nearly the same between 90° to 150° and 210° to 270° (which is marked as $S_3$ but is actually $S_4$). The oscillator strength is low in this area, but the energy of A2M in $S_4$ in this area (which is marked as $S_3$) is only 10 kJ/mol lower than $S_4$ at 179.991° where the oscillator strength has the maximum amount. Alternatively, at 150° and 210° the $S_4$ state could transit into the $S_3$ state. The $S_3$ state has the minimum at 179.99° similar to the ground state $S_0$. So, for A2M in the $S_4$ state the dichloro-triazin ring is rotating out-of-plane until 150° (or 210°). There, $S_4$ is transitioning into $S_3$ and then the dichloro-triazin ring rotates back to be in-plane again. That is very similar to what happens in A1M and A3M, with one major difference: The $S_3$ state of A2M has a large oscillator strength at its minimum (179.991°). So, there are two reasons for the higher fluorescence of A2M: (i) The geometries of A2M in $S_4$ that have a low oscillator strength are energetically close to the geometries with a large oscillator strength. So, at room temperature a considerable fraction of the A2M molecules in state $S_4$ would be in a geometry with a large oscillator strength. (ii) If the $S_4$ state transits to $S_3$ it will get stuck in a geometry which has a large oscillator strength. On the other hand, A1M and A3M excite into the $S_2$ state and rotate out-of-plane to a deep energy hole (at 90° and 270°) where they cannot get out and the oscillator strength is low. Alternatively, they can transit to the $S_1$ state and rotate back in-plane into a deep energy hole where A1M and A3M cannot get out and the oscillator strength is low.

It is worth to note that in A2M, $S_3$ and $S_2$ have conical intersections and $S_2$ has a low oscillator strength at 179.991°. However, $S_3$ and $S_2$ are very close in energy and in equilibrium at room temperature. Therefore, their abundance is almost equal. Furthermore, $S_2$ has a minimum at 179.991° and cannot rotate easily out-of-plane (Figure S30). States $S_1$ and $S_2$ of A1M and A3M as well as $S_3$ and $S_4$ of A2M are shown together in figure S32.

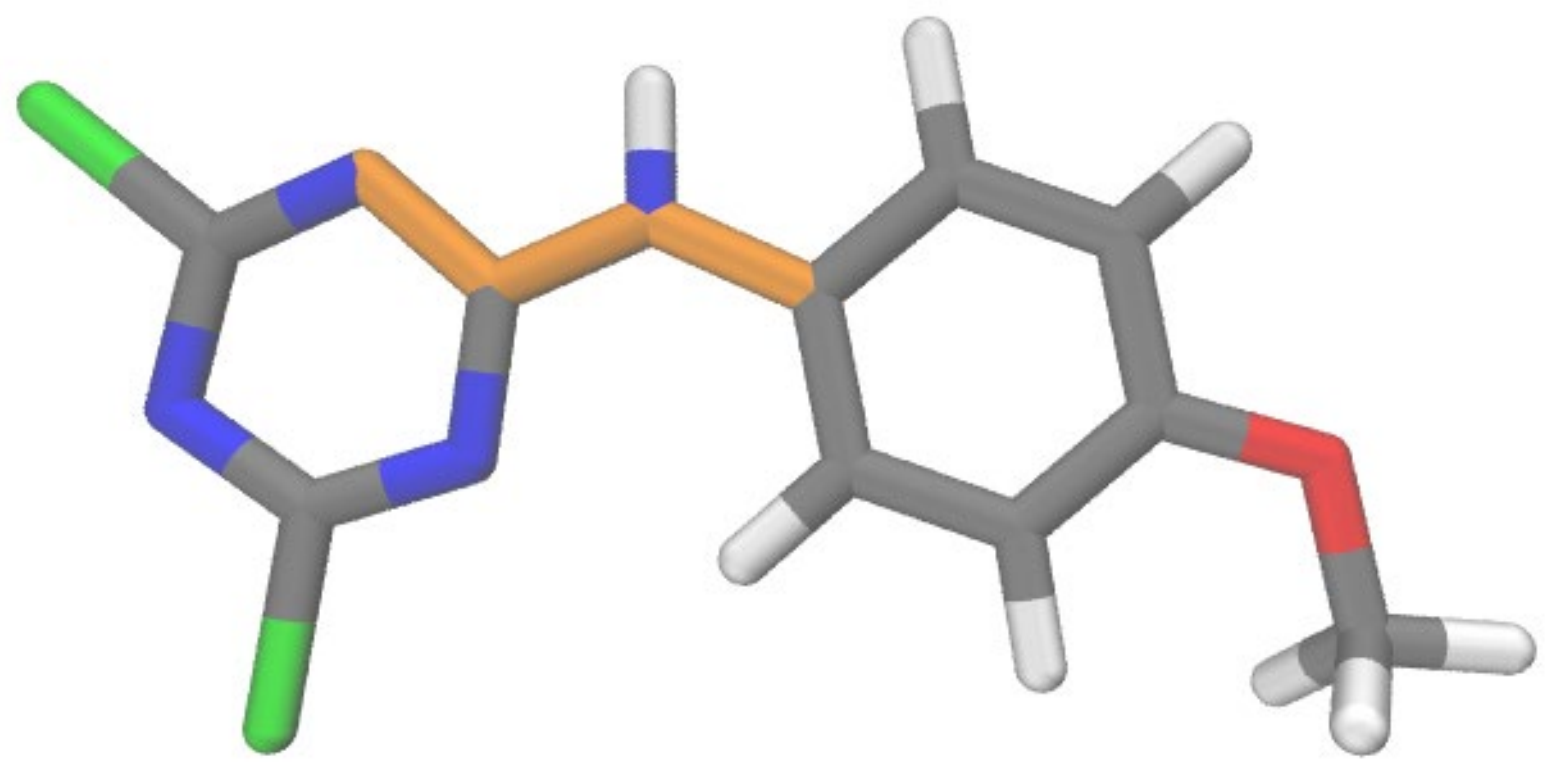

**Figure S26**. Equilibrium ground state structure of A1M. The atoms in the dihedral angle (179.94°) for the rotation are shown in orange. White: hydrogen, gray: carbon, blue: nitrogen, green: chlorine and red: oxygen.

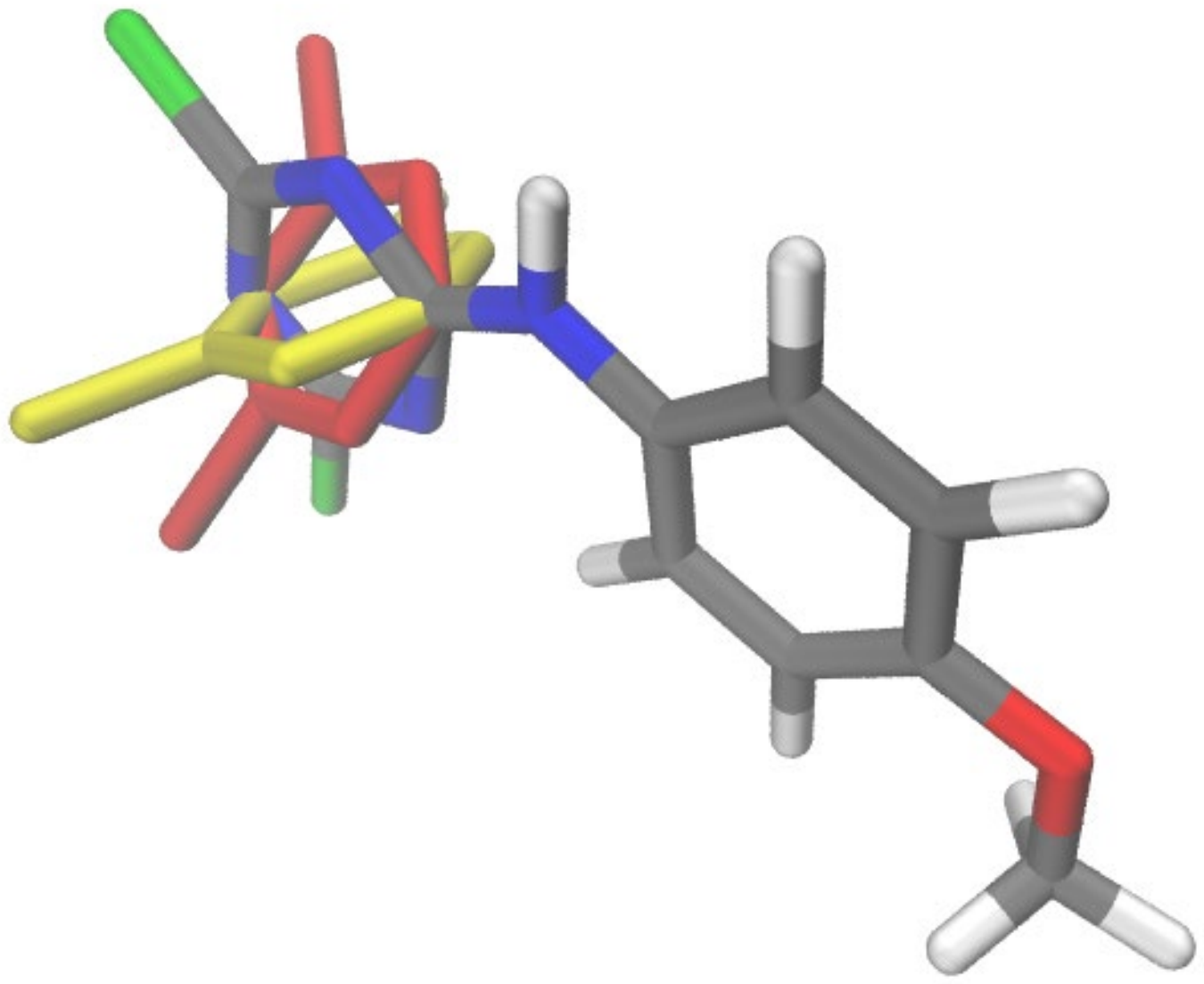

**Figure S27.** The structure of A1M is shown with three different angles for the dihedral angle shown in Figure : 179.94°, 225° and 270°. The equilibrium ground state structure with 179.94° is shown in the “normal” color scheme: White: hydrogen, gray: carbon, blue: nitrogen, green: chlorine and red: oxygen. For the structures with 225° the dichloro-triazin ring is depicted in red and for the structures with 270° in yellow.

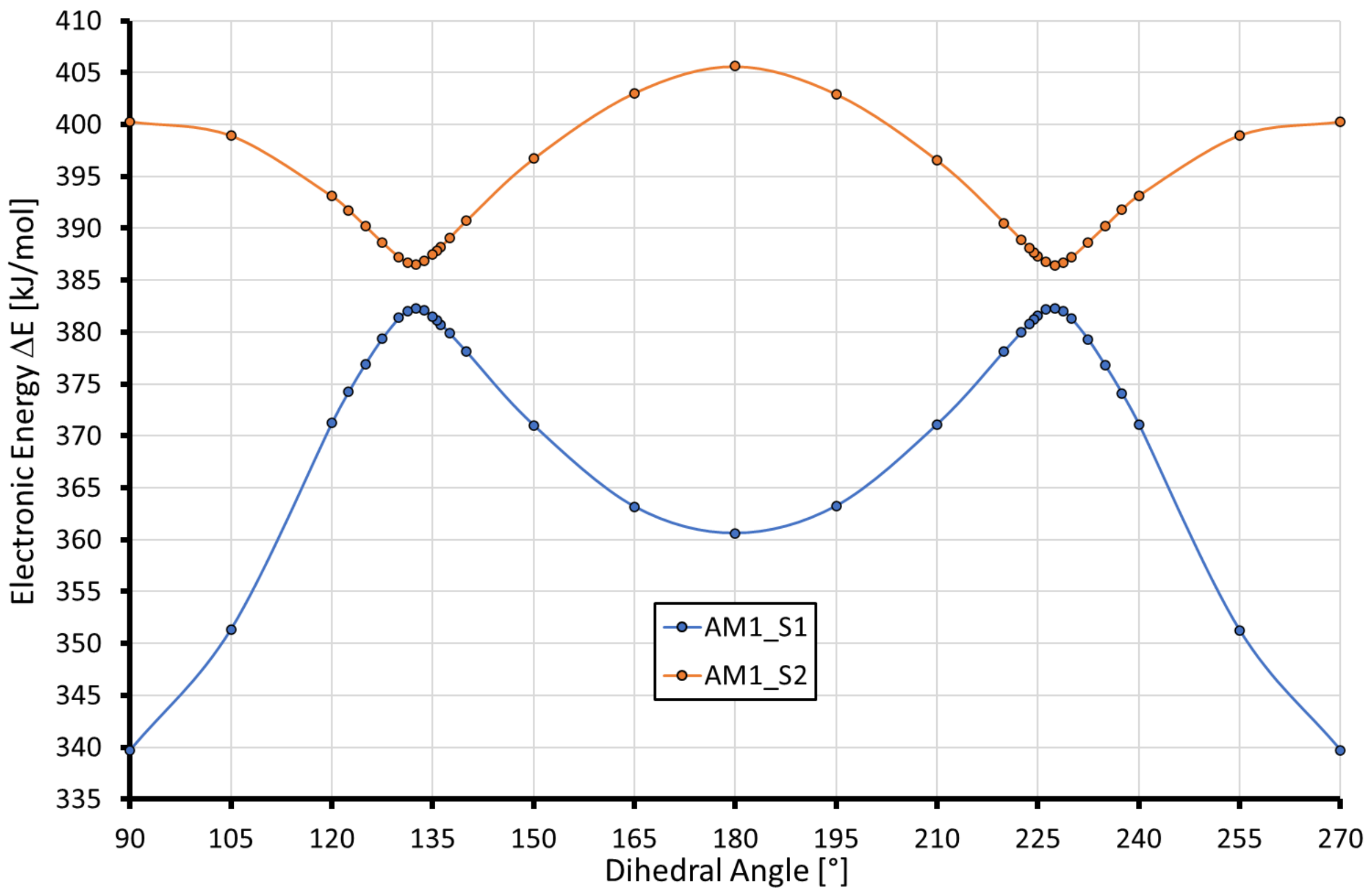

**Figure S28.** The potential energy surface of $S_1$ and $S_2$ of A1M for the rotation around the dihedral angle shown in figure S26 and figure S27. The electronic energy is given relative to A1M in state $S_0$ at equilibrium geometry (179.94°). The energy is given in kJ/mol the angle in °.

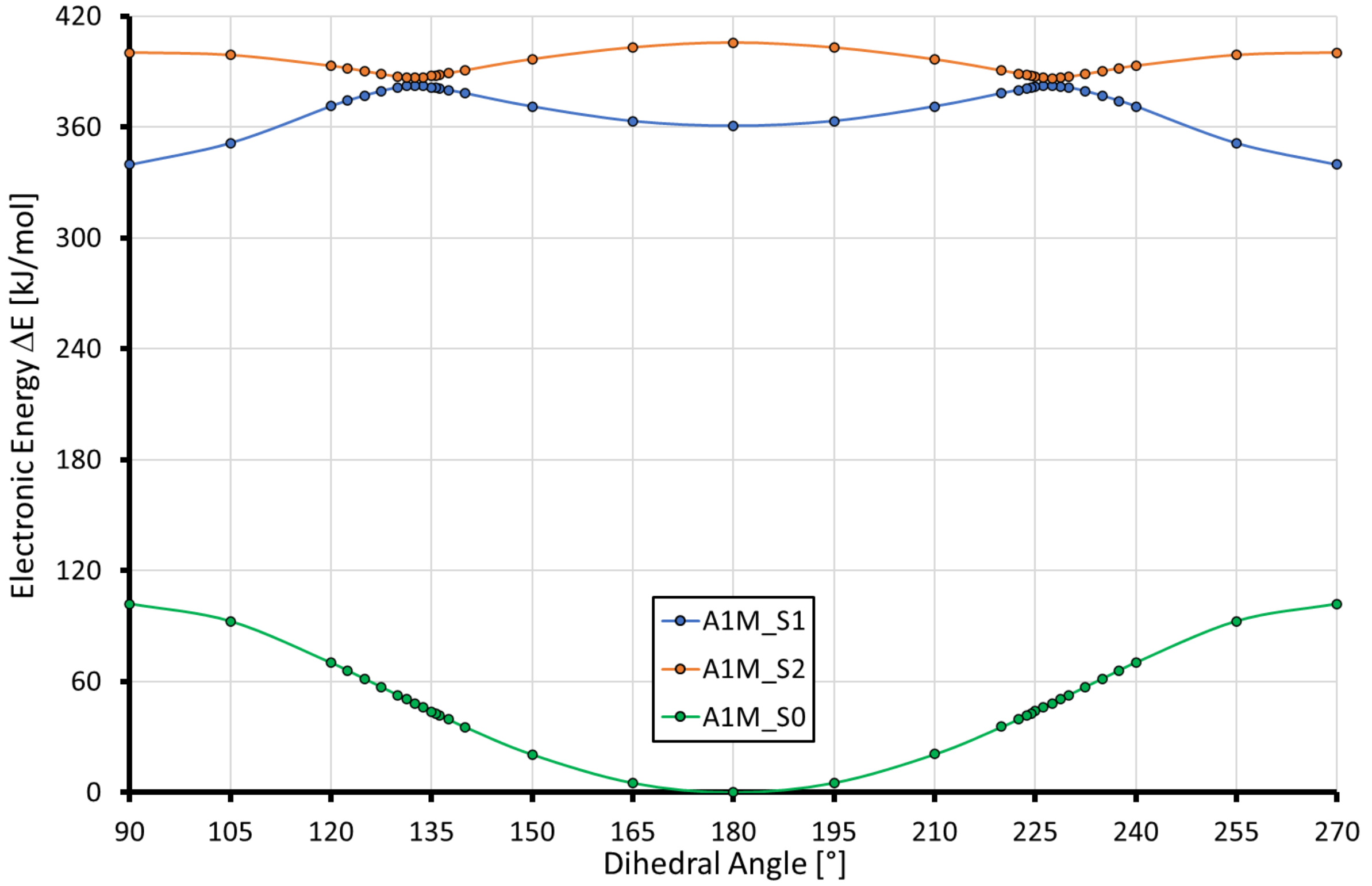

**Figure S29.** The potential energy surface of $S_0$, $S_1$ and $S_2$ of A1M for the rotation around the dihedral angle shown in figure S26 and figure S27. The electronic energy is given relative to A1M in state $S_0$ at equilibrium geometry (179.94°). The energy is given in kJ/mol the angle in °.

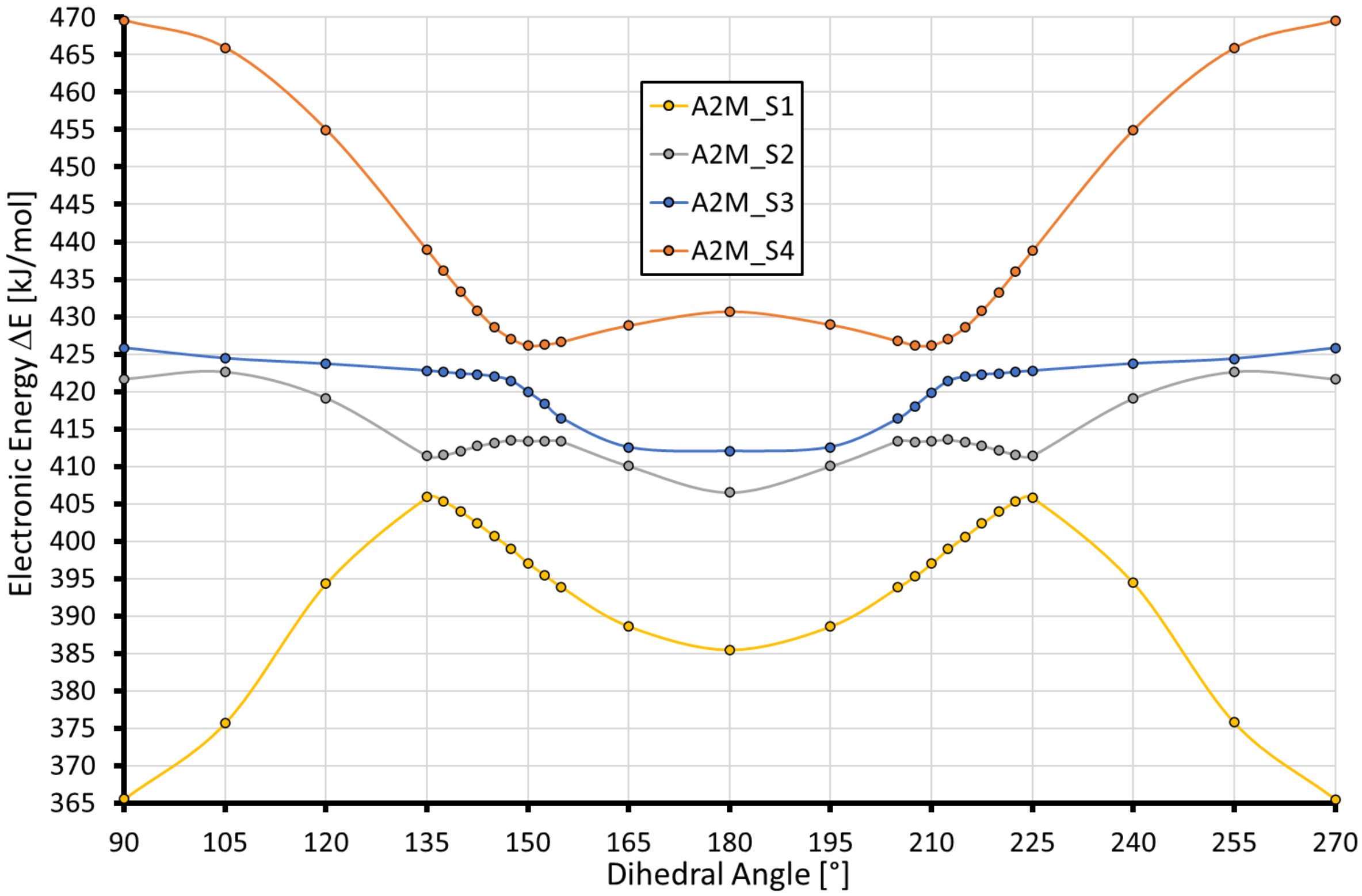

**Figure S30.** The potential energy surface of $S_1$, $S_2$, $S_3$ and $S_4$ of A2M for the rotation around the dihedral angle shown in figure S26 and figure S27. The electronic energy is given relative to A2M in state $S_0$ at equilibrium geometry (179.991°). The energy is given in kJ/mol the angle in °.

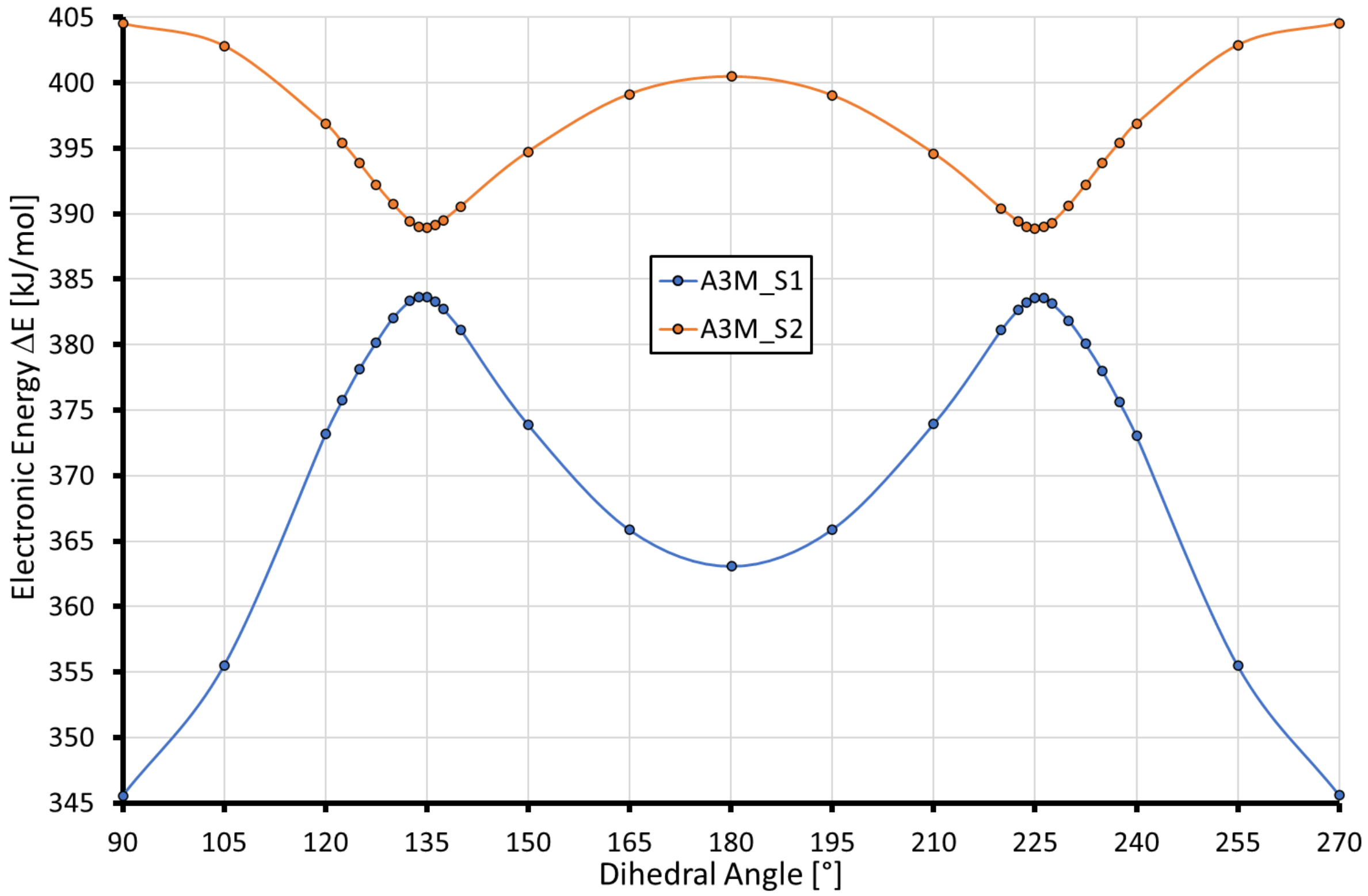

**Figure S31:** The potential energy surface of $S_1$ and $S_2$ of A3M for the rotation around the dihedral angle shown in figure S26 and figure S27. The electronic energy is given relative to A3M in state $S_0$ at equilibrium geometry (180.084°). The energy is given in kJ/mol the angle in °.

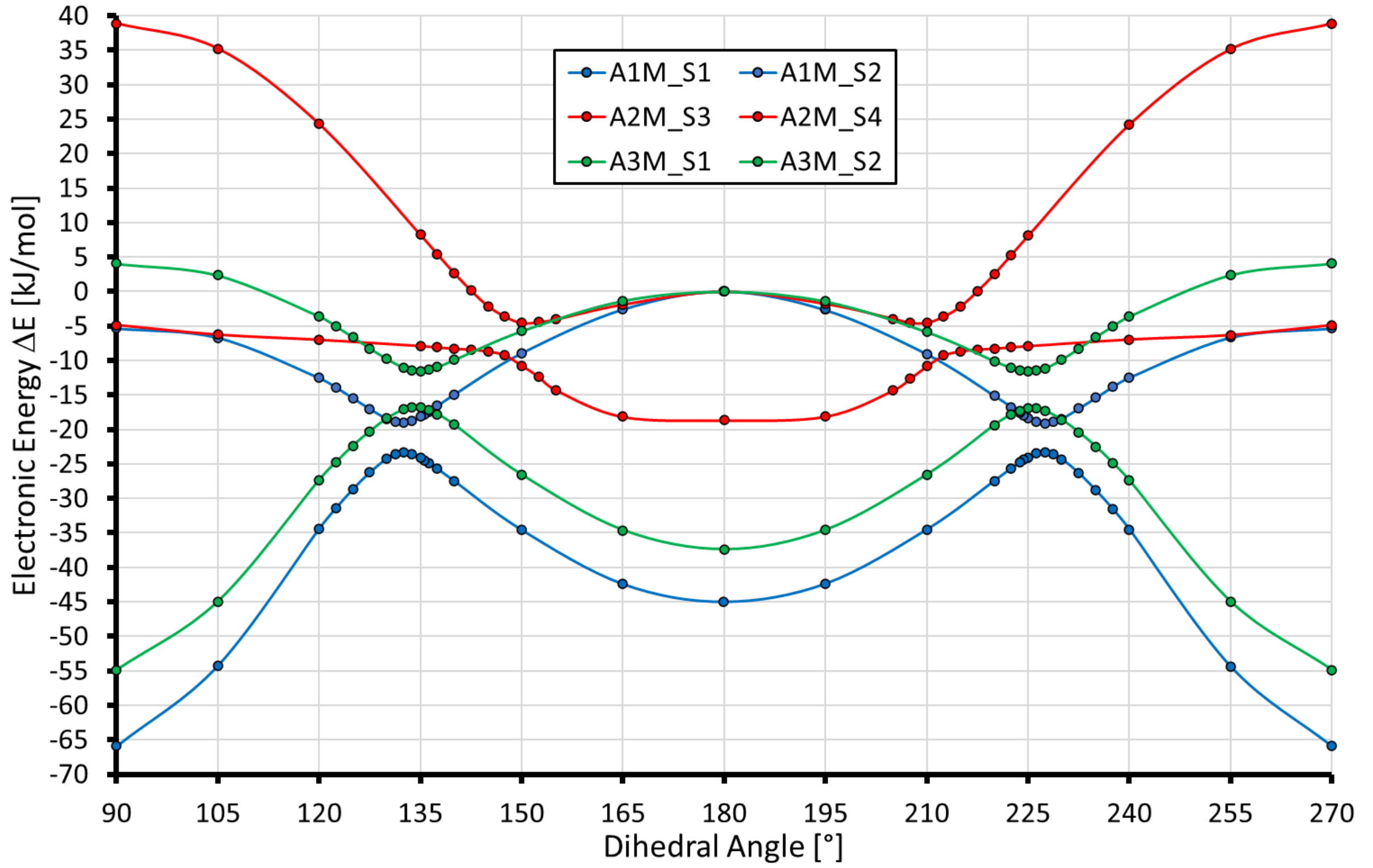

**Figure S32.** The potential energy surface of $S_1$ and $S_2$ of A1M, $S_3$ and $S_4$ of A2M, and $S_1$ and $S_2$ of A3M for the rotation around the dihedral angle shown in figure S26 and figure S27. The electronic energy of A1M is given relative to A1M in state $S_2$ at equilibrium geometry of the $S_0$ state (179.94°). The electronic energy of A2M is given relative to A2M in state $S_4$ at equilibrium geometry of the $S_0$ state (179.991°). The electronic energy of A3M is given relative to A3M in state $S_2$ at equilibrium geometry of the $S_0$ state (180.084°). For example, $S_2$ of A1M as an energy 0 kJ/mol at 179.94°. This makes the PES of the three molecules better comparable. The energy is given in kJ/mol the angle in °.

**Table S22.** Electronic energy and oscillator strength of states $S_1$ and $S_2$ of A1M for the rotation around the dihedral angle shown in figure S26 and figure S27.

| State: | $S_1$ | | | $S_2$ | | |
|---|---|---|---|---|---|---|
| Angle [°] | Energy [kJ/mol] | Oscillator strengths | Orbital | Energy [kJ/mol] | Oscillator strengths | Orbital |
| 270 | 339.70 | 0.00004 | H → L | 400.28 | 0.01021 | H → L+1 |
| 255 | 351.25 | 0.07141 | H → L | 398.96 | 0.00743 | H → L+1 |
| 240 | 371.12 | 0.2357 | H → L | 393.15 | 0.00169 | H → L+1 |
| 237.5 | 374.11 | 0.26642 | H → L | 391.79 | 0.00069 | H → L+1 |
| 235 | 376.80 | 0.29681 | H → L | 390.20 | 0.00045 | H → L+1 |
| 232.5 | 379.27 | 0.32355 | H → L | 388.66 | 0.0038 | H → L+1 |
| 230 | 381.30 | 0.33021 | H → L | 387.19 | 0.02706 | H → L+1 |
| 228.75 | 382.03 | 0.30344 | H → L | 386.68 | 0.06866 | H → L+1 |
| 227.5 | 382.30 | 0.2343 | H → L (85,9%), H → L+1 (12,8%) | 386.46 | 0.15258 | H → L +1 (85,8%), H → L (12,8%) |
| 226.25 | 382.16 | 0.15087 | H → L (73,9%), H → L+1 (24,8%) | 386.77 | 0.25068 | H → L +1 (73,8%), H → L (24,8%) |
| 225 | 381.55 | 0.09409 | H → L (67,9%), H → L+1 (30,8%) | 387.32 | 0.32188 | H → L +1 (67,8%), H → L (30,8%) |
| 224.375 | 381.19 | 0.07648 | H → L (67,8%), H → L+1 (30,9%) | 387.69 | 0.34668 | H → L +1 (67,7%), H → L (30,9%) |
| 223.75 | 380.80 | 0.06341 | H → L (69,2%), H → L+1 (29,5%) | 388.07 | 0.36688 | H → L +1 (69,1%), H → L (29,5%) |
| 222.5 | 379.97 | 0.04615 | H → L (74,9%), H → L+1 (23,7%) | 388.89 | 0.39827 | H → L +1 (74,9%), H → L (23,8%) |
| 220 | 378.17 | 0.02881 | H → L (87,6%), H → L+1 (11,0%) | 390.51 | 0.44318 | H → L +1 (87,6%), H → L (11,1%) |
| 210 | 371.12 | 0.01087 | H → L | 396.54 | 0.56211 | H → L+1 |
| 195 | 363.26 | 0.00675 | H → L | 402.93 | 0.67853 | H → L+1 |
| 179.940 | 360.63 | 0.00664 | H → L | 405.61 | 0.72207 | H → L+1 |
| 165 | 363.18 | 0.0068 | H → L | 403.04 | 0.67935 | H → L+1 |
| 150 | 371.02 | 0.01095 | H → L | 396.70 | 0.56276 | H → L+1 |
| 140 | 378.11 | 0.02893 | H → L (88,3%), H → L+1 (10,3%) | 390.70 | 0.44354 | H → L +1 (88,4%), H → L (10,3%) |
| 137.5 | 379.89 | 0.04587 | H → L (76,3%), H → L+1 (22,3%) | 389.04 | 0.39898 | H → L +1 (76,2%), H → L (22,4%) |
| 136.25 | 380.72 | 0.06244 | H → L (70,5%), H → L+1 (28,2%) | 388.23 | 0.36826 | H → L +1 (70,4%), H → L (28,2%) |
| 135.625 | 381.10 | 0.07484 | H → L (68,9%), H → L+1 (29,8%) | 387.83 | 0.34872 | H → L +1 (68,8%), H → L (29,9%) |
| 135 | 381.46 | 0.09144 | H → L (68,6%), H → L+1 (30,0%) | 387.46 | 0.32492 | H → L +1 (68,6%), H → L (30,1%) |
| 133.75 | 382.06 | 0.14423 | H → L (73,6%), H → L+1 (25,1%) | 386.87 | 0.2577 | H → L +1 (73,6%), H → L (25,1%) |
| 132.5 | 382.28 | 0.22407 | H → L (84,9%), H → L+1 (13,8%) | 386.55 | 0.16319 | H → L +1 (84,8%), H → L (13,8%) |
| 131.25 | 382.05 | 0.29553 | H → L | 386.70 | 0.07692 | H → L+1 |
| 130 | 381.37 | 0.32643 | H → L | 387.18 | 0.03117 | H → L+1 |
| 127.5 | 379.36 | 0.32305 | H → L | 388.61 | 0.00461 | H → L+1 |
| 125 | 376.94 | 0.29706 | H → L | 390.18 | 0.00054 | H → L+1 |
| 122.5 | 374.22 | 0.2669 | H → L | 391.73 | 0.00058 | H → L+1 |
| 120 | 371.22 | 0.23617 | H → L | 393.09 | 0.00155 | H → L+1 |
| 105 | 351.37 | 0.07191 | H → L | 398.92 | 0.00732 | H → L+1 |
| 90 | 339.71 | 0.00005 | H → L | 400.28 | 0.01023 | H → L+1 |

**Table S23.** Electronic energy and oscillator strength of states $S_3$ and $S_4$ of A2M for the rotation around the dihedral angle shown in figure S26 and figure S27.

| State | | $S_3$ | | | $S_4$ | |
|---|---|---|---|---|---|---|
| **Angle [°]** | **Energy [kJ/mol]** | **Oscillator strengths** | **Orbital** | **Energy [kJ/mol]** | **Oscillator strengths** | **Orbital** |
| 270 | 425.84 | 0.00461 | H→L+1(96,8%) | 469.56 | 0.00823 | H-1→L+1(97,4%) |
| 255 | 424.39 | 0.00252 | H→L+1(94,9%) | 465.86 | 0.00726 | H-1→L+1(96,9%) |
| 240 | 423.74 | 0.00481 | H-1→L(94,3%) | 454.92 | 0.00464 | H-1→L+1(95,1%) |
| 225 | 422.78 | 0.03922 | H-1→L(90,6%) | 438.78 | 0.0002 | H-1→L+1(86,6%),H→L+1(6,8%) |
| 222.5 | 422.61 | 0.058 | H-1→L(83,9%),H-1→L+1(13,0%) | 436.02 | 0.00052 | H-1→L+1(80,2%),H-1→L(11,0%) |
| 220 | 422.41 | 0.08404 | H-1→L(71,1%),H-1→L+1(25,4%) | 433.26 | 0.00293 | H-1→L+1(69,1%),H-1→L(21,6%) |
| 217.5 | 422.28 | 0.11454 | H-1→L(58,2%),H-1→L+1(36,9%) | 430.76 | 0.01128 | H-1→L+1(58,7%),H-1→L(31,8%) |
| 215 | 422.01 | 0.13695 | H-1→L(55,4%),H-1→L+1(36,9%) | 428.55 | 0.03715 | H-1→L+1(58,6%),H-1→L(32,3%) |
| 212.5 | 421.42 | 0.12493 | H-1→L(64,5%),H-1→L+1(23,7%), H→L+1(6,6%) | 427.07 | 0.10292 | H-1→L+1(70,2%),H-1→L(21,6%) |
| 210 | 419.85 | 0.08113 | H-1→L(75,0%),H-1→L+1(10,0%), H→L(7,0%) | 426.17 | 0.20029 | H-1→L+1(81,9%),H-1→L(9,9%) |
| 207.5 | 418.03 | 0.04744 | H-1→L(79,5%),H→L(9,5%), H→L+1(5,9%) | 426.15 | 0.28336 | H-1→L+1(85,8%),H→L+1(4,8%) |
| 205 | 416.35 | 0.02852 | H-1→L(81,0%),H→L(11,0%) | 426.75 | 0.34289 | H-1→L+1(86,1%),H→L+1(7,2%) |
| 195 | 412.57 | 0.1802 | H→L+1(77,8%),H-1→L+1(12,1%),H-1→L(5,8%) | 428.91 | 0.48511 | H-1→L+1(84,1%),H→L+1(11,8%) |
| 179.991 | 412.02 | 0.16794 | H→L+1(77,2%),H-1→L+1(12,4%),H-1→L(5,7%) | 430.68 | 0.54795 | H-1→L+1(83,8%),H→L+1(12,4%) |
| 165 | 412.53 | 0.18051 | H→L+1(77,9%),H-1→L+1(12,1%) | 428.81 | 0.48416 | H-1→L+1(84,0%),H→L+1(11,8%) |
| 155 | 416.42 | 0.02975 | H-1→L(80,9%),H→L(11,0%) | 426.64 | 0.34088 | H-1→L+1(86,1%),H→L+1(7,1%) |
| 152.5 | 418.34 | 0.04858 | H-1→L(79,5%),H→L(9,3%), H→L+1(5,8%) | 426.26 | 0.27814 | H-1→L+1(85,7%),H-1→L(4,7%) |
| 150 | 419.94 | 0.08444 | H-1→L(74,6%),H-1→L+1(10,4%), H→L+1(7,0%) | 426.13 | 0.19573 | H-1→L+1(81,5%),H-1→L(10,3%) |
| 147.5 | 421.45 | 0.12857 | H-1→L(63,7%),H-1→L+1(24,6%), H→L+1(6,7%) | 427.06 | 0.09794 | H-1→L+1(69,4%),H-1→L(22,4%) |
| 145 | 422.01 | 0.1383 | H-1→L(54,9%),H-1→L+1(37,4%) | 428.58 | 0.03448 | H-1→L+1(58,0%),H-1→L(32,8%) |
| 142.5 | 422.27 | 0.11454 | H-1→L(58,4%),H-1→L+1(36,8%) | 430.81 | 0.01024 | H-1→L+1(58,8%),H-1→L(31,7%) |
| 140 | 422.40 | 0.08373 | H-1→L(71,7%),H-1→L+1(24,9%) | 433.36 | 0.00255 | H-1→L+1(69,5%),H-1→L(21,1%) |
| 137.5 | 422.58 | 0.05779 | H-1→L(84,3%),H-1→L+1(12,6%) | 436.10 | 0.00042 | H-1→L+1(80,6%),H-1→L(10,6%) |
| 135 | 422.79 | 0.03913 | H-1→L(90,8%) | 438.91 | 0.00024 | H-1→L+1(86,9%),H→L+1(6,8%) |
| 120 | 423.70 | 0.00489 | H-1→L(94,2%) | 454.99 | 0.00479 | H-1→L+1(95,1%) |
| 105 | 424.45 | 0.00256 | H→L+1(95,0%) | 465.93 | 0.00733 | H-1→L+1(96,9%) |
| 90 | 425.84 | 0.00465 | H→L+1(96,8%) | 469.58 | 0.00825 | H-1→L+1(97,4%) |

**Table S24.** Electronic energy and oscillator strength of states $S_1$ and $S_2$ of A3M for the rotation around the dihedral angle shown in figure S26 and figure S27.

| State: | | $S_1$ | | | $S_2$ | |
|---|---|---|---|---|---|---|
| **Angle [°]** | **Energy [kJ/mol]** | **Oscillator strengths** | **Orbital** | **Energy [kJ/mol]** | **Oscillator strengths** | **Orbital** |
| 270 | 345.59 | 0.00003 | H → L | 404.56 | 0.00812 | H → L+1 |
| 255 | 355.48 | 0.06603 | H → L | 402.88 | 0.00518 | H → L+1 |
| 240 | 373.05 | 0.2148 | H → L | 396.89 | 0.00052 | H → L+1 |
| 237.5 | 375.62 | 0.24166 | H → L | 395.44 | 0.0003 | H → L+1 |
| 235 | 378.00 | 0.26771 | H → L | 393.89 | 0.00115 | H → L+1 |
| 232.5 | 380.06 | 0.29074 | H → L | 392.22 | 0.00489 | H → L+1 |
| 230 | 381.86 | 0.30431 | H → L | 390.64 | 0.01782 | H → L+1 |
| 227.5 | 383.17 | 0.285 | H → L | 389.32 | 0.0633 | H → L+1 |
| 226.25 | 383.54 | 0.24725 | H → L | 388.98 | 0.11399 | H → L+1 |
| 225 | 383.57 | 0.19114 | H → L (85,4%), H → L+1 (12,9%) | 388.88 | 0.18278 | H → L +1 (85,2%), H → L (13,0%) |
| 223.75 | 383.22 | 0.13525 | H → L (79,3%), H → L+1 (19,0%) | 389.01 | 0.25125 | H → L +1 (79,2%), H → L (19,1%) |
| 222.5 | 382.67 | 0.09406 | H → L (77,0%), H → L+1 (21,2%) | 389.40 | 0.30483 | H → L +1 (76,9%), H → L (21,4%) |
| 220 | 381.12 | 0.0498 | H → L (82,8%), H → L+1 (15,3%) | 390.40 | 0.37313 | H → L +1 (82,9%), H → L (15,5%) |
| 210 | 373.94 | 0.0118 | H → L | 394.61 | 0.49653 | H → L+1 |
| 195 | 365.89 | 0.00566 | H → L | 399.04 | 0.57569 | H → L+1 |
| 180.084 | 363.08 | 0.00527 | H → L | 400.47 | 0.57637 | H → L+1 |
| 165 | 365.86 | 0.00581 | H → L | 399.14 | 0.57434 | H → L+1 |
| 150 | 373.89 | 0.01202 | H → L | 394.74 | 0.49657 | H → L+1 |
| 140 | 381.16 | 0.0487 | H → L (83,2%), H → L+1 (15,0%) | 390.56 | 0.37518 | H → L +1 (83,2%), H → L (15,2%) |
| 137.5 | 382.70 | 0.09127 | H → L (76,8%), H → L+1 (21,4%) | 389.51 | 0.30868 | H → L +1 (76,7%), H → L (21,6%) |
| 136.25 | 383.29 | 0.13151 | H → L (78,8%), H → L+1 (19,5%) | 389.12 | 0.25611 | H → L +1 (78,6%), H → L (19,6%) |
| 135 | 383.64 | 0.18658 | H → L (84,7%), H → L+1 (13,6%) | 388.94 | 0.18849 | H → L +1 (84,5%), H → L (13,7%) |
| 133.75 | 383.66 | 0.244 | H → L | 389.03 | 0.11844 | H → L+1 |
| 132.5 | 383.36 | 0.28367 | H → L | 389.43 | 0.06588 | H → L+1 |
| 130 | 382.06 | 0.30507 | H → L | 390.72 | 0.01832 | H → L+1 |
| 127.5 | 380.19 | 0.29192 | H → L | 392.23 | 0.00498 | H → L+1 |
| 125 | 378.12 | 0.26902 | H → L | 393.88 | 0.00116 | H → L+1 |
| 122.5 | 375.74 | 0.24298 | H → L | 395.43 | 0.0003 | H → L+1 |
| 120 | 373.16 | 0.21614 | H → L | 396.88 | 0.00052 | H → L+1 |
| 105 | 355.51 | 0.06699 | H → L | 402.82 | 0.00521 | H → L+1 |
| 90 | 345.56 | 0.00006 | H → L | 404.51 | 0.00815 | H → L+1 |

# Cartesian Coordinates and Ground State Energies of the Calculated Molecules

All coordinates are given in Angstrom, all energies are given in Hartree

## *Molecules calculated in gas phase with PBE0-D3(BJ)/def2-TZVP*

### A1M

```
Energy = -1599.757476450
N    -2.9802999    0.9412467    0.2025716
C    -2.3315125    2.0118761    0.6485647
N    -1.1131836    2.3806234    0.3589857
C    -0.4741901    1.5529396   -0.4939690
N    -1.0137059    0.4442966   -1.0143748
C    -2.2461856    0.2170135   -0.6213439
Cl   -3.2059286    3.0188057    1.7290798
Cl   -2.9935840   -1.1928873   -1.2559930
N     0.7779785    1.9218263   -0.7987931
H     1.0472772    2.7743687   -0.3326483
C     1.7520464    1.3463783   -1.6278847
C     1.5805252    0.1804426   -2.3615456
C     2.9756880    2.0162033   -1.7035649
C     2.6138645   -0.3033080   -3.1539789
C     3.8283600    0.3663437   -3.2253366
C     3.9992898    1.5368875   -2.4882322
H     0.6430881   -0.3513512   -2.3162370
H     2.4487467   -1.2139307   -3.7135062
O     4.8851253   -0.0310273   -3.9678305
H     4.9470910    2.0575334   -2.5452637
H     3.1248339    2.9292354   -1.1356256
C     4.7468746   -1.2087626   -4.7248083
H     5.6940337   -1.3512922   -5.2415112
H     3.9431255   -1.1184116   -5.4634215
H     4.5511918   -2.0758899   -4.0848330
```

### A2M

```
Energy = -1714.208991716
N    -3.4391215    1.1683346    0.1711039
C    -2.6590982    1.9726800    0.8854355
N    -1.3835029    2.1970777    0.7171731
C    -0.8324053    1.5162137   -0.3080899
N    -1.5074599    0.6769979   -1.1014702
C    -2.7802422    0.5631600   -0.7988752
Cl   -3.4255363    2.8069223    2.1736755
Cl   -3.6973687   -0.5025385   -1.7856368
N     0.4789400    1.7428989   -0.4859649
H     0.8459775    2.3976134    0.1874195
C     1.4043012    1.2470270   -1.4132236
C     1.0672834    0.3322264   -2.4110110
C     2.7000071    1.7212925   -1.2861103
C     2.0656484   -0.0919173   -3.2754249
C     3.3766479    0.3717698   -3.1666447
```

| | | | |
|---|---|---|---|
| C | 3.6840216 | 1.2794871 | -2.1672857 |
| H | 0.0547830 | -0.0215171 | -2.4900274 |
| O | 1.8482929 | -0.9729290 | -4.2735093 |
| H | 4.1110508 | 0.0055043 | -3.8689754 |
| O | 4.9154870 | 1.7945896 | -1.9683599 |
| H | 2.9714411 | 2.4317823 | -0.5144435 |
| C | 0.5423417 | -1.4780944 | -4.4341950 |
| C | 5.9476651 | 1.3803639 | -2.8313114 |
| H | 0.5853635 | -2.1604817 | -5.2808400 |
| H | 0.2122489 | -2.0231128 | -3.5441933 |
| H | -0.1725449 | -0.6773668 | -4.6484684 |
| H | 6.8440108 | 1.9041056 | -2.5047402 |
| H | 5.7308123 | 1.6476631 | -3.8706931 |
| H | 6.1150158 | 0.3003979 | -2.7644634 |

**A3M**

Energy = -1828.643259377

| | | | |
|---|---|---|---|
| N | -3.1168888 | 0.8744602 | 0.5163873 |
| C | -2.2344903 | 1.5227398 | 1.2697088 |
| N | -0.9850963 | 1.7775271 | 0.9885119 |
| C | -0.5828360 | 1.3136182 | -0.2127718 |
| N | -1.3716087 | 0.6464058 | -1.0632403 |
| C | -2.5989081 | 0.4735510 | -0.6291719 |
| Cl | -2.8145754 | 2.0848459 | 2.7835455 |
| Cl | -3.6573943 | -0.3776938 | -1.6806517 |
| N | 0.6997590 | 1.5719853 | -0.5093771 |
| H | 1.1638063 | 2.0804144 | 0.2271290 |
| C | 1.4868214 | 1.2696688 | -1.6295905 |
| C | 1.0076018 | 0.5841132 | -2.7360195 |
| C | 2.8091960 | 1.7035306 | -1.5733161 |
| C | 1.8716645 | 0.3371210 | -3.7994551 |
| C | 3.1981864 | 0.7702487 | -3.7693805 |
| C | 3.6628629 | 1.4481457 | -2.6397447 |
| H | -0.0168308 | 0.2561235 | -2.7571038 |
| O | 1.4996755 | -0.3223941 | -4.9131854 |
| O | 4.0161599 | 0.5824460 | -4.8366252 |
| O | 4.9581350 | 1.8184112 | -2.6726866 |
| H | 3.1637313 | 2.2324109 | -0.6985120 |
| C | 0.1620167 | -0.7561484 | -5.0054828 |
| C | 5.4679659 | 2.5309819 | -1.5732834 |
| H | -0.5352326 | 0.0861971 | -4.9599146 |
| H | -0.0850604 | -1.4682101 | -4.2115390 |
| H | 0.0722960 | -1.2471624 | -5.9724827 |
| H | 4.9357812 | 3.4770914 | -1.4271356 |
| H | 6.5104013 | 2.7399941 | -1.8055867 |
| H | 5.4162550 | 1.9407709 | -0.6516601 |
| C | 4.6656654 | -0.6761868 | -4.8535660 |
| H | 5.2798218 | -0.6988310 | -5.7531119 |
| H | 3.9374349 | -1.4920332 | -4.8908716 |
| H | 5.3067937 | -0.7952729 | -3.9740455 |

**A1D**

Energy = -1540.944384900

| | | | |
|---|---|---|---|
| N | -3.3124421 | 0.9483233 | -0.1564255 |
| C | -2.8228756 | 2.1948323 | -0.0442944 |
| N | -1.5520172 | 2.5377139 | -0.2549384 |
| C | -0.7458227 | 1.5341865 | -0.6013543 |
| N | -1.1279076 | 0.2534689 | -0.7421478 |
| C | -2.4043282 | 0.0669232 | -0.5026635 |
| N | -3.6380866 | 3.2099577 | 0.3048836 |
| Cl | -2.9650392 | -1.5583711 | -0.6656287 |
| N | 0.5388675 | 1.8814825 | -0.8157272 |
| H | 0.6928736 | 2.8675168 | -0.6784453 |
| C | 1.6649683 | 1.1390254 | -1.1927876 |
| C | 1.6743720 | -0.2295395 | -1.4287540 |
| C | 2.8591995 | 1.8506399 | -1.3364647 |
| C | 2.8516332 | -0.8673728 | -1.8002726 |
| C | 4.0340639 | -0.1538687 | -1.9417018 |
| C | 4.0255419 | 1.2187581 | -1.7043660 |
| H | 0.7629092 | -0.7968907 | -1.3232992 |
| H | 2.8230046 | -1.9341384 | -1.9766307 |
| O | 5.2235132 | -0.6924608 | -2.2995262 |
| H | 4.9482458 | 1.7749689 | -1.8146951 |
| H | 2.8706371 | 2.9211556 | -1.1558533 |
| C | 5.2634855 | -2.0755968 | -2.5470407 |
| H | 6.2920417 | -2.3076449 | -2.8174048 |
| H | 4.6017588 | -2.3549565 | -3.3743307 |
| H | 4.9862840 | -2.6512388 | -1.6570504 |
| H | -3.1492624 | 4.0895144 | 0.3523402 |
| C | -5.0066943 | 3.2607037 | 0.5978977 |
| C | -5.8631209 | 2.1674349 | 0.5942002 |
| C | -7.2084994 | 2.3315601 | 0.9004890 |
| C | -5.5283162 | 4.5177584 | 0.9156275 |
| C | -6.8611097 | 4.6803482 | 1.2192911 |
| C | -7.7197586 | 3.5834395 | 1.2145606 |
| H | -5.4800442 | 1.1885505 | 0.3517180 |
| H | -7.8472496 | 1.4587570 | 0.8877655 |
| O | -9.0136734 | 3.8343661 | 1.5238420 |
| H | -7.2603307 | 5.6567751 | 1.4645347 |
| H | -4.8734384 | 5.3836805 | 0.9238455 |
| C | -9.9044430 | 2.7469608 | 1.5279814 |
| H | -10.8786078 | 3.1514339 | 1.7968401 |
| H | -9.6122747 | 1.9910544 | 2.2652759 |
| H | -9.9695469 | 2.2771484 | 0.5403798 |

**A2D**

Energy = -1769.848354558

| | | | |
|---|---|---|---|
| N | -3.3952418 | 1.0858361 | 0.1030644 |
| C | -2.8623654 | 2.3053868 | 0.2869934 |
| N | -1.6216919 | 2.6525592 | -0.0536918 |
| C | -0.8940988 | 1.6862571 | -0.6129407 |
| N | -1.3269790 | 0.4350974 | -0.8416007 |
| C | -2.5650823 | 0.2387595 | -0.4564897 |
| N | -3.5915756 | 3.2893981 | 0.8540184 |
| Cl | -3.1868118 | -1.3513595 | -0.7224384 |
| N | 0.3610018 | 2.0462923 | -0.9544415 |
| H | 0.5526458 | 3.0094580 | -0.7290645 |
| C | 1.4166080 | 1.3465330 | -1.5465533 |
| C | 1.3316500 | 0.0036849 | -1.9173343 |
| C | 2.5831368 | 2.0675746 | -1.7526644 |
| C | 2.4426262 | -0.5902215 | -2.4961753 |
| C | 3.6259174 | 0.1160095 | -2.7127271 |
| C | 3.6845123 | 1.4470648 | -2.3360754 |
| H | 0.4162109 | -0.5349232 | -1.7480454 |
| O | 2.4637596 | -1.8817671 | -2.8895582 |
| H | 4.4580578 | -0.4006397 | -3.1680239 |
| O | 4.7749477 | 2.2289342 | -2.4963936 |
| H | 2.6619665 | 3.1105382 | -1.4696023 |
| H | -3.0707096 | 4.1482576 | 0.9333627 |
| C | -4.9000955 | 3.3331280 | 1.3443057 |
| C | -5.7667050 | 2.2396697 | 1.3102100 |
| C | -7.0442398 | 2.3968059 | 1.8251311 |
| C | -5.3078349 | 4.5457745 | 1.8794626 |
| C | -6.5964329 | 4.6768773 | 2.3900992 |
| C | -7.4751814 | 3.6071634 | 2.3683641 |
| H | -5.4279822 | 1.3091374 | 0.8904681 |
| O | -7.9576526 | 1.4024545 | 1.8399177 |
| H | -8.4820575 | 3.6687900 | 2.7545106 |
| O | -6.9024992 | 5.8943031 | 2.8904457 |
| H | -4.6447991 | 5.4021783 | 1.9120139 |
| C | 5.9158053 | 1.6495467 | -3.0812388 |
| C | 1.2961783 | -2.6465161 | -2.6967231 |
| C | -7.5798131 | 0.1552122 | 1.3041892 |
| C | -8.1931554 | 6.0793975 | 3.4188380 |
| H | -8.4462742 | -0.4945184 | 1.4128904 |
| H | -6.7324076 | -0.2739804 | 1.8481021 |
| H | -7.3173060 | 0.2376347 | 0.2447205 |
| H | -8.2350970 | 7.1110706 | 3.7633152 |
| H | -8.9639715 | 5.9209282 | 2.6573025 |
| H | -8.3798085 | 5.4082424 | 4.2637250 |
| H | 1.5215205 | -3.6415858 | -3.0758965 |
| H | 1.0312574 | -2.7123077 | -1.6366779 |
| H | 0.4486122 | -2.2294875 | -3.2497882 |
| H | 6.6692734 | 2.4341651 | -3.1190202 |
| H | 5.7095518 | 1.2971107 | -4.0973117 |
| H | 6.2929995 | 0.8147056 | -2.4812734 |

**A3D**

Energy = -1998.716456565

| | | | |
|---|---|---|---|
| N | -2.9910172 | 1.0431825 | 0.0325059 |
| C | -2.4546935 | 2.2703363 | 0.1455402 |
| N | -1.1969147 | 2.5830903 | -0.1661216 |
| C | -0.4557616 | 1.5715259 | -0.6182105 |
| N | -0.8909282 | 0.3090199 | -0.7700210 |
| C | -2.1460789 | 0.1505664 | -0.4246860 |
| N | -3.2004128 | 3.2970662 | 0.6024959 |
| Cl | -2.7712089 | -1.4512176 | -0.5952745 |
| N | 0.8162815 | 1.8927753 | -0.9310046 |
| H | 1.0102791 | 2.8677748 | -0.7689030 |
| C | 1.8865766 | 1.1367405 | -1.4233309 |
| C | 1.7981233 | -0.2167620 | -1.7156437 |
| C | 3.0820116 | 1.8283979 | -1.6114948 |
| C | 2.9223669 | -0.8759235 | -2.2044491 |
| C | 4.1265298 | -0.2009930 | -2.4091443 |
| C | 4.1980941 | 1.1582518 | -2.0974656 |
| H | 0.8662349 | -0.7315509 | -1.5600584 |
| O | 2.9306414 | -2.1883576 | -2.5112389 |
| O | 5.2009290 | -0.8376783 | -2.9457295 |
| O | 5.3974313 | 1.7419070 | -2.3009115 |
| H | 3.1337825 | 2.8826173 | -1.3732656 |
| H | -2.6777777 | 4.1574470 | 0.6366611 |
| C | -4.5311432 | 3.3802875 | 1.0287387 |
| C | -5.3959786 | 2.2956301 | 1.0567728 |
| C | -6.7009531 | 2.4850914 | 1.5022193 |
| C | -4.9599742 | 4.6425735 | 1.4356162 |
| C | -6.2662065 | 4.8192897 | 1.8754299 |
| C | -7.1485405 | 3.7380845 | 1.9230119 |
| H | -5.0446470 | 1.3307031 | 0.7356575 |
| O | -7.6103865 | 1.4919423 | 1.5601051 |
| O | -8.4048264 | 3.9012389 | 2.4157128 |
| O | -6.7713948 | 6.0038463 | 2.2780280 |
| H | -4.2729371 | 5.4778336 | 1.4024102 |
| C | 5.5152451 | 3.1150015 | -2.0283728 |
| C | 1.7305426 | -2.9080547 | -2.3473677 |
| C | -7.1987946 | 0.2018304 | 1.1711663 |
| C | -5.9163273 | 7.1184994 | 2.2751865 |
| H | -6.8835931 | 0.1792034 | 0.1230441 |
| H | -8.0665327 | -0.4419000 | 1.3023510 |
| H | -6.3783023 | -0.1620792 | 1.7974308 |
| H | -5.5544094 | 7.3468146 | 1.2664031 |
| H | -5.0589065 | 6.9686508 | 2.9403638 |
| H | -6.5087742 | 7.9555723 | 2.6399410 |
| H | 1.3998395 | -2.9063691 | -1.3037817 |
| H | 1.9484292 | -3.9285968 | -2.6565440 |
| H | 0.9312678 | -2.5010961 | -2.9745496 |
| H | 5.3216419 | 3.3348162 | -0.9724461 |
| H | 4.8355876 | 3.7076703 | -2.6503637 |
| H | 6.5427143 | 3.3839289 | -2.2664719 |
| C | 6.0065424 | -1.5086901 | -1.9945492 |
| C | -9.3861700 | 4.1380144 | 1.4233267 |
| H | -10.3379377 | 4.2454192 | 1.9428994 |
| H | -9.4474665 | 3.2979401 | 0.7246354 |
| H | -9.1689147 | 5.0588740 | 0.8719557 |
| H | 6.8207439 | -1.9763726 | -2.5474414 |
| H | 5.4338735 | -2.2803952 | -1.4708812 |
| H | 6.4213704 | -0.8026391 | -1.2674964 |

*Molecules calculated in solution with PBE0-D3(BJ)/def2-TZVP COSMO(THF)*

**A1M**

```
Energy = -1599.769766698
N    -2.9807909    0.9390715    0.2012273
C    -2.3310023    2.0096429    0.6484583
N    -1.1160203    2.3819801    0.3627580
C    -0.4683969    1.5585307   -0.4930592
N    -1.0120216    0.4484125   -1.0144878
C    -2.2408342    0.2191150   -0.6242394
Cl   -3.2116211    3.0132000    1.7324200
Cl   -2.9885740   -1.1922761   -1.2620105
N     0.7786402    1.9250239   -0.7971807
H     1.0602693    2.7775814   -0.3352424
C     1.7500781    1.3460483   -1.6277820
C     1.5807063    0.1792245   -2.3622549
C     2.9729904    2.0187541   -1.7019112
C     2.6148186   -0.3049074   -3.1548774
C     3.8285132    0.3678585   -3.2245642
C     3.9976836    1.5397226   -2.4869012
H     0.6459069   -0.3573611   -2.3213279
H     2.4510368   -1.2159307   -3.7142510
O     4.8848337   -0.0306238   -3.9673521
H     4.9434726    2.0656104   -2.5401233
H     3.1184435    2.9314559   -1.1332682
C     4.7440930   -1.2150156   -4.7268728
H     5.6908213   -1.3574680   -5.2437344
H     3.9392106   -1.1208218   -5.4620499
H     4.5482932   -2.0776678   -4.0828731
```

**A2M**

```
Energy = -1714.222033094
N    -3.4395384    1.1671591    0.1663873
C    -2.6603957    1.9709089    0.8829745
N    -1.3868926    2.1981053    0.7212602
C    -0.8265875    1.5201542   -0.3041794
N    -1.5029841    0.6807407   -1.0996458
C    -2.7735626    0.5644016   -0.8029876
Cl   -3.4352271    2.8029535    2.1717127
Cl   -3.6883701   -0.5014809   -1.7936907
N     0.4809752    1.7455524   -0.4804237
H     0.8582384    2.3997579    0.1897983
C     1.4021699    1.2473135   -1.4101241
C     1.0660864    0.3320722   -2.4091695
C     2.6984984    1.7226393   -1.2824661
C     2.0654385   -0.0913559   -3.2741183
C     3.3766485    0.3729273   -3.1654787
C     3.6828461    1.2811830   -2.1645310
H     0.0550094   -0.0244057   -2.4915761
O     1.8467182   -0.9727942   -4.2726841
```

| | | | |
|---|---|---|---|
| H | 4.1152268 | 0.0102305 | -3.8659665 |
| O | 4.9149416 | 1.7950206 | -1.9680167 |
| H | 2.9619774 | 2.4330780 | -0.5078376 |
| C | 0.5361012 | -1.4812886 | -4.4343791 |
| C | 5.9508180 | 1.3783087 | -2.8369858 |
| H | 0.5805398 | -2.1632942 | -5.2808089 |
| H | 0.2108604 | -2.0270060 | -3.5439026 |
| H | -0.1756553 | -0.6791758 | -4.6502284 |
| H | 6.8467835 | 1.9019426 | -2.5104352 |
| H | 5.7291105 | 1.6475199 | -3.8737075 |
| H | 6.1142852 | 0.2989826 | -2.7679395 |

**A3M**

Energy = -1828.659070364

| | | | |
|---|---|---|---|
| C | 5.4683950 | 2.5125750 | -1.5754180 |
| O | 4.9458200 | 1.8244010 | -2.6949920 |
| C | 3.6524140 | 1.4518080 | -2.6564730 |
| C | 3.1808330 | 0.7769950 | -3.7857420 |
| O | 3.9979160 | 0.5768010 | -4.8567230 |
| C | 4.6952750 | -0.6629040 | -4.8195430 |
| C | 1.8534270 | 0.3475960 | -3.8165120 |
| O | 1.4761590 | -0.3040750 | -4.9331690 |
| C | 0.1386360 | -0.7588640 | -5.0133190 |
| C | 0.9937040 | 0.5929730 | -2.7486100 |
| C | 1.4804640 | 1.2716530 | -1.6392750 |
| C | 2.8041440 | 1.7030910 | -1.5850830 |
| N | 0.7039630 | 1.5736390 | -0.5122160 |
| H | 1.1832390 | 2.0794850 | 0.2183140 |
| C | -0.5735230 | 1.3166010 | -0.2112320 |
| N | -0.9790020 | 1.7754780 | 0.9935930 |
| C | -2.2256940 | 1.5184570 | 1.2737550 |
| N | -3.1112290 | 0.8721970 | 0.5217240 |
| C | -2.5919700 | 0.4758980 | -0.6271520 |
| N | -1.3684280 | 0.6510860 | -1.0607870 |
| Cl | -2.8062310 | 2.0757510 | 2.7925320 |
| Cl | -3.6541110 | -0.3729760 | -1.6788980 |
| H | -0.0310050 | 0.2666690 | -2.7710990 |
| H | 3.1600890 | 2.2278570 | -0.7088060 |
| H | -0.5671040 | 0.0752180 | -4.9622680 |
| H | -0.0849840 | -1.4748360 | -4.2171060 |
| H | 0.0473730 | -1.2512810 | -5.9790790 |
| H | 4.9416490 | 3.4569480 | -1.4097440 |
| H | 6.5111400 | 2.7170840 | -1.8085370 |
| H | 5.4117640 | 1.8991470 | -0.6713140 |
| H | 5.3073450 | -0.7068550 | -5.7196190 |
| H | 3.9931190 | -1.5014540 | -4.8142700 |
| H | 5.3395240 | -0.7172920 | -3.9371590 |

**A1D**

Energy = -1540.962010704

| | | | |
|---|---|---|---|
| N | -3.3109271 | 0.9578945 | -0.1606248 |
| C | -2.8220015 | 2.2081511 | -0.0458916 |
| N | -1.5496144 | 2.5471435 | -0.2571453 |
| C | -0.7401084 | 1.5460522 | -0.6050232 |
| N | -1.1258448 | 0.2629328 | -0.7471790 |
| C | -2.4012011 | 0.0808531 | -0.5075273 |
| N | -3.6381225 | 3.2184388 | 0.3041490 |
| Cl | -2.9640015 | -1.5492385 | -0.6725738 |
| N | 0.5428286 | 1.8888633 | -0.8191391 |
| H | 0.7099100 | 2.8744795 | -0.6835936 |
| C | 1.6655189 | 1.1402561 | -1.1949252 |
| C | 1.6755230 | -0.2294262 | -1.4305956 |
| C | 2.8610361 | 1.8519410 | -1.3375413 |
| C | 2.8527714 | -0.8706369 | -1.8006679 |
| C | 4.0357307 | -0.1566610 | -1.9406906 |
| C | 4.0275360 | 1.2174339 | -1.7039137 |
| H | 0.7664182 | -0.8007102 | -1.3270736 |
| H | 2.8232738 | -1.9375953 | -1.9762363 |
| O | 5.2242240 | -0.6988772 | -2.2966675 |
| H | 4.9487167 | 1.7777886 | -1.8120867 |
| H | 2.8713673 | 2.9221022 | -1.1572050 |
| C | 5.2602860 | -2.0896309 | -2.5440936 |
| H | 6.2885009 | -2.3237037 | -2.8125404 |
| H | 4.5979571 | -2.3629643 | -3.3713495 |
| H | 4.9792915 | -2.6589801 | -1.6527007 |
| H | -3.1595167 | 4.1049757 | 0.3562265 |
| C | -5.0070921 | 3.2620728 | 0.5975234 |
| C | -5.8645102 | 2.1682367 | 0.5941172 |
| C | -7.2111186 | 2.3294136 | 0.9018220 |
| C | -5.5290421 | 4.5196913 | 0.9170016 |
| C | -6.8627226 | 4.6800259 | 1.2221950 |
| C | -7.7218194 | 3.5817535 | 1.2174013 |
| H | -5.4861388 | 1.1876807 | 0.3514470 |
| H | -7.8490858 | 1.4559351 | 0.8890170 |
| O | -9.0162631 | 3.8289694 | 1.5279950 |
| H | -7.2575687 | 5.6586993 | 1.4683074 |
| H | -4.8736390 | 5.3847107 | 0.9247891 |
| C | -9.9087666 | 2.7334512 | 1.5296073 |
| H | -10.8832836 | 3.1360856 | 1.7984765 |
| H | -9.6115406 | 1.9802563 | 2.2659007 |
| H | -9.9684511 | 2.2684957 | 0.5406791 |

**A2D**

Energy = -1769.867779424

| | | | |
|---|---|---|---|
| N | -3.3882999 | 1.0980402 | 0.1097680 |
| C | -2.8548167 | 2.3186614 | 0.2984489 |
| N | -1.6128518 | 2.6630903 | -0.0416424 |
| C | -0.8842723 | 1.6991257 | -0.6044176 |
| N | -1.3208687 | 0.4479815 | -0.8371336 |
| C | -2.5584360 | 0.2537372 | -0.4522842 |
| N | -3.5849915 | 3.2984655 | 0.8679200 |
| Cl | -3.1829560 | -1.3371697 | -0.7237867 |
| N | 0.3703427 | 2.0551800 | -0.9459211 |
| H | 0.5740936 | 3.0176989 | -0.7223410 |
| C | 1.4187159 | 1.3495527 | -1.5416867 |
| C | 1.3314226 | 0.0054835 | -1.9126077 |
| C | 2.5871683 | 2.0687141 | -1.7524159 |
| C | 2.4400975 | -0.5907578 | -2.4957857 |
| C | 3.6244423 | 0.1129422 | -2.7169790 |
| C | 3.6856204 | 1.4452459 | -2.3399491 |
| H | 0.4176249 | -0.5350483 | -1.7416585 |
| O | 2.4557773 | -1.8833514 | -2.8888322 |
| H | 4.4574289 | -0.4007491 | -3.1750746 |
| O | 4.7779117 | 2.2224586 | -2.5056828 |
| H | 2.6615318 | 3.1116008 | -1.4673822 |
| H | -3.0720419 | 4.1629264 | 0.9548303 |
| C | -4.8953867 | 3.3348193 | 1.3511046 |
| C | -5.7624500 | 2.2401372 | 1.3126335 |
| C | -7.0433813 | 2.3955548 | 1.8222472 |
| C | -5.3077077 | 4.5472896 | 1.8866302 |
| C | -6.5995685 | 4.6758197 | 2.3916697 |
| C | -7.4788457 | 3.6048313 | 2.3651394 |
| H | -5.4248796 | 1.3088539 | 0.8940585 |
| O | -7.9551543 | 1.3984931 | 1.8308414 |
| H | -8.4870760 | 3.6703350 | 2.7482831 |
| O | -6.9120726 | 5.8908487 | 2.8924536 |
| H | -4.6402140 | 5.4003680 | 1.9196517 |
| C | 5.9208522 | 1.6365293 | -3.0977604 |
| C | 1.2836440 | -2.6504309 | -2.6914857 |
| C | -7.5735501 | 0.1467209 | 1.2934173 |
| C | -8.2128827 | 6.0737230 | 3.4164599 |
| H | -8.4406075 | -0.5023015 | 1.3973855 |
| H | -6.7306596 | -0.2805965 | 1.8443557 |
| H | -7.3087250 | 0.2339046 | 0.2356521 |
| H | -8.2572438 | 7.1048727 | 3.7608068 |
| H | -8.9751491 | 5.9136203 | 2.6483169 |
| H | -8.3986372 | 5.4006340 | 4.2585256 |
| H | 1.5073675 | -3.6450344 | -3.0716606 |
| H | 1.0268453 | -2.7155791 | -1.6301386 |
| H | 0.4379756 | -2.2309864 | -3.2442198 |
| H | 6.6750238 | 2.4197515 | -3.1377116 |
| H | 5.7058897 | 1.2863680 | -4.1115883 |
| H | 6.2943199 | 0.8022555 | -2.4967538 |

**A3D**

Energy = -1998.741444042

| | | | |
|---|---|---|---|
| N | -2.9916635 | 1.0382916 | 0.0382655 |
| C | -2.4540792 | 2.2664807 | 0.1586214 |
| N | -1.1948763 | 2.5760252 | -0.1513285 |
| C | -0.4527337 | 1.5671902 | -0.6084454 |
| N | -0.8918455 | 0.3045165 | -0.7658046 |
| C | -2.1467735 | 0.1485916 | -0.4220814 |
| N | -3.1999847 | 3.2890846 | 0.6198217 |
| Cl | -2.7753507 | -1.4540995 | -0.6008719 |
| N | 0.8185807 | 1.8850366 | -0.9207746 |
| H | 1.0251427 | 2.8592186 | -0.7600097 |
| C | 1.8805782 | 1.1253638 | -1.4221761 |
| C | 1.7929029 | -0.2318383 | -1.7066018 |
| C | 3.0714326 | 1.8209083 | -1.6294781 |
| C | 2.9139785 | -0.8907427 | -2.2047638 |
| C | 4.1119960 | -0.2107724 | -2.4258213 |
| C | 4.1832286 | 1.1520578 | -2.1263948 |
| H | 0.8662487 | -0.7519198 | -1.5389695 |
| O | 2.9236594 | -2.2044752 | -2.5063231 |
| O | 5.1875149 | -0.8551906 | -2.9612125 |
| O | 5.3745495 | 1.7398784 | -2.3541717 |
| H | 3.1192993 | 2.8768297 | -1.3996594 |
| H | -2.6850417 | 4.1553896 | 0.6636584 |
| C | -4.5312725 | 3.3649529 | 1.0420128 |
| C | -5.3996465 | 2.2804130 | 1.0579133 |
| C | -6.7067204 | 2.4688861 | 1.4996076 |
| C | -4.9579289 | 4.6250226 | 1.4604140 |
| C | -6.2647195 | 4.8003303 | 1.8985205 |
| C | -7.1503031 | 3.7201283 | 1.9286284 |
| H | -5.0529802 | 1.3162366 | 0.7301121 |
| O | -7.6192085 | 1.4778691 | 1.5467610 |
| O | -8.4171120 | 3.8816000 | 2.4057966 |
| O | -6.7657536 | 5.9797465 | 2.3164304 |
| H | -4.2666287 | 5.4566898 | 1.4383944 |
| C | 5.4956260 | 3.1187283 | -2.0672211 |
| C | 1.7317357 | -2.9378269 | -2.2999281 |
| C | -7.2166661 | 0.1893357 | 1.1239566 |
| C | -5.9071579 | 7.1026565 | 2.3050050 |
| H | -6.9067109 | 0.1947250 | 0.0749161 |
| H | -8.0890717 | -0.4502368 | 1.2406059 |
| H | -6.3995942 | -0.1936323 | 1.7421747 |
| H | -5.5576242 | 7.3246041 | 1.2922516 |
| H | -5.0461287 | 6.9503473 | 2.9625439 |
| H | -6.4993028 | 7.9376088 | 2.6737309 |
| H | 1.4347008 | -2.9241761 | -1.2472127 |
| H | 1.9530221 | -3.9597725 | -2.6006048 |
| H | 0.9144852 | -2.5471213 | -2.9130183 |
| H | 5.3157863 | 3.3206747 | -1.0070309 |
| H | 4.8049434 | 3.7122211 | -2.6735570 |
| H | 6.5192555 | 3.3891621 | -2.3177409 |
| C | 6.0357410 | -1.4524903 | -1.9885195 |
| C | -9.3679062 | 4.2047587 | 1.3988580 |
| H | -10.3322659 | 4.3052715 | 1.8958009 |
| H | -9.4245159 | 3.4090913 | 0.6503725 |
| H | -9.1088494 | 5.1493252 | 0.9113492 |
| H | 6.8482995 | -1.9342560 | -2.5313721 |
| H | 5.4905779 | -2.2019986 | -1.4075277 |
| H | 6.4463013 | -0.6939190 | -1.3155122 |

*Transition in A1D*

**Calculated in solution with PBE0-D3(BJ)/def2-TZVP COSMO(THF)**

**Position 0**

```
Energy = -1540.961822757
N    -3.3124400    0.9483200   -0.1564300
C    -2.8228800    2.1948300   -0.0442900
N    -1.5520200    2.5377100   -0.2549400
N    -3.6380900    3.2099600    0.3048800
H    -3.1492600    4.0895100    0.3523400
C    -5.0066900    3.2607000    0.5979000
C    -5.8631200    2.1674300    0.5942000
C    -7.2085000    2.3315600    0.9004900
C    -5.5283200    4.5177600    0.9156300
C    -6.8611100    4.6803500    1.2192900
C    -7.7197600    3.5834400    1.2145600
H    -5.4800400    1.1885500    0.3517200
H    -7.8472500    1.4587600    0.8877700
O    -9.0136700    3.8343700    1.5238400
H    -7.2603300    5.6567800    1.4645300
H    -4.8734400    5.3836800    0.9238500
C    -9.9044400    2.7469600    1.5279800
H   -10.8786100    3.1514300    1.7968400
H    -9.6122700    1.9910500    2.2652800
H    -9.9695500    2.2771500    0.5403800
C    -0.7458200    1.5341900   -0.6013500
N    -1.1279100    0.2534700   -0.7421500
C    -2.4043300    0.0669200   -0.5026600
Cl   -2.9650400   -1.5583700   -0.6656300
N     0.5388700    1.8814800   -0.8157300
H     0.6928700    2.8675200   -0.6784500
C     1.6649700    1.1390300   -1.1927900
C     1.6743700   -0.2295400   -1.4287500
C     2.8592000    1.8506400   -1.3364600
C     2.8516300   -0.8673700   -1.8002700
C     4.0340600   -0.1538700   -1.9417000
C     4.0255400    1.2187600   -1.7043700
H     0.7629100   -0.7968900   -1.3233000
H     2.8230000   -1.9341400   -1.9766300
O     5.2235100   -0.6924600   -2.2995300
H     4.9482500    1.7749700   -1.8147000
H     2.8706400    2.9211600   -1.1558500
C     5.2634900   -2.0756000   -2.5470400
H     6.2920400   -2.3076400   -2.8174000
H     4.6017600   -2.3549600   -3.3743300
H     4.9862800   -2.6512400   -1.6570500
```

**Position 1**

```
Energy = -1540.961585929
N    -3.3180160    0.9371760   -0.1584440
C    -2.8251940    2.1857100   -0.0463860
N    -1.5553400    2.5288920   -0.2567380
N    -3.6423940    3.2100700    0.3045420
```

| | | | |
|---|---|---|---|
| H | -3.1484800 | 4.0868580 | 0.3503400 |
| C | -5.0023240 | 3.2636840 | 0.5969600 |
| C | -5.8621720 | 2.1666800 | 0.5934300 |
| C | -7.2053160 | 2.3309380 | 0.8999300 |
| C | -5.5271860 | 4.5232960 | 0.9164220 |
| C | -6.8581900 | 4.6838020 | 1.2199700 |
| C | -7.7172220 | 3.5841380 | 1.2149120 |
| H | -5.4758200 | 1.1891400 | 0.3500340 |
| H | -7.8448420 | 1.4586160 | 0.8875300 |
| O | -9.0094480 | 3.8351080 | 1.5244860 |
| H | -7.2593240 | 5.6592660 | 1.4659320 |
| H | -4.8722940 | 5.3890900 | 0.9245160 |
| C | -9.9031280 | 2.7494300 | 1.5293480 |
| H | -10.8759560 | 3.1565720 | 1.7986120 |
| H | -9.6114920 | 1.9935180 | 2.2666920 |
| H | -9.9690080 | 2.2800720 | 0.5417100 |
| C | -0.7446780 | 1.5365620 | -0.6020840 |
| N | -1.1274960 | 0.2553620 | -0.7431560 |
| C | -2.4037040 | 0.0697060 | -0.5036340 |
| Cl | -2.9648320 | -1.5565440 | -0.6670180 |
| N | 0.5396620 | 1.8829560 | -0.8164120 |
| H | 0.6962780 | 2.8689120 | -0.6794780 |
| C | 1.6650800 | 1.1392760 | -1.1932180 |
| C | 1.6746000 | -0.2295180 | -1.4291200 |
| C | 2.8595680 | 1.8509000 | -1.3366760 |
| C | 2.8518580 | -0.8680240 | -1.8003500 |
| C | 4.0343940 | -0.1544280 | -1.9414980 |
| C | 4.0259400 | 1.2184940 | -1.7042780 |
| H | 0.7636120 | -0.7976540 | -1.3240540 |
| H | 2.8230540 | -1.9348320 | -1.9765520 |
| O | 5.2236520 | -0.6937440 | -2.2989580 |
| H | 4.9483440 | 1.7755340 | -1.8141780 |
| H | 2.8707860 | 2.9213480 | -1.1561220 |
| C | 5.2628500 | -2.0784060 | -2.5464500 |
| H | 6.2913320 | -2.3108520 | -2.8164280 |
| H | 4.6010000 | -2.3565600 | -3.3737340 |
| H | 4.9848820 | -2.6527880 | -1.6561800 |

**Position 2**

Energy = -1540.960739324

| | | | |
|---|---|---|---|
| N | -3.3235920 | 0.9260320 | -0.1604580 |
| C | -2.8275080 | 2.1765900 | -0.0484820 |
| N | -1.5586600 | 2.5200740 | -0.2585360 |
| N | -3.6466980 | 3.2101800 | 0.3042040 |
| H | -3.1477000 | 4.0842060 | 0.3483400 |
| C | -4.9979580 | 3.2666680 | 0.5960200 |
| C | -5.8612240 | 2.1659300 | 0.5926600 |
| C | -7.2021320 | 2.3303160 | 0.8993700 |
| C | -5.5260520 | 4.5288320 | 0.9172140 |
| C | -6.8552700 | 4.6872540 | 1.2206500 |
| C | -7.7146840 | 3.5848360 | 1.2152640 |
| H | -5.4716000 | 1.1897300 | 0.3483480 |
| H | -7.8424340 | 1.4584720 | 0.8872900 |
| O | -9.0052260 | 3.8358460 | 1.5251320 |
| H | -7.2583180 | 5.6617520 | 1.4673340 |
| H | -4.8711480 | 5.3945000 | 0.9251820 |
| C | -9.9018160 | 2.7519000 | 1.5307160 |
| H | -10.8733020 | 3.1617140 | 1.8003840 |
| H | -9.6107140 | 1.9959860 | 2.2681040 |
| H | -9.9684660 | 2.2829940 | 0.5430400 |
| C | -0.7435360 | 1.5389340 | -0.6028180 |
| N | -1.1270820 | 0.2572540 | -0.7441620 |
| C | -2.4030780 | 0.0724920 | -0.5046080 |
| Cl | -2.9646240 | -1.5547180 | -0.6684060 |
| N | 0.5404540 | 1.8844320 | -0.8170940 |
| H | 0.6996860 | 2.8703040 | -0.6805060 |
| C | 1.6651900 | 1.1395220 | -1.1936460 |
| C | 1.6748300 | -0.2294960 | -1.4294900 |
| C | 2.8599360 | 1.8511600 | -1.3368920 |
| C | 2.8520860 | -0.8686780 | -1.8004300 |
| C | 4.0347280 | -0.1549860 | -1.9412960 |
| C | 4.0263400 | 1.2182280 | -1.7041860 |
| H | 0.7643140 | -0.7984180 | -1.3248080 |
| H | 2.8231080 | -1.9355240 | -1.9764740 |
| O | 5.2237940 | -0.6950280 | -2.2983860 |
| H | 4.9484380 | 1.7760980 | -1.8136560 |
| H | 2.8709320 | 2.9215360 | -1.1563940 |
| C | 5.2622100 | -2.0812120 | -2.5458600 |
| H | 6.2906240 | -2.3140640 | -2.8154560 |
| H | 4.6002400 | -2.3581600 | -3.3731380 |
| H | 4.9834840 | -2.6543360 | -1.6553100 |

**Position 3**

Energy = -1540.959303633

| | | | |
|---|---|---|---|
| N | -3.3291680 | 0.9148880 | -0.1624720 |
| C | -2.8298220 | 2.1674700 | -0.0505780 |
| N | -1.5619800 | 2.5112560 | -0.2603340 |
| N | -3.6510020 | 3.2102900 | 0.3038660 |
| H | -3.1469200 | 4.0815540 | 0.3463400 |
| C | -4.9935920 | 3.2696520 | 0.5950800 |
| C | -5.8602760 | 2.1651800 | 0.5918900 |
| C | -7.1989480 | 2.3296940 | 0.8988100 |
| C | -5.5249180 | 4.5343680 | 0.9180060 |
| C | -6.8523500 | 4.6907060 | 1.2213300 |
| C | -7.7121460 | 3.5855340 | 1.2156160 |
| H | -5.4673800 | 1.1903200 | 0.3466620 |
| H | -7.8400260 | 1.4583280 | 0.8870500 |
| O | -9.0010040 | 3.8365840 | 1.5257780 |
| H | -7.2573120 | 5.6642380 | 1.4687360 |
| H | -4.8700020 | 5.3999100 | 0.9258480 |
| C | -9.9005040 | 2.7543700 | 1.5320840 |
| H | -10.8706480 | 3.1668560 | 1.8021560 |
| H | -9.6099360 | 1.9984540 | 2.2695160 |
| H | -9.9679240 | 2.2859160 | 0.5443700 |

```
C     -0.7423940    1.5413060   -0.6035520
N     -1.1266680    0.2591460   -0.7451680
C     -2.4024520    0.0752780   -0.5055820
Cl    -2.9644160   -1.5528920   -0.6697940
N      0.5412460    1.8859080   -0.8177760
H      0.7030940    2.8716960   -0.6815340
C      1.6653000    1.1397680   -1.1940740
C      1.6750600   -0.2294740   -1.4298600
C      2.8603040    1.8514200   -1.3371080
C      2.8523140   -0.8693320   -1.8005100
C      4.0350620   -0.1555440   -1.9410940
C      4.0267400    1.2179620   -1.7040940
H      0.7650160   -0.7991820   -1.3255620
H      2.8231620   -1.9362160   -1.9763960
O      5.2239360   -0.6963120   -2.2978140
H      4.9485320    1.7766620   -1.8131340
H      2.8710780    2.9217240   -1.1566660
C      5.2615700   -2.0840180   -2.5452700
H      6.2899160   -2.3172760   -2.8144840
H      4.5994800   -2.3597600   -3.3725420
H      4.9820860   -2.6558840   -1.6544400
```

**Position 4**

```
Energy = -1540.957266124
N     -3.3347440    0.9037440   -0.1644860
C     -2.8321360    2.1583500   -0.0526740
N     -1.5653000    2.5024380   -0.2621320
N     -3.6553060    3.2104000    0.3035280
H     -3.1461400    4.0789020    0.3443400
C     -4.9892260    3.2726360    0.5941400
C     -5.8593280    2.1644300    0.5911200
C     -7.1957640    2.3290720    0.8982500
C     -5.5237840    4.5399040    0.9187980
C     -6.8494300    4.6941580    1.2220100
C     -7.7096080    3.5862320    1.2159680
H     -5.4631600    1.1909100    0.3449760
H     -7.8376180    1.4581840    0.8868100
O     -8.9967820    3.8373220    1.5264240
H     -7.2563060    5.6667240    1.4701380
H     -4.8688560    5.4053200    0.9265140
C     -9.8991920    2.7568400    1.5334520
H    -10.8679940    3.1719980    1.8039280
H     -9.6091580    2.0009220    2.2709280
H     -9.9673820    2.2888380    0.5457000
C     -0.7412520    1.5436780   -0.6042860
N     -1.1262540    0.2610380   -0.7461740
C     -2.4018260    0.0780640   -0.5065560
Cl    -2.9642080   -1.5510660   -0.6711820
N      0.5420380    1.8873840   -0.8184580
H      0.7065020    2.8730880   -0.6825620
C      1.6654100    1.1400140   -1.1945020
C      1.6752900   -0.2294520   -1.4302300
```

| | | | |
|---|---|---|---|
| C | 2.8606720 | 1.8516800 | -1.3373240 |
| C | 2.8525420 | -0.8699860 | -1.8005900 |
| C | 4.0353960 | -0.1561020 | -1.9408920 |
| C | 4.0271400 | 1.2176960 | -1.7040020 |
| H | 0.7657180 | -0.7999460 | -1.3263160 |
| H | 2.8232160 | -1.9369080 | -1.9763180 |
| O | 5.2240780 | -0.6975960 | -2.2972420 |
| H | 4.9486260 | 1.7772260 | -1.8126120 |
| H | 2.8712240 | 2.9219120 | -1.1569380 |
| C | 5.2609300 | -2.0868240 | -2.5446800 |
| H | 6.2892080 | -2.3204880 | -2.8135120 |
| H | 4.5987200 | -2.3613600 | -3.3719460 |
| H | 4.9806880 | -2.6574320 | -1.6535700 |

**Position 5**

Energy = -1540.954638723

| | | | |
|---|---|---|---|
| N | -3.3403200 | 0.8926000 | -0.1665000 |
| C | -2.8344500 | 2.1492300 | -0.0547700 |
| N | -1.5686200 | 2.4936200 | -0.2639300 |
| N | -3.6596100 | 3.2105100 | 0.3031900 |
| H | -3.1453600 | 4.0762500 | 0.3423400 |
| C | -4.9848600 | 3.2756200 | 0.5932000 |
| C | -5.8583800 | 2.1636800 | 0.5903500 |
| C | -7.1925800 | 2.3284500 | 0.8976900 |
| C | -5.5226500 | 4.5454400 | 0.9195900 |
| C | -6.8465100 | 4.6976100 | 1.2226900 |
| C | -7.7070700 | 3.5869300 | 1.2163200 |
| H | -5.4589400 | 1.1915000 | 0.3432900 |
| H | -7.8352100 | 1.4580400 | 0.8865700 |
| O | -8.9925600 | 3.8380600 | 1.5270700 |
| H | -7.2553000 | 5.6692100 | 1.4715400 |
| H | -4.8677100 | 5.4107300 | 0.9271800 |
| C | -9.8978800 | 2.7593100 | 1.5348200 |
| H | -10.8653400 | 3.1771400 | 1.8057000 |
| H | -9.6083800 | 2.0033900 | 2.2723400 |
| H | -9.9668400 | 2.2917600 | 0.5470300 |
| C | -0.7401100 | 1.5460500 | -0.6050200 |
| N | -1.1258400 | 0.2629300 | -0.7471800 |
| C | -2.4012000 | 0.0808500 | -0.5075300 |
| Cl | -2.9640000 | -1.5492400 | -0.6725700 |
| N | 0.5428300 | 1.8888600 | -0.8191400 |
| H | 0.7099100 | 2.8744800 | -0.6835900 |
| C | 1.6655200 | 1.1402600 | -1.1949300 |
| C | 1.6755200 | -0.2294300 | -1.4306000 |
| C | 2.8610400 | 1.8519400 | -1.3375400 |
| C | 2.8527700 | -0.8706400 | -1.8006700 |
| C | 4.0357300 | -0.1566600 | -1.9406900 |
| C | 4.0275400 | 1.2174300 | -1.7039100 |
| H | 0.7664200 | -0.8007100 | -1.3270700 |
| H | 2.8232700 | -1.9376000 | -1.9762400 |
| O | 5.2242200 | -0.6988800 | -2.2966700 |
| H | 4.9487200 | 1.7777900 | -1.8120900 |

```
H      2.8713700    2.9221000   -1.1572100
C      5.2602900   -2.0896300   -2.5440900
H      6.2885000   -2.3237000   -2.8125400
H      4.5979600   -2.3629600   -3.3713500
H      4.9792900   -2.6589800   -1.6527000
```

*Rotation around the dihedral angle shown in figure S26 and figure S27 for A1M*

**Calculated in solution with PBE0-D3(BJ)/def2-TZVP COSMO(THF)**

**<u>Angle 90°</u>**

```
Energy = -1599.730932073
C      4.7440930   -1.2150160   -4.7268730
O      4.8848340   -0.0306240   -3.9673520
C      3.8285130    0.3678580   -3.2245640
C      2.6148190   -0.3049070   -3.1548770
C      1.5807060    0.1792250   -2.3622550
C      1.7500780    1.3460480   -1.6277820
C      2.9729910    2.0187540   -1.7019110
C      3.9976840    1.5397220   -2.4869010
N      0.7786410    1.9250240   -0.7971810
H      1.0602700    2.7775810   -0.3352420
C     -0.4683970    1.5585310   -0.4930590
N     -0.5820810    0.7500330    0.5855120
C     -1.8051090    0.4023640    0.8679490
N     -2.9181800    0.7479340    0.2276230
C     -2.6719270    1.5367040   -0.8041090
N     -1.5125410    1.9780470   -1.2235020
Cl    -2.0208520   -0.6263860    2.2290940
Cl    -4.0472680    2.0433830   -1.7038900
H      0.6459070   -0.3573610   -2.3213280
H      2.4510370   -1.2159310   -3.7142510
H      4.9434730    2.0656100   -2.5401240
H      3.1184440    2.9314560   -1.1332680
H      5.6908210   -1.3574680   -5.2437340
H      3.9392110   -1.1208220   -5.4620500
H      4.5482930   -2.0776680   -4.0828730
```

**<u>Angle 105°</u>**

```
Energy = -1599.734521702
C      4.7440930   -1.2150160   -4.7268730
O      4.8848340   -0.0306240   -3.9673520
C      3.8285130    0.3678580   -3.2245640
C      2.6148190   -0.3049070   -3.1548770
C      1.5807060    0.1792250   -2.3622550
C      1.7500780    1.3460480   -1.6277820
C      2.9729910    2.0187540   -1.7019110
C      3.9976840    1.5397220   -2.4869010
N      0.7786410    1.9250240   -0.7971810
H      1.0602700    2.7775810   -0.3352420
C     -0.4683970    1.5585310   -0.4930590
N     -0.6254440    1.0262940    0.7406220
```

| | | | |
|---|---|---|---|
| C | -1.8478230 | 0.6744600 | 1.0207000 |
| N | -2.9232790 | 0.7803240 | 0.2457470 |
| C | -2.6369160 | 1.3136560 | -0.9293370 |
| N | -1.4718810 | 1.7190810 | -1.3688560 |
| Cl | -2.1175530 | -0.0102790 | 2.5750380 |
| Cl | -3.9612770 | 1.4956160 | -2.0113940 |
| H | 0.6459070 | -0.3573610 | -2.3213280 |
| H | 2.4510370 | -1.2159310 | -3.7142510 |
| H | 4.9434730 | 2.0656100 | -2.5401240 |
| H | 3.1184440 | 2.9314560 | -1.1332680 |
| H | 5.6908210 | -1.3574680 | -5.2437340 |
| H | 3.9392110 | -1.1208220 | -5.4620500 |
| H | 4.5482930 | -2.0776680 | -4.0828730 |

**Angle 120°**

Energy = -1599.743078795

| | | | |
|---|---|---|---|
| C | 4.7440930 | -1.2150160 | -4.7268730 |
| O | 4.8848340 | -0.0306240 | -3.9673520 |
| C | 3.8285130 | 0.3678580 | -3.2245640 |
| C | 2.6148190 | -0.3049070 | -3.1548770 |
| C | 1.5807060 | 0.1792250 | -2.3622550 |
| C | 1.7500780 | 1.3460480 | -1.6277820 |
| C | 2.9729910 | 2.0187540 | -1.7019110 |
| C | 3.9976840 | 1.5397220 | -2.4869010 |
| N | 0.7786410 | 1.9250240 | -0.7971810 |
| H | 1.0602700 | 2.7775810 | -0.3352420 |
| C | -0.4683970 | 1.5585310 | -0.4930590 |
| N | -0.6946420 | 1.3280880 | 0.8205690 |
| C | -1.9159810 | 0.9717000 | 1.0994210 |
| N | -2.9314020 | 0.8156930 | 0.2550600 |
| C | -2.5810460 | 1.0699940 | -0.9938800 |
| N | -1.4070080 | 1.4361920 | -1.4437530 |
| Cl | -2.2718720 | 0.6627780 | 2.7533550 |
| Cl | -3.8240660 | 0.8972340 | -2.1698670 |
| H | 0.6459070 | -0.3573610 | -2.3213280 |
| H | 2.4510370 | -1.2159310 | -3.7142510 |
| H | 4.9434730 | 2.0656100 | -2.5401240 |
| H | 3.1184440 | 2.9314560 | -1.1332680 |
| H | 5.6908210 | -1.3574680 | -5.2437340 |
| H | 3.9392110 | -1.1208220 | -5.4620500 |
| H | 4.5482930 | -2.0776680 | -4.0828730 |

**Angle 122.5°**

Energy = -1599.744705393

| | | | |
|---|---|---|---|
| O | 4.8848340 | -0.0306240 | -3.9673520 |
| C | 3.8285130 | 0.3678580 | -3.2245640 |
| C | 2.6148190 | -0.3049070 | -3.1548770 |
| C | 1.5807060 | 0.1792250 | -2.3622550 |
| C | 1.7500780 | 1.3460480 | -1.6277820 |
| C | 2.9729910 | 2.0187540 | -1.7019110 |
| C | 3.9976840 | 1.5397220 | -2.4869010 |
| N | 0.7786410 | 1.9250240 | -0.7971820 |
| C | -0.4683980 | 1.5585310 | -0.4930600 |
| N | -0.7083810 | 1.3794390 | 0.8261230 |
| N | -1.3941320 | 1.3880600 | -1.4489540 |
| C | -2.5699550 | 1.0285360 | -0.9983660 |
| N | -2.9330150 | 0.8217100 | 0.2556990 |
| C | -1.9295130 | 1.0222760 | 1.1048870 |
| C | 4.7440930 | -1.2150160 | -4.7268730 |
| Cl | -3.7968290 | 0.7954220 | -2.1808760 |
| Cl | -2.3025100 | 0.7773010 | 2.7657440 |
| H | 1.0602690 | 2.7775810 | -0.3352420 |
| H | 0.6459070 | -0.3573610 | -2.3213280 |
| H | 2.4510370 | -1.2159310 | -3.7142510 |
| H | 4.9434730 | 2.0656100 | -2.5401240 |
| H | 3.1184440 | 2.9314560 | -1.1332680 |
| H | 5.6908210 | -1.3574680 | -5.2437340 |
| H | 3.9392110 | -1.1208220 | -5.4620500 |
| H | 4.5482930 | -2.0776680 | -4.0828730 |

**Angle 125°**

Energy = -1599.746384722

| | | | |
|---|---|---|---|
| O | 4.8848340 | -0.0306240 | -3.9673520 |
| C | 3.8285130 | 0.3678580 | -3.2245640 |
| C | 2.6148190 | -0.3049070 | -3.1548770 |
| C | 1.5807060 | 0.1792250 | -2.3622550 |
| C | 1.7500780 | 1.3460480 | -1.6277820 |
| C | 2.9729910 | 2.0187540 | -1.7019110 |
| C | 3.9976840 | 1.5397220 | -2.4869010 |
| N | 0.7786410 | 1.9250240 | -0.7971820 |
| C | -0.4683980 | 1.5585310 | -0.4930600 |
| N | -0.7226810 | 1.4308300 | 0.8294150 |
| N | -1.3807270 | 1.3398900 | -1.4520340 |
| C | -2.5584100 | 0.9870440 | -1.0010230 |
| N | -2.9346930 | 0.8277300 | 0.2560750 |
| C | -1.9435980 | 1.0728910 | 1.1081270 |
| C | 4.7440930 | -1.2150160 | -4.7268730 |
| Cl | -3.7684740 | 0.6935280 | -2.1873980 |
| Cl | -2.3344010 | 0.8919130 | 2.7730910 |
| H | 1.0602690 | 2.7775810 | -0.3352420 |
| H | 0.6459070 | -0.3573610 | -2.3213280 |
| H | 2.4510370 | -1.2159310 | -3.7142510 |
| H | 4.9434730 | 2.0656100 | -2.5401240 |
| H | 3.1184440 | 2.9314560 | -1.1332680 |
| H | 5.6908210 | -1.3574680 | -5.2437340 |
| H | 3.9392110 | -1.1208220 | -5.4620500 |
| H | 4.5482930 | -2.0776680 | -4.0828730 |

**Angle 127.5°**

Energy = -1599.748076837

| | | | |
|---|---|---|---|
| O | 4.8848340 | -0.0306240 | -3.9673520 |
| C | 3.8285130 | 0.3678580 | -3.2245640 |
| C | 2.6148190 | -0.3049070 | -3.1548770 |
| C | 1.5807060 | 0.1792250 | -2.3622550 |
| C | 1.7500780 | 1.3460480 | -1.6277820 |
| C | 2.9729910 | 2.0187540 | -1.7019110 |
| C | 3.9976840 | 1.5397220 | -2.4869010 |
| N | 0.7786410 | 1.9250240 | -0.7971820 |
| C | -0.4683980 | 1.5585310 | -0.4930600 |
| N | -0.7375180 | 1.4821640 | 0.8304400 |
| N | -1.3668180 | 1.2917730 | -1.4529870 |
| C | -2.5464310 | 0.9455980 | -1.0018500 |
| N | -2.9364330 | 0.8337440 | 0.2561860 |
| C | -1.9582110 | 1.1234490 | 1.1091330 |
| C | 4.7440930 | -1.2150160 | -4.7268730 |
| Cl | -3.7390550 | 0.5917470 | -2.1894220 |
| Cl | -2.3674900 | 1.0063980 | 2.7753790 |
| H | 1.0602690 | 2.7775810 | -0.3352420 |
| H | 0.6459070 | -0.3573610 | -2.3213280 |
| H | 2.4510370 | -1.2159310 | -3.7142510 |
| H | 4.9434730 | 2.0656100 | -2.5401240 |
| H | 3.1184440 | 2.9314560 | -1.1332680 |
| H | 5.6908210 | -1.3574680 | -5.2437340 |
| H | 3.9392110 | -1.1208220 | -5.4620500 |
| H | 4.5482930 | -2.0776680 | -4.0828730 |

**Angle 130°**

Energy = -1599.749771294

| | | | |
|---|---|---|---|
| O | 4.8848340 | -0.0306240 | -3.9673520 |
| C | 3.8285130 | 0.3678580 | -3.2245640 |
| C | 2.6148190 | -0.3049070 | -3.1548770 |
| C | 1.5807060 | 0.1792250 | -2.3622550 |
| C | 1.7500780 | 1.3460480 | -1.6277820 |
| C | 2.9729910 | 2.0187540 | -1.7019110 |
| C | 3.9976840 | 1.5397220 | -2.4869010 |
| N | 0.7786410 | 1.9250240 | -0.7971820 |
| C | -0.4683980 | 1.5585310 | -0.4930600 |
| N | -0.7528630 | 1.5333430 | 0.8291930 |
| N | -1.3524330 | 1.2438000 | -1.4518130 |
| C | -2.5340410 | 0.9042770 | -1.0008430 |
| N | -2.9382330 | 0.8397390 | 0.2560310 |
| C | -1.9733250 | 1.1738550 | 1.1079020 |
| C | 4.7440930 | -1.2150160 | -4.7268730 |
| Cl | -2.4017110 | 1.1205380 | 2.7726030 |
| Cl | -3.7086280 | 0.4902720 | -2.1869450 |
| H | 1.0602690 | 2.7775810 | -0.3352420 |
| H | 0.6459070 | -0.3573610 | -2.3213280 |
| H | 2.4510370 | -1.2159310 | -3.7142510 |
| H | 4.9434730 | 2.0656100 | -2.5401240 |
| H | 3.1184440 | 2.9314560 | -1.1332680 |
| H | 5.6908210 | -1.3574680 | -5.2437340 |
| H | 3.9392110 | -1.1208220 | -5.4620500 |
| H | 4.5482930 | -2.0776680 | -4.0828730 |

**Angle 131.25°**

```
Energy = -1599.750613414
O      4.8848340   -0.0306240   -3.9673520
C      3.8285130    0.3678580   -3.2245640
C      2.6148190   -0.3049070   -3.1548770
C      1.5807060    0.1792250   -2.3622550
C      1.7500780    1.3460480   -1.6277820
C      2.9729910    2.0187540   -1.7019110
C      3.9976840    1.5397220   -2.4869010
N      0.7786410    1.9250240   -0.7971820
C     -0.4683980    1.5585310   -0.4930600
N     -0.7607170    1.5588450    0.8277200
N     -1.3450710    1.2199000   -1.4504260
C     -2.5277000    0.8836890   -0.9996510
N     -2.9391540    0.8427270    0.2558540
C     -1.9810610    1.1989720    1.1064490
C      4.7440930   -1.2150160   -4.7268730
Cl    -3.6930560    0.4397120   -2.1840180
Cl    -2.4192270    1.1774100    2.7693170
H      1.0602690    2.7775810   -0.3352420
H      0.6459070   -0.3573610   -2.3213280
H      2.4510370   -1.2159310   -3.7142510
H      4.9434730    2.0656100   -2.5401240
H      3.1184440    2.9314560   -1.1332680
H      5.6908210   -1.3574680   -5.2437340
H      3.9392110   -1.1208220   -5.4620500
H      4.5482930   -2.0776680   -4.0828730
```

**Angle 132.5°**

```
Energy = -1599.751467901
O      4.8848340   -0.0306240   -3.9673520
C      3.8285130    0.3678580   -3.2245640
C      2.6148190   -0.3049070   -3.1548770
C      1.5807060    0.1792250   -2.3622550
C      1.7500780    1.3460480   -1.6277820
C      2.9729910    2.0187540   -1.7019110
C      3.9976840    1.5397220   -2.4869010
N      0.7786410    1.9250240   -0.7971820
C     -0.4683980    1.5585310   -0.4930600
N     -0.7686870    1.5842710    0.8256790
N     -1.3375990    1.1960660   -1.4485110
C     -2.5212650    0.8631600   -0.9980040
N     -2.9400890    0.8457050    0.2556100
C     -1.9889110    1.2240140    1.1044380
C      4.7440930   -1.2150160   -4.7268730
Cl    -3.6772520    0.3892980   -2.1799690
Cl    -2.4370020    1.2341170    2.7647690
H      1.0602690    2.7775810   -0.3352420
H      0.6459070   -0.3573610   -2.3213280
H      2.4510370   -1.2159310   -3.7142510
H      4.9434730    2.0656100   -2.5401240
```

| | | | |
|---|---|---|---|
| H | 3.1184440 | 2.9314560 | -1.1332680 |
| H | 5.6908210 | -1.3574680 | -5.2437340 |
| H | 3.9392110 | -1.1208220 | -5.4620500 |
| H | 4.5482930 | -2.0776680 | -4.0828730 |

**Angle 133.75°**

Energy = -1599.752300162

| | | | |
|---|---|---|---|
| O | 4.8848340 | -0.0306240 | -3.9673520 |
| C | 3.8285130 | 0.3678580 | -3.2245640 |
| C | 2.6148190 | -0.3049070 | -3.1548770 |
| C | 1.5807060 | 0.1792250 | -2.3622550 |
| C | 1.7500780 | 1.3460480 | -1.6277820 |
| C | 2.9729910 | 2.0187540 | -1.7019110 |
| C | 3.9976840 | 1.5397220 | -2.4869010 |
| N | 0.7786410 | 1.9250240 | -0.7971810 |
| C | -0.4683980 | 1.5585310 | -0.4930600 |
| N | -0.7767680 | 1.6096090 | 0.8230730 |
| N | -1.3300220 | 1.1723160 | -1.4460650 |
| C | -2.5147390 | 0.8427030 | -0.9958990 |
| N | -2.9410360 | 0.8486730 | 0.2553010 |
| C | -1.9968710 | 1.2489690 | 1.1018700 |
| C | 4.7440930 | -1.2150160 | -4.7268730 |
| Cl | -2.4550250 | 1.2906270 | 2.7589610 |
| Cl | -3.6612260 | 0.3390600 | -2.1748000 |
| H | 1.0602700 | 2.7775810 | -0.3352420 |
| H | 0.6459070 | -0.3573610 | -2.3213280 |
| H | 2.4510370 | -1.2159310 | -3.7142510 |
| H | 4.9434730 | 2.0656100 | -2.5401240 |
| H | 3.1184440 | 2.9314560 | -1.1332680 |
| H | 5.6908210 | -1.3574680 | -5.2437340 |
| H | 3.9392110 | -1.1208220 | -5.4620500 |
| H | 4.5482930 | -2.0776680 | -4.0828730 |

**Angle 135°**

Energy = -1599.753125900

| | | | |
|---|---|---|---|
| C | 4.7440930 | -1.2150160 | -4.7268730 |
| O | 4.8848340 | -0.0306240 | -3.9673520 |
| C | 3.8285130 | 0.3678580 | -3.2245640 |
| C | 2.6148190 | -0.3049070 | -3.1548770 |
| C | 1.5807060 | 0.1792250 | -2.3622550 |
| C | 1.7500780 | 1.3460480 | -1.6277820 |
| C | 2.9729910 | 2.0187540 | -1.7019110 |
| C | 3.9976840 | 1.5397220 | -2.4869010 |
| N | 0.7786410 | 1.9250240 | -0.7971810 |
| H | 1.0602700 | 2.7775810 | -0.3352420 |
| C | -0.4683970 | 1.5585310 | -0.4930590 |
| N | -0.7849580 | 1.6348480 | 0.8199040 |
| C | -2.0049370 | 1.2738260 | 1.0987480 |
| N | -2.9419960 | 0.8516280 | 0.2549270 |
| C | -2.5081260 | 0.8223240 | -0.9933390 |
| N | -1.3223430 | 1.1486580 | -1.4430890 |
| Cl | -2.4732900 | 1.3469150 | 2.7518940 |
| Cl | -3.6449850 | 0.2890150 | -2.1685090 |
| H | 0.6459070 | -0.3573610 | -2.3213280 |
| H | 2.4510370 | -1.2159310 | -3.7142510 |
| H | 4.9434730 | 2.0656100 | -2.5401240 |
| H | 3.1184440 | 2.9314560 | -1.1332680 |
| H | 5.6908210 | -1.3574680 | -5.2437340 |
| H | 3.9392110 | -1.1208220 | -5.4620500 |
| H | 4.5482930 | -2.0776680 | -4.0828730 |

**Angle 135.635°**

Energy = -1599.753540271

| | | | |
|---|---|---|---|
| O | 4.8848340 | -0.0306240 | -3.9673520 |
| C | 3.8285130 | 0.3678580 | -3.2245640 |
| C | 2.6148190 | -0.3049070 | -3.1548770 |
| C | 1.5807060 | 0.1792250 | -2.3622550 |
| C | 1.7500780 | 1.3460480 | -1.6277820 |
| C | 2.9729910 | 2.0187540 | -1.7019110 |
| C | 3.9976840 | 1.5397220 | -2.4869010 |
| N | 0.7786410 | 1.9250240 | -0.7971810 |
| C | -0.4683980 | 1.5585310 | -0.4930610 |
| N | -0.7890960 | 1.6474270 | 0.8181070 |
| N | -1.3184670 | 1.1368680 | -1.4414070 |
| C | -2.5047870 | 0.8121680 | -0.9918920 |
| N | -2.9424820 | 0.8531010 | 0.2547090 |
| C | -2.0090130 | 1.2862140 | 1.0969750 |
| C | 4.7440930 | -1.2150160 | -4.7268730 |
| Cl | -3.6367840 | 0.2640760 | -2.1649520 |
| Cl | -2.4825170 | 1.3749670 | 2.7478860 |
| H | 1.0602700 | 2.7775810 | -0.3352420 |
| H | 0.6459070 | -0.3573610 | -2.3213280 |
| H | 2.4510370 | -1.2159310 | -3.7142510 |
| H | 4.9434730 | 2.0656100 | -2.5401240 |
| H | 3.1184440 | 2.9314560 | -1.1332680 |
| H | 5.6908210 | -1.3574680 | -5.2437340 |
| H | 3.9392110 | -1.1208220 | -5.4620500 |
| H | 4.5482930 | -2.0776680 | -4.0828730 |

**Angle 136.25°**

Energy = -1599.753947759

| | | | |
|---|---|---|---|
| O | 4.8848340 | -0.0306240 | -3.9673520 |
| C | 3.8285130 | 0.3678580 | -3.2245640 |
| C | 2.6148190 | -0.3049070 | -3.1548770 |
| C | 1.5807060 | 0.1792250 | -2.3622550 |
| C | 1.7500780 | 1.3460480 | -1.6277820 |
| C | 2.9729910 | 2.0187540 | -1.7019110 |
| C | 3.9976840 | 1.5397220 | -2.4869010 |
| N | 0.7786410 | 1.9250240 | -0.7971810 |
| C | -0.4683980 | 1.5585310 | -0.4930600 |
| N | -0.7932540 | 1.6599770 | 0.8161710 |
| N | -1.3145670 | 1.1251060 | -1.4395880 |
| C | -2.5014290 | 0.8020370 | -0.9903260 |
| N | -2.9429700 | 0.8545730 | 0.2544840 |
| C | -2.0131090 | 1.2985750 | 1.0950690 |
| C | 4.7440930 | -1.2150160 | -4.7268730 |
| Cl | -2.4917910 | 1.4029580 | 2.7435700 |
| Cl | -3.6285370 | 0.2391950 | -2.1611070 |
| H | 1.0602700 | 2.7775810 | -0.3352420 |
| H | 0.6459070 | -0.3573610 | -2.3213280 |
| H | 2.4510370 | -1.2159310 | -3.7142510 |
| H | 4.9434730 | 2.0656100 | -2.5401240 |
| H | 3.1184440 | 2.9314560 | -1.1332680 |
| H | 5.6908210 | -1.3574680 | -5.2437340 |
| H | 3.9392110 | -1.1208220 | -5.4620500 |
| H | 4.5482930 | -2.0776680 | -4.0828730 |

**Angle 137.5°**

Energy = -1599.754754981

| | | | |
|---|---|---|---|
| O | 4.8848340 | -0.0306240 | -3.9673520 |
| C | 3.8285130 | 0.3678580 | -3.2245640 |
| C | 2.6148190 | -0.3049070 | -3.1548770 |
| C | 1.5807060 | 0.1792250 | -2.3622550 |
| C | 1.7500780 | 1.3460480 | -1.6277820 |
| C | 2.9729910 | 2.0187540 | -1.7019110 |
| C | 3.9976840 | 1.5397220 | -2.4869010 |
| N | 0.7786410 | 1.9250240 | -0.7971810 |
| C | -0.4683980 | 1.5585310 | -0.4930600 |
| N | -0.8016500 | 1.6849820 | 0.8118770 |
| N | -1.3066970 | 1.1016680 | -1.4355600 |
| C | -2.4946500 | 0.7818490 | -0.9868590 |
| N | -2.9439540 | 0.8575010 | 0.2539760 |
| C | -2.0213780 | 1.3232020 | 1.0908390 |
| C | 4.7440930 | -1.2150160 | -4.7268730 |
| Cl | -2.5105150 | 1.4587240 | 2.7339970 |
| Cl | -3.6118900 | 0.1896180 | -2.1525910 |
| H | 1.0602700 | 2.7775810 | -0.3352420 |
| H | 0.6459070 | -0.3573610 | -2.3213280 |
| H | 2.4510370 | -1.2159310 | -3.7142510 |
| H | 4.9434730 | 2.0656100 | -2.5401240 |
| H | 3.1184440 | 2.9314560 | -1.1332680 |
| H | 5.6908210 | -1.3574680 | -5.2437340 |
| H | 3.9392110 | -1.1208220 | -5.4620500 |
| H | 4.5482930 | -2.0776680 | -4.0828730 |

**Angle 140°**

Energy = -1599.756328317

| | | | |
|---|---|---|---|
| C | 4.7440930 | -1.2150160 | -4.7268730 |
| O | 4.8848340 | -0.0306240 | -3.9673520 |
| C | 3.8285130 | 0.3678580 | -3.2245640 |
| C | 2.6148190 | -0.3049070 | -3.1548770 |
| C | 1.5807060 | 0.1792250 | -2.3622550 |
| C | 1.7500780 | 1.3460480 | -1.6277820 |
| C | 2.9729910 | 2.0187540 | -1.7019110 |
| C | 3.9976840 | 1.5397220 | -2.4869010 |
| N | 0.7786410 | 1.9250240 | -0.7971820 |
| C | -0.4683980 | 1.5585310 | -0.4930600 |
| N | -0.8187290 | 1.7345750 | 0.8016160 |
| N | -1.2906900 | 1.0551850 | -1.4259360 |
| C | -2.4808620 | 0.7418080 | -0.9785740 |
| N | -2.9459560 | 0.8633080 | 0.2527640 |
| C | -2.0381990 | 1.3720460 | 1.0807280 |
| Cl | -3.5780300 | 0.0912900 | -2.1322390 |
| Cl | -2.5486030 | 1.5693280 | 2.7111130 |
| H | 1.0602680 | 2.7775810 | -0.3352420 |
| H | 0.6459070 | -0.3573610 | -2.3213280 |
| H | 2.4510370 | -1.2159310 | -3.7142510 |
| H | 4.9434730 | 2.0656100 | -2.5401240 |
| H | 3.1184440 | 2.9314560 | -1.1332680 |
| H | 5.6908210 | -1.3574680 | -5.2437340 |
| H | 3.9392110 | -1.1208220 | -5.4620500 |
| H | 4.5482930 | -2.0776680 | -4.0828730 |

## Angle 150°

Energy = -1599.761973619

| | | | |
|---|---|---|---|
| C | 4.7440930 | -1.2150160 | -4.7268730 |
| O | 4.8848340 | -0.0306240 | -3.9673520 |
| C | 3.8285130 | 0.3678580 | -3.2245640 |
| C | 2.6148190 | -0.3049070 | -3.1548770 |
| C | 1.5807060 | 0.1792250 | -2.3622550 |
| C | 1.7500780 | 1.3460480 | -1.6277820 |
| C | 2.9729910 | 2.0187540 | -1.7019110 |
| C | 3.9976840 | 1.5397220 | -2.4869010 |
| N | 0.7786410 | 1.9250240 | -0.7971810 |
| H | 1.0602700 | 2.7775810 | -0.3352420 |
| C | -0.4683970 | 1.5585310 | -0.4930590 |
| N | -0.8902380 | 1.9256700 | 0.7386740 |
| C | -2.1086300 | 1.5602490 | 1.0187270 |
| N | -2.9543380 | 0.8856830 | 0.2453560 |
| C | -2.4231250 | 0.5875230 | -0.9277510 |
| N | -1.2236550 | 0.8760740 | -1.3669100 |
| Cl | -2.7080820 | 1.9955120 | 2.5707550 |
| Cl | -3.4362380 | -0.2875910 | -2.0074130 |
| H | 0.6459070 | -0.3573610 | -2.3213280 |
| H | 2.4510370 | -1.2159310 | -3.7142510 |
| H | 4.9434730 | 2.0656100 | -2.5401240 |
| H | 3.1184440 | 2.9314560 | -1.1332680 |
| H | 5.6908210 | -1.3574680 | -5.2437340 |
| H | 3.9392110 | -1.1208220 | -5.4620500 |
| H | 4.5482930 | -2.0776680 | -4.0828730 |

## Angle 165°

Energy = -1599.767822140

| | | | |
|---|---|---|---|
| C | 4.7440930 | -1.2150160 | -4.7268730 |
| O | 4.8848340 | -0.0306240 | -3.9673520 |
| C | 3.8285130 | 0.3678580 | -3.2245640 |
| C | 2.6148190 | -0.3049070 | -3.1548770 |
| C | 1.5807060 | 0.1792250 | -2.3622550 |
| C | 1.7500780 | 1.3460480 | -1.6277820 |
| C | 2.9729910 | 2.0187540 | -1.7019110 |
| C | 3.9976840 | 1.5397220 | -2.4869010 |
| N | 0.7786410 | 1.9250240 | -0.7971810 |
| H | 1.0602700 | 2.7775810 | -0.3352420 |
| C | -0.4683970 | 1.5585310 | -0.4930590 |
| N | -1.0033070 | 2.1807340 | 0.5824140 |
| C | -2.2199920 | 1.8114500 | 0.8648100 |
| N | -2.9675880 | 0.9155350 | 0.2270010 |
| C | -2.3318360 | 0.3815930 | -0.8015870 |
| N | -1.1176690 | 0.6370170 | -1.2204070 |
| Cl | -2.9602470 | 2.5643660 | 2.2222820 |
| Cl | -3.2120510 | -0.7932900 | -1.6975560 |
| H | 0.6459070 | -0.3573610 | -2.3213280 |
| H | 2.4510370 | -1.2159310 | -3.7142510 |
| H | 4.9434730 | 2.0656100 | -2.5401240 |
| H | 3.1184440 | 2.9314560 | -1.1332680 |
| H | 5.6908210 | -1.3574680 | -5.2437340 |
| H | 3.9392110 | -1.1208220 | -5.4620500 |
| H | 4.5482930 | -2.0776680 | -4.0828730 |

**Angle 179.9404°**

Energy = -1599.769766697

| | | | |
|---|---|---|---|
| N | -2.9807909 | 0.9390715 | 0.2012273 |
| C | -2.3310023 | 2.0096429 | 0.6484583 |
| N | -1.1160203 | 2.3819801 | 0.3627580 |
| C | -0.4683969 | 1.5585307 | -0.4930592 |
| N | -1.0120216 | 0.4484125 | -1.0144878 |
| C | -2.2408342 | 0.2191150 | -0.6242394 |
| Cl | -3.2116211 | 3.0132000 | 1.7324200 |
| Cl | -2.9885740 | -1.1922761 | -1.2620105 |
| N | 0.7786402 | 1.9250239 | -0.7971807 |
| H | 1.0602693 | 2.7775814 | -0.3352424 |
| C | 1.7500781 | 1.3460483 | -1.6277820 |
| C | 1.5807063 | 0.1792245 | -2.3622549 |
| C | 2.9729904 | 2.0187541 | -1.7019112 |
| C | 2.6148186 | -0.3049074 | -3.1548774 |
| C | 3.8285132 | 0.3678585 | -3.2245642 |
| C | 3.9976836 | 1.5397226 | -2.4869012 |
| H | 0.6459069 | -0.3573611 | -2.3213279 |
| H | 2.4510368 | -1.2159307 | -3.7142510 |
| O | 4.8848337 | -0.0306238 | -3.9673521 |
| H | 4.9434726 | 2.0656104 | -2.5401233 |
| H | 3.1184435 | 2.9314559 | -1.1332682 |
| C | 4.7440930 | -1.2150156 | -4.7268728 |
| H | 5.6908213 | -1.3574680 | -5.2437344 |
| H | 3.9392106 | -1.1208218 | -5.4620499 |
| H | 4.5482932 | -2.0776678 | -4.0828731 |

**Angle 195°**

```
Energy = -1599.767785192
C     4.7440930   -1.2150160   -4.7268730
O     4.8848340   -0.0306240   -3.9673520
C     3.8285130    0.3678580   -3.2245640
C     2.6148190   -0.3049070   -3.1548770
C     1.5807060    0.1792250   -2.3622550
C     1.7500780    1.3460480   -1.6277820
C     2.9729910    2.0187540   -1.7019110
C     3.9976840    1.5397220   -2.4869010
N     0.7786410    1.9250240   -0.7971810
H     1.0602700    2.7775810   -0.3352420
C    -0.4683970    1.5585310   -0.4930590
N    -1.2219840    2.5176810    0.0917870
C    -2.4353630    2.1432780    0.3815700
N    -2.9931980    0.9549210    0.1694530
C    -2.1552810    0.1095580   -0.4054610
N    -0.9127030    0.3212530   -0.7604760
Cl   -3.4479450    3.3158580    1.1281100
Cl   -2.7784810   -1.4612960   -0.7247270
H     0.6459070   -0.3573610   -2.3213280
H     2.4510370   -1.2159310   -3.7142510
H     4.9434730    2.0656100   -2.5401240
H     3.1184440    2.9314560   -1.1332680
H     5.6908210   -1.3574680   -5.2437340
H     3.9392110   -1.1208220   -5.4620500
H     4.5482930   -2.0776680   -4.0828730
```

**Angle 210°**

```
Energy = -1599.761911739
C     4.7440930   -1.2150160   -4.7268730
O     4.8848340   -0.0306240   -3.9673520
C     3.8285130    0.3678580   -3.2245640
C     2.6148190   -0.3049070   -3.1548770
C     1.5807060    0.1792250   -2.3622550
C     1.7500780    1.3460480   -1.6277820
C     2.9729910    2.0187540   -1.7019110
C     3.9976840    1.5397220   -2.4869010
N     0.7786410    1.9250240   -0.7971820
C    -0.4683980    1.5585310   -0.4930600
N    -1.3126920    2.5766020   -0.2091470
C    -2.5246960    2.2012910    0.0851770
Cl   -3.6502440    3.4472830    0.4569760
N    -3.0038130    0.9617710    0.1341820
C    -2.0820490    0.0619920   -0.1624950
N    -0.8276920    0.2660640   -0.4783930
Cl   -2.5986460   -1.5780800   -0.1280530
H     1.0602690    2.7775810   -0.3352420
H     0.6459070   -0.3573610   -2.3213280
H     2.4510370   -1.2159310   -3.7142510
```

```
H      4.9434730    2.0656100   -2.5401240
H      3.1184440    2.9314560   -1.1332680
H      5.6908210   -1.3574680   -5.2437340
H      3.9392110   -1.1208220   -5.4620500
H      4.5482930   -2.0776680   -4.0828730
```

**Angle 220°**

```
Energy = -1599.756272753
C      4.7440930   -1.2150160   -4.7268730
O      4.8848340   -0.0306240   -3.9673520
C      3.8285130    0.3678580   -3.2245640
C      2.6148190   -0.3049070   -3.1548770
C      1.5807060    0.1792250   -2.3622550
C      1.7500780    1.3460480   -1.6277820
C      2.9729910    2.0187540   -1.7019110
C      3.9976840    1.5397220   -2.4869010
N      0.7786410    1.9250240   -0.7971820
C     -0.4683980    1.5585310   -0.4930600
N     -1.3618240    2.5713970   -0.4168790
N     -0.7816490    0.2709710   -0.2836790
C     -2.0423830    0.0661980    0.0052220
N     -3.0095590    0.9611250    0.1098420
C     -2.5730820    2.1961520   -0.1194190
Cl    -2.5012420   -1.5677310    0.2838230
Cl    -3.7598190    3.4356870   -0.0063070
H      1.0602680    2.7775810   -0.3352420
H      0.6459070   -0.3573610   -2.3213280
H      2.4510370   -1.2159310   -3.7142510
H      4.9434730    2.0656100   -2.5401240
H      3.1184440    2.9314560   -1.1332680
H      5.6908210   -1.3574680   -5.2437340
H      3.9392110   -1.1208220   -5.4620500
H      4.5482930   -2.0776680   -4.0828730
```

**Angle 222.5°**

```
Energy = -1599.754689549
O      4.8848340   -0.0306240   -3.9673520
C      3.8285130    0.3678580   -3.2245640
C      2.6148190   -0.3049070   -3.1548770
C      1.5807060    0.1792250   -2.3622550
C      1.7500780    1.3460480   -1.6277820
C      2.9729910    2.0187540   -1.7019110
C      3.9976840    1.5397220   -2.4869010
N      0.7786410    1.9250240   -0.7971810
C     -0.4683980    1.5585310   -0.4930600
N     -1.3724600    2.5645080   -0.4687990
N     -0.7716780    0.2774340   -0.2350120
C     -2.0337950    0.0717600    0.0471410
N     -3.0108020    0.9603090    0.1037600
C     -2.5835570    2.1893640   -0.1705530
C      4.7440930   -1.2150160   -4.7268730
```

| | | | |
|---|---|---|---|
| Cl | -3.7835440 | 3.4203290 | -0.1220980 |
| Cl | -2.4801510 | -1.5540670 | 0.3867660 |
| H | 1.0602700 | 2.7775810 | -0.3352420 |
| H | 0.6459070 | -0.3573610 | -2.3213280 |
| H | 2.4510370 | -1.2159310 | -3.7142510 |
| H | 4.9434730 | 2.0656100 | -2.5401240 |
| H | 3.1184440 | 2.9314560 | -1.1332680 |
| H | 5.6908210 | -1.3574680 | -5.2437340 |
| H | 3.9392110 | -1.1208220 | -5.4620500 |
| H | 4.5482930 | -2.0776680 | -4.0828730 |

**Angle 223.75°**

Energy = -1599.753881801

| | | | |
|---|---|---|---|
| O | 4.8848340 | -0.0306240 | -3.9673520 |
| C | 3.8285130 | 0.3678580 | -3.2245640 |
| C | 2.6148190 | -0.3049070 | -3.1548770 |
| C | 1.5807060 | 0.1792250 | -2.3622550 |
| C | 1.7500780 | 1.3460480 | -1.6277820 |
| C | 2.9729910 | 2.0187540 | -1.7019110 |
| C | 3.9976840 | 1.5397220 | -2.4869010 |
| N | 0.7786410 | 1.9250240 | -0.7971810 |
| C | -0.4683980 | 1.5585310 | -0.4930600 |
| N | -1.3775180 | 2.5602330 | -0.4946890 |
| N | -0.7669390 | 0.2814450 | -0.2107440 |
| C | -2.0297110 | 0.0752130 | 0.0680440 |
| N | -3.0113930 | 0.9598030 | 0.1007270 |
| C | -2.5885380 | 2.1851520 | -0.1960530 |
| C | 4.7440930 | -1.2150160 | -4.7268730 |
| Cl | -3.7948240 | 3.4107950 | -0.1798390 |
| Cl | -2.4701240 | -1.5455860 | 0.4380990 |
| H | 1.0602700 | 2.7775810 | -0.3352420 |
| H | 0.6459070 | -0.3573610 | -2.3213280 |
| H | 2.4510370 | -1.2159310 | -3.7142510 |
| H | 4.9434730 | 2.0656100 | -2.5401240 |
| H | 3.1184440 | 2.9314560 | -1.1332680 |
| H | 5.6908210 | -1.3574680 | -5.2437340 |
| H | 3.9392110 | -1.1208220 | -5.4620500 |
| H | 4.5482930 | -2.0776680 | -4.0828730 |

**Angle 224.375°**

Energy = -1599.753472397

| | | | |
|---|---|---|---|
| O | 4.8848340 | -0.0306240 | -3.9673520 |
| C | 3.8285130 | 0.3678580 | -3.2245640 |
| C | 2.6148190 | -0.3049070 | -3.1548770 |
| C | 1.5807060 | 0.1792250 | -2.3622550 |
| C | 1.7500780 | 1.3460480 | -1.6277820 |
| C | 2.9729910 | 2.0187540 | -1.7019110 |
| C | 3.9976840 | 1.5397220 | -2.4869010 |
| N | 0.7786410 | 1.9250240 | -0.7971810 |
| C | -0.4683980 | 1.5585310 | -0.4930610 |
| N | -1.3799810 | 2.5578890 | -0.5076120 |
| N | -0.7646320 | 0.2836450 | -0.1986320 |
| C | -2.0277240 | 0.0771070 | 0.0784770 |
| N | -3.0116810 | 0.9595270 | 0.0992130 |
| C | -2.5909630 | 2.1828420 | -0.2087810 |
| C | 4.7440930 | -1.2150160 | -4.7268730 |
| Cl | -2.4652430 | -1.5409340 | 0.4637200 |
| Cl | -3.8003160 | 3.4055670 | -0.2086590 |
| H | 1.0602700 | 2.7775810 | -0.3352420 |
| H | 0.6459070 | -0.3573610 | -2.3213280 |
| H | 2.4510370 | -1.2159310 | -3.7142510 |
| H | 4.9434730 | 2.0656100 | -2.5401240 |
| H | 3.1184440 | 2.9314560 | -1.1332680 |
| H | 5.6908210 | -1.3574680 | -5.2437340 |
| H | 3.9392110 | -1.1208220 | -5.4620500 |
| H | 4.5482930 | -2.0776680 | -4.0828730 |

## Angle 225°

Energy = -1599.753057688

| | | | |
|---|---|---|---|
| C | 4.7440930 | -1.2150160 | -4.7268730 |
| O | 4.8848340 | -0.0306240 | -3.9673520 |
| C | 3.8285130 | 0.3678580 | -3.2245640 |
| C | 2.6148190 | -0.3049070 | -3.1548770 |
| C | 1.5807060 | 0.1792250 | -2.3622550 |
| C | 1.7500780 | 1.3460480 | -1.6277820 |
| C | 2.9729910 | 2.0187540 | -1.7019110 |
| C | 3.9976840 | 1.5397220 | -2.4869010 |
| N | 0.7786410 | 1.9250240 | -0.7971810 |
| H | 1.0602700 | 2.7775810 | -0.3352420 |
| C | -0.4683970 | 1.5585310 | -0.4930590 |
| N | -1.3823970 | 2.5554050 | -0.5205160 |
| C | -2.5933430 | 2.1803950 | -0.2214890 |
| N | -3.0119630 | 0.9592330 | 0.0977030 |
| C | -2.0257710 | 0.0791100 | 0.0888980 |
| N | -0.7623650 | 0.2859730 | -0.1865350 |
| Cl | -3.8057070 | 3.4000300 | -0.2374390 |
| Cl | -2.4604490 | -1.5360120 | 0.4893090 |
| H | 0.6459070 | -0.3573610 | -2.3213280 |
| H | 2.4510370 | -1.2159310 | -3.7142510 |
| H | 4.9434730 | 2.0656100 | -2.5401240 |
| H | 3.1184440 | 2.9314560 | -1.1332680 |
| H | 5.6908210 | -1.3574680 | -5.2437340 |
| H | 3.9392110 | -1.1208220 | -5.4620500 |
| H | 4.5482930 | -2.0776680 | -4.0828730 |

## Angle 226.25°

Energy = -1599.752225480

| | | | |
|---|---|---|---|
| O | 4.8848340 | -0.0306240 | -3.9673520 |
| C | 3.8285130 | 0.3678580 | -3.2245640 |
| C | 2.6148190 | -0.3049070 | -3.1548770 |
| C | 1.5807060 | 0.1792250 | -2.3622550 |
| C | 1.7500780 | 1.3460480 | -1.6277820 |
| C | 2.9729910 | 2.0187540 | -1.7019110 |
| C | 3.9976840 | 1.5397220 | -2.4869010 |
| N | 0.7786410 | 1.9250240 | -0.7971810 |
| C | -0.4683980 | 1.5585310 | -0.4930600 |
| N | -1.3870980 | 2.5500290 | -0.5462690 |
| N | -0.7579610 | 0.2910170 | -0.1623980 |
| C | -2.0219770 | 0.0834520 | 0.1096890 |
| N | -3.0125130 | 0.9585990 | 0.0946860 |
| C | -2.5979720 | 2.1750980 | -0.2468530 |
| C | 4.7440930 | -1.2150160 | -4.7268730 |
| Cl | -3.8161900 | 3.3880410 | -0.2948730 |
| Cl | -2.4511320 | -1.5253470 | 0.5403670 |
| H | 1.0602700 | 2.7775810 | -0.3352420 |
| H | 0.6459070 | -0.3573610 | -2.3213280 |
| H | 2.4510370 | -1.2159310 | -3.7142510 |
| H | 4.9434730 | 2.0656100 | -2.5401240 |
| H | 3.1184440 | 2.9314560 | -1.1332680 |
| H | 5.6908210 | -1.3574680 | -5.2437340 |
| H | 3.9392110 | -1.1208220 | -5.4620500 |
| H | 4.5482930 | -2.0776680 | -4.0828730 |

**Angle 227.5°**

Energy = -1599.751403034

| | | | |
|---|---|---|---|
| O | 4.8848340 | -0.0306240 | -3.9673520 |
| C | 3.8285130 | 0.3678580 | -3.2245640 |
| C | 2.6148190 | -0.3049070 | -3.1548770 |
| C | 1.5807060 | 0.1792250 | -2.3622550 |
| C | 1.7500780 | 1.3460480 | -1.6277820 |
| C | 2.9729910 | 2.0187540 | -1.7019110 |
| C | 3.9976840 | 1.5397220 | -2.4869010 |
| N | 0.7786410 | 1.9250240 | -0.7971820 |
| C | -0.4683980 | 1.5585310 | -0.4930600 |
| N | -1.3916170 | 2.5441050 | -0.5719350 |
| N | -0.7537290 | 0.2965730 | -0.1383430 |
| C | -2.0183300 | 0.0882360 | 0.1304100 |
| N | -3.0130410 | 0.9579010 | 0.0916800 |
| C | -2.6024220 | 2.1692630 | -0.2721310 |
| C | 4.7440930 | -1.2150160 | -4.7268730 |
| Cl | -2.4421760 | -1.5135980 | 0.5912530 |
| Cl | -3.8262670 | 3.3748300 | -0.3521120 |
| H | 1.0602690 | 2.7775810 | -0.3352420 |
| H | 0.6459070 | -0.3573610 | -2.3213280 |
| H | 2.4510370 | -1.2159310 | -3.7142510 |
| H | 4.9434730 | 2.0656100 | -2.5401240 |
| H | 3.1184440 | 2.9314560 | -1.1332680 |
| H | 5.6908210 | -1.3574680 | -5.2437340 |
| H | 3.9392110 | -1.1208220 | -5.4620500 |
| H | 4.5482930 | -2.0776680 | -4.0828730 |

**Angle 228.75°**

Energy = -1599.750546871

| | | | |
|---|---|---|---|
| O | 4.8848340 | -0.0306240 | -3.9673520 |
| C | 3.8285130 | 0.3678580 | -3.2245640 |
| C | 2.6148190 | -0.3049070 | -3.1548770 |
| C | 1.5807060 | 0.1792250 | -2.3622550 |
| C | 1.7500780 | 1.3460480 | -1.6277820 |
| C | 2.9729910 | 2.0187540 | -1.7019110 |
| C | 3.9976840 | 1.5397220 | -2.4869010 |
| N | 0.7786410 | 1.9250240 | -0.7971820 |
| C | -0.4683980 | 1.5585310 | -0.4930600 |
| N | -1.3959510 | 2.5376380 | -0.5975000 |
| N | -0.7496680 | 0.3026390 | -0.1143820 |
| C | -2.0148320 | 0.0934580 | 0.1510500 |
| N | -3.0135470 | 0.9571390 | 0.0886850 |
| C | -2.6066900 | 2.1628920 | -0.2973090 |
| C | 4.7440930 | -1.2150160 | -4.7268730 |
| Cl | -2.4335850 | -1.5007710 | 0.6419390 |
| Cl | -3.8359330 | 3.3604080 | -0.4091260 |
| H | 1.0602690 | 2.7775810 | -0.3352420 |
| H | 0.6459070 | -0.3573610 | -2.3213280 |
| H | 2.4510370 | -1.2159310 | -3.7142510 |
| H | 4.9434730 | 2.0656100 | -2.5401240 |
| H | 3.1184440 | 2.9314560 | -1.1332680 |
| H | 5.6908210 | -1.3574680 | -5.2437340 |
| H | 3.9392110 | -1.1208220 | -5.4620500 |
| H | 4.5482930 | -2.0776680 | -4.0828730 |

## Angle 230°

Energy = -1599.749706089

| | | | |
|---|---|---|---|
| O | 4.8848340 | -0.0306240 | -3.9673520 |
| C | 3.8285130 | 0.3678580 | -3.2245640 |
| C | 2.6148190 | -0.3049070 | -3.1548770 |
| C | 1.5807060 | 0.1792250 | -2.3622550 |
| C | 1.7500780 | 1.3460480 | -1.6277820 |
| C | 2.9729910 | 2.0187540 | -1.7019110 |
| C | 3.9976840 | 1.5397220 | -2.4869010 |
| N | 0.7786410 | 1.9250240 | -0.7971820 |
| C | -0.4683980 | 1.5585310 | -0.4930600 |
| N | -1.4000980 | 2.5306280 | -0.6229520 |
| N | -0.7457810 | 0.3092110 | -0.0905230 |
| C | -2.0114820 | 0.0991170 | 0.1716010 |
| N | -3.0140310 | 0.9563130 | 0.0857050 |
| C | -2.6107740 | 2.1559870 | -0.3223770 |
| C | 4.7440930 | -1.2150160 | -4.7268730 |
| Cl | -3.8451830 | 3.3447780 | -0.4658920 |
| Cl | -2.4253620 | -1.4868710 | 0.6924060 |
| H | 1.0602690 | 2.7775810 | -0.3352420 |
| H | 0.6459070 | -0.3573610 | -2.3213280 |
| H | 2.4510370 | -1.2159310 | -3.7142510 |
| H | 4.9434730 | 2.0656100 | -2.5401240 |
| H | 3.1184440 | 2.9314560 | -1.1332680 |
| H | 5.6908210 | -1.3574680 | -5.2437340 |
| H | 3.9392110 | -1.1208220 | -5.4620500 |

```
H     4.5482930   -2.0776680   -4.0828730
```

**Angle 232.5°**

```
Energy = -1599.748007886
O     4.8848340   -0.0306240   -3.9673520
C     3.8285130    0.3678580   -3.2245640
C     2.6148190   -0.3049070   -3.1548770
C     1.5807060    0.1792250   -2.3622550
C     1.7500780    1.3460480   -1.6277820
C     2.9729910    2.0187540   -1.7019110
C     3.9976840    1.5397220   -2.4869010
N     0.7786410    1.9250240   -0.7971820
C    -0.4683980    1.5585310   -0.4930600
N    -1.4078270    2.5150010   -0.6734740
N    -0.7385400    0.3238650   -0.0431690
C    -2.0052430    0.1117350    0.2123900
N    -3.0149340    0.9544740    0.0797880
C    -2.6183850    2.1405930   -0.3721350
C     4.7440930   -1.2150160   -4.7268730
Cl   -2.4100420   -1.4558800    0.7925740
Cl   -3.8624200    3.3099300   -0.5785670
H     1.0602690    2.7775810   -0.3352420
H     0.6459070   -0.3573610   -2.3213280
H     2.4510370   -1.2159310   -3.7142510
H     4.9434730    2.0656100   -2.5401240
H     3.1184440    2.9314560   -1.1332680
H     5.6908210   -1.3574680   -5.2437340
H     3.9392110   -1.1208220   -5.4620500
H     4.5482930   -2.0776680   -4.0828730
```

**Angle 235°**

```
Energy = -1599.746320982
O     4.8848340   -0.0306240   -3.9673520
C     3.8285130    0.3678580   -3.2245640
C     2.6148190   -0.3049070   -3.1548770
C     1.5807060    0.1792250   -2.3622550
C     1.7500780    1.3460480   -1.6277820
C     2.9729910    2.0187540   -1.7019110
C     3.9976840    1.5397220   -2.4869010
N     0.7786410    1.9250240   -0.7971820
C    -0.4683980    1.5585310   -0.4930600
N    -1.4147870    2.4972530   -0.7234030
N    -0.7320170    0.3405070    0.0036290
C    -1.9996230    0.1260650    0.2527020
N    -3.0157460    0.9523860    0.0739410
C    -2.6252390    2.1231110   -0.4213100
C     4.7440930   -1.2150160   -4.7268730
Cl   -2.3962430   -1.4206850    0.8915690
Cl   -3.8779450    3.2703520   -0.6899200
H     1.0602690    2.7775810   -0.3352420
H     0.6459070   -0.3573610   -2.3213280
```

| | | | |
|---|---|---|---|
| H | 2.4510370 | -1.2159310 | -3.7142510 |
| H | 4.9434730 | 2.0656100 | -2.5401240 |
| H | 3.1184440 | 2.9314560 | -1.1332680 |
| H | 5.6908210 | -1.3574680 | -5.2437340 |
| H | 3.9392110 | -1.1208220 | -5.4620500 |
| H | 4.5482930 | -2.0776680 | -4.0828730 |

**Angle 237.5°**

Energy = -1599.744634772

| | | | |
|---|---|---|---|
| O | 4.8848340 | -0.0306240 | -3.9673520 |
| C | 3.8285130 | 0.3678580 | -3.2245640 |
| C | 2.6148190 | -0.3049070 | -3.1548770 |
| C | 1.5807060 | 0.1792250 | -2.3622550 |
| C | 1.7500780 | 1.3460480 | -1.6277820 |
| C | 2.9729910 | 2.0187540 | -1.7019110 |
| C | 3.9976840 | 1.5397220 | -2.4869010 |
| N | 0.7786410 | 1.9250240 | -0.7971820 |
| C | -0.4683980 | 1.5585310 | -0.4930600 |
| N | -1.4209670 | 2.4774190 | -0.7726450 |
| N | -0.7262270 | 0.3591050 | 0.0497830 |
| C | -1.9946340 | 0.1420790 | 0.2924580 |
| N | -3.0164670 | 0.9500540 | 0.0681740 |
| C | -2.6313250 | 2.1035730 | -0.4698070 |
| C | 4.7440930 | -1.2150160 | -4.7268730 |
| Cl | -2.3839930 | -1.3813520 | 0.9892000 |
| Cl | -3.8917280 | 3.2261200 | -0.7997400 |
| H | 1.0602690 | 2.7775810 | -0.3352420 |
| H | 0.6459070 | -0.3573610 | -2.3213280 |
| H | 2.4510370 | -1.2159310 | -3.7142510 |
| H | 4.9434730 | 2.0656100 | -2.5401240 |
| H | 3.1184440 | 2.9314560 | -1.1332680 |
| H | 5.6908210 | -1.3574680 | -5.2437340 |
| H | 3.9392110 | -1.1208220 | -5.4620500 |
| H | 4.5482930 | -2.0776680 | -4.0828730 |

**Angle 240°**

Energy = -1599.743002403

| | | | |
|---|---|---|---|
| C | 4.7440930 | -1.2150160 | -4.7268730 |
| O | 4.8848340 | -0.0306240 | -3.9673520 |
| C | 3.8285130 | 0.3678580 | -3.2245640 |
| C | 2.6148190 | -0.3049070 | -3.1548770 |
| C | 1.5807060 | 0.1792250 | -2.3622550 |
| C | 1.7500780 | 1.3460480 | -1.6277820 |
| C | 2.9729910 | 2.0187540 | -1.7019110 |
| C | 3.9976840 | 1.5397220 | -2.4869010 |
| N | 0.7786410 | 1.9250240 | -0.7971810 |
| H | 1.0602700 | 2.7775810 | -0.3352420 |
| C | -0.4683970 | 1.5585310 | -0.4930590 |
| N | -1.4263520 | 2.4555350 | -0.8211040 |
| C | -2.6366280 | 2.0820160 | -0.5175320 |
| N | -3.0170940 | 0.9474800 | 0.0625010 |
| C | -1.9902840 | 0.1597460 | 0.3315840 |
| N | -0.7211780 | 0.3796230 | 0.0952060 |
| Cl | -3.9037420 | 3.1773200 | -0.9078160 |
| Cl | -2.3733120 | -1.3379590 | 1.0852820 |
| H | 0.6459070 | -0.3573610 | -2.3213280 |
| H | 2.4510370 | -1.2159310 | -3.7142510 |
| H | 4.9434730 | 2.0656100 | -2.5401240 |
| H | 3.1184440 | 2.9314560 | -1.1332680 |
| H | 5.6908210 | -1.3574680 | -5.2437340 |
| H | 3.9392110 | -1.1208220 | -5.4620500 |
| H | 4.5482930 | -2.0776680 | -4.0828730 |

**Angle 255°**

Energy = -1599.734478624

| | | | |
|---|---|---|---|
| C | 4.7440930 | -1.2150160 | -4.7268730 |
| O | 4.8848340 | -0.0306240 | -3.9673520 |
| C | 3.8285130 | 0.3678580 | -3.2245640 |
| C | 2.6148190 | -0.3049070 | -3.1548770 |
| C | 1.5807060 | 0.1792250 | -2.3622550 |
| C | 1.7500780 | 1.3460480 | -1.6277820 |
| C | 2.9729910 | 2.0187540 | -1.7019110 |
| C | 3.9976840 | 1.5397220 | -2.4869010 |
| N | 0.7786410 | 1.9250240 | -0.7971810 |
| H | 1.0602700 | 2.7775810 | -0.3352420 |
| C | -0.4683970 | 1.5585310 | -0.4930590 |
| N | -1.4415610 | 2.2837980 | -1.0904260 |
| C | -2.6516010 | 1.9128570 | -0.7827770 |
| N | -3.0188560 | 0.9273140 | 0.0309750 |
| C | -1.9780070 | 0.2984060 | 0.5490260 |
| N | -0.7069370 | 0.5406310 | 0.3476290 |
| Cl | -3.9376690 | 2.7943280 | -1.5084670 |
| Cl | -2.3431720 | -0.9974170 | 1.6192530 |
| H | 0.6459070 | -0.3573610 | -2.3213280 |
| H | 2.4510370 | -1.2159310 | -3.7142510 |
| H | 4.9434730 | 2.0656100 | -2.5401240 |
| H | 3.1184440 | 2.9314560 | -1.1332680 |
| H | 5.6908210 | -1.3574680 | -5.2437340 |
| H | 3.9392110 | -1.1208220 | -5.4620500 |
| H | 4.5482930 | -2.0776680 | -4.0828730 |

**Angle 270°**

Energy = -1599.730931942

| | | | |
|---|---|---|---|
| C | 4.7440930 | -1.2150160 | -4.7268730 |
| O | 4.8848340 | -0.0306240 | -3.9673520 |
| C | 3.8285130 | 0.3678580 | -3.2245640 |
| C | 2.6148190 | -0.3049070 | -3.1548770 |
| C | 1.5807060 | 0.1792250 | -2.3622550 |
| C | 1.7500780 | 1.3460480 | -1.6277820 |
| C | 2.9729910 | 2.0187540 | -1.7019110 |
| C | 3.9976840 | 1.5397220 | -2.4869010 |
| N | 0.7786410 | 1.9250240 | -0.7971810 |
| H | 1.0602700 | 2.7775810 | -0.3352420 |
| C | -0.4683970 | 1.5585310 | -0.4930590 |
| N | -1.4269870 | 2.0518970 | -1.3101280 |
| C | -2.6372400 | 1.6844470 | -0.9991480 |
| N | -3.0171280 | 0.9001090 | 0.0052740 |
| C | -1.9897750 | 0.4856400 | 0.7264050 |
| N | -0.7206120 | 0.7580260 | 0.5535340 |
| Cl | -3.9051740 | 2.2771570 | -1.9984610 |
| Cl | -2.3720820 | -0.5375950 | 2.0548330 |
| H | 0.6459070 | -0.3573610 | -2.3213280 |
| H | 2.4510370 | -1.2159310 | -3.7142510 |
| H | 4.9434730 | 2.0656100 | -2.5401240 |
| H | 3.1184440 | 2.9314560 | -1.1332680 |
| H | 5.6908210 | -1.3574680 | -5.2437340 |
| H | 3.9392110 | -1.1208220 | -5.4620500 |
| H | 4.5482930 | -2.0776680 | -4.0828730 |

*Rotation around the dihedral angle shown in figure S26 and figure S27 for A2M*

**Calculated in solution with PBE0-D3(BJ)/def2-TZVP COSMO(THF)**

**<u>Angle 90°</u>**

Energy = -1714.185615178

| | | | |
|---|---|---|---|
| C | 5.9508180 | 1.3783090 | -2.8369860 |
| O | 4.9149420 | 1.7950210 | -1.9680170 |
| C | 3.6828460 | 1.2811830 | -2.1645310 |
| C | 3.3766480 | 0.3729270 | -3.1654790 |
| C | 2.0654380 | -0.0913560 | -3.2741180 |
| O | 1.8467180 | -0.9727940 | -4.2726840 |
| C | 0.5361010 | -1.4812880 | -4.4343780 |
| C | 1.0660860 | 0.3320730 | -2.4091690 |
| C | 1.4021700 | 1.2473140 | -1.4101240 |
| C | 2.6984990 | 1.7226400 | -1.2824660 |
| N | 0.4809760 | 1.7455530 | -0.4804230 |
| H | 0.8582390 | 2.3997580 | 0.1897990 |
| C | -0.8265870 | 1.5201550 | -0.3041790 |
| N | -1.1197190 | 0.4966370 | 0.5274280 |
| C | -2.3958880 | 0.2868210 | 0.6915860 |
| N | -3.4059750 | 0.9542530 | 0.1431030 |
| C | -2.9875020 | 1.9273700 | -0.6471190 |
| N | -1.7523220 | 2.2687450 | -0.9186070 |
| Cl | -2.8376340 | -1.0023900 | 1.7386310 |
| Cl | -4.2155920 | 2.8567330 | -1.4103680 |
| H | 0.0550090 | -0.0244050 | -2.4915760 |
| H | 4.1152260 | 0.0102300 | -3.8659660 |
| H | 2.9619780 | 2.4330780 | -0.5078380 |
| H | 0.5805390 | -2.1632940 | -5.2808080 |
| H | 0.2108600 | -2.0270060 | -3.5439020 |
| H | -0.1756560 | -0.6791750 | -4.6502280 |
| H | 6.8467840 | 1.9019430 | -2.5104350 |
| H | 5.7291110 | 1.6475200 | -3.8737080 |
| H | 6.1142850 | 0.2989830 | -2.7679400 |

**Angle 105°**

Energy = -1714.188921708

| | | | |
|---|---|---|---|
| C | 5.9508180 | 1.3783090 | -2.8369860 |
| O | 4.9149420 | 1.7950210 | -1.9680170 |
| C | 3.6828460 | 1.2811830 | -2.1645310 |
| C | 3.3766480 | 0.3729270 | -3.1654790 |
| C | 2.0654380 | -0.0913560 | -3.2741180 |
| O | 1.8467180 | -0.9727940 | -4.2726840 |
| C | 0.5361010 | -1.4812880 | -4.4343780 |
| C | 1.0660860 | 0.3320730 | -2.4091690 |
| C | 1.4021700 | 1.2473140 | -1.4101240 |
| C | 2.6984990 | 1.7226400 | -1.2824660 |
| N | 0.4809760 | 1.7455530 | -0.4804230 |
| H | 0.8582390 | 2.3997580 | 0.1897990 |
| C | -0.8265870 | 1.5201550 | -0.3041790 |
| N | -1.1300080 | 0.7285230 | 0.7476480 |
| C | -2.4060910 | 0.5163900 | 0.9094900 |
| N | -3.4073000 | 0.9833770 | 0.1705200 |
| C | -2.9792830 | 1.7416830 | -0.8236130 |
| N | -1.7427240 | 2.0523370 | -1.1241600 |
| Cl | -2.8606630 | -0.4837290 | 2.2310930 |
| Cl | -4.1953130 | 2.3991410 | -1.8451270 |
| H | 0.0550090 | -0.0244050 | -2.4915760 |
| H | 4.1152260 | 0.0102300 | -3.8659660 |
| H | 2.9619780 | 2.4330780 | -0.5078380 |
| H | 0.5805390 | -2.1632940 | -5.2808080 |
| H | 0.2108600 | -2.0270060 | -3.5439020 |
| H | -0.1756560 | -0.6791750 | -4.6502280 |
| H | 6.8467840 | 1.9019430 | -2.5104350 |
| H | 5.7291110 | 1.6475200 | -3.8737080 |
| H | 6.1142850 | 0.2989830 | -2.7679400 |

**Angle 120°**

Energy = -1714.196804867

| | | | |
|---|---|---|---|
| C | 5.9508180 | 1.3783090 | -2.8369860 |
| O | 4.9149420 | 1.7950210 | -1.9680170 |
| C | 3.6828460 | 1.2811830 | -2.1645310 |
| C | 3.3766480 | 0.3729270 | -3.1654790 |
| C | 2.0654380 | -0.0913560 | -3.2741180 |
| O | 1.8467180 | -0.9727940 | -4.2726840 |
| C | 0.5361010 | -1.4812880 | -4.4343780 |
| C | 1.0660860 | 0.3320730 | -2.4091690 |
| C | 1.4021700 | 1.2473140 | -1.4101240 |
| C | 2.6984990 | 1.7226400 | -1.2824660 |
| N | 0.4809760 | 1.7455530 | -0.4804230 |
| H | 0.8582390 | 2.3997580 | 0.1897990 |
| C | -0.8265870 | 1.5201550 | -0.3041790 |
| N | -1.1574470 | 1.0078360 | 0.9012870 |
| C | -2.4332660 | 0.7928830 | 1.0614800 |
| N | -3.4107680 | 1.0183960 | 0.1895810 |
| C | -2.9573230 | 1.5179800 | -0.9467860 |
| N | -1.7171190 | 1.7916600 | -1.2675750 |
| Cl | -2.9220460 | 0.1409880 | 2.5746340 |
| Cl | -4.1411820 | 1.8479080 | -2.1484920 |
| H | 0.0550090 | -0.0244050 | -2.4915760 |
| H | 4.1152260 | 0.0102300 | -3.8659660 |
| H | 2.9619780 | 2.4330780 | -0.5078380 |
| H | 0.5805390 | -2.1632940 | -5.2808080 |
| H | 0.2108600 | -2.0270060 | -3.5439020 |
| H | -0.1756560 | -0.6791750 | -4.6502280 |
| H | 6.8467840 | 1.9019430 | -2.5104350 |
| H | 5.7291110 | 1.6475200 | -3.8737080 |
| H | 6.1142850 | 0.2989830 | -2.7679400 |

**Angle 135°**

Energy = -1714.206203715

| | | | |
|---|---|---|---|
| C | 5.9508180 | 1.3783090 | -2.8369860 |
| O | 4.9149420 | 1.7950210 | -1.9680170 |
| C | 3.6828460 | 1.2811830 | -2.1645310 |
| C | 3.3766480 | 0.3729270 | -3.1654790 |
| C | 2.0654380 | -0.0913560 | -3.2741180 |
| O | 1.8467180 | -0.9727940 | -4.2726840 |
| C | 0.5361010 | -1.4812880 | -4.4343780 |
| C | 1.0660860 | 0.3320730 | -2.4091690 |
| C | 1.4021700 | 1.2473140 | -1.4101240 |
| C | 2.6984990 | 1.7226400 | -1.2824660 |
| N | 0.4809760 | 1.7455530 | -0.4804230 |
| H | 0.8582390 | 2.3997580 | 0.1897990 |
| C | -0.8265870 | 1.5201550 | -0.3041790 |
| N | -1.2001670 | 1.3155430 | 0.9778730 |
| C | -2.4755620 | 1.0974580 | 1.1372000 |
| N | -3.4161410 | 1.0569230 | 0.1989880 |
| C | -2.9231190 | 1.2715050 | -1.0082450 |
| N | -1.6772520 | 1.5044760 | -1.3390780 |
| Cl | -3.0176010 | 0.8291850 | 2.7458430 |
| Cl | -4.0568870 | 1.2406010 | -2.2997900 |
| H | 0.0550090 | -0.0244050 | -2.4915760 |
| H | 4.1152260 | 0.0102300 | -3.8659660 |
| H | 2.9619780 | 2.4330780 | -0.5078380 |
| H | 0.5805390 | -2.1632940 | -5.2808080 |
| H | 0.2108600 | -2.0270060 | -3.5439020 |
| H | -0.1756560 | -0.6791750 | -4.6502280 |
| H | 6.8467840 | 1.9019430 | -2.5104350 |
| H | 5.7291110 | 1.6475200 | -3.8737080 |
| H | 6.1142850 | 0.2989830 | -2.7679400 |

**Angle 137.5°**

Energy = -1714.207739074

| | | | |
|---|---|---|---|
| O | 4.9149420 | 1.7950210 | -1.9680170 |
| C | 3.6828460 | 1.2811830 | -2.1645310 |
| C | 3.3766480 | 0.3729270 | -3.1654790 |
| C | 2.0654380 | -0.0913560 | -3.2741180 |
| O | 1.8467180 | -0.9727940 | -4.2726840 |
| C | 1.0660860 | 0.3320730 | -2.4091690 |
| C | 1.4021700 | 1.2473140 | -1.4101240 |
| C | 2.6984990 | 1.7226400 | -1.2824660 |
| N | 0.4809760 | 1.7455530 | -0.4804230 |
| C | -0.8265870 | 1.5201550 | -0.3041790 |
| N | -1.2085820 | 1.3681290 | 0.9826990 |
| N | -1.6693990 | 1.4553940 | -1.3435860 |
| C | -2.9163810 | 1.2293760 | -1.0121300 |
| N | -3.4171980 | 1.0635000 | 0.1995610 |
| C | -2.4838940 | 1.1495060 | 1.1419620 |
| C | 5.9508180 | 1.3783090 | -2.8369860 |
| C | 0.5361010 | -1.4812880 | -4.4343780 |
| Cl | -4.0402820 | 1.1368050 | -2.3093370 |
| Cl | -3.0364220 | 0.9467940 | 2.7566220 |
| H | 0.8582390 | 2.3997580 | 0.1897990 |
| H | 0.0550090 | -0.0244050 | -2.4915760 |
| H | 4.1152260 | 0.0102300 | -3.8659660 |
| H | 2.9619780 | 2.4330780 | -0.5078380 |
| H | 0.5805390 | -2.1632940 | -5.2808080 |
| H | 0.2108600 | -2.0270060 | -3.5439020 |
| H | -0.1756560 | -0.6791750 | -4.6502280 |
| H | 6.8467840 | 1.9019430 | -2.5104350 |
| H | 5.7291110 | 1.6475200 | -3.8737080 |
| H | 6.1142850 | 0.2989830 | -2.7679400 |

**Angle 140°**

Energy = -1714.209225547

| | | | |
|---|---|---|---|
| O | 4.9149420 | 1.7950210 | -1.9680170 |
| C | 3.6828460 | 1.2811830 | -2.1645310 |
| C | 3.3766480 | 0.3729270 | -3.1654790 |
| C | 2.0654380 | -0.0913560 | -3.2741180 |
| O | 1.8467180 | -0.9727940 | -4.2726840 |
| C | 1.0660860 | 0.3320730 | -2.4091690 |
| C | 1.4021700 | 1.2473140 | -1.4101240 |
| C | 2.6984990 | 1.7226400 | -1.2824660 |
| N | 0.4809760 | 1.7455530 | -0.4804230 |
| C | -0.8265870 | 1.5201550 | -0.3041790 |
| N | -1.2173260 | 1.4208240 | 0.9852180 |
| N | -1.6612390 | 1.4062130 | -1.3459400 |
| C | -2.9093790 | 1.1871640 | -1.0141660 |
| N | -3.4182960 | 1.0700920 | 0.1998470 |
| C | -2.4925500 | 1.2016610 | 1.1444410 |
| C | 5.9508180 | 1.3783090 | -2.8369860 |
| C | 0.5361010 | -1.4812880 | -4.4343780 |
| Cl | -4.0230260 | 1.0327990 | -2.3143310 |
| Cl | -3.0559800 | 1.0646460 | 2.7622420 |
| H | 0.8582390 | 2.3997580 | 0.1897990 |
| H | 0.0550090 | -0.0244050 | -2.4915760 |
| H | 4.1152260 | 0.0102300 | -3.8659660 |
| H | 2.9619780 | 2.4330780 | -0.5078380 |
| H | 0.5805390 | -2.1632940 | -5.2808080 |
| H | 0.2108600 | -2.0270060 | -3.5439020 |
| H | -0.1756560 | -0.6791750 | -4.6502280 |
| H | 6.8467840 | 1.9019430 | -2.5104350 |
| H | 5.7291110 | 1.6475200 | -3.8737080 |
| H | 6.1142850 | 0.2989830 | -2.7679400 |

**Angle 142.5°**

Energy = -1714.210663734

| | | | |
|---|---|---|---|
| O | 4.9149420 | 1.7950210 | -1.9680170 |
| C | 3.6828460 | 1.2811830 | -2.1645310 |
| C | 3.3766480 | 0.3729270 | -3.1654790 |
| C | 2.0654380 | -0.0913560 | -3.2741180 |
| O | 1.8467180 | -0.9727940 | -4.2726840 |
| C | 1.0660860 | 0.3320730 | -2.4091690 |
| C | 1.4021700 | 1.2473140 | -1.4101240 |
| C | 2.6984990 | 1.7226400 | -1.2824660 |
| N | 0.4809760 | 1.7455530 | -0.4804230 |
| C | -0.8265870 | 1.5201550 | -0.3041790 |
| N | -1.2263830 | 1.4735260 | 0.9854240 |
| N | -1.6527860 | 1.3570250 | -1.3461370 |
| C | -2.9021260 | 1.1449450 | -1.0143490 |
| N | -3.4194330 | 1.0766830 | 0.1998430 |
| C | -2.5015160 | 1.2538230 | 1.1446310 |
| C | 5.9508180 | 1.3783090 | -2.8369860 |
| C | 0.5361010 | -1.4812880 | -4.4343780 |
| Cl | -4.0051530 | 0.9287780 | -2.3147610 |
| Cl | -3.0762370 | 1.1825130 | 2.7626910 |
| H | 0.8582390 | 2.3997580 | 0.1897990 |
| H | 0.0550090 | -0.0244050 | -2.4915760 |
| H | 4.1152260 | 0.0102300 | -3.8659660 |
| H | 2.9619780 | 2.4330780 | -0.5078380 |
| H | 0.5805390 | -2.1632940 | -5.2808080 |
| H | 0.2108600 | -2.0270060 | -3.5439020 |
| H | -0.1756560 | -0.6791750 | -4.6502280 |
| H | 6.8467840 | 1.9019430 | -2.5104350 |
| H | 5.7291110 | 1.6475200 | -3.8737080 |
| H | 6.1142850 | 0.2989830 | -2.7679400 |

**Angle 145°**

Energy = -1714.212047353

| | | | |
|---|---|---|---|
| O | 4.9149420 | 1.7950210 | -1.9680170 |
| C | 3.6828460 | 1.2811830 | -2.1645310 |
| C | 3.3766480 | 0.3729270 | -3.1654790 |
| C | 2.0654380 | -0.0913560 | -3.2741180 |
| O | 1.8467180 | -0.9727940 | -4.2726840 |
| C | 1.0660860 | 0.3320730 | -2.4091690 |
| C | 1.4021700 | 1.2473140 | -1.4101240 |
| C | 2.6984990 | 1.7226400 | -1.2824660 |
| N | 0.4809760 | 1.7455530 | -0.4804230 |
| C | -0.8265870 | 1.5201550 | -0.3041790 |

```
N     -1.2357350    1.5261350    0.9833190
N     -1.6440580    1.3079260   -1.3441750
C     -2.8946360    1.1028020   -1.0126800
N     -3.4206060    1.0832630    0.1995500
C     -2.5107740    1.3058930    1.1425330
C      5.9508180    1.3783090   -2.8369860
C      0.5361010   -1.4812880   -4.4343780
Cl    -3.0971550    1.3001690    2.7579680
Cl    -3.9866970    0.8249430   -2.3106280
H      0.8582390    2.3997580    0.1897990
H      0.0550090   -0.0244050   -2.4915760
H      4.1152260    0.0102300   -3.8659660
H      2.9619780    2.4330780   -0.5078380
H      0.5805390   -2.1632940   -5.2808080
H      0.2108600   -2.0270060   -3.5439020
H     -0.1756560   -0.6791750   -4.6502280
H      6.8467840    1.9019430   -2.5104350
H      5.7291110    1.6475200   -3.8737080
H      6.1142850    0.2989830   -2.7679400
```

**Angle 147.5°**

```
Energy = -1714.213333427
O      4.9149420    1.7950210   -1.9680170
C      3.6828460    1.2811830   -2.1645310
C      3.3766480    0.3729270   -3.1654790
C      2.0654380   -0.0913560   -3.2741180
O      1.8467180   -0.9727940   -4.2726840
C      1.0660860    0.3320730   -2.4091690
C      1.4021700    1.2473140   -1.4101240
C      2.6984990    1.7226400   -1.2824660
N      0.4809760    1.7455530   -0.4804230
C     -0.8265870    1.5201550   -0.3041790
N     -1.2453650    1.5785500    0.9789040
N     -1.6350700    1.2590040   -1.3400590
C     -2.8869230    1.0608090   -1.0091620
N     -3.4218140    1.0898150    0.1989680
C     -2.5203080    1.3577690    1.1381490
C      5.9508180    1.3783090   -2.8369860
C      0.5361010   -1.4812880   -4.4343780
Cl    -3.9676910    0.7214830   -2.3019380
Cl    -3.1186950    1.4173940    2.7480820
H      0.8582390    2.3997580    0.1897990
H      0.0550090   -0.0244050   -2.4915760
H      4.1152260    0.0102300   -3.8659660
H      2.9619780    2.4330780   -0.5078380
H      0.5805390   -2.1632940   -5.2808080
H      0.2108600   -2.0270060   -3.5439020
H     -0.1756560   -0.6791750   -4.6502280
H      6.8467840    1.9019430   -2.5104350
H      5.7291110    1.6475200   -3.8737080
H      6.1142850    0.2989830   -2.7679400
```

**Angle 150°**

```
Energy = -1714.214624260
C      5.9508180    1.3783090   -2.8369860
O      4.9149420    1.7950210   -1.9680170
C      3.6828460    1.2811830   -2.1645310
C      3.3766480    0.3729270   -3.1654790
C      2.0654380   -0.0913560   -3.2741180
O      1.8467180   -0.9727940   -4.2726840
C      0.5361010   -1.4812880   -4.4343780
C      1.0660860    0.3320730   -2.4091690
C      1.4021700    1.2473140   -1.4101240
C      2.6984990    1.7226400   -1.2824660
N      0.4809760    1.7455530   -0.4804230
H      0.8582390    2.3997580    0.1897990
C     -0.8265870    1.5201550   -0.3041790
N     -1.2552550    1.6306730    0.9721890
C     -2.5300980    1.4093570    1.1314890
N     -3.4230540    1.0963330    0.1980990
C     -2.8790030    1.0190550   -1.0038020
N     -1.6258400    1.2103580   -1.3337970
Cl    -3.1408150    1.5339650    2.7330530
Cl    -3.9481740    0.6186070   -2.2887090
H      0.0550090   -0.0244050   -2.4915760
H      4.1152260    0.0102300   -3.8659660
H      2.9619780    2.4330780   -0.5078380
H      0.5805390   -2.1632940   -5.2808080
H      0.2108600   -2.0270060   -3.5439020
H     -0.1756560   -0.6791750   -4.6502280
H      6.8467840    1.9019430   -2.5104350
H      5.7291110    1.6475200   -3.8737080
H      6.1142850    0.2989830   -2.7679400
```

**Angle 152.5°**

```
Energy = -1714.215758501
O      4.9149420    1.7950210   -1.9680170
C      3.6828460    1.2811830   -2.1645310
C      3.3766480    0.3729270   -3.1654790
C      2.0654380   -0.0913560   -3.2741180
O      1.8467180   -0.9727940   -4.2726840
C      1.0660860    0.3320730   -2.4091690
C      1.4021700    1.2473140   -1.4101240
C      2.6984990    1.7226400   -1.2824660
N      0.4809760    1.7455530   -0.4804230
C     -0.8265870    1.5201550   -0.3041790
N     -1.2653850    1.6824010    0.9631880
N     -1.6163860    1.1620780   -1.3253990
C     -2.8708900    0.9776120   -0.9966090
N     -3.4243240    1.1027960    0.1969440
C     -2.5401260    1.4605520    1.1225660
C      5.9508180    1.3783090   -2.8369860
C      0.5361010   -1.4812880   -4.4343780
```

| | | | |
|---|---|---|---|
| Cl | -3.1634720 | 1.6496500 | 2.7129090 |
| Cl | -3.9281820 | 0.5165030 | -2.2709640 |
| H | 0.8582390 | 2.3997580 | 0.1897990 |
| H | 0.0550090 | -0.0244050 | -2.4915760 |
| H | 4.1152260 | 0.0102300 | -3.8659660 |
| H | 2.9619780 | 2.4330780 | -0.5078380 |
| H | 0.5805390 | -2.1632940 | -5.2808080 |
| H | 0.2108600 | -2.0270060 | -3.5439020 |
| H | -0.1756560 | -0.6791750 | -4.6502280 |
| H | 6.8467840 | 1.9019430 | -2.5104350 |
| H | 5.7291110 | 1.6475200 | -3.8737080 |
| H | 6.1142850 | 0.2989830 | -2.7679400 |

**<u>Angle 155°</u>**

Energy = -1714.216807871

| | | | |
|---|---|---|---|
| O | 4.9149420 | 1.7950210 | -1.9680170 |
| C | 3.6828460 | 1.2811830 | -2.1645310 |
| C | 3.3766480 | 0.3729270 | -3.1654790 |
| C | 2.0654380 | -0.0913560 | -3.2741180 |
| O | 1.8467180 | -0.9727940 | -4.2726840 |
| C | 1.0660860 | 0.3320730 | -2.4091690 |
| C | 1.4021700 | 1.2473140 | -1.4101240 |
| C | 2.6984990 | 1.7226400 | -1.2824660 |
| N | 0.4809760 | 1.7455530 | -0.4804230 |
| C | -0.8265870 | 1.5201550 | -0.3041790 |
| N | -1.2757370 | 1.7336390 | 0.9519160 |
| N | -1.6067250 | 1.1142540 | -1.3148830 |
| C | -2.8625980 | 0.9365590 | -0.9875970 |
| N | -3.4256220 | 1.1091970 | 0.1955050 |
| C | -2.5503730 | 1.5112620 | 1.1113960 |
| C | 5.9508180 | 1.3783090 | -2.8369860 |
| C | 0.5361010 | -1.4812880 | -4.4343780 |
| Cl | -3.9077510 | 0.4153630 | -2.2487390 |
| Cl | -3.1866250 | 1.7642400 | 2.6876860 |
| H | 0.8582390 | 2.3997580 | 0.1897990 |
| H | 0.0550090 | -0.0244050 | -2.4915760 |
| H | 4.1152260 | 0.0102300 | -3.8659660 |
| H | 2.9619780 | 2.4330780 | -0.5078380 |
| H | 0.5805390 | -2.1632940 | -5.2808080 |
| H | 0.2108600 | -2.0270060 | -3.5439020 |
| H | -0.1756560 | -0.6791750 | -4.6502280 |
| H | 6.8467840 | 1.9019430 | -2.5104350 |
| H | 5.7291110 | 1.6475200 | -3.8737080 |
| H | 6.1142850 | 0.2989830 | -2.7679400 |

**<u>Angle 165°</u>**

Energy = -1714.220153330

| | | | |
|---|---|---|---|
| C | 5.9508180 | 1.3783090 | -2.8369860 |
| O | 4.9149420 | 1.7950210 | -1.9680170 |
| C | 3.6828460 | 1.2811830 | -2.1645310 |
| C | 3.3766480 | 0.3729270 | -3.1654790 |
| C | 2.0654380 | -0.0913560 | -3.2741180 |
| O | 1.8467180 | -0.9727940 | -4.2726840 |
| C | 0.5361010 | -1.4812880 | -4.4343780 |
| C | 1.0660860 | 0.3320730 | -2.4091690 |
| C | 1.4021700 | 1.2473140 | -1.4101240 |
| C | 2.6984990 | 1.7226400 | -1.2824660 |
| N | 0.4809760 | 1.7455530 | -0.4804230 |
| H | 0.8582390 | 2.3997580 | 0.1897990 |
| C | -0.8265870 | 1.5201550 | -0.3041790 |
| N | -1.3189580 | 1.9317500 | 0.8846210 |
| C | -2.5931550 | 1.7073270 | 1.0447360 |
| N | -3.4310360 | 1.1339390 | 0.1869740 |
| C | -2.8279800 | 0.7778340 | -0.9337580 |
| N | -1.5663870 | 0.9293490 | -1.2520920 |
| Cl | -3.2832920 | 2.2072960 | 2.5371330 |
| Cl | -3.8224510 | 0.0243140 | -2.1160040 |
| H | 0.0550090 | -0.0244050 | -2.4915760 |
| H | 4.1152260 | 0.0102300 | -3.8659660 |
| H | 2.9619780 | 2.4330780 | -0.5078380 |
| H | 0.5805390 | -2.1632940 | -5.2808080 |
| H | 0.2108600 | -2.0270060 | -3.5439020 |
| H | -0.1756560 | -0.6791750 | -4.6502280 |
| H | 6.8467840 | 1.9019430 | -2.5104350 |
| H | 5.7291110 | 1.6475200 | -3.8737080 |
| H | 6.1142850 | 0.2989830 | -2.7679400 |

## **Angle 179.990646°**

Energy = -1714.222042491

| | | | |
|---|---|---|---|
| C | 5.9508180 | 1.3783090 | -2.8369860 |
| O | 4.9149420 | 1.7950210 | -1.9680170 |
| C | 3.6828460 | 1.2811830 | -2.1645310 |
| C | 3.3766480 | 0.3729270 | -3.1654790 |
| C | 2.0654380 | -0.0913560 | -3.2741180 |
| O | 1.8467180 | -0.9727940 | -4.2726840 |
| C | 0.5361010 | -1.4812880 | -4.4343780 |
| C | 1.0660860 | 0.3320730 | -2.4091690 |
| C | 1.4021700 | 1.2473140 | -1.4101240 |
| C | 2.6984990 | 1.7226400 | -1.2824660 |
| N | 0.4809760 | 1.7455530 | -0.4804230 |
| H | 0.8582390 | 2.3997580 | 0.1897990 |
| C | -0.8265870 | 1.5201550 | -0.3041790 |
| N | -1.3868920 | 2.1981060 | 0.7212610 |
| C | -2.6603950 | 1.9709100 | 0.8829760 |
| N | -3.4395380 | 1.1671600 | 0.1663890 |
| C | -2.7735620 | 0.5644030 | -0.8029870 |
| N | -1.5029840 | 0.6807420 | -1.0996450 |
| Cl | -3.4352260 | 2.8029540 | 2.1717140 |
| Cl | -3.6883700 | -0.5014800 | -1.7936900 |
| H | 0.0550090 | -0.0244050 | -2.4915760 |
| H | 4.1152260 | 0.0102300 | -3.8659660 |
| H | 2.9619780 | 2.4330780 | -0.5078380 |
| H | 0.5805390 | -2.1632940 | -5.2808080 |

```
H      0.2108600   -2.0270060   -3.5439020
H     -0.1756560   -0.6791750   -4.6502280
H      6.8467840    1.9019430   -2.5104350
H      5.7291110    1.6475200   -3.8737080
H      6.1142850    0.2989830   -2.7679400
```

**Angle 195°**

```
Energy = -1714.220162136
C      5.9508180    1.3783090   -2.8369860
O      4.9149420    1.7950210   -1.9680170
C      3.6828460    1.2811830   -2.1645310
C      3.3766480    0.3729270   -3.1654790
C      2.0654380   -0.0913560   -3.2741180
O      1.8467180   -0.9727940   -4.2726840
C      0.5361010   -1.4812880   -4.4343780
C      1.0660860    0.3320730   -2.4091690
C      1.4021700    1.2473140   -1.4101240
C      2.6984990    1.7226400   -1.2824660
N      0.4809760    1.7455530   -0.4804230
H      0.8582390    2.3997580    0.1897990
C     -0.8265870    1.5201550   -0.3041790
N     -1.4545520    2.4120330    0.4928790
C     -2.7273590    2.1825830    0.6568730
N     -3.4479950    1.1937880    0.1376990
C     -2.7193570    0.3929490   -0.6201090
N     -1.4398350    0.4810620   -0.8865080
Cl    -3.5865440    3.2813400    1.6608870
Cl    -3.5548220   -0.9238190   -1.3430150
H      0.0550090   -0.0244050   -2.4915760
H      4.1152260    0.0102300   -3.8659660
H      2.9619780    2.4330780   -0.5078380
H      0.5805390   -2.1632940   -5.2808080
H      0.2108600   -2.0270060   -3.5439020
H     -0.1756560   -0.6791750   -4.6502280
H      6.8467840    1.9019430   -2.5104350
H      5.7291110    1.6475200   -3.8737080
H      6.1142850    0.2989830   -2.7679400
```

**Angle 205°**

```
Energy = -1714.216827283
O      4.9149420    1.7950210   -1.9680170
C      3.6828460    1.2811830   -2.1645310
C      3.3766480    0.3729270   -3.1654790
C      2.0654380   -0.0913560   -3.2741180
O      1.8467180   -0.9727940   -4.2726840
C      1.0660860    0.3320730   -2.4091690
C      1.4021700    1.2473140   -1.4101240
C      2.6984990    1.7226400   -1.2824660
N      0.4809760    1.7455530   -0.4804230
C     -0.8265870    1.5201550   -0.3041790
N     -1.4971440    2.5178790    0.3122510
```

```
N    -1.4000810    0.3822590   -0.7179320
C    -2.6852310    0.3080920   -0.4754470
N    -3.4533130    1.2069230    0.1150450
C    -2.7695120    2.2872960    0.4780660
C     5.9508180    1.3783090   -2.8369860
C     0.5361010   -1.4812880   -4.4343780
Cl   -3.4707470   -1.1328140   -0.9865460
Cl   -3.6817970    3.5180180    1.2568900
H     0.8582390    2.3997580    0.1897990
H     0.0550090   -0.0244050   -2.4915760
H     4.1152260    0.0102300   -3.8659660
H     2.9619780    2.4330780   -0.5078380
H     0.5805390   -2.1632940   -5.2808080
H     0.2108600   -2.0270060   -3.5439020
H    -0.1756560   -0.6791750   -4.6502280
H     6.8467840    1.9019430   -2.5104350
H     5.7291110    1.6475200   -3.8737080
H     6.1142850    0.2989830   -2.7679400
```

**Angle 207.5°**

```
Energy = -1714.215816682
O     4.9149420    1.7950210   -1.9680170
C     3.6828460    1.2811830   -2.1645310
C     3.3766480    0.3729270   -3.1654790
C     2.0654380   -0.0913560   -3.2741180
O     1.8467180   -0.9727940   -4.2726840
C     1.0660860    0.3320730   -2.4091690
C     1.4021700    1.2473140   -1.4101240
C     2.6984990    1.7226400   -1.2824660
N     0.4809760    1.7455530   -0.4804230
C    -0.8265870    1.5201550   -0.3041790
N    -1.5072920    2.5392540    0.2642960
N    -1.3906090    0.3623070   -0.6731770
C    -2.6771000    0.2909540   -0.4370410
N    -3.4545800    1.2095710    0.1090330
C    -2.7795540    2.3084400    0.4305950
C     5.9508180    1.3783090   -2.8369860
C     0.5361010   -1.4812880   -4.4343780
Cl   -3.7044920    3.5658100    1.1496340
Cl   -3.4507150   -1.1750220   -0.8919070
H     0.8582390    2.3997580    0.1897990
H     0.0550090   -0.0244050   -2.4915760
H     4.1152260    0.0102300   -3.8659660
H     2.9619780    2.4330780   -0.5078380
H     0.5805390   -2.1632940   -5.2808080
H     0.2108600   -2.0270060   -3.5439020
H    -0.1756560   -0.6791750   -4.6502280
H     6.8467840    1.9019430   -2.5104350
H     5.7291110    1.6475200   -3.8737080
H     6.1142850    0.2989830   -2.7679400
```

**Angle 210°**

Energy = -1714.214656224

| | | | |
|---|---|---|---|
| C | 5.9508180 | 1.3783090 | -2.8369860 |
| O | 4.9149420 | 1.7950210 | -1.9680170 |
| C | 3.6828460 | 1.2811830 | -2.1645310 |
| C | 3.3766480 | 0.3729270 | -3.1654790 |
| C | 2.0654380 | -0.0913560 | -3.2741180 |
| O | 1.8467180 | -0.9727940 | -4.2726840 |
| C | 0.5361010 | -1.4812880 | -4.4343780 |
| C | 1.0660860 | 0.3320730 | -2.4091690 |
| C | 1.4021700 | 1.2473140 | -1.4101240 |
| C | 2.6984990 | 1.7226400 | -1.2824660 |
| N | 0.4809760 | 1.7455530 | -0.4804230 |
| H | 0.8582390 | 2.3997580 | 0.1897990 |
| C | -0.8265870 | 1.5201550 | -0.3041790 |
| N | -1.5172020 | 2.5585070 | 0.2154000 |
| C | -2.7893610 | 2.3274830 | 0.3821940 |
| N | -3.4558160 | 1.2119520 | 0.1029050 |
| C | -2.6691600 | 0.2755140 | -0.3978790 |
| N | -1.3813600 | 0.3443340 | -0.6275440 |
| Cl | -3.7266530 | 3.6088580 | 1.0402740 |
| Cl | -3.4311540 | -1.2130440 | -0.7954090 |
| H | 0.0550090 | -0.0244050 | -2.4915760 |
| H | 4.1152260 | 0.0102300 | -3.8659660 |
| H | 2.9619780 | 2.4330780 | -0.5078380 |
| H | 0.5805390 | -2.1632940 | -5.2808080 |
| H | 0.2108600 | -2.0270060 | -3.5439020 |
| H | -0.1756560 | -0.6791750 | -4.6502280 |
| H | 6.8467840 | 1.9019430 | -2.5104350 |
| H | 5.7291110 | 1.6475200 | -3.8737080 |
| H | 6.1142850 | 0.2989830 | -2.7679400 |

**Angle 212.5°**

Energy = -1714.213361918

| | | | |
|---|---|---|---|
| O | 4.9149420 | 1.7950210 | -1.9680170 |
| C | 3.6828460 | 1.2811830 | -2.1645310 |
| C | 3.3766480 | 0.3729270 | -3.1654790 |
| C | 2.0654380 | -0.0913560 | -3.2741180 |
| O | 1.8467180 | -0.9727940 | -4.2726840 |
| C | 1.0660860 | 0.3320730 | -2.4091690 |
| C | 1.4021700 | 1.2473140 | -1.4101240 |
| C | 2.6984990 | 1.7226400 | -1.2824660 |
| N | 0.4809760 | 1.7455530 | -0.4804230 |
| C | -0.8265870 | 1.5201550 | -0.3041790 |
| N | -1.5268530 | 2.5756000 | 0.1656600 |
| N | -1.3723520 | 0.3283750 | -0.5811200 |
| C | -2.6614260 | 0.2618020 | -0.3580360 |
| N | -3.4570200 | 1.2140610 | 0.0966750 |
| C | -2.7989120 | 2.3443870 | 0.3329590 |
| C | 5.9508180 | 1.3783090 | -2.8369860 |
| C | 0.5361010 | -1.4812880 | -4.4343780 |
| Cl | -3.4121020 | -1.2468060 | -0.6972370 |
| Cl | -3.7482360 | 3.6470750 | 0.9290260 |
| H | 0.8582390 | 2.3997580 | 0.1897990 |
| H | 0.0550090 | -0.0244050 | -2.4915760 |
| H | 4.1152260 | 0.0102300 | -3.8659660 |
| H | 2.9619780 | 2.4330780 | -0.5078380 |
| H | 0.5805390 | -2.1632940 | -5.2808080 |
| H | 0.2108600 | -2.0270060 | -3.5439020 |
| H | -0.1756560 | -0.6791750 | -4.6502280 |
| H | 6.8467840 | 1.9019430 | -2.5104350 |
| H | 5.7291110 | 1.6475200 | -3.8737080 |
| H | 6.1142850 | 0.2989830 | -2.7679400 |

**Angle 215°**

Energy = -1714.212077863

| | | | |
|---|---|---|---|
| O | 4.9149420 | 1.7950210 | -1.9680170 |
| C | 3.6828460 | 1.2811830 | -2.1645310 |
| C | 3.3766480 | 0.3729270 | -3.1654790 |
| C | 2.0654380 | -0.0913560 | -3.2741180 |
| O | 1.8467180 | -0.9727940 | -4.2726840 |
| C | 1.0660860 | 0.3320730 | -2.4091690 |
| C | 1.4021700 | 1.2473140 | -1.4101240 |
| C | 2.6984990 | 1.7226400 | -1.2824660 |
| N | 0.4809760 | 1.7455530 | -0.4804230 |
| C | -0.8265870 | 1.5201550 | -0.3041790 |
| N | -1.5362280 | 2.5905030 | 0.1151670 |
| N | -1.3636020 | 0.3144630 | -0.5339960 |
| C | -2.6539140 | 0.2498470 | -0.3175940 |
| N | -3.4581890 | 1.2158970 | 0.0903490 |
| C | -2.8081890 | 2.3591240 | 0.2829780 |
| C | 5.9508180 | 1.3783090 | -2.8369860 |
| C | 0.5361010 | -1.4812880 | -4.4343780 |
| Cl | -3.7692010 | 3.6803920 | 0.8160950 |
| Cl | -3.3935960 | -1.2762410 | -0.5975860 |
| H | 0.8582390 | 2.3997580 | 0.1897990 |
| H | 0.0550090 | -0.0244050 | -2.4915760 |
| H | 4.1152260 | 0.0102300 | -3.8659660 |
| H | 2.9619780 | 2.4330780 | -0.5078380 |
| H | 0.5805390 | -2.1632940 | -5.2808080 |
| H | 0.2108600 | -2.0270060 | -3.5439020 |
| H | -0.1756560 | -0.6791750 | -4.6502280 |
| H | 6.8467840 | 1.9019430 | -2.5104350 |
| H | 5.7291110 | 1.6475200 | -3.8737080 |
| H | 6.1142850 | 0.2989830 | -2.7679400 |

**Angle 217.5°**

Energy = -1714.210699040

| | | | |
|---|---|---|---|
| O | 4.9149420 | 1.7950210 | -1.9680170 |
| C | 3.6828460 | 1.2811830 | -2.1645310 |
| C | 3.3766480 | 0.3729270 | -3.1654790 |
| C | 2.0654380 | -0.0913560 | -3.2741180 |
| O | 1.8467180 | -0.9727940 | -4.2726840 |
| C | 1.0660860 | 0.3320730 | -2.4091690 |
| C | 1.4021700 | 1.2473140 | -1.4101240 |
| C | 2.6984990 | 1.7226400 | -1.2824660 |
| N | 0.4809760 | 1.7455530 | -0.4804230 |
| C | -0.8265870 | 1.5201550 | -0.3041790 |
| N | -1.5453090 | 2.6031860 | 0.0640170 |
| N | -1.3551260 | 0.3026210 | -0.4862570 |
| C | -2.6466370 | 0.2396680 | -0.2766220 |
| N | -3.4593210 | 1.2174540 | 0.0839450 |
| C | -2.8171750 | 2.3716630 | 0.2323490 |
| C | 5.9508180 | 1.3783090 | -2.8369860 |
| C | 0.5361010 | -1.4812880 | -4.4343780 |
| Cl | -3.3756700 | -1.3012980 | -0.4966330 |
| Cl | -3.7895080 | 3.7087450 | 0.7016960 |
| H | 0.8582390 | 2.3997580 | 0.1897990 |
| H | 0.0550090 | -0.0244050 | -2.4915760 |
| H | 4.1152260 | 0.0102300 | -3.8659660 |
| H | 2.9619780 | 2.4330780 | -0.5078380 |
| H | 0.5805390 | -2.1632940 | -5.2808080 |
| H | 0.2108600 | -2.0270060 | -3.5439020 |
| H | -0.1756560 | -0.6791750 | -4.6502280 |
| H | 6.8467840 | 1.9019430 | -2.5104350 |
| H | 5.7291110 | 1.6475200 | -3.8737080 |
| H | 6.1142850 | 0.2989830 | -2.7679400 |

**Angle 220°**

Energy = -1714.209268896

| | | | |
|---|---|---|---|
| O | 4.9149420 | 1.7950210 | -1.9680170 |
| C | 3.6828460 | 1.2811830 | -2.1645310 |
| C | 3.3766480 | 0.3729270 | -3.1654790 |
| C | 2.0654380 | -0.0913560 | -3.2741180 |
| O | 1.8467180 | -0.9727940 | -4.2726840 |
| C | 1.0660860 | 0.3320730 | -2.4091690 |
| C | 1.4021700 | 1.2473140 | -1.4101240 |
| C | 2.6984990 | 1.7226400 | -1.2824660 |
| N | 0.4809760 | 1.7455530 | -0.4804230 |
| C | -0.8265870 | 1.5201550 | -0.3041790 |
| N | -1.5540780 | 2.6136250 | 0.0123100 |
| N | -1.3469410 | 0.2928740 | -0.4379980 |
| C | -2.6396100 | 0.2312880 | -0.2352040 |
| N | -3.4604140 | 1.2187310 | 0.0774700 |
| C | -2.8258520 | 2.3819820 | 0.1811670 |
| C | 5.9508180 | 1.3783090 | -2.8369860 |
| C | 0.5361010 | -1.4812880 | -4.4343780 |
| Cl | -3.3583580 | -1.3219250 | -0.3945800 |
| Cl | -3.8091180 | 3.7320800 | 0.5860500 |
| H | 0.8582390 | 2.3997580 | 0.1897990 |
| H | 0.0550090 | -0.0244050 | -2.4915760 |
| H | 4.1152260 | 0.0102300 | -3.8659660 |
| H | 2.9619780 | 2.4330780 | -0.5078380 |
| H | 0.5805390 | -2.1632940 | -5.2808080 |
| H | 0.2108600 | -2.0270060 | -3.5439020 |
| H | -0.1756560 | -0.6791750 | -4.6502280 |
| H | 6.8467840 | 1.9019430 | -2.5104350 |
| H | 5.7291110 | 1.6475200 | -3.8737080 |
| H | 6.1142850 | 0.2989830 | -2.7679400 |

**Angle 222.5°**

Energy = -1714.207775475

| | | | |
|---|---|---|---|
| O | 4.9149420 | 1.7950210 | -1.9680170 |
| C | 3.6828460 | 1.2811830 | -2.1645310 |
| C | 3.3766480 | 0.3729270 | -3.1654790 |
| C | 2.0654380 | -0.0913560 | -3.2741180 |
| O | 1.8467180 | -0.9727940 | -4.2726840 |
| C | 1.0660860 | 0.3320730 | -2.4091690 |
| C | 1.4021700 | 1.2473140 | -1.4101240 |
| C | 2.6984990 | 1.7226400 | -1.2824660 |
| N | 0.4809760 | 1.7455530 | -0.4804230 |
| C | -0.8265870 | 1.5201550 | -0.3041790 |
| N | -1.5625190 | 2.6218010 | -0.0398580 |
| N | -1.3390620 | 0.2852390 | -0.3893100 |
| C | -2.6328450 | 0.2247200 | -0.1934160 |
| N | -3.4614650 | 1.2197250 | 0.0709400 |
| C | -2.8342050 | 2.3900600 | 0.1295310 |
| C | 5.9508180 | 1.3783090 | -2.8369860 |
| C | 0.5361010 | -1.4812880 | -4.4343780 |
| Cl | -3.3416950 | -1.3380840 | -0.2916180 |
| Cl | -3.8279940 | 3.7503520 | 0.4693760 |
| H | 0.8582390 | 2.3997580 | 0.1897990 |
| H | 0.0550090 | -0.0244050 | -2.4915760 |
| H | 4.1152260 | 0.0102300 | -3.8659660 |
| H | 2.9619780 | 2.4330780 | -0.5078380 |
| H | 0.5805390 | -2.1632940 | -5.2808080 |
| H | 0.2108600 | -2.0270060 | -3.5439020 |
| H | -0.1756560 | -0.6791750 | -4.6502280 |
| H | 6.8467840 | 1.9019430 | -2.5104350 |
| H | 5.7291110 | 1.6475200 | -3.8737080 |
| H | 6.1142850 | 0.2989830 | -2.7679400 |

**Angle 225°**

Energy = -1714.206246541

| | | | |
|---|---|---|---|
| C | 5.9508180 | 1.3783090 | -2.8369860 |
| O | 4.9149420 | 1.7950210 | -1.9680170 |
| C | 3.6828460 | 1.2811830 | -2.1645310 |
| C | 3.3766480 | 0.3729270 | -3.1654790 |
| C | 2.0654380 | -0.0913560 | -3.2741180 |
| O | 1.8467180 | -0.9727940 | -4.2726840 |
| C | 0.5361010 | -1.4812880 | -4.4343780 |
| C | 1.0660860 | 0.3320730 | -2.4091690 |
| C | 1.4021700 | 1.2473140 | -1.4101240 |
| C | 2.6984990 | 1.7226400 | -1.2824660 |
| N | 0.4809760 | 1.7455530 | -0.4804230 |
| H | 0.8582390 | 2.3997580 | 0.1897990 |
| C | -0.8265870 | 1.5201550 | -0.3041790 |
| N | -1.5706160 | 2.6276980 | -0.0923870 |
| C | -2.8422160 | 2.3958830 | 0.0775370 |
| N | -3.4624730 | 1.2204340 | 0.0643640 |
| C | -2.6263560 | 0.2199800 | -0.1513410 |
| N | -1.3315050 | 0.2797320 | -0.3402850 |
| Cl | -3.8461010 | 3.7635280 | 0.3518930 |
| Cl | -3.3257100 | -1.3497450 | -0.1879450 |
| H | 0.0550090 | -0.0244050 | -2.4915760 |
| H | 4.1152260 | 0.0102300 | -3.8659660 |
| H | 2.9619780 | 2.4330780 | -0.5078380 |
| H | 0.5805390 | -2.1632940 | -5.2808080 |
| H | 0.2108600 | -2.0270060 | -3.5439020 |
| H | -0.1756560 | -0.6791750 | -4.6502280 |
| H | 6.8467840 | 1.9019430 | -2.5104350 |
| H | 5.7291110 | 1.6475200 | -3.8737080 |
| H | 6.1142850 | 0.2989830 | -2.7679400 |

**Angle 240°**

Energy = -1714.196838243

| | | | |
|---|---|---|---|
| C | 5.9508180 | 1.3783090 | -2.8369860 |
| O | 4.9149420 | 1.7950210 | -1.9680170 |
| C | 3.6828460 | 1.2811830 | -2.1645310 |
| C | 3.3766480 | 0.3729270 | -3.1654790 |
| C | 2.0654380 | -0.0913560 | -3.2741180 |
| O | 1.8467180 | -0.9727940 | -4.2726840 |
| C | 0.5361010 | -1.4812880 | -4.4343780 |
| C | 1.0660860 | 0.3320730 | -2.4091690 |
| C | 1.4021700 | 1.2473140 | -1.4101240 |
| C | 2.6984990 | 1.7226400 | -1.2824660 |
| N | 0.4809760 | 1.7455530 | -0.4804230 |
| H | 0.8582390 | 2.3997580 | 0.1897990 |
| C | -0.8265870 | 1.5201550 | -0.3041790 |
| N | -1.6111520 | 2.6148920 | -0.4095100 |
| C | -2.8823230 | 2.3831230 | -0.2363360 |
| N | -3.4675120 | 1.2186550 | 0.0247010 |
| C | -2.5938640 | 0.2301300 | 0.1027020 |
| N | -1.2936670 | 0.2916610 | -0.0443080 |
| Cl | -3.9367470 | 3.7348090 | -0.3573450 |
| Cl | -3.2456760 | -1.3246060 | 0.4379790 |
| H | 0.0550090 | -0.0244050 | -2.4915760 |
| H | 4.1152260 | 0.0102300 | -3.8659660 |
| H | 2.9619780 | 2.4330780 | -0.5078380 |
| H | 0.5805390 | -2.1632940 | -5.2808080 |
| H | 0.2108600 | -2.0270060 | -3.5439020 |
| H | -0.1756560 | -0.6791750 | -4.6502280 |
| H | 6.8467840 | 1.9019430 | -2.5104350 |
| H | 5.7291110 | 1.6475200 | -3.8737080 |
| H | 6.1142850 | 0.2989830 | -2.7679400 |

**Angle 255°**

```
Energy = -1714.188943001
C      5.9508180    1.3783090   -2.8369860
O      4.9149420    1.7950210   -1.9680170
C      3.6828460    1.2811830   -2.1645310
C      3.3766480    0.3729270   -3.1654790
C      2.0654380   -0.0913560   -3.2741180
O      1.8467180   -0.9727940   -4.2726840
C      0.5361010   -1.4812880   -4.4343780
C      1.0660860    0.3320730   -2.4091690
C      1.4021700    1.2473140   -1.4101240
C      2.6984990    1.7226400   -1.2824660
N      0.4809760    1.7455530   -0.4804230
H      0.8582390    2.3997580    0.1897990
C     -0.8265870    1.5201550   -0.3041790
N     -1.6360500    2.5209600   -0.7143550
C     -2.9069490    2.2900720   -0.5380370
N     -3.4705910    1.2067360   -0.0133800
C     -2.5738970    0.3052730    0.3469400
N     -1.2704240    0.3793060    0.2402170
Cl    -3.9924150    3.5246580   -1.0391050
Cl    -3.1965060   -1.1393390    1.0397070
H      0.0550090   -0.0244050   -2.4915760
H      4.1152260    0.0102300   -3.8659660
H      2.9619780    2.4330780   -0.5078380
H      0.5805390   -2.1632940   -5.2808080
H      0.2108600   -2.0270060   -3.5439020
H     -0.1756560   -0.6791750   -4.6502280
H      6.8467840    1.9019430   -2.5104350
H      5.7291110    1.6475200   -3.8737080
H      6.1142850    0.2989830   -2.7679400
```

**Angle 270°**

```
Energy = -1714.185622583
C      5.9508180    1.3783090   -2.8369860
O      4.9149420    1.7950210   -1.9680170
C      3.6828460    1.2811830   -2.1645310
C      3.3766480    0.3729270   -3.1654790
C      2.0654380   -0.0913560   -3.2741180
O      1.8467180   -0.9727940   -4.2726840
C      0.5361010   -1.4812880   -4.4343780
C      1.0660860    0.3320730   -2.4091690
C      1.4021700    1.2473140   -1.4101240
C      2.6984990    1.7226400   -1.2824660
N      0.4809760    1.7455530   -0.4804230
H      0.8582390    2.3997580    0.1897990
C     -0.8265870    1.5201550   -0.3041790
N     -1.6436120    2.3523040   -0.9861490
C     -2.9144150    2.1230720   -0.8070040
N     -3.4714980    1.1854910   -0.0472840
C     -2.5678150    0.4402880    0.5647260
```

| | | | |
|---|---|---|---|
| N | -1.2633610 | 0.5366950 | 0.4939000 |
| Cl | -4.0093100 | 3.1473970 | -1.6469270 |
| Cl | -3.1815520 | -0.8065700 | 1.5762320 |
| H | 0.0550090 | -0.0244050 | -2.4915760 |
| H | 4.1152260 | 0.0102300 | -3.8659660 |
| H | 2.9619780 | 2.4330780 | -0.5078380 |
| H | 0.5805390 | -2.1632940 | -5.2808080 |
| H | 0.2108600 | -2.0270060 | -3.5439020 |
| H | -0.1756560 | -0.6791750 | -4.6502280 |
| H | 6.8467840 | 1.9019430 | -2.5104350 |
| H | 5.7291110 | 1.6475200 | -3.8737080 |
| H | 6.1142850 | 0.2989830 | -2.7679400 |

*Rotation around the dihedral angle shown in figure S26 and figure S27 for A3M*

**Calculated in solution with PBE0-D3(BJ)/def2-TZVP COSMO(THF)**

**<u>Angle 90°</u>**

Energy = -1828.621715892

| | | | |
|---|---|---|---|
| C | 5.4683950 | 2.5125750 | -1.5754180 |
| O | 4.9458200 | 1.8244010 | -2.6949920 |
| C | 3.6524140 | 1.4518080 | -2.6564730 |
| C | 3.1808330 | 0.7769950 | -3.7857420 |
| O | 3.9979160 | 0.5768010 | -4.8567230 |
| C | 4.6952750 | -0.6629040 | -4.8195430 |
| C | 1.8534270 | 0.3475960 | -3.8165120 |
| O | 1.4761590 | -0.3040750 | -4.9331690 |
| C | 0.1386360 | -0.7588640 | -5.0133190 |
| C | 0.9937040 | 0.5929730 | -2.7486100 |
| C | 1.4804640 | 1.2716530 | -1.6392750 |
| C | 2.8041440 | 1.7030910 | -1.5850830 |
| N | 0.7039630 | 1.5736390 | -0.5122160 |
| H | 1.1832390 | 2.0794850 | 0.2183140 |
| C | -0.5735230 | 1.3166010 | -0.2112320 |
| N | -0.7836220 | 0.1465710 | 0.4317860 |
| C | -2.0325270 | -0.0909440 | 0.7192120 |
| N | -3.0872800 | 0.6734950 | 0.4536840 |
| C | -2.7490850 | 1.7851940 | -0.1758770 |
| N | -1.5512170 | 2.1740020 | -0.5360530 |
| Cl | -2.3693360 | -1.5637510 | 1.5387740 |
| Cl | -4.0408800 | 2.8489820 | -0.5689670 |
| H | -0.0310050 | 0.2666690 | -2.7710990 |
| H | 3.1600890 | 2.2278570 | -0.7088060 |
| H | -0.5671040 | 0.0752180 | -4.9622680 |
| H | -0.0849840 | -1.4748360 | -4.2171060 |
| H | 0.0473730 | -1.2512810 | -5.9790790 |
| H | 4.9416490 | 3.4569480 | -1.4097440 |
| H | 6.5111400 | 2.7170840 | -1.8085370 |
| H | 5.4117640 | 1.8991470 | -0.6713140 |
| H | 5.3073450 | -0.7068550 | -5.7196190 |
| H | 3.9931190 | -1.5014540 | -4.8142700 |
| H | 5.3395240 | -0.7172920 | -3.9371590 |

**Angle 105°**

```
Energy = -1828.625127494
C      5.4683950    2.5125750   -1.5754180
O      4.9458200    1.8244010   -2.6949920
C      3.6524140    1.4518080   -2.6564730
C      3.1808330    0.7769950   -3.7857420
O      3.9979160    0.5768010   -4.8567230
C      4.6952750   -0.6629040   -4.8195430
C      1.8534270    0.3475960   -3.8165120
O      1.4761590   -0.3040750   -4.9331690
C      0.1386360   -0.7588640   -5.0133190
C      0.9937040    0.5929730   -2.7486100
C      1.4804640    1.2716530   -1.6392750
C      2.8041440    1.7030910   -1.5850830
N      0.7039630    1.5736390   -0.5122160
H      1.1832390    2.0794850    0.2183140
C     -0.5735230    1.3166010   -0.2112320
N     -0.7588580    0.3294370    0.6930590
C     -2.0080900    0.0897930    0.9772800
N     -3.0842870    0.6958580    0.4854860
C     -2.7689730    1.6381750   -0.3858430
N     -1.5743410    2.0029760   -0.7802510
Cl    -2.3140890   -1.1550010    2.1223290
Cl    -4.0897890    2.4871270   -1.0855750
H     -0.0310050    0.2666690   -2.7710990
H      3.1600890    2.2278570   -0.7088060
H     -0.5671040    0.0752180   -4.9622680
H     -0.0849840   -1.4748360   -4.2171060
H      0.0473730   -1.2512810   -5.9790790
H      4.9416490    3.4569480   -1.4097440
H      6.5111400    2.7170840   -1.8085370
H      5.4117640    1.8991470   -0.6713140
H      5.3073450   -0.7068550   -5.7196190
H      3.9931190   -1.5014540   -4.8142700
H      5.3395240   -0.7172920   -3.9371590
```

**Angle 120°**

```
Energy = -1828.633187106
C      5.4683950    2.5125750   -1.5754180
O      4.9458200    1.8244010   -2.6949920
C      3.6524140    1.4518080   -2.6564730
C      3.1808330    0.7769950   -3.7857420
O      3.9979160    0.5768010   -4.8567230
C      4.6952750   -0.6629040   -4.8195430
C      1.8534270    0.3475960   -3.8165120
O      1.4761590   -0.3040750   -4.9331690
C      0.1386360   -0.7588640   -5.0133190
C      0.9937040    0.5929730   -2.7486100
C      1.4804640    1.2716530   -1.6392750
C      2.8041440    1.7030910   -1.5850830
```

| | | | |
|---|---|---|---|
| N | 0.7039630 | 1.5736390 | -0.5122160 |
| H | 1.1832390 | 2.0794850 | 0.2183140 |
| C | -0.5735230 | 1.3166010 | -0.2112320 |
| N | -0.7585860 | 0.5721070 | 0.9014510 |
| C | -2.0078500 | 0.3295950 | 1.1830870 |
| N | -3.0842800 | 0.7254950 | 0.5108250 |
| C | -2.7691760 | 1.4430980 | -0.5532960 |
| N | -1.5745670 | 1.7760600 | -0.9749980 |
| Cl | -2.3135620 | -0.6126910 | 2.5876950 |
| Cl | -4.0902560 | 2.0070350 | -1.4975530 |
| H | -0.0310050 | 0.2666690 | -2.7710990 |
| H | 3.1600890 | 2.2278570 | -0.7088060 |
| H | -0.5671040 | 0.0752180 | -4.9622680 |
| H | -0.0849840 | -1.4748360 | -4.2171060 |
| H | 0.0473730 | -1.2512810 | -5.9790790 |
| H | 4.9416490 | 3.4569480 | -1.4097440 |
| H | 6.5111400 | 2.7170840 | -1.8085370 |
| H | 5.4117640 | 1.8991470 | -0.6713140 |
| H | 5.3073450 | -0.7068550 | -5.7196190 |
| H | 3.9931190 | -1.5014540 | -4.8142700 |
| H | 5.3395240 | -0.7172920 | -3.9371590 |

**Angle 122.5°**

Energy = -1828.634755626

| | | | |
|---|---|---|---|
| O | 4.9458200 | 1.8244010 | -2.6949920 |
| C | 3.6524140 | 1.4518080 | -2.6564730 |
| C | 3.1808330 | 0.7769950 | -3.7857420 |
| O | 3.9979160 | 0.5768010 | -4.8567230 |
| C | 1.8534270 | 0.3475960 | -3.8165120 |
| O | 1.4761590 | -0.3040750 | -4.9331690 |
| C | 0.9937040 | 0.5929730 | -2.7486100 |
| C | 1.4804640 | 1.2716530 | -1.6392750 |
| C | 2.8041440 | 1.7030910 | -1.5850830 |
| N | 0.7039630 | 1.5736390 | -0.5122160 |
| C | -0.5735230 | 1.3166010 | -0.2112320 |
| N | -0.7609470 | 0.6172540 | 0.9299830 |
| N | -1.5723550 | 1.7338460 | -1.0016560 |
| C | -2.7672750 | 1.4068080 | -0.5762190 |
| N | -3.0845720 | 0.7310060 | 0.5142920 |
| C | -2.0101860 | 0.3742070 | 1.2112620 |
| C | 4.6952750 | -0.6629040 | -4.8195430 |
| C | 0.1386360 | -0.7588640 | -5.0133190 |
| C | 5.4683950 | 2.5125750 | -1.5754180 |
| Cl | -4.0855740 | 1.9177240 | -1.5539460 |
| Cl | -2.3188510 | -0.5118050 | 2.6514000 |
| H | 1.1832390 | 2.0794850 | 0.2183140 |
| H | -0.0310050 | 0.2666690 | -2.7710990 |
| H | 3.1600890 | 2.2278570 | -0.7088060 |
| H | -0.5671040 | 0.0752180 | -4.9622680 |
| H | -0.0849840 | -1.4748360 | -4.2171060 |
| H | 0.0473730 | -1.2512810 | -5.9790790 |
| H | 4.9416500 | 3.4569480 | -1.4097440 |
| H | 6.5111410 | 2.7170830 | -1.8085370 |
| H | 5.4117640 | 1.8991470 | -0.6713140 |
| H | 5.3073450 | -0.7068550 | -5.7196190 |
| H | 3.9931190 | -1.5014540 | -4.8142700 |
| H | 5.3395240 | -0.7172920 | -3.9371590 |

**Angle 125°**

Energy = -1828.636358311

| | | | |
|---|---|---|---|
| O | 4.9458200 | 1.8244010 | -2.6949920 |
| C | 3.6524140 | 1.4518080 | -2.6564730 |
| C | 3.1808330 | 0.7769950 | -3.7857420 |
| O | 3.9979160 | 0.5768010 | -4.8567230 |
| C | 1.8534270 | 0.3475960 | -3.8165120 |
| O | 1.4761590 | -0.3040750 | -4.9331690 |
| C | 0.9937040 | 0.5929730 | -2.7486100 |
| C | 1.4804640 | 1.2716530 | -1.6392750 |
| C | 2.8041440 | 1.7030910 | -1.5850830 |
| N | 0.7039630 | 1.5736390 | -0.5122160 |
| C | -0.5735230 | 1.3166010 | -0.2112320 |
| N | -0.7639890 | 0.6635240 | 0.9565860 |
| N | -1.5695070 | 1.6905830 | -1.0265130 |
| C | -2.7648270 | 1.3696150 | -0.5975940 |
| N | -3.0849470 | 0.7366540 | 0.5175230 |
| C | -2.0131960 | 0.4199260 | 1.2375310 |
| C | 4.6952750 | -0.6629040 | -4.8195430 |
| C | 0.1386360 | -0.7588640 | -5.0133190 |
| C | 5.4683950 | 2.5125750 | -1.5754180 |
| Cl | -4.0795460 | 1.8261930 | -1.6065280 |
| Cl | -2.3256610 | -0.4084120 | 2.7107950 |
| H | 1.1832390 | 2.0794850 | 0.2183140 |
| H | -0.0310050 | 0.2666690 | -2.7710990 |
| H | 3.1600890 | 2.2278570 | -0.7088060 |
| H | -0.5671040 | 0.0752180 | -4.9622680 |
| H | -0.0849840 | -1.4748360 | -4.2171060 |
| H | 0.0473730 | -1.2512810 | -5.9790790 |
| H | 4.9416500 | 3.4569480 | -1.4097440 |
| H | 6.5111410 | 2.7170830 | -1.8085370 |
| H | 5.4117640 | 1.8991470 | -0.6713140 |
| H | 5.3073450 | -0.7068550 | -5.7196190 |
| H | 3.9931190 | -1.5014540 | -4.8142700 |
| H | 5.3395240 | -0.7172920 | -3.9371590 |

**Angle 127.5°**

Energy = -1828.637984407

| | | | |
|---|---|---|---|
| O | 4.9458200 | 1.8244010 | -2.6949920 |
| C | 3.6524140 | 1.4518080 | -2.6564730 |
| C | 3.1808330 | 0.7769950 | -3.7857420 |
| O | 3.9979160 | 0.5768010 | -4.8567230 |
| C | 1.8534270 | 0.3475960 | -3.8165120 |
| O | 1.4761590 | -0.3040750 | -4.9331690 |
| C | 0.9937040 | 0.5929730 | -2.7486100 |
| C | 1.4804640 | 1.2716530 | -1.6392750 |
| C | 2.8041440 | 1.7030910 | -1.5850830 |
| N | 0.7039630 | 1.5736390 | -0.5122160 |
| C | -0.5735230 | 1.3166010 | -0.2112320 |
| N | -0.7677060 | 0.7108300 | 0.9812110 |
| N | -1.5660280 | 1.6463530 | -1.0495200 |
| C | -2.7618380 | 1.3315900 | -0.6173780 |
| N | -3.0854050 | 0.7424270 | 0.5205120 |
| C | -2.0168730 | 0.4666680 | 1.2618450 |
| C | 4.6952750 | -0.6629040 | -4.8195430 |
| C | 0.1386360 | -0.7588640 | -5.0133190 |
| C | 5.4683950 | 2.5125750 | -1.5754180 |
| Cl | -4.0721840 | 1.7326180 | -1.6551950 |
| Cl | -2.3339760 | -0.3027100 | 2.7657690 |
| H | 1.1832390 | 2.0794850 | 0.2183140 |
| H | -0.0310050 | 0.2666690 | -2.7710990 |
| H | 3.1600890 | 2.2278570 | -0.7088060 |
| H | -0.5671040 | 0.0752180 | -4.9622680 |
| H | -0.0849840 | -1.4748360 | -4.2171060 |
| H | 0.0473730 | -1.2512810 | -5.9790790 |
| H | 4.9416500 | 3.4569480 | -1.4097440 |
| H | 6.5111410 | 2.7170830 | -1.8085370 |
| H | 5.4117640 | 1.8991470 | -0.6713140 |
| H | 5.3073450 | -0.7068550 | -5.7196190 |
| H | 3.9931190 | -1.5014540 | -4.8142700 |
| H | 5.3395240 | -0.7172920 | -3.9371590 |

**Angle 130°**

Energy = -1828.639593579

| | | | |
|---|---|---|---|
| O | 3.9979160 | 0.5768010 | -4.8567230 |
| C | 3.1808330 | 0.7769950 | -3.7857420 |
| C | 1.8534270 | 0.3475960 | -3.8165120 |
| O | 1.4761590 | -0.3040750 | -4.9331690 |
| C | 0.9937040 | 0.5929730 | -2.7486100 |
| C | 1.4804640 | 1.2716530 | -1.6392750 |
| C | 2.8041440 | 1.7030910 | -1.5850830 |
| C | 3.6524140 | 1.4518080 | -2.6564730 |
| O | 4.9458200 | 1.8244010 | -2.6949920 |
| N | 0.7039630 | 1.5736390 | -0.5122160 |
| C | -0.5735230 | 1.3166010 | -0.2112320 |
| N | -0.7720890 | 0.7590790 | 1.0038100 |
| N | -1.5619260 | 1.6012420 | -1.0706330 |
| C | -2.7583120 | 1.2928060 | -0.6355330 |
| N | -3.0859430 | 0.7483130 | 0.5232530 |
| C | -2.0212080 | 0.5143410 | 1.2841560 |
| C | 5.4683950 | 2.5125750 | -1.5754180 |
| C | 4.6952750 | -0.6629040 | -4.8195430 |
| C | 0.1386360 | -0.7588640 | -5.0133190 |
| Cl | -2.3437820 | -0.1949010 | 2.8162170 |
| Cl | -4.0635030 | 1.6371760 | -1.6998550 |
| H | 1.1832390 | 2.0794850 | 0.2183140 |
| H | -0.0310050 | 0.2666690 | -2.7710990 |
| H | 3.1600890 | 2.2278570 | -0.7088060 |
| H | -0.5671040 | 0.0752180 | -4.9622680 |
| H | -0.0849840 | -1.4748360 | -4.2171060 |
| H | 0.0473730 | -1.2512810 | -5.9790790 |
| H | 4.9416500 | 3.4569480 | -1.4097440 |
| H | 6.5111410 | 2.7170830 | -1.8085370 |
| H | 5.4117640 | 1.8991470 | -0.6713140 |
| H | 5.3073450 | -0.7068550 | -5.7196190 |
| H | 3.9931190 | -1.5014540 | -4.8142700 |
| H | 5.3395240 | -0.7172920 | -3.9371590 |

**Angle 132.5°**

Energy = -1828.641209836

| | | | |
|---|---|---|---|
| O | 4.9458200 | 1.8244010 | -2.6949920 |
| C | 3.6524140 | 1.4518080 | -2.6564730 |
| C | 3.1808330 | 0.7769950 | -3.7857420 |
| O | 3.9979160 | 0.5768010 | -4.8567230 |
| C | 1.8534270 | 0.3475960 | -3.8165120 |
| O | 1.4761590 | -0.3040750 | -4.9331690 |
| C | 0.9937040 | 0.5929730 | -2.7486100 |
| C | 1.4804640 | 1.2716530 | -1.6392750 |
| C | 2.8041440 | 1.7030910 | -1.5850830 |
| N | 0.7039630 | 1.5736390 | -0.5122160 |
| C | -0.5735230 | 1.3166010 | -0.2112320 |
| N | -0.7771320 | 0.8081810 | 1.0243400 |
| N | -1.5572070 | 1.5553330 | -1.0898110 |
| C | -2.7542560 | 1.2533380 | -0.6520270 |
| N | -3.0865620 | 0.7543050 | 0.5257420 |
| C | -2.0261940 | 0.5628570 | 1.3044250 |
| C | 4.6952750 | -0.6629040 | -4.8195430 |
| C | 0.1386360 | -0.7588640 | -5.0133190 |
| C | 5.4683950 | 2.5125750 | -1.5754180 |
| Cl | -4.0535190 | 1.5400500 | -1.7404240 |
| Cl | -2.3550610 | -0.0851860 | 2.8620420 |
| H | 1.1832390 | 2.0794850 | 0.2183140 |
| H | -0.0310050 | 0.2666690 | -2.7710990 |
| H | 3.1600890 | 2.2278570 | -0.7088060 |
| H | -0.5671040 | 0.0752180 | -4.9622680 |
| H | -0.0849840 | -1.4748360 | -4.2171060 |
| H | 0.0473730 | -1.2512810 | -5.9790790 |
| H | 4.9416500 | 3.4569480 | -1.4097440 |
| H | 6.5111410 | 2.7170830 | -1.8085370 |
| H | 5.4117640 | 1.8991470 | -0.6713140 |
| H | 5.3073450 | -0.7068550 | -5.7196190 |
| H | 3.9931190 | -1.5014540 | -4.8142700 |
| H | 5.3395240 | -0.7172920 | -3.9371590 |

**Angle 133.75°**

Energy = -1828.642013173

| | | | |
|---|---|---|---|
| O | 4.9458200 | 1.8244010 | -2.6949920 |
| C | 3.6524140 | 1.4518080 | -2.6564730 |
| C | 3.1808330 | 0.7769950 | -3.7857420 |
| O | 3.9979160 | 0.5768010 | -4.8567230 |
| C | 1.8534270 | 0.3475960 | -3.8165120 |
| O | 1.4761590 | -0.3040750 | -4.9331690 |
| C | 0.9937040 | 0.5929730 | -2.7486100 |
| C | 1.4804640 | 1.2716530 | -1.6392750 |
| C | 2.8041440 | 1.7030910 | -1.5850830 |
| N | 0.7039630 | 1.5736390 | -0.5122160 |
| C | -0.5735230 | 1.3166010 | -0.2112320 |
| N | -0.7798980 | 0.8330240 | 1.0338170 |
| N | -1.5546200 | 1.5321070 | -1.0986640 |
| C | -2.7520330 | 1.2333690 | -0.6596380 |
| N | -3.0869010 | 0.7573340 | 0.5268910 |
| C | -2.0289290 | 0.5874010 | 1.3137800 |
| C | 5.4683950 | 2.5125750 | -1.5754180 |
| C | 4.6952750 | -0.6629040 | -4.8195430 |
| C | 0.1386360 | -0.7588640 | -5.0133190 |
| Cl | -2.3612440 | -0.0296820 | 2.8831940 |
| Cl | -4.0480450 | 1.4909110 | -1.7591500 |
| H | 1.1832390 | 2.0794850 | 0.2183140 |
| H | -0.0310050 | 0.2666690 | -2.7710990 |
| H | 3.1600890 | 2.2278570 | -0.7088060 |
| H | -0.5671040 | 0.0752180 | -4.9622680 |
| H | -0.0849840 | -1.4748360 | -4.2171060 |
| H | 0.0473730 | -1.2512810 | -5.9790790 |
| H | 4.9416490 | 3.4569480 | -1.4097440 |
| H | 6.5111400 | 2.7170840 | -1.8085370 |
| H | 5.4117640 | 1.8991470 | -0.6713140 |
| H | 5.3073450 | -0.7068550 | -5.7196190 |
| H | 3.9931190 | -1.5014540 | -4.8142700 |
| H | 5.3395240 | -0.7172920 | -3.9371590 |

**Angle 135°**

Energy = -1828.642798773

| | | | |
|---|---|---|---|
| C | 5.4683950 | 2.5125750 | -1.5754180 |
| O | 4.9458200 | 1.8244010 | -2.6949920 |
| C | 3.6524140 | 1.4518080 | -2.6564730 |
| C | 3.1808330 | 0.7769950 | -3.7857420 |
| O | 3.9979160 | 0.5768010 | -4.8567230 |
| C | 4.6952750 | -0.6629040 | -4.8195430 |
| C | 1.8534270 | 0.3475960 | -3.8165120 |
| O | 1.4761590 | -0.3040750 | -4.9331690 |
| C | 0.1386360 | -0.7588640 | -5.0133190 |
| C | 0.9937040 | 0.5929730 | -2.7486100 |
| C | 1.4804640 | 1.2716530 | -1.6392750 |
| C | 2.8041440 | 1.7030910 | -1.5850830 |
| N | 0.7039630 | 1.5736390 | -0.5122160 |
| H | 1.1832390 | 2.0794850 | 0.2183140 |
| C | -0.5735230 | 1.3166010 | -0.2112320 |
| N | -0.7828250 | 0.8580440 | 1.0427630 |
| C | -2.0318230 | 0.6121210 | 1.3226090 |
| N | -3.0872600 | 0.7603870 | 0.5279730 |
| C | -2.7496790 | 1.2132580 | -0.6668260 |
| N | -1.5518810 | 1.5087160 | -1.1070200 |
| Cl | -2.3677900 | 0.0262210 | 2.9031570 |
| Cl | -4.0422500 | 1.4414230 | -1.7768260 |
| H | -0.0310050 | 0.2666690 | -2.7710990 |
| H | 3.1600890 | 2.2278570 | -0.7088060 |
| H | -0.5671040 | 0.0752180 | -4.9622680 |
| H | -0.0849840 | -1.4748360 | -4.2171060 |
| H | 0.0473730 | -1.2512810 | -5.9790790 |
| H | 4.9416490 | 3.4569480 | -1.4097440 |
| H | 6.5111400 | 2.7170840 | -1.8085370 |
| H | 5.4117640 | 1.8991470 | -0.6713140 |
| H | 5.3073450 | -0.7068550 | -5.7196190 |
| H | 3.9931190 | -1.5014540 | -4.8142700 |
| H | 5.3395240 | -0.7172920 | -3.9371590 |

## Angle 136.25°

Energy = -1828.643598121

| | | | |
|---|---|---|---|
| O | 4.9458200 | 1.8244010 | -2.6949920 |
| C | 3.6524140 | 1.4518080 | -2.6564730 |
| C | 3.1808330 | 0.7769950 | -3.7857420 |
| O | 3.9979160 | 0.5768010 | -4.8567230 |
| C | 1.8534270 | 0.3475960 | -3.8165120 |
| O | 1.4761590 | -0.3040750 | -4.9331690 |
| C | 0.9937040 | 0.5929730 | -2.7486100 |
| C | 1.4804640 | 1.2716530 | -1.6392750 |
| C | 2.8041440 | 1.7030910 | -1.5850830 |
| N | 0.7039630 | 1.5736390 | -0.5122160 |
| C | -0.5735230 | 1.3166010 | -0.2112320 |
| N | -0.7859100 | 0.8832290 | 1.0511730 |
| N | -1.5489950 | 1.4851690 | -1.1148750 |
| C | -2.7471980 | 1.1930140 | -0.6735790 |
| N | -3.0876380 | 0.7634590 | 0.5289930 |
| C | -2.0348730 | 0.6370050 | 1.3309110 |
| C | 5.4683950 | 2.5125750 | -1.5754180 |
| C | 4.6952750 | -0.6629040 | -4.8195430 |
| C | 0.1386360 | -0.7588640 | -5.0133190 |
| Cl | -2.3746890 | 0.0824930 | 2.9219250 |
| Cl | -4.0361430 | 1.3916060 | -1.7934380 |
| H | 1.1832390 | 2.0794850 | 0.2183140 |
| H | -0.0310050 | 0.2666690 | -2.7710990 |
| H | 3.1600890 | 2.2278570 | -0.7088060 |
| H | -0.5671040 | 0.0752180 | -4.9622680 |
| H | -0.0849840 | -1.4748360 | -4.2171060 |
| H | 0.0473730 | -1.2512810 | -5.9790790 |
| H | 4.9416490 | 3.4569480 | -1.4097440 |
| H | 6.5111400 | 2.7170840 | -1.8085370 |
| H | 5.4117640 | 1.8991470 | -0.6713140 |
| H | 5.3073450 | -0.7068550 | -5.7196190 |
| H | 3.9931190 | -1.5014540 | -4.8142700 |
| H | 5.3395240 | -0.7172920 | -3.9371590 |

**Angle 137.5°**

```
Energy = -1828.644376112
O      4.9458200    1.8244010   -2.6949920
C      3.6524140    1.4518080   -2.6564730
C      3.1808330    0.7769950   -3.7857420
O      3.9979160    0.5768010   -4.8567230
C      1.8534270    0.3475960   -3.8165120
O      1.4761590   -0.3040750   -4.9331690
C      0.9937040    0.5929730   -2.7486100
C      1.4804640    1.2716530   -1.6392750
C      2.8041440    1.7030910   -1.5850830
N      0.7039630    1.5736390   -0.5122160
C     -0.5735230    1.3166010   -0.2112320
N     -0.7891540    0.9085690    1.0590430
N     -1.5459600    1.4614780   -1.1222250
C     -2.7445890    1.1726460   -0.6799010
N     -3.0880360    0.7665500    0.5299460
C     -2.0380810    0.6620420    1.3386780
C      5.4683950    2.5125750   -1.5754180
C      4.6952750   -0.6629040   -4.8195430
C      0.1386360   -0.7588640   -5.0133190
Cl    -2.3819440    0.1391100    2.9394860
Cl    -4.0297210    1.3414840   -1.8089850
H      1.1832390    2.0794850    0.2183140
H     -0.0310050    0.2666690   -2.7710990
H      3.1600890    2.2278570   -0.7088060
H     -0.5671040    0.0752180   -4.9622680
H     -0.0849840   -1.4748360   -4.2171060
H      0.0473730   -1.2512810   -5.9790790
H      4.9416490    3.4569480   -1.4097440
H      6.5111400    2.7170840   -1.8085370
H      5.4117640    1.8991470   -0.6713140
H      5.3073450   -0.7068550   -5.7196190
H      3.9931190   -1.5014540   -4.8142700
H      5.3395240   -0.7172920   -3.9371590
```

**Angle 140°**

```
Energy = -1828.645895208
O      4.9458200    1.8244010   -2.6949920
C      3.6524140    1.4518080   -2.6564730
C      3.1808330    0.7769950   -3.7857420
O      3.9979160    0.5768010   -4.8567230
C      1.8534270    0.3475960   -3.8165120
O      1.4761590   -0.3040750   -4.9331690
C      0.9937040    0.5929730   -2.7486100
C      1.4804640    1.2716530   -1.6392750
C      2.8041440    1.7030910   -1.5850830
N      0.7039630    1.5736390   -0.5122160
C     -0.5735230    1.3166010   -0.2112320
N     -0.7961110    0.9596620    1.0731490
N     -1.5394520    1.4137100   -1.1353970
```

| | | | |
|---|---|---|---|
| C | -2.7389950 | 1.1315800 | -0.6912310 |
| N | -3.0888880 | 0.7727830 | 0.5316510 |
| C | -2.0449580 | 0.7125230 | 1.3525990 |
| C | 4.6952750 | -0.6629040 | -4.8195430 |
| C | 0.1386360 | -0.7588640 | -5.0133190 |
| C | 5.4683950 | 2.5125750 | -1.5754180 |
| Cl | -4.0159500 | 1.2404260 | -1.8368460 |
| Cl | -2.3974980 | 0.2532680 | 2.9709580 |
| H | 1.1832390 | 2.0794850 | 0.2183140 |
| H | -0.0310050 | 0.2666690 | -2.7710990 |
| H | 3.1600890 | 2.2278570 | -0.7088060 |
| H | -0.5671040 | 0.0752180 | -4.9622680 |
| H | -0.0849840 | -1.4748360 | -4.2171060 |
| H | 0.0473730 | -1.2512810 | -5.9790790 |
| H | 4.9416500 | 3.4569480 | -1.4097440 |
| H | 6.5111410 | 2.7170830 | -1.8085370 |
| H | 5.4117640 | 1.8991470 | -0.6713140 |
| H | 5.3073450 | -0.7068550 | -5.7196190 |
| H | 3.9931190 | -1.5014540 | -4.8142700 |
| H | 5.3395240 | -0.7172920 | -3.9371590 |

**Angle 150°**

Energy = -1828.651428963

| | | | |
|---|---|---|---|
| C | 5.4683950 | 2.5125750 | -1.5754180 |
| O | 4.9458200 | 1.8244010 | -2.6949920 |
| C | 3.6524140 | 1.4518080 | -2.6564730 |
| C | 3.1808330 | 0.7769950 | -3.7857420 |
| O | 3.9979160 | 0.5768010 | -4.8567230 |
| C | 4.6952750 | -0.6629040 | -4.8195430 |
| C | 1.8534270 | 0.3475960 | -3.8165120 |
| O | 1.4761590 | -0.3040750 | -4.9331690 |
| C | 0.1386360 | -0.7588640 | -5.0133190 |
| C | 0.9937040 | 0.5929730 | -2.7486100 |
| C | 1.4804640 | 1.2716530 | -1.6392750 |
| C | 2.8041440 | 1.7030910 | -1.5850830 |
| N | 0.7039630 | 1.5736390 | -0.5122160 |
| H | 1.1832390 | 2.0794850 | 0.2183140 |
| C | -0.5735230 | 1.3166010 | -0.2112320 |
| N | -0.8299220 | 1.1677610 | 1.1073620 |
| C | -2.0783770 | 0.9181170 | 1.3863360 |
| N | -3.0930240 | 0.7981560 | 0.5357630 |
| C | -2.7118110 | 0.9643180 | -0.7186960 |
| N | -1.5078290 | 1.2191630 | -1.1673220 |
| Cl | -2.4730770 | 0.7181940 | 3.0472180 |
| Cl | -3.9490430 | 0.8288370 | -1.9043610 |
| H | -0.0310050 | 0.2666690 | -2.7710990 |
| H | 3.1600890 | 2.2278570 | -0.7088060 |
| H | -0.5671040 | 0.0752180 | -4.9622680 |
| H | -0.0849840 | -1.4748360 | -4.2171060 |
| H | 0.0473730 | -1.2512810 | -5.9790790 |
| H | 4.9416490 | 3.4569480 | -1.4097440 |
| H | 6.5111400 | 2.7170840 | -1.8085370 |
| H | 5.4117640 | 1.8991470 | -0.6713140 |
| H | 5.3073450 | -0.7068550 | -5.7196190 |
| H | 3.9931190 | -1.5014540 | -4.8142700 |
| H | 5.3395240 | -0.7172920 | -3.9371590 |

**Angle 165°**

Energy = -1828.657112328

| | | | |
|---|---|---|---|
| C | 5.4683950 | 2.5125750 | -1.5754180 |
| O | 4.9458200 | 1.8244010 | -2.6949920 |
| C | 3.6524140 | 1.4518080 | -2.6564730 |
| C | 3.1808330 | 0.7769950 | -3.7857420 |
| O | 3.9979160 | 0.5768010 | -4.8567230 |
| C | 4.6952750 | -0.6629040 | -4.8195430 |
| C | 1.8534270 | 0.3475960 | -3.8165120 |
| O | 1.4761590 | -0.3040750 | -4.9331690 |
| C | 0.1386360 | -0.7588640 | -5.0133190 |
| C | 0.9937040 | 0.5929730 | -2.7486100 |
| C | 1.4804640 | 1.2716530 | -1.6392750 |
| C | 2.8041440 | 1.7030910 | -1.5850830 |
| N | 0.7039630 | 1.5736390 | -0.5122160 |
| H | 1.1832390 | 2.0794850 | 0.2183140 |
| C | -0.5735230 | 1.3166010 | -0.2112320 |
| N | -0.8966670 | 1.4801530 | 1.0908480 |
| C | -2.1443380 | 1.2267300 | 1.3699270 |
| N | -3.1011790 | 0.8362280 | 0.5336630 |
| C | -2.6581540 | 0.7132420 | -0.7053700 |
| N | -1.4454130 | 0.9271350 | -1.1517930 |
| Cl | -2.6222490 | 1.4160720 | 3.0100600 |
| Cl | -3.8169850 | 0.2110230 | -1.8714680 |
| H | -0.0310050 | 0.2666690 | -2.7710990 |
| H | 3.1600890 | 2.2278570 | -0.7088060 |
| H | -0.5671040 | 0.0752180 | -4.9622680 |
| H | -0.0849840 | -1.4748360 | -4.2171060 |
| H | 0.0473730 | -1.2512810 | -5.9790790 |
| H | 4.9416490 | 3.4569480 | -1.4097440 |
| H | 6.5111400 | 2.7170840 | -1.8085370 |
| H | 5.4117640 | 1.8991470 | -0.6713140 |
| H | 5.3073450 | -0.7068550 | -5.7196190 |
| H | 3.9931190 | -1.5014540 | -4.8142700 |
| H | 5.3395240 | -0.7172920 | -3.9371590 |

**Angle 180.083664°**

Energy = -1828.659070364

| | | | |
|---|---|---|---|
| C | 5.4683950 | 2.5125750 | -1.5754180 |
| O | 4.9458200 | 1.8244010 | -2.6949920 |
| C | 3.6524140 | 1.4518080 | -2.6564730 |
| C | 3.1808330 | 0.7769950 | -3.7857420 |
| O | 3.9979160 | 0.5768010 | -4.8567230 |
| C | 4.6952750 | -0.6629040 | -4.8195430 |
| C | 1.8534270 | 0.3475960 | -3.8165120 |
| O | 1.4761590 | -0.3040750 | -4.9331690 |
| C | 0.1386360 | -0.7588640 | -5.0133190 |
| C | 0.9937040 | 0.5929730 | -2.7486100 |
| C | 1.4804640 | 1.2716530 | -1.6392750 |
| C | 2.8041440 | 1.7030910 | -1.5850830 |
| N | 0.7039630 | 1.5736390 | -0.5122160 |
| H | 1.1832390 | 2.0794850 | 0.2183140 |
| C | -0.5735230 | 1.3166010 | -0.2112320 |
| N | -0.9790020 | 1.7754780 | 0.9935930 |
| C | -2.2256940 | 1.5184570 | 1.2737550 |
| N | -3.1112290 | 0.8721970 | 0.5217240 |
| C | -2.5919700 | 0.4758980 | -0.6271520 |
| N | -1.3684280 | 0.6510860 | -1.0607870 |
| Cl | -2.8062310 | 2.0757510 | 2.7925320 |
| Cl | -3.6541110 | -0.3729760 | -1.6788980 |
| H | -0.0310050 | 0.2666690 | -2.7710990 |
| H | 3.1600890 | 2.2278570 | -0.7088060 |
| H | -0.5671040 | 0.0752180 | -4.9622680 |
| H | -0.0849840 | -1.4748360 | -4.2171060 |
| H | 0.0473730 | -1.2512810 | -5.9790790 |
| H | 4.9416490 | 3.4569480 | -1.4097440 |
| H | 6.5111400 | 2.7170840 | -1.8085370 |
| H | 5.4117640 | 1.8991470 | -0.6713140 |
| H | 5.3073450 | -0.7068550 | -5.7196190 |
| H | 3.9931190 | -1.5014540 | -4.8142700 |
| H | 5.3395240 | -0.7172920 | -3.9371590 |

**Angle 195°**

Energy = -1828.657123743

| | | | |
|---|---|---|---|
| C | 5.4683950 | 2.5125750 | -1.5754180 |
| O | 4.9458200 | 1.8244010 | -2.6949920 |
| C | 3.6524140 | 1.4518080 | -2.6564730 |
| C | 3.1808330 | 0.7769950 | -3.7857420 |
| O | 3.9979160 | 0.5768010 | -4.8567230 |
| C | 4.6952750 | -0.6629040 | -4.8195430 |
| C | 1.8534270 | 0.3475960 | -3.8165120 |
| O | 1.4761590 | -0.3040750 | -4.9331690 |
| C | 0.1386360 | -0.7588640 | -5.0133190 |
| C | 0.9937040 | 0.5929730 | -2.7486100 |
| C | 1.4804640 | 1.2716530 | -1.6392750 |
| C | 2.8041440 | 1.7030910 | -1.5850830 |
| N | 0.7039630 | 1.5736390 | -0.5122160 |
| H | 1.1832390 | 2.0794850 | 0.2183140 |
| C | -0.5735230 | 1.3166010 | -0.2112320 |
| N | -1.0698820 | 2.0290710 | 0.8244310 |
| C | -2.3154850 | 1.7689360 | 1.1065550 |
| N | -3.1223150 | 0.9030600 | 0.5010310 |
| C | -2.5189220 | 0.2721060 | -0.4911460 |
| N | -1.2834630 | 0.4140690 | -0.9025710 |
| Cl | -3.0092840 | 2.6421410 | 2.4143930 |
| Cl | -3.4743530 | -0.8743870 | -1.3441420 |
| H | -0.0310050 | 0.2666690 | -2.7710990 |
| H | 3.1600890 | 2.2278570 | -0.7088060 |
| H | -0.5671040 | 0.0752180 | -4.9622680 |
| H | -0.0849840 | -1.4748360 | -4.2171060 |
| H | 0.0473730 | -1.2512810 | -5.9790790 |
| H | 4.9416490 | 3.4569480 | -1.4097440 |
| H | 6.5111400 | 2.7170840 | -1.8085370 |
| H | 5.4117640 | 1.8991470 | -0.6713140 |
| H | 5.3073450 | -0.7068550 | -5.7196190 |
| H | 3.9931190 | -1.5014540 | -4.8142700 |
| H | 5.3395240 | -0.7172920 | -3.9371590 |

**Angle 210°**

Energy = -1828.651442076

| | | | |
|---|---|---|---|
| C | 5.4683950 | 2.5125750 | -1.5754180 |
| O | 4.9458200 | 1.8244010 | -2.6949920 |
| C | 3.6524140 | 1.4518080 | -2.6564730 |
| C | 3.1808330 | 0.7769950 | -3.7857420 |
| O | 3.9979160 | 0.5768010 | -4.8567230 |
| C | 4.6952750 | -0.6629040 | -4.8195430 |
| C | 1.8534270 | 0.3475960 | -3.8165120 |
| O | 1.4761590 | -0.3040750 | -4.9331690 |
| C | 0.1386360 | -0.7588640 | -5.0133190 |
| C | 0.9937040 | 0.5929730 | -2.7486100 |
| C | 1.4804640 | 1.2716530 | -1.6392750 |
| C | 2.8041440 | 1.7030910 | -1.5850830 |
| N | 0.7039630 | 1.5736390 | -0.5122160 |
| H | 1.1832390 | 2.0794850 | 0.2183140 |
| C | -0.5735230 | 1.3166010 | -0.2112320 |
| N | -1.1645470 | 2.2281900 | 0.5926840 |
| C | -2.4090070 | 1.9655780 | 0.8775410 |
| N | -3.1338540 | 0.9272650 | 0.4727230 |
| C | -2.4428370 | 0.1121090 | -0.3048470 |
| N | -1.1949660 | 0.2279950 | -0.6858630 |
| Cl | -3.2207720 | 3.0867790 | 1.8964780 |
| Cl | -3.2871280 | -1.2680150 | -0.8856460 |
| H | -0.0310050 | 0.2666690 | -2.7710990 |
| H | 3.1600890 | 2.2278570 | -0.7088060 |
| H | -0.5671040 | 0.0752180 | -4.9622680 |
| H | -0.0849840 | -1.4748360 | -4.2171060 |
| H | 0.0473730 | -1.2512810 | -5.9790790 |
| H | 4.9416490 | 3.4569480 | -1.4097440 |
| H | 6.5111400 | 2.7170840 | -1.8085370 |
| H | 5.4117640 | 1.8991470 | -0.6713140 |
| H | 5.3073450 | -0.7068550 | -5.7196190 |
| H | 3.9931190 | -1.5014540 | -4.8142700 |
| H | 5.3395240 | -0.7172920 | -3.9371590 |

**Angle 220°**

Energy = -1828.645923792

| | | | |
|---|---|---|---|
| O | 4.9458200 | 1.8244010 | -2.6949920 |
| C | 3.6524140 | 1.4518080 | -2.6564730 |
| C | 3.1808330 | 0.7769950 | -3.7857420 |
| O | 3.9979160 | 0.5768010 | -4.8567230 |
| C | 1.8534270 | 0.3475960 | -3.8165120 |
| O | 1.4761590 | -0.3040750 | -4.9331690 |
| C | 0.9937040 | 0.5929730 | -2.7486100 |
| C | 1.4804640 | 1.2716530 | -1.6392750 |
| C | 2.8041440 | 1.7030910 | -1.5850830 |
| N | 0.7039630 | 1.5736390 | -0.5122160 |
| C | -0.5735230 | 1.3166010 | -0.2112320 |
| N | -1.2262980 | 2.3229240 | 0.4114920 |
| N | -1.1372430 | 0.1394890 | -0.5164470 |
| C | -2.3932070 | 0.0359990 | -0.1592000 |
| N | -3.1413790 | 0.9387610 | 0.4506050 |
| C | -2.4700090 | 2.0591120 | 0.6985030 |
| C | 4.6952750 | -0.6629040 | -4.8195430 |
| C | 0.1386360 | -0.7588640 | -5.0133190 |
| C | 5.4683950 | 2.5125750 | -1.5754180 |
| Cl | -3.1650090 | -1.4552360 | -0.5272160 |
| Cl | -3.3587170 | 3.2982610 | 1.4915940 |
| H | 1.1832390 | 2.0794850 | 0.2183140 |
| H | -0.0310050 | 0.2666690 | -2.7710990 |
| H | 3.1600890 | 2.2278570 | -0.7088060 |
| H | -0.5671040 | 0.0752180 | -4.9622680 |
| H | -0.0849840 | -1.4748360 | -4.2171060 |
| H | 0.0473730 | -1.2512810 | -5.9790790 |
| H | 4.9416500 | 3.4569480 | -1.4097440 |
| H | 6.5111410 | 2.7170830 | -1.8085370 |
| H | 5.4117640 | 1.8991470 | -0.6713140 |
| H | 5.3073450 | -0.7068550 | -5.7196190 |
| H | 3.9931190 | -1.5014540 | -4.8142700 |
| H | 5.3395240 | -0.7172920 | -3.9371590 |

**Angle 222.5°**

Energy = -1828.644394681

| | | | |
|---|---|---|---|
| O | 4.9458200 | 1.8244010 | -2.6949920 |
| C | 3.6524140 | 1.4518080 | -2.6564730 |
| C | 3.1808330 | 0.7769950 | -3.7857420 |
| O | 3.9979160 | 0.5768010 | -4.8567230 |
| C | 1.8534270 | 0.3475960 | -3.8165120 |
| O | 1.4761590 | -0.3040750 | -4.9331690 |
| C | 0.9937040 | 0.5929730 | -2.7486100 |
| C | 1.4804640 | 1.2716530 | -1.6392750 |
| C | 2.8041440 | 1.7030910 | -1.5850830 |
| N | 0.7039630 | 1.5736390 | -0.5122160 |
| C | -0.5735230 | 1.3166010 | -0.2112320 |
| N | -1.2412950 | 2.3414000 | 0.3636200 |
| N | -1.1232250 | 0.1222310 | -0.4716880 |
| C | -2.3811560 | 0.0211590 | -0.1207180 |
| N | -3.1432050 | 0.9410000 | 0.4447650 |
| C | -2.4848230 | 2.0773500 | 0.6512030 |
| C | 5.4683950 | 2.5125750 | -1.5754180 |
| C | 4.6952750 | -0.6629040 | -4.8195430 |
| C | 0.1386360 | -0.7588640 | -5.0133190 |
| Cl | -3.3922150 | 3.3394960 | 1.3846280 |
| Cl | -3.1353550 | -1.4917400 | -0.4325200 |
| H | 1.1832390 | 2.0794850 | 0.2183140 |
| H | -0.0310050 | 0.2666690 | -2.7710990 |
| H | 3.1600890 | 2.2278570 | -0.7088060 |
| H | -0.5671040 | 0.0752180 | -4.9622680 |
| H | -0.0849840 | -1.4748360 | -4.2171060 |
| H | 0.0473730 | -1.2512810 | -5.9790790 |
| H | 4.9416490 | 3.4569480 | -1.4097440 |
| H | 6.5111400 | 2.7170840 | -1.8085370 |
| H | 5.4117640 | 1.8991470 | -0.6713140 |
| H | 5.3073450 | -0.7068550 | -5.7196190 |
| H | 3.9931190 | -1.5014540 | -4.8142700 |
| H | 5.3395240 | -0.7172920 | -3.9371590 |

**Angle 223.75°**

Energy = -1828.643620603

| | | | |
|---|---|---|---|
| O | 4.9458200 | 1.8244010 | -2.6949920 |
| C | 3.6524140 | 1.4518080 | -2.6564730 |
| C | 3.1808330 | 0.7769950 | -3.7857420 |
| O | 3.9979160 | 0.5768010 | -4.8567230 |
| C | 1.8534270 | 0.3475960 | -3.8165120 |
| O | 1.4761590 | -0.3040750 | -4.9331690 |
| C | 0.9937040 | 0.5929730 | -2.7486100 |
| C | 1.4804640 | 1.2716530 | -1.6392750 |
| C | 2.8041440 | 1.7030910 | -1.5850830 |
| N | 0.7039630 | 1.5736390 | -0.5122160 |
| C | -0.5735230 | 1.3166010 | -0.2112320 |
| N | -1.2487070 | 2.3498300 | 0.3393600 |
| N | -1.1162970 | 0.1143580 | -0.4490060 |
| C | -2.3751990 | 0.0143880 | -0.1012180 |
| N | -3.1441080 | 0.9420210 | 0.4418050 |
| C | -2.4921450 | 2.0856710 | 0.6272330 |
| C | 5.4683950 | 2.5125750 | -1.5754180 |
| C | 4.6952750 | -0.6629040 | -4.8195430 |
| C | 0.1386360 | -0.7588640 | -5.0133190 |
| Cl | -3.4087720 | 3.3583090 | 1.3304210 |
| Cl | -3.1206980 | -1.5083950 | -0.3845320 |
| H | 1.1832390 | 2.0794850 | 0.2183140 |
| H | -0.0310050 | 0.2666690 | -2.7710990 |
| H | 3.1600890 | 2.2278570 | -0.7088060 |
| H | -0.5671040 | 0.0752180 | -4.9622680 |
| H | -0.0849840 | -1.4748360 | -4.2171060 |
| H | 0.0473730 | -1.2512810 | -5.9790790 |
| H | 4.9416490 | 3.4569480 | -1.4097440 |
| H | 6.5111400 | 2.7170840 | -1.8085370 |
| H | 5.4117640 | 1.8991470 | -0.6713140 |
| H | 5.3073450 | -0.7068550 | -5.7196190 |
| H | 3.9931190 | -1.5014540 | -4.8142700 |
| H | 5.3395240 | -0.7172920 | -3.9371590 |

**Angle 225°**

```
Energy = -1828.642813340
C      5.4683950    2.5125750   -1.5754180
O      4.9458200    1.8244010   -2.6949920
C      3.6524140    1.4518080   -2.6564730
C      3.1808330    0.7769950   -3.7857420
O      3.9979160    0.5768010   -4.8567230
C      4.6952750   -0.6629040   -4.8195430
C      1.8534270    0.3475960   -3.8165120
O      1.4761590   -0.3040750   -4.9331690
C      0.1386360   -0.7588640   -5.0133190
C      0.9937040    0.5929730   -2.7486100
C      1.4804640    1.2716530   -1.6392750
C      2.8041440    1.7030910   -1.5850830
N      0.7039630    1.5736390   -0.5122160
H      1.1832390    2.0794850    0.2183140
C     -0.5735230    1.3166010   -0.2112320
N     -1.2560570    2.3577160    0.3148980
C     -2.4994050    2.0934540    0.6030640
N     -3.1450030    0.9429740    0.4388210
C     -2.3692910    0.0080520   -0.0815560
N     -1.1094260    0.1069920   -0.4261350
Cl    -3.4251900    3.3759060    1.2757650
Cl    -3.1061620   -1.5239740   -0.3361450
H     -0.0310050    0.2666690   -2.7710990
H      3.1600890    2.2278570   -0.7088060
H     -0.5671040    0.0752180   -4.9622680
H     -0.0849840   -1.4748360   -4.2171060
H      0.0473730   -1.2512810   -5.9790790
H      4.9416490    3.4569480   -1.4097440
H      6.5111400    2.7170840   -1.8085370
H      5.4117640    1.8991470   -0.6713140
H      5.3073450   -0.7068550   -5.7196190
H      3.9931190   -1.5014540   -4.8142700
H      5.3395240   -0.7172920   -3.9371590
```

**Angle 226.25°**

```
Energy = -1828.642036055
O      4.9458200    1.8244010   -2.6949920
C      3.6524140    1.4518080   -2.6564730
C      3.1808330    0.7769950   -3.7857420
O      3.9979160    0.5768010   -4.8567230
C      1.8534270    0.3475960   -3.8165120
O      1.4761590   -0.3040750   -4.9331690
C      0.9937040    0.5929730   -2.7486100
C      1.4804640    1.2716530   -1.6392750
C      2.8041440    1.7030910   -1.5850830
N      0.7039630    1.5736390   -0.5122160
C     -0.5735230    1.3166010   -0.2112320
N     -1.2633410    2.3650550    0.2902480
N     -1.1026170    0.1001390   -0.4030890
```

| | | | |
|---|---|---|---|
| C | -2.3634380 | 0.0021590 | -0.0617420 |
| N | -3.1458900 | 0.9438630 | 0.4358140 |
| C | -2.5066010 | 2.1006970 | 0.5787080 |
| C | 5.4683950 | 2.5125750 | -1.5754180 |
| C | 4.6952750 | -0.6629040 | -4.8195430 |
| C | 0.1386360 | -0.7588640 | -5.0133190 |
| Cl | -3.4414610 | 3.3922790 | 1.2206870 |
| Cl | -3.0917590 | -1.5384690 | -0.2873870 |
| H | 1.1832390 | 2.0794850 | 0.2183140 |
| H | -0.0310050 | 0.2666690 | -2.7710990 |
| H | 3.1600890 | 2.2278570 | -0.7088060 |
| H | -0.5671040 | 0.0752180 | -4.9622680 |
| H | -0.0849840 | -1.4748360 | -4.2171060 |
| H | 0.0473730 | -1.2512810 | -5.9790790 |
| H | 4.9416490 | 3.4569480 | -1.4097440 |
| H | 6.5111400 | 2.7170840 | -1.8085370 |
| H | 5.4117640 | 1.8991470 | -0.6713140 |
| H | 5.3073450 | -0.7068550 | -5.7196190 |
| H | 3.9931190 | -1.5014540 | -4.8142700 |
| H | 5.3395240 | -0.7172920 | -3.9371590 |

**Angle 227.5°**

Energy = -1828.641249512

| | | | |
|---|---|---|---|
| O | 4.9458200 | 1.8244010 | -2.6949920 |
| C | 3.6524140 | 1.4518080 | -2.6564730 |
| C | 3.1808330 | 0.7769950 | -3.7857420 |
| O | 3.9979160 | 0.5768010 | -4.8567230 |
| C | 1.8534270 | 0.3475960 | -3.8165120 |
| O | 1.4761590 | -0.3040750 | -4.9331690 |
| C | 0.9937040 | 0.5929730 | -2.7486100 |
| C | 1.4804640 | 1.2716530 | -1.6392750 |
| C | 2.8041440 | 1.7030910 | -1.5850830 |
| N | 0.7039630 | 1.5736390 | -0.5122160 |
| C | -0.5735230 | 1.3166010 | -0.2112320 |
| N | -1.2705570 | 2.3718430 | 0.2654190 |
| N | -1.0958730 | 0.0938020 | -0.3798760 |
| C | -2.3576380 | -0.0032930 | -0.0417870 |
| N | -3.1467690 | 0.9446820 | 0.4327830 |
| C | -2.5137280 | 2.1073950 | 0.5541760 |
| C | 4.6952750 | -0.6629040 | -4.8195430 |
| C | 0.1386360 | -0.7588640 | -5.0133190 |
| C | 5.4683950 | 2.5125750 | -1.5754180 |
| Cl | -3.0774900 | -1.5518730 | -0.2382780 |
| Cl | -3.4575790 | 3.4074210 | 1.1652110 |
| H | 1.1832390 | 2.0794850 | 0.2183140 |
| H | -0.0310050 | 0.2666690 | -2.7710990 |
| H | 3.1600890 | 2.2278570 | -0.7088060 |
| H | -0.5671040 | 0.0752180 | -4.9622680 |
| H | -0.0849840 | -1.4748360 | -4.2171060 |
| H | 0.0473730 | -1.2512810 | -5.9790790 |
| H | 4.9416500 | 3.4569480 | -1.4097440 |
| H | 6.5111410 | 2.7170830 | -1.8085370 |
| H | 5.4117640 | 1.8991470 | -0.6713140 |
| H | 5.3073450 | -0.7068550 | -5.7196190 |
| H | 3.9931190 | -1.5014540 | -4.8142700 |
| H | 5.3395240 | -0.7172920 | -3.9371590 |

**Angle 230°**

Energy = -1828.639627240

| | | | |
|---|---|---|---|
| O | 3.9979160 | 0.5768010 | -4.8567230 |
| C | 3.1808330 | 0.7769950 | -3.7857420 |
| C | 1.8534270 | 0.3475960 | -3.8165120 |
| O | 1.4761590 | -0.3040750 | -4.9331690 |
| C | 0.9937040 | 0.5929730 | -2.7486100 |
| C | 1.4804640 | 1.2716530 | -1.6392750 |
| C | 2.8041440 | 1.7030910 | -1.5850830 |
| C | 3.6524140 | 1.4518080 | -2.6564730 |
| O | 4.9458200 | 1.8244010 | -2.6949920 |
| N | 0.7039630 | 1.5736390 | -0.5122160 |
| C | -0.5735230 | 1.3166010 | -0.2112320 |
| N | -1.2847660 | 2.3837510 | 0.2152780 |
| N | -1.0825910 | 0.0826840 | -0.3329970 |
| C | -2.3462180 | -0.0128560 | -0.0014830 |
| N | -3.1484990 | 0.9461190 | 0.4266680 |
| C | -2.5277640 | 2.1191440 | 0.5046360 |
| C | 5.4683950 | 2.5125750 | -1.5754180 |
| C | 4.6952750 | -0.6629040 | -4.8195430 |
| C | 0.1386360 | -0.7588640 | -5.0133190 |
| Cl | -3.4893180 | 3.4339820 | 1.0531800 |
| Cl | -3.0493910 | -1.5753880 | -0.1390980 |
| H | 1.1832390 | 2.0794850 | 0.2183140 |
| H | -0.0310050 | 0.2666690 | -2.7710990 |
| H | 3.1600890 | 2.2278570 | -0.7088060 |
| H | -0.5671040 | 0.0752180 | -4.9622680 |
| H | -0.0849840 | -1.4748360 | -4.2171060 |
| H | 0.0473730 | -1.2512810 | -5.9790790 |
| H | 4.9416500 | 3.4569480 | -1.4097440 |
| H | 6.5111410 | 2.7170830 | -1.8085370 |
| H | 5.4117640 | 1.8991470 | -0.6713140 |
| H | 5.3073450 | -0.7068550 | -5.7196190 |
| H | 3.9931190 | -1.5014540 | -4.8142700 |
| H | 5.3395240 | -0.7172920 | -3.9371590 |

**Angle 232.5°**

Energy = -1828.637998996

| | | | |
|---|---|---|---|
| O | 4.9458200 | 1.8244010 | -2.6949920 |
| C | 3.6524140 | 1.4518080 | -2.6564730 |
| C | 3.1808330 | 0.7769950 | -3.7857420 |
| O | 3.9979160 | 0.5768010 | -4.8567230 |
| C | 1.8534270 | 0.3475960 | -3.8165120 |
| O | 1.4761590 | -0.3040750 | -4.9331690 |
| C | 0.9937040 | 0.5929730 | -2.7486100 |
| C | 1.4804640 | 1.2716530 | -1.6392750 |
| C | 2.8041440 | 1.7030910 | -1.5850830 |
| N | 0.7039630 | 1.5736390 | -0.5122160 |
| C | -0.5735230 | 1.3166010 | -0.2112320 |
| N | -1.2986590 | 2.3934200 | 0.1645680 |
| N | -1.0696060 | 0.0736600 | -0.2855880 |
| C | -2.3350530 | -0.0206180 | 0.0392750 |
| N | -3.1501910 | 0.9472830 | 0.4204820 |
| C | -2.5414870 | 2.1286810 | 0.4545340 |
| C | 4.6952750 | -0.6629040 | -4.8195430 |
| C | 0.1386360 | -0.7588640 | -5.0133190 |
| C | 5.4683950 | 2.5125750 | -1.5754180 |
| Cl | -3.0219200 | -1.5944720 | -0.0387990 |
| Cl | -3.5203500 | 3.4555390 | 0.9398800 |
| H | 1.1832390 | 2.0794850 | 0.2183140 |
| H | -0.0310050 | 0.2666690 | -2.7710990 |
| H | 3.1600890 | 2.2278570 | -0.7088060 |
| H | -0.5671040 | 0.0752180 | -4.9622680 |
| H | -0.0849840 | -1.4748360 | -4.2171060 |
| H | 0.0473730 | -1.2512810 | -5.9790790 |
| H | 4.9416500 | 3.4569480 | -1.4097440 |
| H | 6.5111410 | 2.7170830 | -1.8085370 |
| H | 5.4117640 | 1.8991470 | -0.6713140 |
| H | 5.3073450 | -0.7068550 | -5.7196190 |
| H | 3.9931190 | -1.5014540 | -4.8142700 |
| H | 5.3395240 | -0.7172920 | -3.9371590 |

## Angle 235°

Energy = -1828.636366487

| | | | |
|---|---|---|---|
| O | 4.9458200 | 1.8244010 | -2.6949920 |
| C | 3.6524140 | 1.4518080 | -2.6564730 |
| C | 3.1808330 | 0.7769950 | -3.7857420 |
| O | 3.9979160 | 0.5768010 | -4.8567230 |
| C | 1.8534270 | 0.3475960 | -3.8165120 |
| O | 1.4761590 | -0.3040750 | -4.9331690 |
| C | 0.9937040 | 0.5929730 | -2.7486100 |
| C | 1.4804640 | 1.2716530 | -1.6392750 |
| C | 2.8041440 | 1.7030910 | -1.5850830 |
| N | 0.7039630 | 1.5736390 | -0.5122160 |
| C | -0.5735230 | 1.3166010 | -0.2112320 |
| N | -1.3122090 | 2.4008300 | 0.1133870 |
| N | -1.0569420 | 0.0667480 | -0.2377400 |
| C | -2.3241650 | -0.0265650 | 0.0804130 |
| N | -3.1518400 | 0.9481720 | 0.4142400 |
| C | -2.5548710 | 2.1359860 | 0.4039680 |
| C | 4.6952750 | -0.6629040 | -4.8195430 |
| C | 0.1386360 | -0.7588640 | -5.0133190 |
| C | 5.4683950 | 2.5125750 | -1.5754180 |
| Cl | -2.9951290 | -1.6090890 | 0.0624320 |
| Cl | -3.5506130 | 3.4720490 | 0.8255310 |
| H | 1.1832390 | 2.0794850 | 0.2183140 |
| H | -0.0310050 | 0.2666690 | -2.7710990 |
| H | 3.1600890 | 2.2278570 | -0.7088060 |
| H | -0.5671040 | 0.0752180 | -4.9622680 |
| H | -0.0849840 | -1.4748360 | -4.2171060 |
| H | 0.0473730 | -1.2512810 | -5.9790790 |
| H | 4.9416500 | 3.4569480 | -1.4097440 |
| H | 6.5111410 | 2.7170830 | -1.8085370 |
| H | 5.4117640 | 1.8991470 | -0.6713140 |
| H | 5.3073450 | -0.7068550 | -5.7196190 |
| H | 3.9931190 | -1.5014540 | -4.8142700 |
| H | 5.3395240 | -0.7172920 | -3.9371590 |

**Angle 237.5°**

Energy = -1828.634764902

| | | | |
|---|---|---|---|
| O | 4.9458200 | 1.8244010 | -2.6949920 |
| C | 3.6524140 | 1.4518080 | -2.6564730 |
| C | 3.1808330 | 0.7769950 | -3.7857420 |
| O | 3.9979160 | 0.5768010 | -4.8567230 |
| C | 1.8534270 | 0.3475960 | -3.8165120 |
| O | 1.4761590 | -0.3040750 | -4.9331690 |
| C | 0.9937040 | 0.5929730 | -2.7486100 |
| C | 1.4804640 | 1.2716530 | -1.6392750 |
| C | 2.8041440 | 1.7030910 | -1.5850830 |
| N | 0.7039630 | 1.5736390 | -0.5122160 |
| C | -0.5735230 | 1.3166010 | -0.2112320 |
| N | -1.3253890 | 2.4059680 | 0.0618330 |
| N | -1.0446230 | 0.0619600 | -0.1895430 |
| C | -2.3135730 | -0.0306860 | 0.1218500 |
| N | -3.1534440 | 0.9487840 | 0.4079540 |
| C | -2.5678890 | 2.1410450 | 0.3530330 |
| C | 4.6952750 | -0.6629040 | -4.8195430 |
| C | 0.1386360 | -0.7588640 | -5.0133190 |
| C | 5.4683950 | 2.5125750 | -1.5754180 |
| Cl | -2.9690680 | -1.6192110 | 0.1643990 |
| Cl | -3.5800510 | 3.4834830 | 0.7103490 |
| H | 1.1832390 | 2.0794850 | 0.2183140 |
| H | -0.0310050 | 0.2666690 | -2.7710990 |
| H | 3.1600890 | 2.2278570 | -0.7088060 |
| H | -0.5671040 | 0.0752180 | -4.9622680 |
| H | -0.0849840 | -1.4748360 | -4.2171060 |
| H | 0.0473730 | -1.2512810 | -5.9790790 |
| H | 4.9416500 | 3.4569480 | -1.4097440 |
| H | 6.5111410 | 2.7170830 | -1.8085370 |
| H | 5.4117640 | 1.8991470 | -0.6713140 |
| H | 5.3073450 | -0.7068550 | -5.7196190 |
| H | 3.9931190 | -1.5014540 | -4.8142700 |
| H | 5.3395240 | -0.7172920 | -3.9371590 |

**Angle 240°**

Energy = -1828.633190640

| | | | |
|---|---|---|---|
| C | 5.4683950 | 2.5125750 | -1.5754180 |
| O | 4.9458200 | 1.8244010 | -2.6949920 |
| C | 3.6524140 | 1.4518080 | -2.6564730 |
| C | 3.1808330 | 0.7769950 | -3.7857420 |
| O | 3.9979160 | 0.5768010 | -4.8567230 |
| C | 4.6952750 | -0.6629040 | -4.8195430 |
| C | 1.8534270 | 0.3475960 | -3.8165120 |
| O | 1.4761590 | -0.3040750 | -4.9331690 |
| C | 0.1386360 | -0.7588640 | -5.0133190 |
| C | 0.9937040 | 0.5929730 | -2.7486100 |
| C | 1.4804640 | 1.2716530 | -1.6392750 |
| C | 2.8041440 | 1.7030910 | -1.5850830 |
| N | 0.7039630 | 1.5736390 | -0.5122160 |
| H | 1.1832390 | 2.0794850 | 0.2183140 |
| C | -0.5735230 | 1.3166010 | -0.2112320 |
| N | -1.3381750 | 2.4088230 | 0.0100030 |
| C | -2.5805180 | 2.1438500 | 0.3018280 |
| N | -3.1550000 | 0.9491180 | 0.4016360 |
| C | -2.3032980 | -0.0329710 | 0.1635110 |
| N | -1.0326720 | 0.0593060 | -0.1410870 |
| Cl | -3.6086070 | 3.4898180 | 0.5945540 |
| Cl | -2.9437870 | -1.6248200 | 0.2669130 |
| H | -0.0310050 | 0.2666690 | -2.7710990 |
| H | 3.1600890 | 2.2278570 | -0.7088060 |
| H | -0.5671040 | 0.0752180 | -4.9622680 |
| H | -0.0849840 | -1.4748360 | -4.2171060 |
| H | 0.0473730 | -1.2512810 | -5.9790790 |
| H | 4.9416490 | 3.4569480 | -1.4097440 |
| H | 6.5111400 | 2.7170840 | -1.8085370 |
| H | 5.4117640 | 1.8991470 | -0.6713140 |
| H | 5.3073450 | -0.7068550 | -5.7196190 |
| H | 3.9931190 | -1.5014540 | -4.8142700 |
| H | 5.3395240 | -0.7172920 | -3.9371590 |

**Angle 255°**

Energy = -1828.625111067

| | | | |
|---|---|---|---|
| C | 5.4683950 | 2.5125750 | -1.5754180 |
| O | 4.9458200 | 1.8244010 | -2.6949920 |
| C | 3.6524140 | 1.4518080 | -2.6564730 |
| C | 3.1808330 | 0.7769950 | -3.7857420 |
| O | 3.9979160 | 0.5768010 | -4.8567230 |
| C | 4.6952750 | -0.6629040 | -4.8195430 |
| C | 1.8534270 | 0.3475960 | -3.8165120 |
| O | 1.4761590 | -0.3040750 | -4.9331690 |
| C | 0.1386360 | -0.7588640 | -5.0133190 |
| C | 0.9937040 | 0.5929730 | -2.7486100 |
| C | 1.4804640 | 1.2716530 | -1.6392750 |
| C | 2.8041440 | 1.7030910 | -1.5850830 |
| N | 0.7039630 | 1.5736390 | -0.5122160 |
| H | 1.1832390 | 2.0794850 | 0.2183140 |
| C | -0.5735230 | 1.3166010 | -0.2112320 |
| N | -1.4053060 | 2.3780270 | -0.3012230 |
| C | -2.6468180 | 2.1133310 | -0.0056370 |
| N | -3.1631640 | 0.9452760 | 0.3637020 |
| C | -2.2493540 | -0.0081670 | 0.4136520 |
| N | -0.9699350 | 0.0881870 | 0.1498550 |
| Cl | -3.7585240 | 3.4207530 | -0.1007300 |
| Cl | -2.8110690 | -1.5636800 | 0.8824310 |
| H | -0.0310050 | 0.2666690 | -2.7710990 |
| H | 3.1600890 | 2.2278570 | -0.7088060 |
| H | -0.5671040 | 0.0752180 | -4.9622680 |
| H | -0.0849840 | -1.4748360 | -4.2171060 |
| H | 0.0473730 | -1.2512810 | -5.9790790 |
| H | 4.9416490 | 3.4569480 | -1.4097440 |
| H | 6.5111400 | 2.7170840 | -1.8085370 |
| H | 5.4117640 | 1.8991470 | -0.6713140 |
| H | 5.3073450 | -0.7068550 | -5.7196190 |
| H | 3.9931190 | -1.5014540 | -4.8142700 |
| H | 5.3395240 | -0.7172920 | -3.9371590 |

**Angle 270°**

Energy = -1828.621711218

| | | | |
|---|---|---|---|
| C | 5.4683950 | 2.5125750 | -1.5754180 |
| O | 4.9458200 | 1.8244010 | -2.6949920 |
| C | 3.6524140 | 1.4518080 | -2.6564730 |
| C | 3.1808330 | 0.7769950 | -3.7857420 |
| O | 3.9979160 | 0.5768010 | -4.8567230 |
| C | 4.6952750 | -0.6629040 | -4.8195430 |
| C | 1.8534270 | 0.3475960 | -3.8165120 |
| O | 1.4761590 | -0.3040750 | -4.9331690 |
| C | 0.1386360 | -0.7588640 | -5.0133190 |
| C | 0.9937040 | 0.5929730 | -2.7486100 |
| C | 1.4804640 | 1.2716530 | -1.6392750 |
| C | 2.8041440 | 1.7030910 | -1.5850830 |
| N | 0.7039630 | 1.5736390 | -0.5122160 |
| H | 1.1832390 | 2.0794850 | 0.2183140 |
| C | -0.5735230 | 1.3166010 | -0.2112320 |
| N | -1.4528740 | 2.2674280 | -0.5975700 |
| C | -2.6937880 | 2.0039770 | -0.2983790 |
| N | -3.1689400 | 0.9317120 | 0.3276040 |
| C | -2.2111360 | 0.0807760 | 0.6518200 |
| N | -0.9254910 | 0.1916660 | 0.4268640 |
| Cl | -3.8647250 | 3.1734180 | -0.7627060 |
| Cl | -2.7170520 | -1.3447220 | 1.4684620 |
| H | -0.0310050 | 0.2666690 | -2.7710990 |
| H | 3.1600890 | 2.2278570 | -0.7088060 |
| H | -0.5671040 | 0.0752180 | -4.9622680 |
| H | -0.0849840 | -1.4748360 | -4.2171060 |
| H | 0.0473730 | -1.2512810 | -5.9790790 |
| H | 4.9416490 | 3.4569480 | -1.4097440 |
| H | 6.5111400 | 2.7170840 | -1.8085370 |
| H | 5.4117640 | 1.8991470 | -0.6713140 |
| H | 5.3073450 | -0.7068550 | -5.7196190 |
| H | 3.9931190 | -1.5014540 | -4.8142700 |
| H | 5.3395240 | -0.7172920 | -3.9371590 |

## IR Spectroscopy

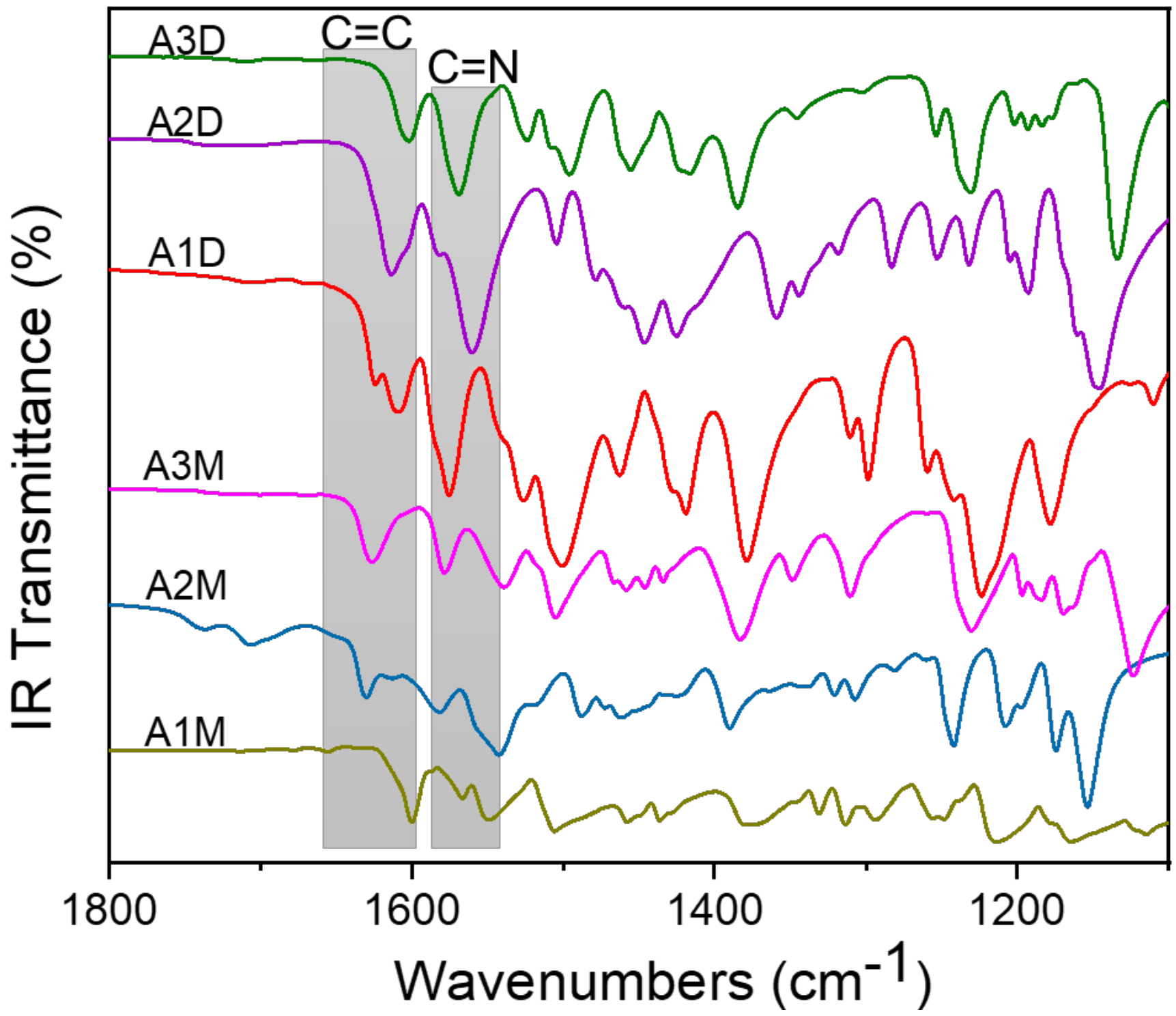

**Figure S33.** Fingerprint region of the IR spectra of our molecular probes.

## Raman Spectroscopy

### Molecular spectra

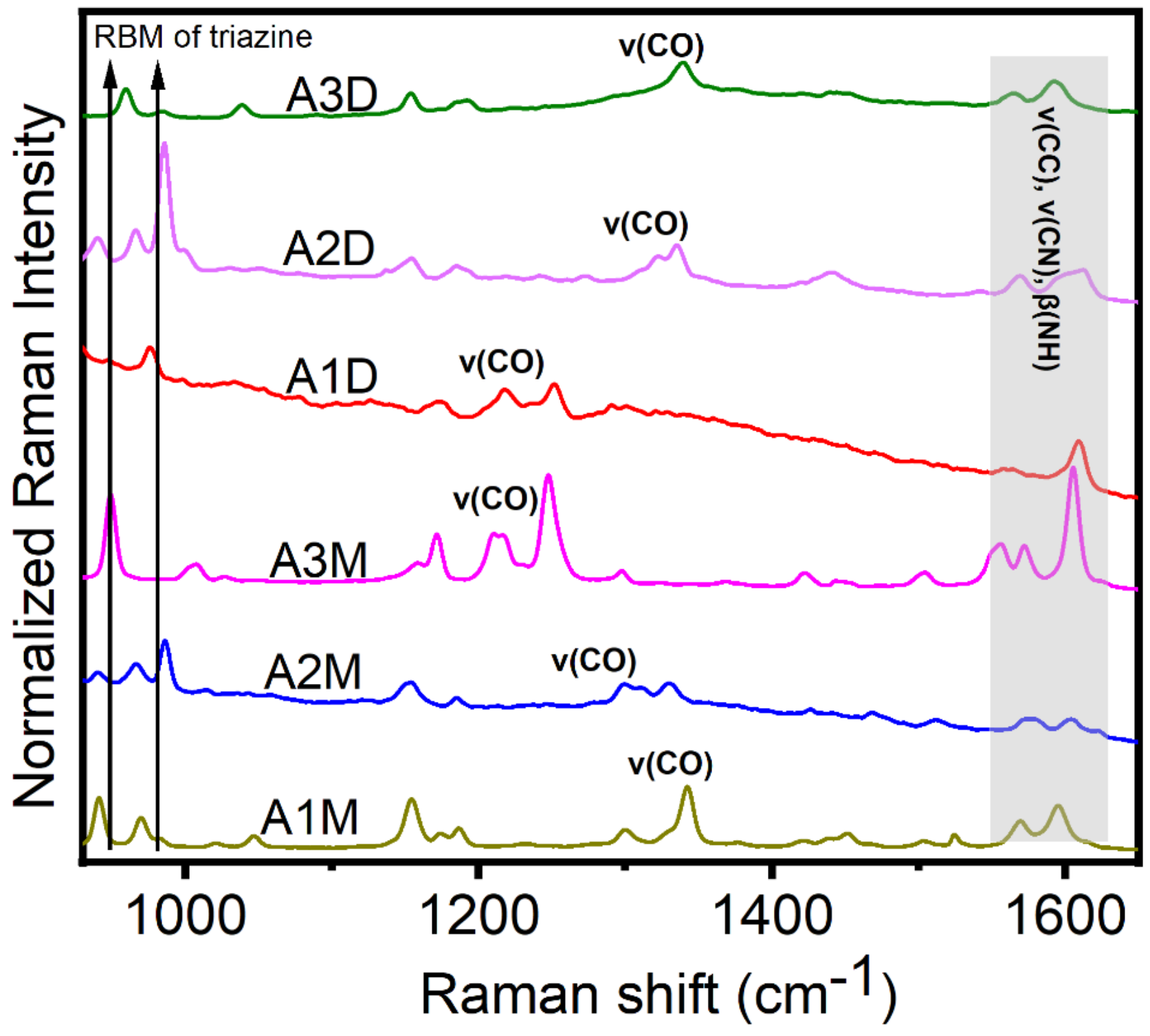

**Figure S34.** Raman spectra of our molecular probes. Labels are added to the most prominent features of the spectra.

## Functionalized SWNTs spectra

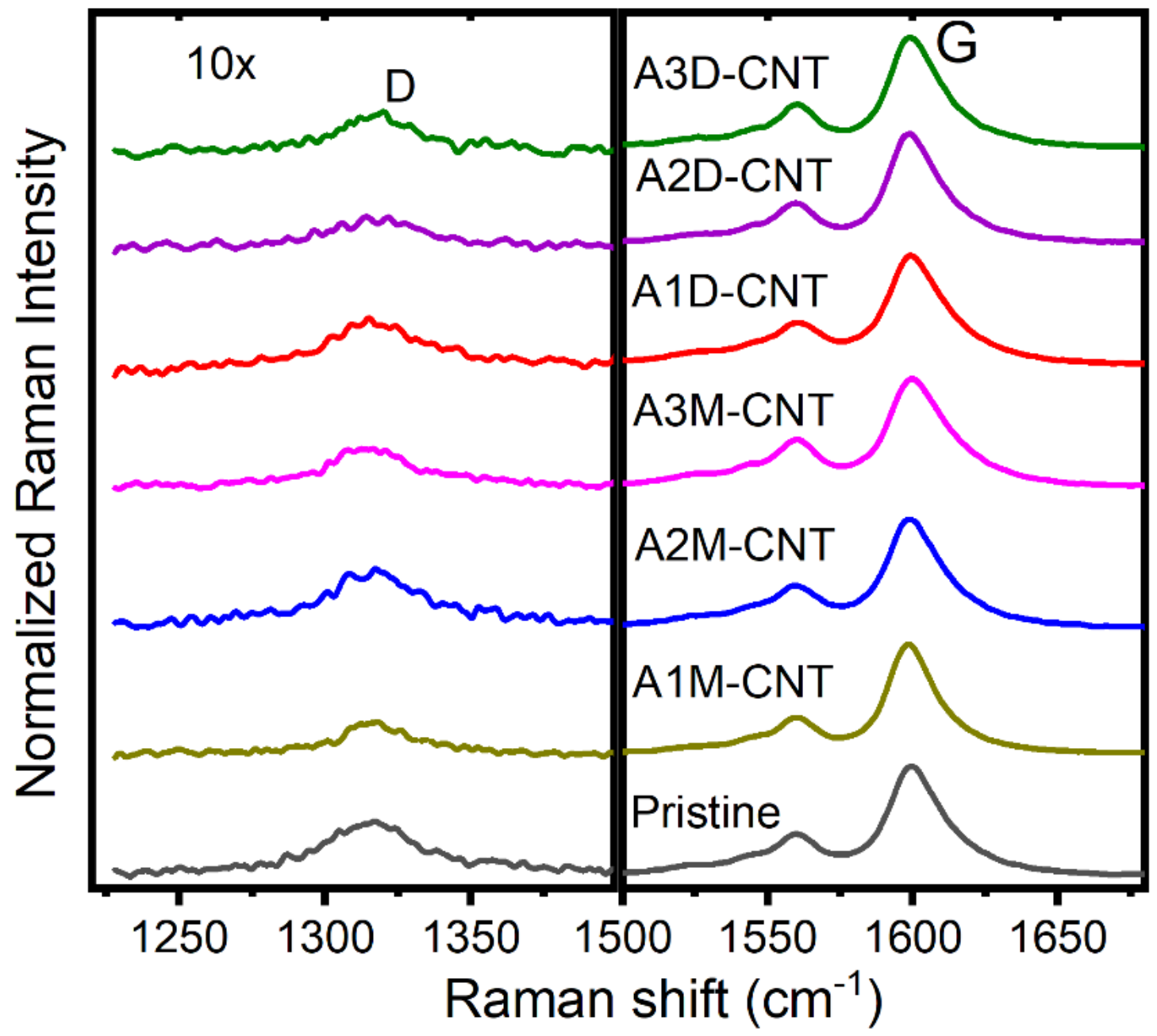

**Figure S35.** Raman spectra of our functionalized SWNTs zoomed to the D and G band spectral regions.

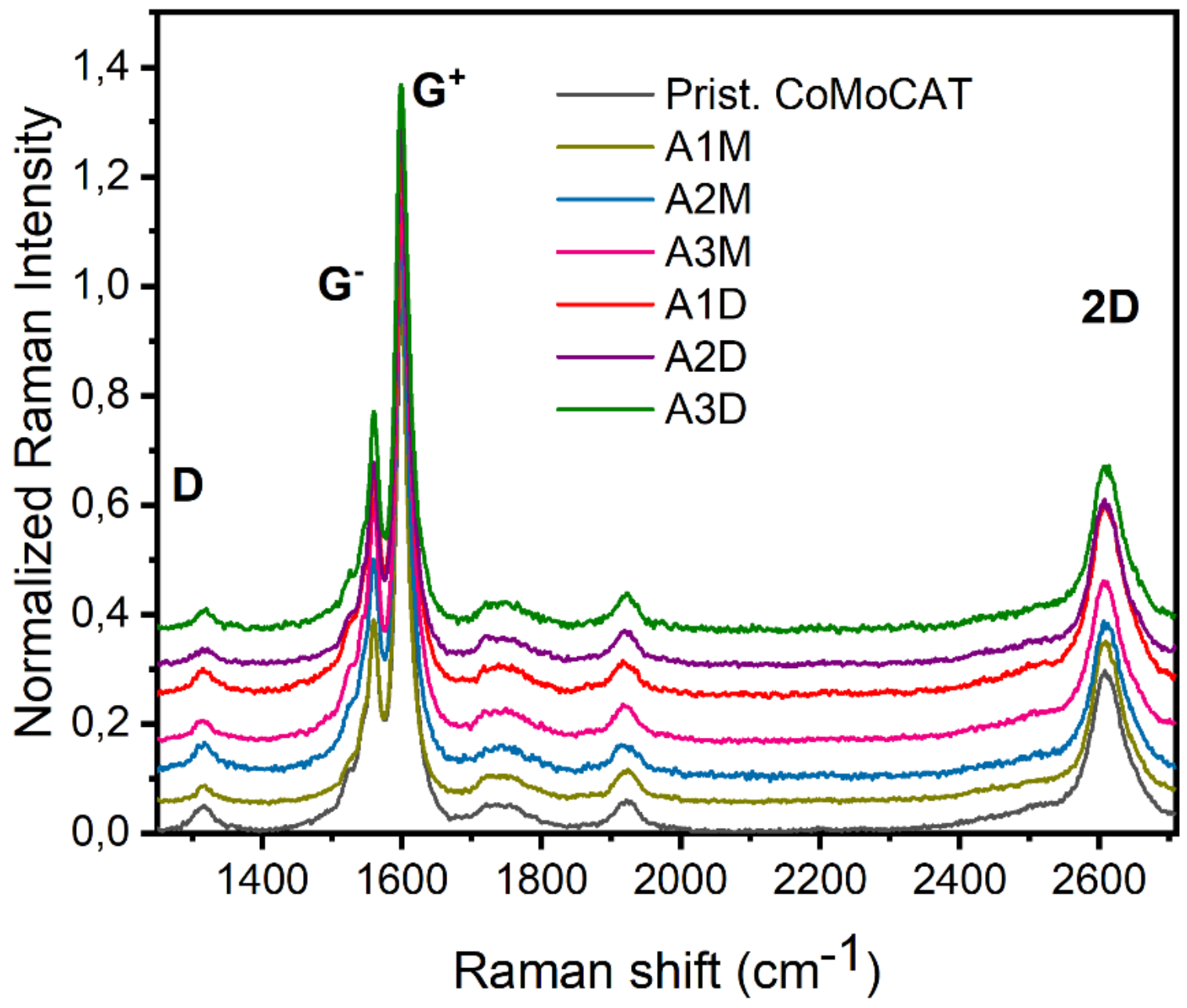

**Figure S36.** Overview Raman spectra of our functionalized SWNTs.